\documentclass[preprintcolumn]{aastex631}
\def\apjl{{ApJ}}                
\def\apjs{{ApJS}}               
\def\ao{{Appl.~Opt.}}           
\def\aap{{A\&A}}                

\usepackage{amsmath}	
\usepackage{amssymb}	

\usepackage{graphicx}
\usepackage{hyperref} 
\usepackage{listings}
\usepackage{color}
\usepackage{graphicx,times}
\usepackage{natbib}
\usepackage{amssymb}
\usepackage[Figureright]{rotating}
\usepackage{newtxtext}
\usepackage[T1]{fontenc}
\usepackage{ae,aecompl}
\usepackage{threeparttable}
\usepackage{multirow}
\usepackage{ulem}
\usepackage{txfonts}
\usepackage{float}
\usepackage{placeins}

\bibpunct{(}{)}{;}{a}{}{,}
\usepackage{hyperref}
\usepackage{longtable}
\usepackage{booktabs}
\definecolor{dkgreen}{rgb}{0,0.6,0}
\definecolor{gray}{rgb}{0.5,0.5,0.5}
\definecolor{mauve}{rgb}{0.58,0,0.82}
\definecolor{golden}{rgb}{0.86,0.65,0.01}

\shorttitle{Gaia DR3 Open Cluster Cepheids: Catalog and Calibrations}
\shortauthors{Deng et al.}

\usepackage[Figureright]{rotating}
\usepackage{ulem}
\usepackage{CJK}
\usepackage{amsmath}
\usepackage{natbib}

\usepackage{threeparttable} 
\usepackage{booktabs} 
\begin{document}
\begin{CJK*}{UTF8}{gbsn}

\setlength{\baselineskip}{15pt}
\title{Gaia DR3 Open Cluster Cepheids: A Unified Catalog with Calibrated Period-Age and Period-Wesenheit Relations}

\correspondingauthor{Anbing Ren}
\email{abren@cwnu.edu.cn}
\author[0009-0008-5286-1060]{Shunhong Deng}
\affiliation{School of Physics and Astronomy, China West Normal University, No. 1 Shida Road,    Nanchong 637009, China}

\author[0000-0002-6989-8192]{Zhihong He}
\affiliation{School of Physics and Astronomy, China West Normal University, No. 1 Shida Road,    Nanchong 637009, China}

\author[0000-0002-2037-2480]{Anbing Ren}
\affiliation{School of Physics and Astronomy, China West Normal University, No. 1 Shida Road,    Nanchong 637009, China}

\author[0000-0003-0089-2005]{Qian Cui}
\affiliation{School of Physics and Astronomy, China West Normal University, No. 1 Shida Road,    Nanchong 637009, China}

\author{Xiaoyue Zhou}
\affiliation{CAS Key Laboratory of Optical Astronomy, National Astronomical Observatories, Chinese Academy of Sciences, Beijing 100101, China}
\affiliation{School of Astronomy and Space Science, University of the Chinese Academy of Sciences, Beijing 101408, China}

\author{Liming Peng}
\affiliation{School of Physics and Astronomy, China West Normal University, No. 1 Shida Road,    Nanchong 637009, China}

\author{Chenxin Wang}
\affiliation{School of Physics and Astronomy, China West Normal University, No. 1 Shida Road,    Nanchong 637009, China}

\author{Ziang Chen}
\affiliation{School of Physics and Astronomy, China West Normal University, No. 1 Shida Road,    Nanchong 637009, China}

\author[0000-0003-3736-6076]{Yangping Luo}
\affiliation{School of Physics and Astronomy, China West Normal University, No. 1 Shida Road,    Nanchong 637009, China}

\author[0000-0002-5745-827X]{Kun Wang}
\affiliation{School of Physics and Astronomy, China West Normal University, No. 1 Shida Road,    Nanchong 637009, China}
\begin{abstract}
Classical Cepheids (CCs) in Galactic open clusters (OCs) provide essential observational constraints for calibrating the period-age relation (PAR) and the period-Wesenheit relation (PWR) of CCs. However, distant and long-period OC Cepheids remain limited, while the confirmed samples still require more precise determinations of their physical properties, such as ages and extinctions. In this work, we present a comprehensive census of OC Cepheids based on an extensive sample of distant OCs from Gaia Data Release 3 (DR3). By combining astrometric and photometric membership analyses, we identified 110 CCs associated with 102 OCs, of which 41 CCs across 37 OCs were classified as OC Cepheids, while the remaining cases were considered candidate or rejected associations. Our results are consistent with previous studies, while 4 of the 41 OC Cepheids are newly reported here. Using updated cluster parameters derived from manual isochrone fitting, we primarily refined the PAR to $\log \textit{Age}$ = (-0.595 $\pm$ 0.044) $\log P$ + (8.430 $\pm$ 0.042) and recalibrated the PWR to $W_G$ = (-3.615 $\pm$ 0.083) $\log P$ + (-2.379 $\pm$ 0.096). This study expands the sample of confirmed and candidate OC Cepheids. The newly longest-period confirmed OC Cepheid is BM Per (CWNU 3123) with $\log P = 1.36$, and two newly discovered OC Cepheid candidates have distances exceeding 6 kpc. Moreover, the PAR and PWR are improved by incorporating refined OC ages and updated parallaxes, respectively.
\end{abstract}
\section{Introduction}
\label{sec:intro}
Classical Cepheids (CCs) are pulsating variable stars that have shaped our understanding of cosmic distance measurement. Their unique period-luminosity relation, first established by \citet{leavitt1912}, provides a fundamental tool for studying stellar populations across different scales. Within the Milky Way, they enable precise mapping of Galactic structure and kinematics \citep[e.g.,][]{chen2019,skowron2019,Lemasle2022}. In cosmology, the Hubble constant is calibrated using CCs \citep{Riess2021}, providing constraints on the universe's expansion history that are independent of early-universe CMB inferences within the $\Lambda\mathrm{CDM}$ framework. The reliability of CCs as standard candles depends on thorough characterization of their properties in various stellar environments. This requirement drives ongoing efforts to improve calibration accuracy through multi-wavelength observations and theoretical modeling \citep{Freedman2001,Sandage2006}. Such work bridges Galactic and extragalactic studies, making CCs indispensable across astrophysical disciplines.

Open clusters (OCs) are young, metal-rich stellar systems concentrated in the Galactic disk, whose members formed nearly simultaneously and share common astrometric and chemical properties. These uniform characteristics allow the clusters' ages, distances, and extinctions to be determined with high precision. For CCs belonging to these OCs, such uniformity allows their physical parameters, particularly ages to be constrained more robustly.
Since the first identification of an OC Cepheid, subsequent studies \citep[e.g.,][]{Anderson2013,Lohr2018,Alonso-Santiago2020,mauricio2023} have steadily expanded this field. Large-scale surveys such as the VISTA Variables in the Via Lactea survey (VVV; \citealt{Minniti2010}), the Gaia astrometric missions \citep{gaia2018,gaia2021,gaia2023}, the All-Sky Automated Survey for Supernovae (ASAS-SN; \citealt{Jayasinghe2018}), the Asteroid Terrestrial-impact Last Alert System (ATLAS; \citealt{Heinze2018}), the Wide-field Infrared Survey Explorer (WISE; \citealt{Wright2010,chen2018}), the Optical Gravitational Lensing Experiment (OGLE; \citealt{Soszynski2017,Udalski2018}), and the Zwicky Transient Facility (ZTF; \citealt{chen2020}) have delivered extensive catalogs of Galactic CCs. Despite these advances, accurately determining CCs' properties remains challenging, which makes OC associations particularly valuable. In particular, Gaia's high-precision astrometry has played a pivotal role in enabling the discovery of both new OCs, substantially increasing the number of confirmed OC Cepheids in recent years.

CCs are FGK-type giants or supergiants, and longer-period variables exhibit higher luminosities and greater masses, which implies younger ages. The clear correlation between period and age is referred to as the period-age relation (PAR) \citep{Bono2005, Anderson2016, DeSomma2021}. \citet[][hereafter Bono05]{Bono2005} developed theoretical PAR models using nonlinear convective pulsation and updated evolutionary tracks. These models were subsequently refined by \citet[][hereafter Anderson16]{Anderson2016} to incorporate the effects of stellar rotation. Since the ages of OCs can be directly determined through isochrone fitting, they offer the most robust empirical constraints on the PAR.

Observational calibrations of the PAR using OCs have a long history \citep[e.g.,][]{Efremov1978}. More recently, \citet[][hereafter ZC21]{zc2021} calibrated the PAR using 33 OC Cepheids with 27 fitted OC ages, finding a small offset relative to the rotating models of Anderson16 at short periods. Subsequent studies have identified additional OC Cepheids \citep{Hao2022,mauricio2023,wang2024}, while reliable cluster age determinations remain limited, leaving the empirical calibration of the PAR open for further refinement.

Recent studies have refined the period-Wesenheit relation (PWR) for OC Cepheids \citep{zc2021, lin2022, mauricio2023, wang2024}. These works have taken into account not only the effects of parallax and extinction on the luminosities of CCs, but also the influence of metallicity, resulting in more precise calibrations. Nevertheless, the derived PWR fits show variations across different studies. Such discrepancies likely arise from differences in sample selection and membership determination, also reflecting the current limitations of the available OC Cepheids. Although a large number of OC Cepheids and candidates have been reported, the membership between many CCs and their host clusters still needs to be confirmed.

In this paper, we compiled a  homogeneous sample of OC Cepheids based on Gaia Data Release 3 (DR3)~\citep{gaia2023}, using astrometric and photometric membership analyses. The host cluster ages, as well as extinction, and distance parameters, were then refined through isochrone fitting. The PAR and PWR were recalibrated using refined OC ages and updated OC parallaxes, respectively, which resulted in more reliable calibrations and improved precision in Cepheid-based distance measurements.

We organize this paper as follows. In Section~\ref{sect:data} and Section ~\ref{sect:method}, we describe the datasets and methodology employed to identify OC Cepheids, followed by the data filtering process and the resulting sample in Section~\ref{sect:results}. Section~\ref{sect:analysis} presents our analysis of the PAR and PWR, and Section~\ref{sect:summary} summarizes the study. 
\section{Data}
\label{sect:data}
\subsection{Classical Cepheid}
The CC sample used in this study is compiled from six catalogs published since 2019, including those by \citet{skowron2019, chen2019, chen2020, pietrukowicz2021, inno2021, Ripepi2022}. \citet{skowron2019} and \citet{chen2019} employed CCs as tracers to map the structure of the Milky Way. \citet{pietrukowicz2021} reported 3,659 CCs from a wide-field survey that combined data from multiple projects. \citet{inno2021} combined Gaia astrometry with ASAS-SN photometry and identified nearly 1,900 CCs, over 90\% of which are considered reliable. \citet{chen2020} performed an extensive search for variable stars using ZTF data and identified about 700 CCs. \citet{Ripepi2022} obtained metallicity measurements for 499 CCs to refine the period-Wesenheit-metallicity (PWZ) relation. 
\FloatBarrier
\begin{figure*}[!htbp]
\centering
\includegraphics[width=0.9\linewidth]{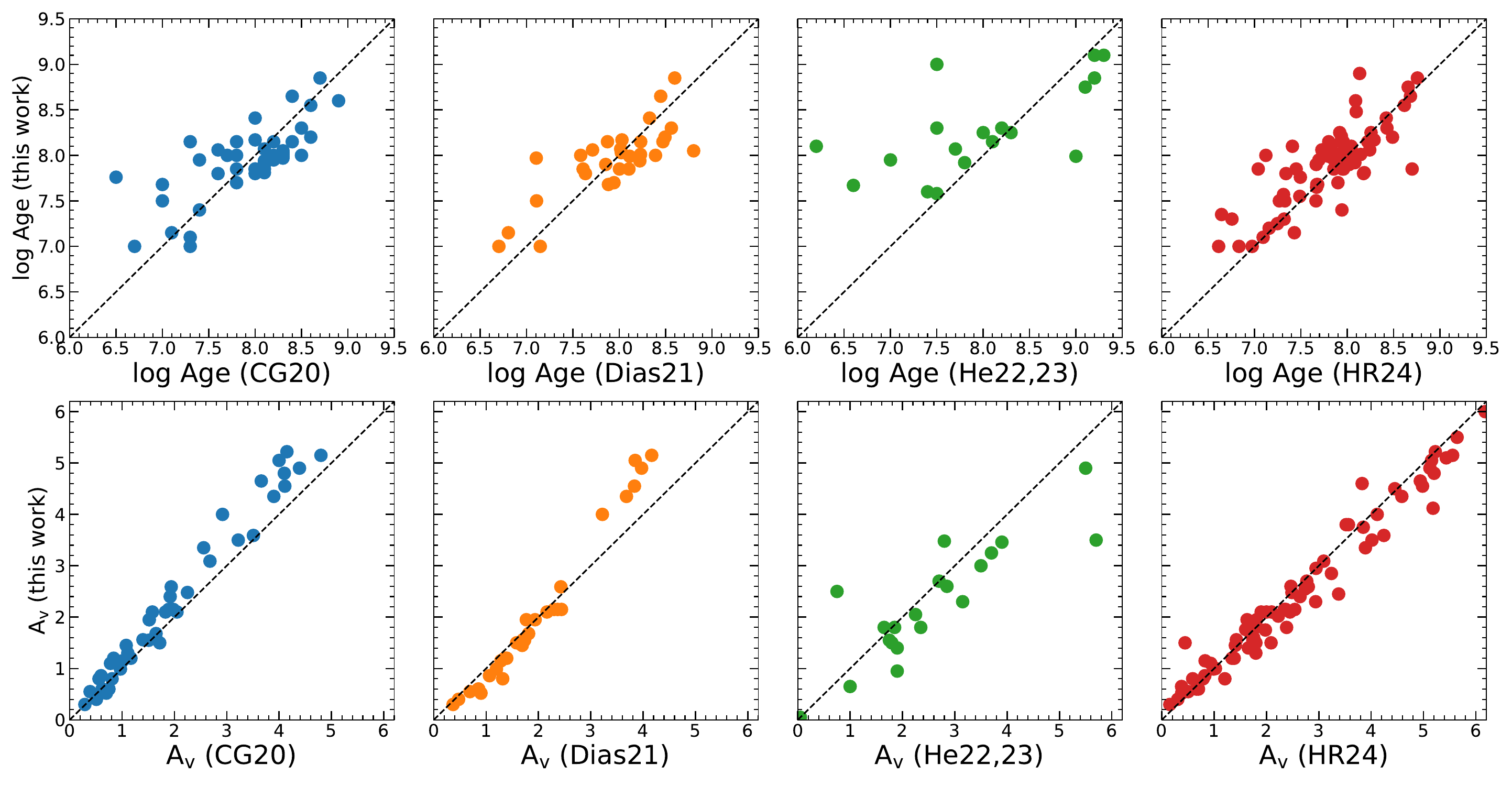} \hspace{0.0cm}
\caption{Comparison of the cluster ages and extinctions obtained from the isochrone fitting with those from the literature, including \citet{CG2020}, \citet{dias2021}, \citet{he2022,he2023}, and \citet{huntemily2024}. Points of different colors correspond to OCs from the respective references.}
\label{fig1:result_1}
\end{figure*}
\subsection{Open Cluster}
The OC data used in this study, collected from the Gaia astrometric mission \citep{gaia2018, gaia2021, gaia2023}, provide stellar positions and proper motions ($l$, $b$, $\varpi$, $\mu_{\alpha^{*}}$, and $\mu_{\delta}$), along with photometry in the Gaia bands ($G$, $G_{BP}$, and $G_{RP}$). With the release of Gaia DR3, the number of sources increased from 1.3 billion in DR2 to 1.8 billion.  In particular, the precision of parallax measurements for sources with $G$ band fainter than 18 mag improved by 30\% in DR3, enhancing the capability to search for distant OCs in the Milky Way.

Our OC sample is compiled from the catalogs of \citet[][hereafter He22]{he2022}, \citet[][hereafter He23]{he2023}, and \citet[][hereafter HR24]{huntemily2024}. He22 employed the unsupervised Density-Based Spatial Clustering of Applications with Noise (DBSCAN) algorithm to identify 1,656 nearby OCs in the Milky Way, limited to Galactic latitudes \(|b| < 20^{\circ}\) and distances beyond 1.2 kpc. Extending this work, He23 restricted the latitude range to \(|b| < 10^{\circ}\), which yielded nearly three times more distant OCs (\(d > 5\) kpc), including systems with extinction above 5 mag and ages older than 100 Myr. Consequently, the catalog expanded to 2,085 OCs, including candidates. An updated catalog of Milky Way OCs based on Gaia DR3 was presented by HR24. Using the Hierarchical Density-Based Spatial Clustering of Applications with Noise (HDBSCAN) algorithm, they identified 7,167 OCs and provided detailed information including positions, parallaxes, proper motions, photometry, ages, and extinction values. Among these, 3,530 OCs are considered high-quality, characterized by well-defined isochrones derived from neural-network fitting and by the exclusion of kinematic outliers through Jacobi radius filtering.

We also supplemented our sample with additional OCs from \citet[][hereafter CG20]{CG2020} and \citet[][hereafter CRA23]{mauricio2023}, incorporating 3 OCs from each catalog.
\subsection{Previous OC Cepheids} 
OC Cepheid samples have expanded with the advent of updated and higher-precision data, enabling the identification of new objects and the refinement of both the PAR and PWR. ZC21 conducted a comprehensive survey of CCs in Galactic OCs and identified 33 OC Cepheids, including 13 new discoveries. Using Gaia EDR3 \citep{gaia2021} parallaxes and proper motions, they confirmed cluster membership for the CCs and re-estimated the cluster ages, extinctions, and distances via isochrone fitting, thereby enabling refined calibrations of the PAR and PWR. 

\citet[][hereafter Hao22]{Hao2022} further expanded the selection of OC Cepheids by cross-matching the CC catalog of \citet{pietrukowicz2021} with several OC catalogs \citep[e.g.,][]{CG2020, Castro-Ginard2022}. This effort resulted in the identification of 50 CCs associated with 45 OCs, of which 39 were confirmed as reliable OC Cepheids. \citet[][hereafter Lin22]{lin2022} focused on calibrating the PWR for 51 OC Cepheids using different distance estimates, including individual CCs' parallaxes, cluster-averaged parallaxes, and distance modulus, based on Gaia photometry.

The impact of metallicity on the PWR has been highlighted in recent studies. CRA23 performed a systematic search for OC Cepheids using Gaia DR3 \citep{gaia2023}, selecting candidates based on spatial position, parallax, and proper motion, further constrained by the color-magnitude diagram (CMD). They identified 34 CCs in 28 OCs as their primary sample, along with 4 candidates. By incorporating cluster-averaged parallaxes, they recalibrated the Leavitt law and further refined it with metallicity constraints. \citet[][hereafter Wang24]{wang2024} employed the OC catalog of \citet{huntemily2024} with the CC catalog from \citet{pietrukowicz2021}, which yielded 43 OC Cepheids used to calibrate the PWZ relation. 

By compiling Galactic OC Cepheids from these five studies (ZC21, Hao22, Lin22, CRA23 and Wang24), we obtained a total of 66 CCs associated with 59 OCs. Additionally, we took into account the OC Cepheid candidates reported by \citet{Majaess2024}.

\section{Method}
\label{sect:method}
\subsection{Cross-Matching and Identifying OC Members with CCs}
The selection of member stars in OCs based on spatial and kinematic properties is a well-established approach, as stars within the same OC generally share similar positions and motions \citep{zc2021,Hao2022,mauricio2023,wang2024}. Accordingly, we performed an initial selection of CCs and OC members using five spatial parameters ($l$, $b$, $\varpi$, $\mu_{\alpha^{*}}$, and $\mu_{\delta}$), with a cross-matching radius of one arcsecond applied in TOPCAT \citep{Taylor2005}.

To further ensure accurate membership, we defined the dispersions $\delta_\varpi$ and $\delta_{\rm pm}$ to quantify the deviation of a CC from its host OC in parallax ($\varpi$) and proper motion ($pm$), with $R$ denoting its angular distance from the cluster center, respectively:
\begin{align}\label{eq_delta}
\delta_\varpi &= \frac{|\varpi - \bar{\varpi}|}{\sqrt{\sigma_\varpi^2 + \sigma_{\bar{\varpi}}^2}} \\[2mm]
\delta_{pm} &= \frac{\sqrt{(\mu_{\alpha*} - \bar{\mu}_{\alpha*})^2 + (\mu_\delta - \bar{\mu}_\delta)^2}}
               {\sqrt{\sigma_{\mu_{\alpha*}}^2 + \sigma_{\bar{\mu}_{\alpha*}}^2 + \sigma_{\mu_\delta}^2 + \sigma_{\bar{\mu}_\delta}^2 + (\bar{\varpi}/4.74)^2}} \\[2mm]
R &= \sqrt{ \big[(l - \bar{l}) \cos b \big]^2 + (b - \bar{b})^2 }
\end{align}
where ($l$, $b$), $\varpi$, and ($\mu_{\alpha^{*}}$, $\mu_{\delta}$) denote the Galactic coordinates, parallax, and proper motions in right ascension and declination of the CCs, respectively. The parameters $\bar{l}$ and $\bar{b}$ are defined as the average Galactic coordinates of the member stars in each OC, and their uncertainties, $\sigma_l$ and $\sigma_b$, are given by the corresponding standard deviations. The values of $\bar{\mu}_{\alpha^{*}}$, and $\bar{\mu}_\delta$ represent the weighted means of the host OC member stars and ($\bar{\varpi}/4.74$) corresponds to the typical intrinsic velocity dispersion of OCs ($\sim$1 km\,s$^{-1}$, \citealp{Mermilliod2009}). The individual member star parallaxes were first corrected for the zero-point (zpt) offset \citep{Lindegren2021a} and then used to calculate the weighted mean parallax ($\bar{\varpi}$) of each host OC. Although the weighted mean of a large number of member stars would nominally benefit from the $\sqrt{N}$ reduction in statistical uncertainty, the angular covariance of Gaia parallaxes and proper motions exhibits many times larger than the weighted mean \citep[e.g.,][]{Lindegren2021b, MaizApellaniz2021, Vasiliev2021, Zinn2021, Riess2022}. To account for this effect, we incorporated the angular covariance between member stars, with values of $\sim 700~\mu\rm as^2$ for parallaxes and $\sim 550~\mu\rm as^2~yr^{-2}$ for proper motions from \citet{Lindegren2021b}, in the calculation the uncertainties of $\sigma_{{\varpi}}$, $\sigma_{\bar{\varpi}}$, $\sigma_{\mu_{\alpha^{*}}}$, $\sigma_{\mu_\delta}$, $\sigma_{\bar\mu_{\alpha^{*}}}$, and $\sigma_{\bar\mu_\delta}$.

Larger dispersions indicate stronger deviations of CCs from their host OCs, thereby reflecting lower astrometric membership likelihood. Accordingly, we adopted a $3\sigma$ threshold in $\delta_\varpi$ and $\delta_{\rm pm}$, and further required that CCs lie within the spatial extent of the cluster core, to identify reliable cluster members.
\subsection{Isochrone Fitting} 
After filtering OC Cepheids based on astrometric dispersions, we further applied isochrone fitting to validate their membership. The distinct position of CCs in the CMD of their host OCs, corresponding to the blue loop evolutionary stage \citep{Turner2006}, enables reliable separation from other member stars. Using theoretical data from the PARSEC models \citet{Bressan2012} and Gaia EDR3 photometry, we then performed isochrone fitting for each OC to re-estimate age, extinction, and distance parameters, thereby establishing the basis for the subsequent PAR and PWR analyses. We note that the precise location of the instability strip boundaries for CCs is sensitive to the adopted input physics in stellar models, including the treatment of convection, turbulent pressure, metallicity, and stellar rotation \citep[e.g.,][]{Espinoza-Arancibia2024,Deka2024}. To account for these uncertainties, we visually inspected the CMDs and adopted an empirical $\sim$0.5 mag range when selecting OC Cepheids.

During the isochrone fitting, we systematically considered the effects of distance, metallicity, extinction, and age. The cluster distances were derived from the weighted mean parallaxes of its member stars. To better constrain the fitting, we incorporated metallicity as well. Given that member stars of an OC typically exhibit homogeneous metallicities, the CC metallicity was taken as representative of the host cluster. For CCs without spectroscopic metallicity, we used the spectroscopic metallicities of their host OCs from APOGEE \citep{Hayden2015, Jonsson2020},  LAMOST \citep{Cui2012}, and Gaia-ESO \citep{Hourihane2023}. When such data were unavailable for both OCs and CCs, metallicities were estimated from the Galactic metallicity gradient of \citet{Ripepi2022}.

We initially constructed CMDs using age and extinction parameters from published OC catalogs. In some clusters, however, the fitted isochrones appeared shifted toward the red edge of the main sequence, likely due to systematic overestimation of BP-band photometry for faint stars \citep{Riello2021}. To address this, we adjusted the extinction values downward, moving the isochrones toward the blue edge and thereby establishing the final extinction estimates. The cluster ages were then derived by locating the main-sequence turn-off of member stars in the CMD. To improve the precision of the isochrone fits, we restricted the age search range to [6.8, 10.1] dex with a step size of 0.01 dex.

As described in our previous studies (e.g., He22; He23), the extinction coefficient $c$ for each OC was calculated using a polynomial function:
\begin{equation}\label{extinction_coefficient}
\begin{split}
c = c_1 + c_2 \ast bp\_rp_0 + c_3 \ast bp\_rp_0^2 + c_4 \ast bp\_rp_0^3 \\+ c_5 \ast A_0
+ c_6 \ast A_0^2 + c_7  \ast A_0^3 + c_8  \ast bp\_rp_0  \ast A_0 \\+ c_9  \ast A_0  \ast bp\_rp_0^2 + c_{10}  \ast bp\_rp_0  \ast A_0^2
\end{split}
\end{equation}
where the coefficients $c_1$-$c_{10}$ were adopted from the auxiliary data provided by ESA/Gaia/DPAC Coordination Unit 5 (CU5; Photometric Processing) and compiled by Carine Babusiaux. The \(bp\_rp_0\) denotes the intrinsic Gaia color index corrected for extinction, and \(A_0\) is derived from our isochrone fitting and serves as the standard measure of interstellar extinction in the Gaia photometric system.

Based on the OC parameters reported in He22, He23, and HR24, we performed isochrone fitting for 102 OCs to refine their age and extinction parameters. The results were then compared with those from literature catalogs \citep[e.g.,][]{CG2020, dias2021, he2022, he2023, huntemily2024}.
Figure~\ref{fig1:result_1} presents the distribution of ages from our isochrone fitting and from the literature. All four references exhibit an overall upward trend within the range $\log Age =$ [6.0, 7.5] dex.
Both \citet[][hereafter Dias21]{dias2021} and HR24 display similar upward discrepancies at $\log Age < 7.5$, followed by downward deviations at older ages. In contrast, CG20 shows both upward and downward deviations at younger ages. Results from He22 and He23 exhibit mainly upward discrepancies at younger ages and downward discrepancies at older ages.

Figure~\ref{fig1:result_1} also shows the extinction distributions. CG20 predominantly exhibit upward discrepancies, while HR24, He22 and He23 generally display downward deviations, with a few upward cases between 0 and 3 dex. Dias21, by comparison, display downward discrepancies in the range of [0, 3] dex and upward deviations from 3 to 5 dex. 
These differences likely arise from the diverse fitting methodologies employed in the literature. In contrast, our isochrone fitting is performed by visual inspections, yielding a homogeneous set of OC ages and extinctions and providing a consistent framework for calibrating OC Cepheids.

\setlength{\tabcolsep}{1.5pt} 
\begin{table*}
\caption{Astrometric parameters of the 41 OC Cepheids.}
\centering
\label{table:table1}
\begin{tabular}{cccccccccccc}
\hline\hline
Index & Cepheid&$\alpha (^{\circ})$ & $\delta (^{\circ})$ & $\varpi$ (mas) &$\mu_{\alpha^{*}}$ (mas/yr) & $\mu_{\delta}$ (mas/yr) &  $\delta_\varpi$&$\delta_{pm}$ & $R $ &Note &Cluster \\ 
\hline
1 & NO Cas & 6.019 & 61.342 & 0.298 $\pm$ 0.029 & -2.83 $\pm$ 0.02 & -1.21 $\pm$ 0.02 & 0.39 & 1.95 & 8.41 & N123456 & NGC 103 \\
2 & J183904-1049.3 & 279.767 & -10.823 & 0.575 $\pm$ 0.033 & 0.24 $\pm$ 0.03 & -0.96 $\pm$ 0.03 & 1.88 & 0.29 & 2.35 & N6 & CWNU 2461 \\
3 & EV Sct & 279.165 & -8.185 & 0.526 $\pm$ 0.032 & -0.21 $\pm$ 0.02 & -2.55 $\pm$ 0.02 & 0.78 & 1.33 & 2.65 & N12345 & NGC 6664 \\
4 & J213533.70+533049.3 & 323.890 & 53.514 & 0.214 $\pm$ 0.028 & -2.88 $\pm$ 0.02 & -3.11 $\pm$ 0.02 & 0.04 & 1.75 & 2.69 & N12356 & Kronberger 84 \\
5 & FZ Car & 163.507 & -59.200 & 0.234 $\pm$ 0.029 & -7.22 $\pm$ 0.02 & 3.16 $\pm$ 0.02 & 0.64 & 1.48 & 1.34 & N6 & CWNU 2232 \\
6 & Gaia DR3 5254518760118884864 & 158.775 & -59.635 & 0.239 $\pm$ 0.029 & -5.56 $\pm$ 0.02 & 2.47 $\pm$ 0.02 & 0.09 & 1.56 & 0.08 & N5 & HSC 2354 \\
7 & QZ Nor & 242.835 & -54.354 & 0.484 $\pm$ 0.033 & -1.90 $\pm$ 0.03 & -3.85 $\pm$ 0.02 & 0.80 & 0.30 & 6.41 & N5 & Theia 3005 \\
8 & GI Car & 168.500 & -57.911 & 0.462 $\pm$ 0.032 & -7.99 $\pm$ 0.02 & 1.76 $\pm$ 0.02 & 0.10 & 1.35 & 17.74 & N6 & CWNU 2478 \\
9 & CE Cas B & 359.538 & 61.214 & 0.333 $\pm$ 0.031 & -3.30 $\pm$ 0.02 & -1.81 $\pm$ 0.02 & 0.46 & 1.37 & 2.21 & N12345 & NGC 7790 \\
10 & V335 Pup & 119.240 & -22.825 & 0.443 $\pm$ 0.032 & -2.97 $\pm$ 0.02 & 2.89 $\pm$ 0.02 & 0.44 & 0.14 & 1.77 & N12356 & UBC 229 \\
11 & CF Cas & 359.575 & 61.221 & 0.316 $\pm$ 0.029 & -3.24 $\pm$ 0.02 & -1.77 $\pm$ 0.02 & 0.09 & 0.52 & 1.28 & N12345 & NGC 7790 \\
12 & CE Cas A & 359.539 & 61.214 & 0.332 $\pm$ 0.031 & -3.30 $\pm$ 0.02 & -1.87 $\pm$ 0.02 & 0.42 & 2.10 & 2.16 & N12345 & NGC 7790 \\
13 & VW Cru & 188.328 & -63.506 & 0.738 $\pm$ 0.031 & -3.90 $\pm$ 0.02 & -1.13 $\pm$ 0.02 & 0.31 & 0.40 & 5.05 & N45 & CWNU 175 \\
14 & CV Mon & 99.270 & 3.064 & 0.601 $\pm$ 0.030 & 0.35 $\pm$ 0.02 & -0.67 $\pm$ 0.02 & 0.35 & 0.63 & 1.61 & N123456 & vdBergh 1 \\
15 & V Cen & 218.138 & -56.888 & 1.409 $\pm$ 0.034 & -6.70 $\pm$ 0.02 & -7.07 $\pm$ 0.02 & 2.25 & 0.84 & 23.69 & N12345 & NGC 5662 \\
16 & OGLE-GD-CEP-1544 & 290.889 & 15.368 & 0.314 $\pm$ 0.038 & -3.02 $\pm$ 0.03 & -6.16 $\pm$ 0.03 & 0.72 & 0.85 & 3.33 & N6 & HSC 413 \\
17 & V724 Pup & 118.438 & -36.970 & 0.318 $\pm$ 0.029 & -2.68 $\pm$ 0.02 & 3.86 $\pm$ 0.02 & 0.04 & 1.45 & 7.13 & N356 & UBC 1429 \\
18 & GI Cyg & 299.890 & 33.746 & 0.273 $\pm$ 0.030 & -3.45 $\pm$ 0.02 & -6.58 $\pm$ 0.02 & 0.07 & 2.37 & 4.30 & N1235 & UBC 135 \\
19 & OGLE-BLG-CEP-098 & 270.519 & -23.707 & 0.395 $\pm$ 0.037 & 0.26 $\pm$ 0.04 & -1.08 $\pm$ 0.03 & 1.11 & 0.47 & 0.84 & Y & Bochum 14 \\
20 & J211659.94+514556.7 & 319.250 & 51.766 & 0.336 $\pm$ 0.032 & -3.91 $\pm$ 0.03 & -4.72 $\pm$ 0.02 & 0.71 & 2.09 & 0.56 & N1236 & Berkeley 55 \\
21 & CS Vel & 145.293 & -53.816 & 0.272 $\pm$ 0.029 & -4.57 $\pm$ 0.02 & 3.13 $\pm$ 0.02 & 0.01 & 1.51 & 1.85 & N123456 & Ruprecht 79 \\
22 & X Cru & 191.593 & -59.125 & 0.654 $\pm$ 0.033 & -5.93 $\pm$ 0.02 & -0.17 $\pm$ 0.02 & 0.33 & 0.25 & 15.72 & N123456 & UBC 290 \\
23 & V367 Sct & 278.397 & -10.427 & 0.473 $\pm$ 0.033 & 0.08 $\pm$ 0.03 & -0.27 $\pm$ 0.02 & 0.92 & 1.44 & 2.91 & N12345 & NGC 6649 \\
24 & X Vul & 299.369 & 26.556 & 0.864 $\pm$ 0.034 & -1.35 $\pm$ 0.02 & -4.25 $\pm$ 0.03 & 0.58 & 2.01 & 16.63 & N12345 & UBC 129 \\
25 & J194806.54+260526.1 & 297.027 & 26.091 & 0.182 $\pm$ 0.032 & -2.87 $\pm$ 0.02 & -5.55 $\pm$ 0.02 & 1.03 & 2.01 & 4.23 & N236 & FSR 0158 \\
26 & U Sgr & 277.972 & -19.125 & 1.605 $\pm$ 0.035 & -1.79 $\pm$ 0.03 & -6.13 $\pm$ 0.02 & 1.55 & 0.37 & 0.81 & N12345 & IC 4725 \\
27 & R Mus & 190.521 & -69.408 & 1.076 $\pm$ 0.032 & -4.34 $\pm$ 0.02 & -2.05 $\pm$ 0.03 & 0.07 & 0.64 & 15.33 & Y & Theia 358 \\
28 & RS Ori & 95.555 & 14.678 & 0.589 $\pm$ 0.040 & 0.20 $\pm$ 0.04 & 0.01 $\pm$ 0.03 & 0.51 & 0.24 & 3.04 & N123456 & FSR 0951 \\
29 & DL Cas & 7.494 & 60.212 & 0.580 $\pm$ 0.038 & -2.71 $\pm$ 0.03 & -1.19 $\pm$ 0.03 & 0.56 & 0.96 & 3.15 & N12345 & NGC 129 \\
30 & IQ Nor & 228.206 & -54.755 & 0.535 $\pm$ 0.032 & -0.90 $\pm$ 0.02 & -1.82 $\pm$ 0.03 & 0.03 & 1.14 & 15.61 & N45 & CWNU 19 \\
31 & S Nor & 244.716 & -57.900 & 1.099 $\pm$ 0.034 & -1.61 $\pm$ 0.03 & -2.14 $\pm$ 0.03 & 0.97 & 1.22 & 0.72 & N12345 & NGC 6087 \\
32 & J201151.18+342447.2 & 302.963 & 34.413 & 0.238 $\pm$ 0.030 & -3.18 $\pm$ 0.02 & -4.91 $\pm$ 0.02 & 1.19 & 2.43 & 2.42 & N1236 & Berkeley 51 \\
33 & CN Sct & 280.627 & -4.331 & 0.390 $\pm$ 0.040 & -1.04 $\pm$ 0.04 & -2.25 $\pm$ 0.03 & 0.47 & 0.64 & 9.52 & N2356 & Trumpler 35 \\
34 & AN Aur & 74.923 & 40.836 & 0.285 $\pm$ 0.032 & -0.43 $\pm$ 0.03 & -1.51 $\pm$ 0.02 & 0.05 & 0.70 & 2.46 & N356 & UBC 1273 \\
35 & TW Nor & 241.230 & -51.954 & 0.360 $\pm$ 0.033 & -1.89 $\pm$ 0.03 & -2.81 $\pm$ 0.02 & 1.58 & 0.67 & 1.51 & N1234 & Lynga 6 \\
36 & V340 Nor & 243.322 & -54.235 & 0.491 $\pm$ 0.036 & -2.07 $\pm$ 0.03 & -2.63 $\pm$ 0.03 & 0.59 & 0.93 & 0.72 & N12345 & NGC 6067 \\
37 & GQ Vul & 296.992 & 26.000 & 0.148 $\pm$ 0.033 & -2.81 $\pm$ 0.02 & -5.47 $\pm$ 0.03 & 2.00 & 0.23 & 1.58 & N1236 & FSR 0158 \\
38 & CD Cyg & 301.111 & 34.112 & 0.394 $\pm$ 0.031 & -1.97 $\pm$ 0.02 & -5.58 $\pm$ 0.02 & 0.10 & 0.47 & 8.21 & N56 & Berkeley 84 \\
39 & NSVS 8389949 & 297.130 & 26.161 & 0.252 $\pm$ 0.036 & -2.88 $\pm$ 0.02 & -5.53 $\pm$ 0.03 & 1.00 & 1.79 & 10.83 & Y & FSR 0158 \\
40 & BM Per & 67.414 & 48.422 & 0.334 $\pm$ 0.034 & 0.04 $\pm$ 0.03 & -0.93 $\pm$ 0.02 & 0.58 & 0.55 & 2.70 & Y & CWNU 3123 \\
41 & SV Vul & 297.879 & 27.460 & 0.402 $\pm$ 0.034 & -2.16 $\pm$ 0.02 & -5.96 $\pm$ 0.03 & 0.66 & 0.85 & 9.15 & N123456 & UBC 130 \\
\hline
\end{tabular}
\tablecomments{Positions, parallaxes with zpt correction, and proper motions are listed for each OC Cepheid. The astrometric dispersions ($\delta_\varpi$, $\delta_{pm}$) denote the deviations of CCs from their host OCs in parallax and proper motion, respectively. $R$ denotes angular distance of CC from cluster center. Y: newly identified OC Cepheids; N: previously identified OC Cepheids. References—(1) \citet{zc2021}; (2) \citet{Hao2022}; (3) \citet{lin2022}; (4) \citet{mauricio2023}; (5) \citet{wang2024}; (6) \citet{Majaess2024}.}
\end{table*}

\setlength{\tabcolsep}{2pt} 
\begin{table*}
\centering
\caption{Pulsation mode, period, Gaia-band magnitudes, and metallicity of CCs for 41 OC Cepheids.}
\label{table:table2}
\begin{tabular}{ccccclllcc}
\hline\hline
Index &Cepheid & Gaia ID & Mode& P (days) & $G$ (mag)	&$G_{BP}$ (mag)	& $G_{RP}$ (mag)&	[M/H]&	Reference (CCs)\\
\hline
1 & NO Cas & 428893303878673408 & 1O & 2.58 & 10.97 & 11.65 & 10.16 &   & ade \\
2 & J183904-1049.3 & 4154758024243120640 & 1O & 3.06 & 9.95 & 10.83 & 9.05 & 0.04 & bcef \\
3 & EV Sct & 4156512638614879104 & 1O & 3.09 & 9.62 & 10.38 & 8.77 & 0.09 & acdef \\
4 & J213533.70+533049.3 & 2174193504173455744 & 1O & 3.20 & 11.90 & 12.75 & 10.98 &   & abcde \\
5 & FZ Car & 5338810654405875328 & 1O & 3.58 & 11.44 & 12.24 & 10.54 &   & abcde \\
6 & Gaia DR3 5254518760118884864 & 5254518760118884864 & 1O & 3.77 & 11.64 & 12.51 & 10.71 &   & e \\
7 & QZ Nor & 5932565900081831040 & 1O & 3.79 & 8.59 & 9.08 & 7.93 & 0.21 & cdef \\
8 & GI Car & 5339554336583678464 & 1O & 4.43 & 8.11 & 8.50 & 7.52 &   & acde \\
9 & CE Cas B & 2011892320749270912 & F & 4.48 & 10.58 & 11.15 & 9.68 &   & cdeg \\
10 & V335 Pup & 5711201189563596416 & 1O & 4.86 & 8.47 & 8.90 & 7.85 & 0.06 & cef \\
11 & CF Cas & 2011892703004353792 & F & 4.88 & 10.66 & 11.39 & 9.82 & 0.02 & acdef \\
12 & CE Cas A & 2011892325047232256 & F & 5.14 & 10.50 & 11.08 & 9.59 & 0.11 & cdef \\
13 & VW Cru & 6053622679932061056 & F & 5.27 & 9.03 & 9.89 & 8.10 & 0.16 & acdef \\
14 & CV Mon & 3127142323596470144 & F & 5.38 & 9.66 & 10.54 & 8.73 & 0.09 & acdef \\
15 & V Cen & 5891675303053080704 & F & 5.49 & 6.53 & 7.03 & 5.87 & 0.12 & acef \\
16 & OGLE-GD-CEP-1544 & 4320062347542878208 & F & 5.54 & 12.72 & 14.79 & 11.41 &   & bcde \\
17 & V724 Pup & 5539107450645759232 & F & 5.56 & 10.57 & 11.37 & 9.70 &   & acde \\
18 & GI Cyg & 2034428980559814400 & F & 5.78 & 11.07 & 11.98 & 10.11 & 0.24 & abcdef \\
19 & OGLE-BLG-CEP-098$^{\mathrm{*}}$ & 4069162310993340288 & F & 5.80 & 11.69 & 13.07 & 10.54 &   & ag \\
20 & J211659.94+514556.7 & 2172278803454142720 & F & 5.85 & 12.37 & 14.01 & 11.15 &   & bce \\
21 & CS Vel & 5308893149150949248 & F & 5.90 & 11.10 & 11.96 & 10.18 & -0.02 & acdef \\
22 & X Cru & 6059764002146656128 & F & 6.22 & 8.08 & 8.63 & 7.38 & 0.12 & cdef \\
23 & V367 Sct & 4155020566971108224 & F1O & 6.29 & 10.51 & 11.81 & 9.40 & 0.05 & abcdef \\
24 & X Vul & 2027263738133623168 & F & 6.31 & 8.24 & 9.12 & 7.30 & 0.13 & acdef \\
25 & J194806.54+260526.1 & 2027003119517288832 & 1O & 6.65 & 12.40 & 14.04 & 11.18 &   & e \\
26 & U Sgr$^{\mathrm{*}}$ & 4092905375639902464 & F & 6.75 & 6.43 & 7.13 & 5.64 & 0.14 & acef \\
27 & R Mus & 5855468247702904704 & F & 7.51 & 6.11 & 6.50 & 5.55 & -0.11 & acef \\
28 & RS Ori & 3368698813404804352 & F & 7.57 & 8.09 & 8.65 & 7.34 & 0.11 & acef \\
29 & DL Cas & 428620663657823232 & F & 8.00 & 8.52 & 9.21 & 7.69 & 0.05 & acdef \\
30 & IQ Nor & 5887482762148283008 & F & 8.22 & 9.03 & 9.86 & 8.13 & 0.22 & acdef \\
31 & S Nor & 5835124087174043136 & F & 9.75 & 6.17 & 6.65 & 5.49 & 0.1 & acef \\
32 & J201151.18+342447.2 & 2055722599436380416 & F & 9.83 & 13.57 & 15.43 & 12.29 &   & abce \\
33 & CN Sct & 4256740751371777408 & F & 9.99 & 11.14 & 12.71 & 9.94 & 0.3 & abcdef \\
34 & AN Aur & 201574982848108416 & F & 10.29 & 9.95 & 10.69 & 9.12 & -0.13 & acdef \\
35 & TW Nor & 5981406000300438016 & F & 10.79 & 10.53 & 11.91 & 9.38 & 0.27 & acdef \\
36 & V340 Nor & 5932569709575669504 & F & 11.29 & 7.99 & 8.62 & 7.22 &   & ace \\
37 & GQ Vul & 2026993563193159168 & F & 12.66 & 12.22 & 13.91 & 10.98 &   & abcde \\
38 & CD Cyg & 2058374144759464064 & F & 17.08 & 8.46 & 9.21 & 7.59 & 0.12 & acdef \\
39 & NSVS 8389949 & 2027002290578474368 & F & 20.76 & 12.41 & 14.56 & 11.04 &   & e \\
40 & BM Per & 258412862056729472 & F & 22.96 & 9.50 & 10.65 & 8.41 & 0.2 & acef \\
41 & SV Vul & 2027951173435143680 & F & 44.99 & 6.66 & 7.48 & 5.77 & 0.11 & acdef \\
\hline
\end{tabular}
\tablecomments{Metallicity values are taken from the CC catalog of \citet{Ripepi2022}. For CCs' photometry, the intensity-averaged magnitudes ($G$, ${G_{\rm BP}}$, and $G_{\rm RP}$) were extracted from \texttt{gaiadr3.vari\_cepheid} \citep{Ripepi2023}. In other cases, indicated with an asterisk next to the CC, Gaia DR3 photometry was adopted. References—a: \citet{skowron2019}; b: \citet{chen2019}; c: \citet{chen2020}; d: \citet{inno2021}; e: \citet{pietrukowicz2021}; f: \citet{Ripepi2022}.}
\end{table*}

\setlength{\tabcolsep}{12pt} 
\begin{table*}
\centering
\caption{Recalibrated parallax, extinction, distance modulus, and age of OCs from isochrone fitting for 41 OC Cepheids}
\label{table:table3}
\begin{tabular}{cccccccccccc}
\hline\hline
Index & Cluster &$\varpi$ (mas) &$A_v$ (mag)& $\mu$ (mag)& log ${Age}$&Referrence (OCs)\\
\hline
1 & NGC 103 & 0.309 $\pm$ 0.003 & 1.45 & 12.55 & 8.17 $\pm$ 0.05 & B \\
2 & CWNU 2461 & 0.511 $\pm$ 0.008 & 3.00 & 11.46 & 7.94 $\pm$ 0.03 & C \\
3 & NGC 6664 & 0.502 $\pm$ 0.003 & 2.15 & 11.50 & 7.97 $\pm$ 0.05 & B \\
4 & Kronberger 84 & 0.215 $\pm$ 0.006 & 2.59 & 13.34 & 8.00 $\pm$ 0.05 & B \\
5 & CWNU 2232 & 0.252 $\pm$ 0.004 & 1.82 & 12.99 & 8.06 $\pm$ 0.03 & C \\
6 & HSC 2354 & 0.237 $\pm$ 0.007 & 2.70 & 13.13 & 8.00 $\pm$ 0.05 & B \\
7 & Theia 3005 & 0.511 $\pm$ 0.005 & 1.00 & 11.46 & 7.96 $\pm$ 0.04 & B \\
8 & CWNU 2478 & 0.466 $\pm$ 0.006 & 0.65 & 11.66 & 7.88 $\pm$ 0.03 & C \\
9 & NGC 7790 & 0.319 $\pm$ 0.004 & 1.68 & 12.48 & 8.07 $\pm$ 0.05 & B \\
10 & UBC 229 & 0.429 $\pm$ 0.005 & 0.80 & 11.84 & 7.90 $\pm$ 0.05 & B \\
11 & NGC 7790 & 0.319 $\pm$ 0.004 & 1.68 & 12.48 & 8.07 $\pm$ 0.05 & B \\
12 & NGC 7790 & 0.319 $\pm$ 0.004 & 1.68 & 12.48 & 8.07 $\pm$ 0.05 & B \\
13 & CWNU 175 & 0.728 $\pm$ 0.006 & 1.50 & 10.69 & 8.25 $\pm$ 0.05 & B \\
14 & vdBergh 1 & 0.590 $\pm$ 0.006 & 2.15 & 11.15 & 8.06 $\pm$ 0.06 & B \\
15 & NGC 5662 & 1.331 $\pm$ 0.003 & 0.86 & 9.38 & 8.03 $\pm$ 0.05 & B \\
16 & HSC 413 & 0.342 $\pm$ 0.012 & 5.20 & 12.33 & 7.90 $\pm$ 0.05 & B \\
17 & UBC 1429 & 0.317 $\pm$ 0.006 & 2.02 & 12.50 & 8.04 $\pm$ 0.04 & B \\
18 & UBC 135 & 0.271 $\pm$ 0.004 & 2.10 & 12.84 & 8.00 $\pm$ 0.05 & B \\
19 & Bochum 14 & 0.353 $\pm$ 0.004 & 4.12 & 12.26 & 8.00 $\pm$ 0.10 & B \\
20 & Berkeley 55 & 0.359 $\pm$ 0.006 & 5.15 & 12.23 & 7.99 $\pm$ 0.06 & B \\
21 & Ruprecht 79 & 0.272 $\pm$ 0.004 & 2.15 & 12.83 & 8.00 $\pm$ 0.05 & B \\
22 & UBC 290 & 0.643 $\pm$ 0.002 & 0.60 & 10.96 & 8.01 $\pm$ 0.05 & B \\
23 & NGC 6649 & 0.504 $\pm$ 0.003 & 4.35 & 11.49 & 7.85 $\pm$ 0.05 & B \\
24 & UBC 129 & 0.884 $\pm$ 0.003 & 1.30 & 10.27 & 8.00 $\pm$ 0.05 & B \\
25 & FSR 0158 & 0.215 $\pm$ 0.005 & 4.90 & 13.34 & 7.73 $\pm$ 0.05 & B \\
26 & IC 4725 & 1.551 $\pm$ 0.002 & 1.20 & 9.05 & 7.94 $\pm$ 0.04 & B \\
27 & Theia 358 & 1.073 $\pm$ 0.005 & 0.56 & 9.85 & 7.92 $\pm$ 0.06 & B \\
28 & FSR 0951 & 0.610 $\pm$ 0.003 & 0.80 & 11.07 & 8.85 $\pm$ 0.05 & B \\
29 & NGC 129 & 0.558 $\pm$ 0.002 & 1.55 & 11.27 & 7.95 $\pm$ 0.05 & B \\
30 & CWNU 19 & 0.534 $\pm$ 0.005 & 1.40 & 11.36 & 7.94 $\pm$ 0.04 & B \\
31 & NGC 6087 & 1.065 $\pm$ 0.003 & 0.55 & 9.86 & 7.85 $\pm$ 0.05 & B \\
32 & Berkeley 51 & 0.201 $\pm$ 0.005 & 5.50 & 13.48 & 7.95 $\pm$ 0.05 & B \\
33 & Trumpler 35 & 0.371 $\pm$ 0.005 & 3.50 & 12.15 & 7.80 $\pm$ 0.05 & B \\
34 & UBC 1273 & 0.284 $\pm$ 0.009 & 1.76 & 12.74 & 7.90 $\pm$ 0.05 & B \\
35 & Lynga 6 & 0.413 $\pm$ 0.003 & 3.59 & 11.92 & 7.79 $\pm$ 0.03 & B \\
36 & NGC 6067 & 0.512 $\pm$ 0.002 & 0.99 & 11.45 & 7.85 $\pm$ 0.05 & B \\
37 & FSR 0158 & 0.215 $\pm$ 0.005 & 4.90 & 13.34 & 7.73 $\pm$ 0.05 & B \\
38 & Berkeley 84 & 0.391 $\pm$ 0.005 & 1.75 & 12.04 & 7.68 $\pm$ 0.05 & B \\
39 & FSR 0158 & 0.215 $\pm$ 0.005 & 4.90 & 13.34 & 7.73 $\pm$ 0.05 & B \\
40 & CWNU 3123 & 0.355 $\pm$ 0.014 & 3.25 & 12.25 & 7.61 $\pm$ 0.04 & A \\
41 & UBC 130 & 0.424 $\pm$ 0.003 & 2.10 & 11.86 & 7.40 $\pm$ 0.05 & B \\
\hline
\end{tabular}
\tablecomments{References—A: \citet{he2023}; B: \citet{huntemily2024}; C: \citet{he2022}.}
\end{table*}

\section{Results}
\label{sect:results}
\subsection{Open Cluster housing Cepheids} 
\label{sect:cross-match}
Using the OC and CC catalogs introduced in Section~2, we applied the method described in Section~3.1 to perform an initial cross-identification of OC members and CCs. This process established preliminary associations between CCs and OCs, resulting in 210 CCs linked to 202 OCs. Since the OCs were drawn from multiple catalogs, occasional discrepancies in naming conventions appeared for clusters with consistent parameters. After resolving such duplications, the sample was refined to 193 CCs associated with 181 OCs. We then evaluated the astrometric membership of CCs in OCs using the dispersions $\delta_\varpi$ and $\delta_{pm}$ relative to the cluster center, as defined in Equation~\ref{eq_delta}, adopting a 3$\sigma$ threshold as the membership criterion. We also took into account the angular distances of CCs relative to those of the member stars within each OC to ensure they are located near the cluster core. This step led to the exclusion of 55 CCs associated with 51 OCs.

After updating the cluster parameters through isochrone fitting, we further removed CCs that exhibited deviations of more than 0.5 mag from their expected position near the blue loop stage in the CMD of their host OCs, indicating that they are unlikely genuine OC Cepheids. This resulted in the exclusion of 43 CCs associated with 39 OCs. We note that our sample does not fully match those of ZC21, Hao22, Lin22, CRA23, and Wang24, with 8, 3, 19, 11, and 11 CCs excluded, respectively. This discrepancy primarily arises from differences in the OC member stars and CCs considered. By merging these datasets, we identified 27 additional CCs associated with 25 OCs. The final sample thus consists of 110 CCs associated with 102 OCs.

\subsection{Classification of OC Cepheids} 
Among the identified OC-Cepheid associations, some CCs exhibit parallaxes, proper motions, and angular distances from the cluster center that deviate significantly from those of the OC members, suggesting they may not be physically associated with the clusters. In addition, the positions of certain CCs in the CMD are inconsistent with the expected location of CCs near the blue loop of their host OCs. These discrepancies indicate that CCs may not be associated with their host OCs. Consequently, we classified the 110 OC-Cepheid pairs, comprising 110 CCs associated with 102 OCs, into three categories: reliable OC Cepheids, OC Cepheid candidates, and rejected samples.

For the reliable OC Cepheids, the astrometric dispersions in parallax and proper motion both fell within 3$\sigma$ of the OC members and they are situated near the cluster core. Simultaneously, their positions in the CMD are consistent with the expected location near the blue loop of their host clusters. These criteria indicate that the CCs are reliably associated with their host OCs. Based on this selection, we identified 41 CCs linked to 37 OCs, including 4 newly discovered OC Cepheids. For comparison, ZC21, Hao22, CRA23, and Wang24 reported 33, 39, 34, and 43 reliable OC Cepheids, respectively. In particular, RS Ori, previously suggested to be associated with FSR 0951, does not fully satisfy our membership criteria. By taking into account the distribution of red giant branch stars during isochrone fitting, we obtained an older age of 8.85 dex for FSR 0951. As a result, RS Ori's position in the CMD deviates from the expected blue loop evolutionary stage by more than 0.5 mag. Previous studies have suggested the presence of a possible hot companion (e.g., \citealt{Evans1990,Shetye2024}), which could make the system appear brighter and bluer, thereby explaining its apparent deviation from the blue loop. Given its cluster membership and these peculiarities, RS Ori is classified as an OC Cepheid.
\begin{figure*}[!htbp]
\centering
\includegraphics[width=0.23\linewidth]{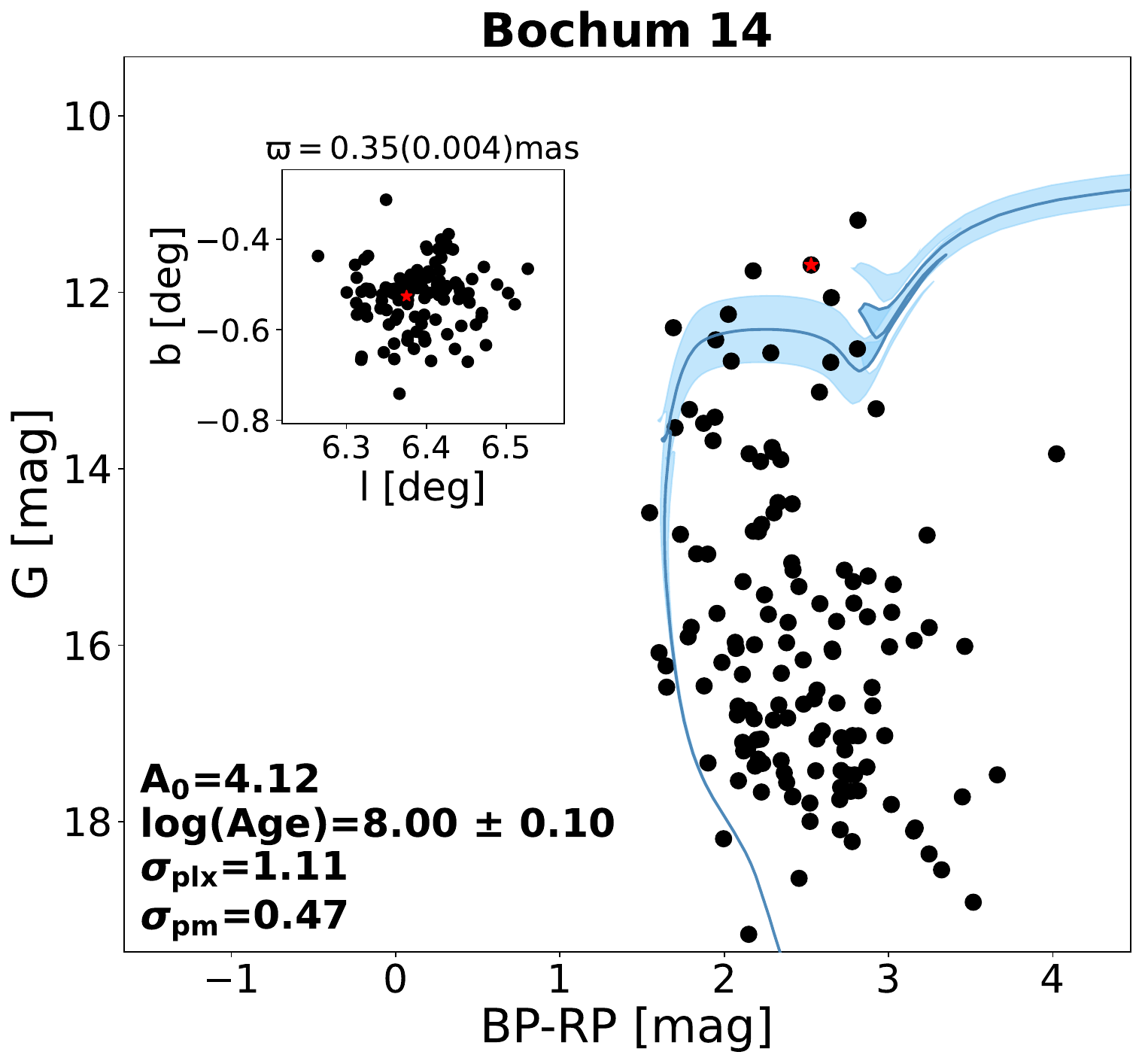} 
\includegraphics[width=0.23\linewidth]{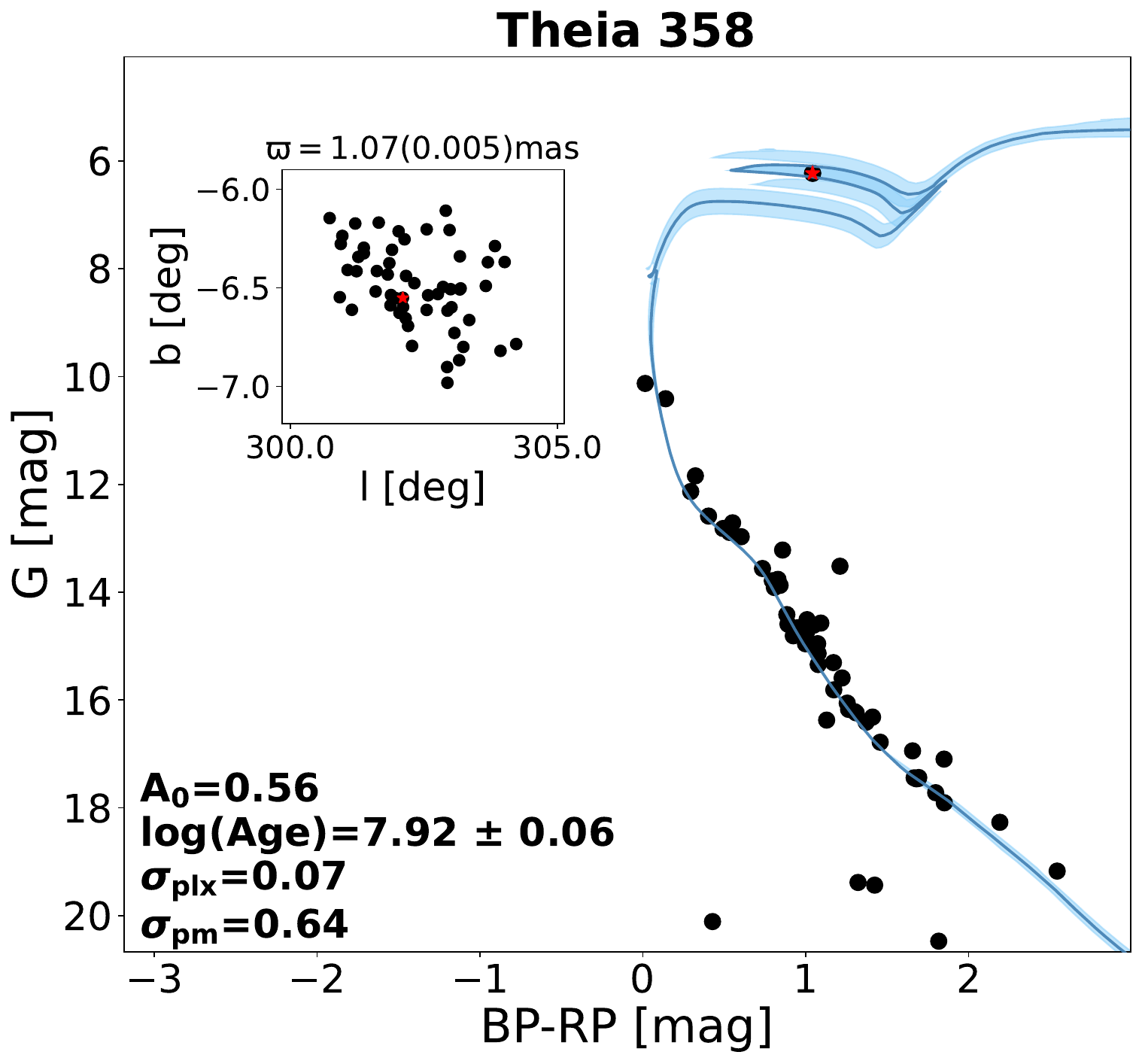} 
\includegraphics[width=0.23\linewidth]{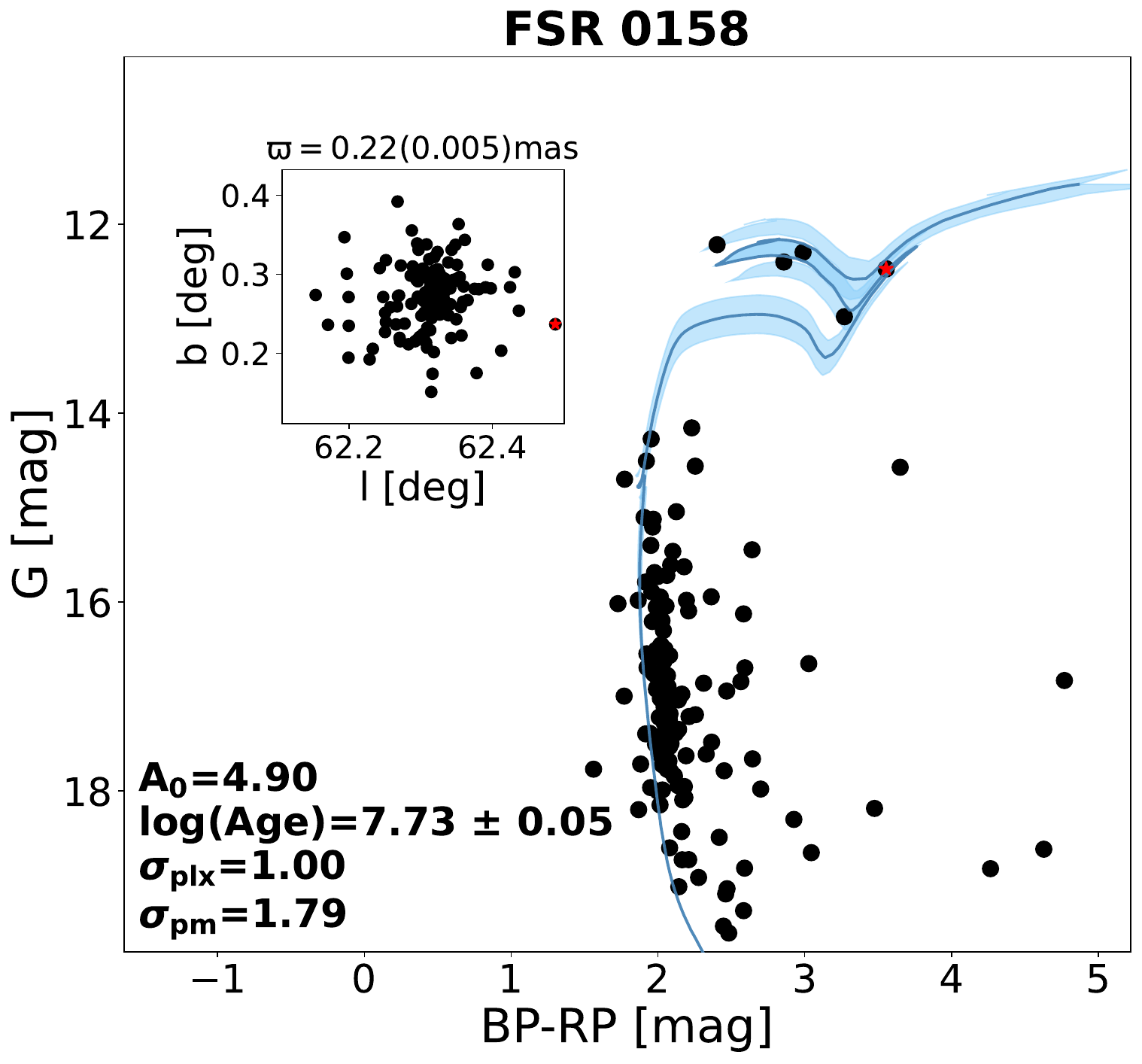}\hspace{0.02\linewidth}
\includegraphics[width=0.23\linewidth]{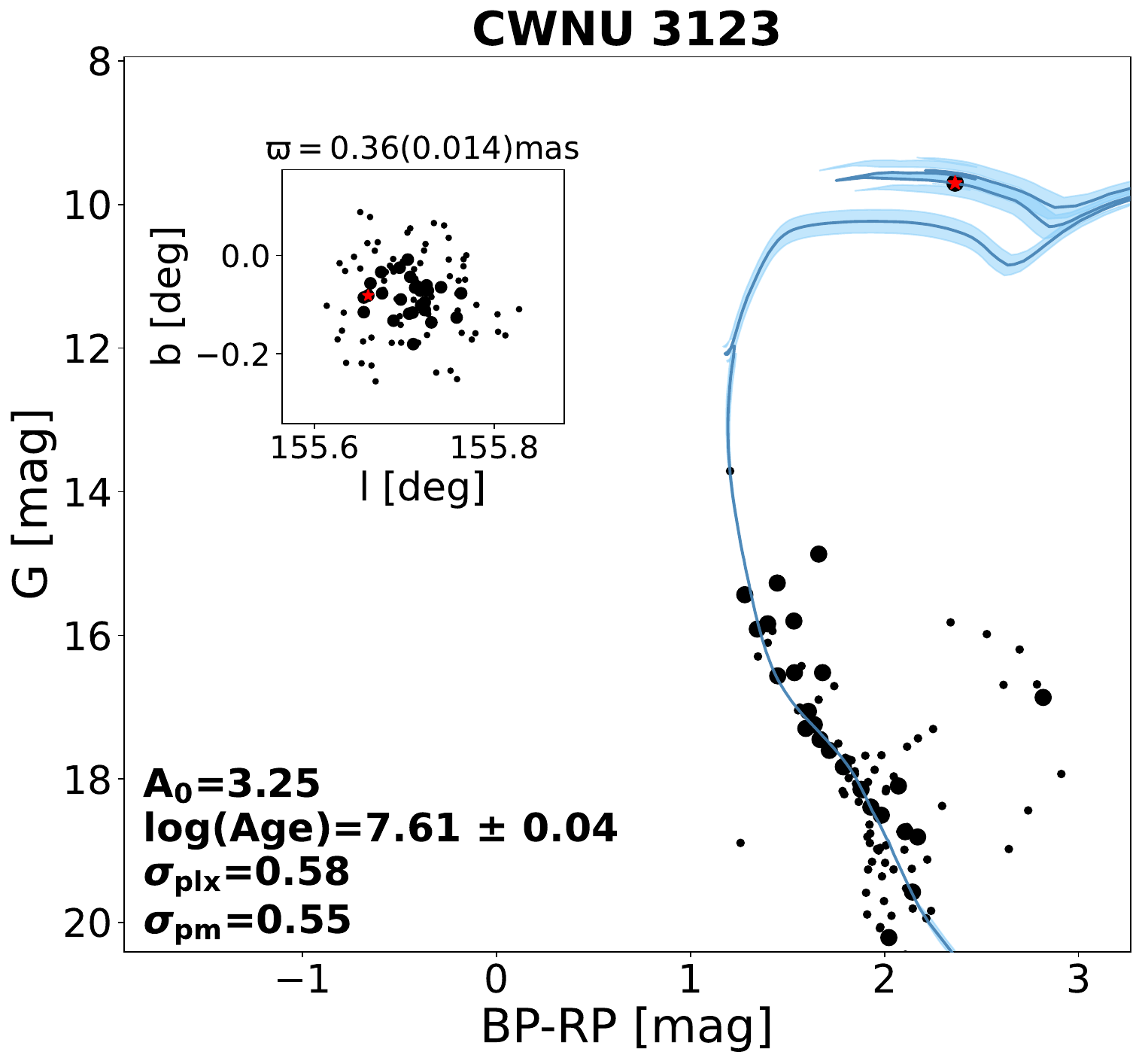}
\caption {CMDs of the 4 newly identified OC Cepheids are shown, including the extinction, age, and distance parameters of their host clusters: Bochum 14, Theia 358, FSR 0158 (NSVS 8389949), and CWNU 3123. In each CMD, black points represent cluster member stars, and the inset shows the spatial distribution of members in Galactic coordinates. The size of the black points indicates the membership reliability of cluster stars as reported by He23: larger points correspond to high-confidence member stars, while smaller points represent candidate members. The blue shaded region indicates the age uncertainty, while the dark blue line corresponds to the $\log{Age}$ derived from isochrone fitting. Astrometric dispersions ($\delta_\varpi$ and $\delta_{pm}$), parallax, extinction, and age values are provided in each panel.}
\label{fig2:newOCCs}
\end{figure*}

The OC Cepheid candidates showed three types of anomalies relative to their host OCs. The first group exhibited $\delta_\varpi$ and $\delta_{pm}$ were within 3$\sigma$, and was located near the cluster core, but the CCs deviated from the blue loop in the CMD by more than 0.5 mag with atypical luminosities.
The second group had $\delta_\varpi$ and $\delta_{pm}$ between 3$\sigma$ and 5$\sigma$, but their CMD positions remained close to the blue loop.
Another group exhibited astrometric dispersions consistent with cluster membership and CMD positions near the blue loop, but with the largest angular distances among OC members, placing them near the outskirts of the cluster (e.g., V0379 Cas associated with NGC 129). Based on these criteria, we identified 19 CCs associated with 19 OCs as OC Cepheid candidates. 

Rejected samples included CCs whose $\delta_\varpi$ and $\delta_{pm}$ differed from their host OCs by more than 5$\sigma$ or with the largest angular distances within OC members, indicating unreliable membership. We also rejected CCs with $\delta_\varpi$ and $\delta_{pm}$ between 3$\sigma$ and 5$\sigma$, and situated near the cluster core if they showed significant deviations by more than 0.5 mag from the blue loop in the CMD. In addition, OCs with fewer than 20 identified members were considered unreliable. As a result, we identified 50 rejected samples, comprising 50 CCs associated with 49 OCs. 

The 41 reliable OC Cepheids and their associated OC and stellar parameters are listed in Tables~\ref{table:table1}, \ref{table:table2}, and \ref{table:table3}. The 19 OC Cepheid candidates and 50 rejected samples are summarized in Appendix~\ref{appdendixa}. The CMDs of the 4 newly identified OC Cepheids are presented in Figures~\ref{fig2:newOCCs}, while those of the remaining reliable, candidate, and rejected samples are presented in Appendix~\ref{appendixc}.

\subsection{Newly found OC Cepheids} 
\subsubsection{Bochum 14 and OGLE-BLG-CEP-098 (5.80 days)}
OGLE-BLG-CEP-098 is associated with Bochum 14 from HR24. Bochum 14 is located at a distance of 2696 $\pm$ 189 pc. This CC is located $\sim$0.84 arcmin from the OC center, well within the cluster core. OGLE-BLG-CEP-098 meets the astrometric membership criteria with $\delta_{\varpi} = 1.11\,\sigma$ and $\delta_{{pm}} = 0.47\,\sigma$. Isochrone fitting refines the cluster parameters to $A_V = 4.12$, DM = 12.41, and $\log Age = 8.00$. Notably, Bochum 14 exhibits a high extinction and significant differential extinction across the cluster field. Together with the case of HSC 413, this underscores the importance of near-infrared observations for more reliable determinations of cluster parameters.

\subsubsection{Theia 358 and R Mus (7.51 days)}
R Mus is associated with Theia 358 (HR24) and is located at approximately 15.33 arcmin from the cluster center, well inside the cluster boundaries. Its host OC is at a distance of $939 \pm 16$ pc. The astrometric  membership criteria for R Mus yield $\delta_{\varpi} = 0.07\sigma$ and $\delta_{pm} = 0.64\,\sigma$. In the CMD, R Mus is positioned near the blue loop. Its agreement with both the PAR and PWR further supports the reliability of our fitting and underscores its significance as a representative sample for studying CC properties in nearby OCs.

\subsubsection{FSR 0158 and NSVS 8389949 (20.76 days)}
OC Cepheid NSVS 8389949 is offset by approximately 10.83 arcmin from the center of FSR 0158 (HR24), yet still lies within the cluster's core region. Its host OC is at a distance of $4220 \pm 445$ pc. The astrometric membership criteria for NSVS 8389949 yield $\delta_{\varpi} = 1.00\,\sigma$ and $\delta_{pm} = 1.79\,\sigma$. This CC is long-period with $\log P = 1.32$, helping to address the scarcity of long-period samples, which are essential for calibrating distances to extragalactic galaxies \citep{riess2019}. Additionally, FSR 0158 hosts one other OC Cepheid and one candidate, further enhancing the cluster's contribution to calibrating the PAR and PWR relations.

\subsubsection{CWNU 3123 and BM Per (22.96 days)}
BM Per, with a period of 22.96 days, is classified in this study as a long-period OC Cepheid ($\log P > 1.1$). Its host OC is at a distance of $3472 \pm 253$ pc. BM Per is located at about 2.70 arcmin from the center of CWNU 3123, well within the cluster core. The CC satisfies the astrometric membership criteria, yielding $\delta_{\varpi} = 0.58\,\sigma$ and $\delta_{pm} = 0.55\,\sigma$. In both the PAR and PWR diagrams, BM Per occupies an underpopulated region of long-period OC Cepheids, highlighting its significance within our sample.
\begin{figure}
\centering
\includegraphics[width=0.75\linewidth]{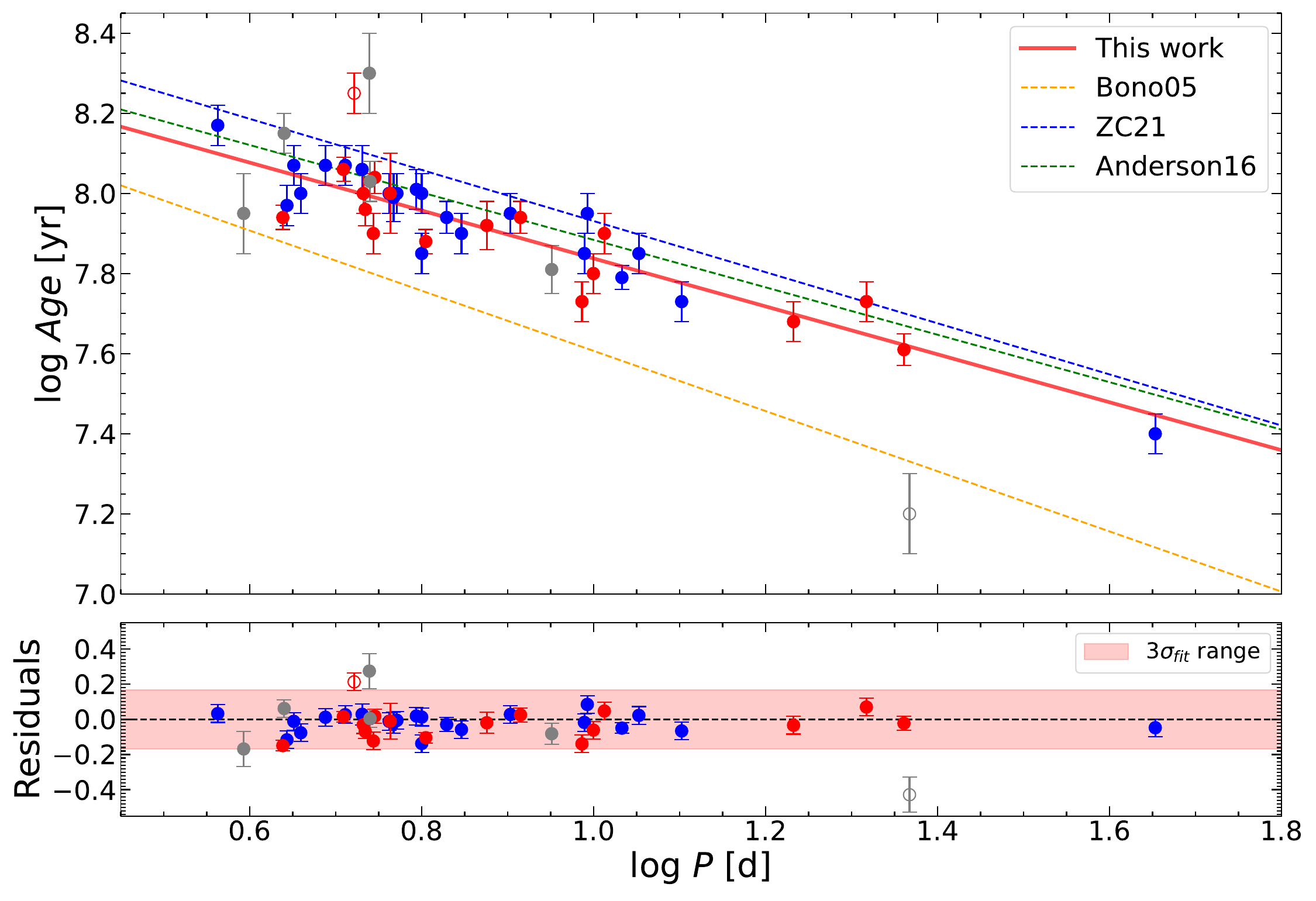} 
\hspace{0.0cm}
\caption{The PAR for the 41 F mode and 1O mode OC Cepheids, compared with those derived by \citet{Bono2005}, \citet{Anderson2016}, and \citet{zc2021}. Red and blue points represent 41 OC Cepheids from this work; blue points appear in the ZC21 PAR sample, while red points are newly identified outside of ZC21. Gray filled points and open circle are candidates or rejected samples in this work that are the other part of the PAR fitting samples in ZC21. One outlier is shown as red open circle.}
\label{fig3:PAR}
\end{figure}

\section{Analyses}
\label{sect:analysis} 
\subsection{Fitting of Period-Age Relations for OC Cepheids} 
Over the past two decades, numerous studies have explored the theoretical PAR models for CCs~\citep[e.g.,][]{Bono2005, Anderson2016, DeSomma2021}.
However, observational calibrations of the PAR remain limited due to the small number of reliable OC Cepheids, especially those with long periods. In addition, deriving accurate age estimates for CCs is inherently challenging. Recently, ZC21 used Gaia EDR3 data to identify 33 OC Cepheids and calibrated the PAR based on cluster ages determined from isochrone fitting.

The PAR derived by ZC21 exhibits differences from the theoretical models of Anderson16 in the short-period range, while showing good agreement in the long-period regime. Compared to Bono05, deviations are present across the full period range. ZC21 includes 3 long-period CCs ($\log P > 1.1$) and 30 short-period CCs ($\log P < 1.1$), which underscores the relative scarcity of long-period CCs in their sample.
\begin{figure}
\centering
\includegraphics[width=0.75\linewidth]{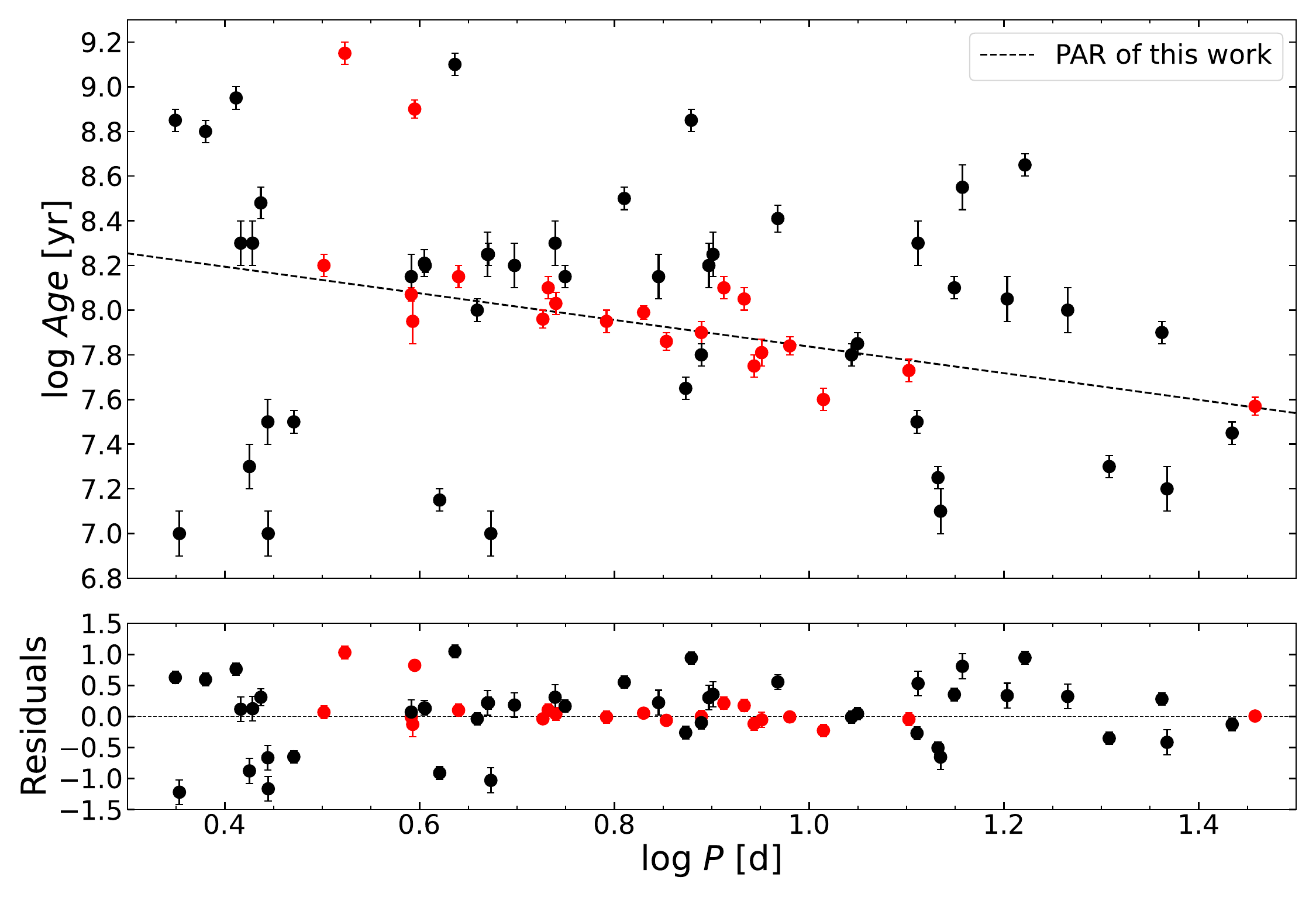} \hspace{0.0cm}
\caption{Period-age distribution of OC Cepheid candidates and rejected samples. Red points represent OC Cepheid candidates, while black points indicate the rejected samples. The black line represents the PAR derived in this work.}
\label{fig4:PAR BC}
\end{figure}

In this study, we performed the search for OC Cepheids using an expanded and updated observational dataset of OCs and CCs. We identified 41 OC Cepheids, including 30 fundamental (F) mode pulsators, 10 first overtone (1O) mode pulsators and 1 multimode pulsator. The number of long-period CCs ($\log P > 1.1$) increased to 5, while the short-period CCs ($\log P < 1.1$) grew to 35. In PAR fitting, the periods of 1O mode pulsators were converted to their F mode equivalents using the relation from \citet{Pilecki2024}:
\begin{equation}
\label{PF1O}
\begin{split}
P_{\text{F}} / P_{\text{1O}} = 1.371 + 0.106 \log P_{\text{1O}}
\end{split}
\end{equation}
allowing for a combined PAR fit of OC Cepheids across different pulsation modes. The final relation, shown in Figure~\ref{fig3:PAR} (red curve), is
\begin{equation}
\begin{split}
\log \textit{Age} = (-0.595 \pm 0.044) \log P + (8.430 \pm 0.042), \ \sigma_{fit} = 0.057
\end{split}
\end{equation}
The 1$\sigma_{fit}$ scatter of 0.057 dex in $\log\textit{Age}$ corresponds to an age uncertainty of approximately 14\%, indicating a high correlation suitable for precise age determination.

Figure \ref{fig3:PAR} compares our PAR with those of Bono05, Anderson16, and ZC21. Similar to ZC21, our results show some discrepancies with Bono05 across the entire period range. 
Across the whole period range, our fit exhibits a shallower slope and smaller intercept compared to ZC21. Interestingly, our fitted slope is nearly identical to Anderson16, although the intercept shows some deviation. Compared with ZC21, we identify three newly discovered long-period CCs ($\log P > 1.1$), all of which closely follow our fitting relation with deviations smaller than 3$\sigma$.
Overall, our fitted PAR lies systematically below previous calibrations of Anderson16 and ZC21 over the entire period range, suggesting slightly younger ages by $\sim$0.06 dex and $\sim$0.15 dex relative to Anderson16 and ZC21, respectively.

Notably, we excluded one long-period sample from ZC21, 2MASS J17184740-3817292 associated with BH 222 (gray open circle), because it was classified as a rejected sample due to a spatial dispersion exceeding 3$\sigma$. In Figure~\ref{fig3:PAR}, this object is indicated by a gray open circle. ZC21 samples considered as candidates are marked by gray filled circles in the same figure and were not included in our PAR fitting. Additionally, one outlier VW Cru linked to CWNU 175 (red open circle) was excluded from the fitting. The outlier may be influenced by binary stars in OCs, which can lead to overestimated extinction values during isochrone fitting and thus biased age estimates. Another potential factor is the effect of stellar rotation on CCs~\citep{Huang2010, Zorec2012}. As noted by Anderson16, rotation modifies the distribution of metallicity both internally and on the stellar surface, significantly affecting stellar evolution.

We further compared the period-age distributions of the OC Cepheid candidates and rejected samples with the PAR fitting results for both F and 1O mode pulsators. As shown in Figure \ref{fig4:PAR BC}, most OC Cepheid candidates exhibit only minor deviations from the fitted curve (less than 0.2 dex). Remarkably, KQ Sco, associated with UBC 1558, shows a discrepancy of just 0.002 dex. Four candidates, however, display larger deviations. In contrast, most rejected samples display substantial discrepancies from the fitting results, although 13 samples show relatively small deviations (less than 0.2 dex). These results indicate that some OC Cepheid candidates may indeed be reliable, while the rejected samples are likely not genuine OC Cepheids or may reflect inaccuracies in the age estimates of their host OCs. With improved astrometric parameters for CCs and OCs from Gaia Data Release 4 (DR4), future work will facilitate a more robust confirmation of the membership for these OC Cepheid candidates.

\begin{figure*}
\centering
\includegraphics[width=1\linewidth]{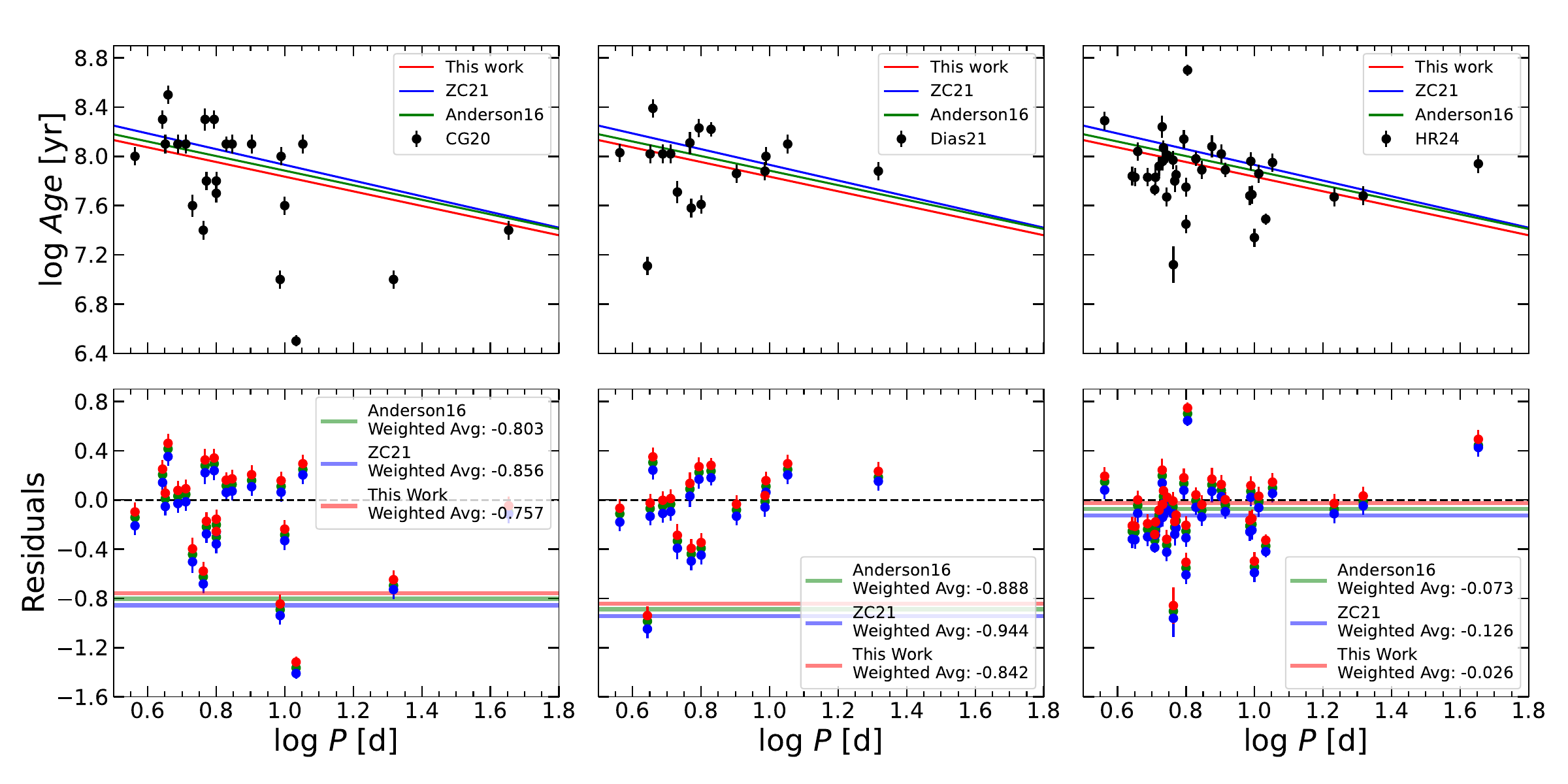} \hspace{0.0cm}
\caption{Comparison of OC age determinations using different PARs. The upper panels compare cluster ages and CC period distributions from \citet{CG2020}, \citet{dias2021}, and ~\citet{huntemily2024}. In the lower panels, red, green and blue points represent the age residuals relative to the PAR fits from this work, \citet{Anderson2016}, and \citet{zc2021}, respectively. The corresponding weighted mean residuals are indicated by the red, green, and blue lines.}
\label{fig5:PAR age match}
\end{figure*}

Finally, we evaluated the performance of our PAR fitting by comparing the age weighted average residuals of OCs from several studies~\citep[e.g.,][]{CG2020, dias2021, huntemily2024, he2022, he2023} with the results from this work, ZC21, and Anderson16, as shown in Figure \ref{fig5:PAR age match}. The comparison indicates that our PAR produces smaller weighted average residuals across both short-period and long-period samples. This improvement highlights the advantage of calibrating OC ages through manual isochrone fitting. Nevertheless, discrepancies in literature ages still remain, suggesting that further refinement is warranted. We expect that more precise age estimates in future studies will enhance the PAR, thereby providing stronger constraints on the formation and evolution of the Galactic disk and its spiral structure.
\begin{figure}
\centering
\includegraphics[width=0.75\linewidth]{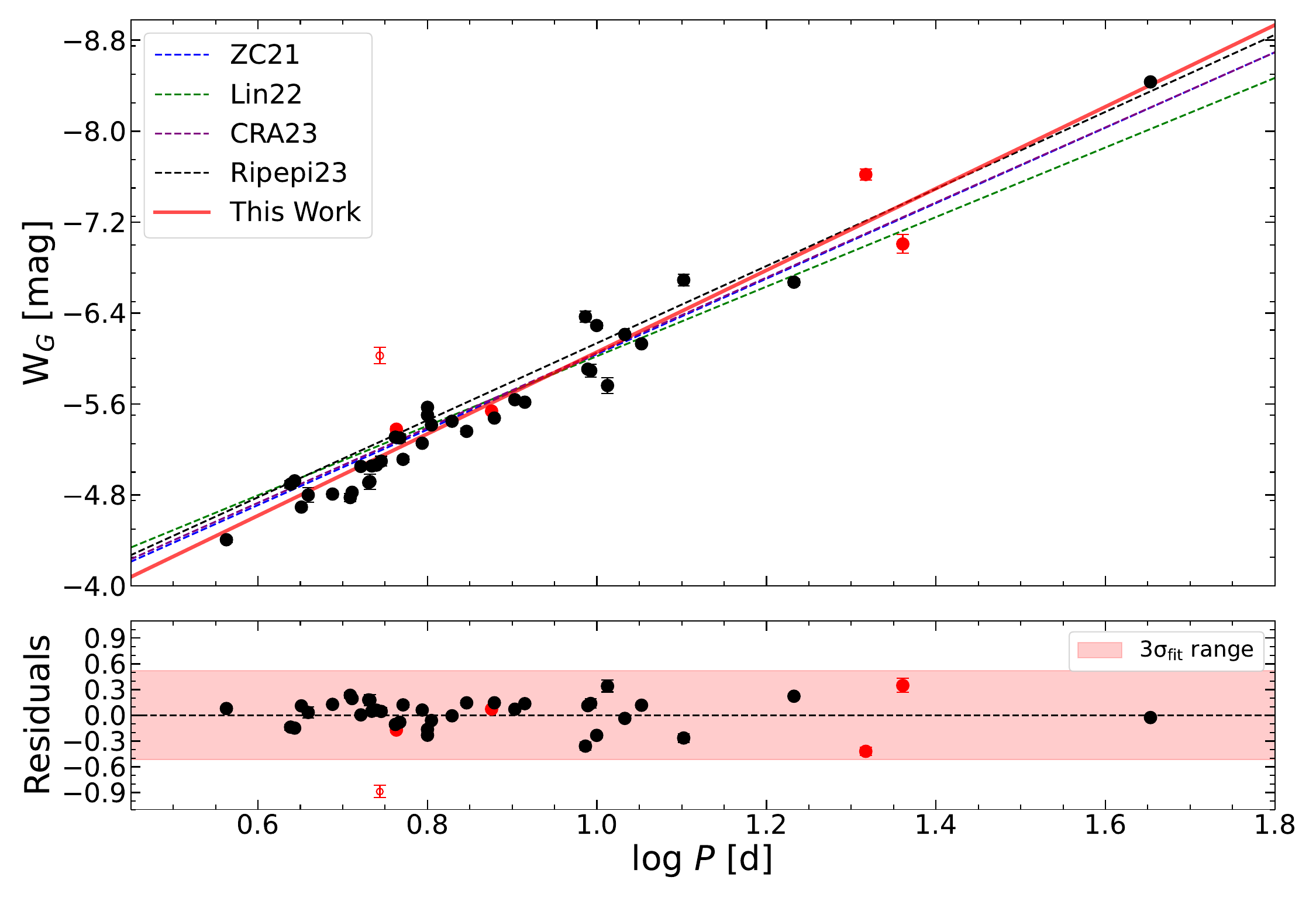}
\caption{The PWR of 41 OC Cepheids, including both F and 1O mode pulsators. Red points represent the newly identified samples from this work, black points denote previously confirmed samples, and one open circle indicates the outlier. The fitted relations are compared with those from \citet{zc2021}, \citet{lin2022}, \citet{mauricio2023} and, \citet{Ripepi2023}.}
\label{fig6:PWR}
\end{figure}

\subsection{Fitting of Period-Wesenheit Relations for Cepheids}
The PWR of CCs is a crucial tool for measuring cosmic distances. In recent years, substantial progress has been made toward refining the PWR. Many studies directly employed individual parameters of CCs for PWR calibration~\citep{Ripepi2019,Ripepi2022,Ripepi2023}, but these results were limited by Gaia's systematic uncertainty in parallax~\citep{Lindegren2021a} and the underestimation of the formal uncertainties~\citep{Fabricius2021}. Subsequent efforts incorporated systematic error estimates and adopted parallax zpt corrections to mitigate these effects~\citep{Riess2022}. For OC Cepheids, the distances derived from their host clusters effectively reduce both random and systematic errors in the individual parallax measurements of the CCs (Lin22).

In this study, we calibrate the PWR using 41 OC Cepheids, including 30 F mode pulsators, 10 1O mode pulsators and 1 multimode pulsator. Consistent with the PAR, the periods of the 1O mode pulsators are converted to their F mode equivalents following Equation~\ref{PF1O} \citep{Pilecki2024}. The PWR fitting is then performed on the combined sample, adopting a linear relation:
\begin{align}
W_G &= a \log P + b \\
W_G &= W(G, \textit{G}_{\textit{BP}}, \textit{G}_{\textit{RP}})
      = G - \lambda \left(\textit{G}_{\textit{BP}} - \textit{G}_{\textit{RP}}\right)
\end{align}
where $W_G$ denotes the Wesenheit absolute magnitude \citep[e.g.,][]{Madore1982, Madore2017}, with $G$, $G_{BP}$, and $G_{RP}$ denoting Gaia band magnitudes. To ensure more reliable photometry for the CCs, we used the intensity-averaged magnitudes from \texttt{gaiadr3.vari\_cepheid} \citep{Ripepi2023} for fitting the PWR. We note that two OC Cepheids do not appear in \texttt{gaiadr3.vari\_cepheid}. For the extinction coefficient, we adopt $\lambda = 1.9$ as recommended by \citet{Ripepi2019}, which is appropriate for Gaia photometric bands. We adopt cluster-averaged parallaxes with zpt corrections instead of individual CCs parallaxes to reduce systematic uncertainties.
As shown in Figure~\ref{fig6:PWR}, the PWR fit for 41 OC Cepheids yields the following result (red curve):
\begin{equation}
\begin{split}
W_\textit{G} = (-3.615 \pm 0.083) \log P + (-2.379 \pm 0.096), \sigma_{fit} = 0.153
\end{split}
\end{equation}
which appears steeper than those from previous studies~\citep[e.g.,][]{zc2021,lin2022,mauricio2023,Ripepi2023}. This difference is likely due to the different samples used for the fitting. Simultaneously, we adopt cluster-averaged parallaxes obtained from the weighted average of the parallaxes of member stars, which effectively reduces both random and systematic errors. In contrast, when using the parallaxes of individual CCs for the fitting, the relation becomes steeper. Using cluster-averaged parallaxes, on the other hand, reduces the scatter.

Further, we analyze the effect of applying the parallax zpt correction in PWR fitting. Using the same cluster-averaged parallaxes and $\lambda$, the zpt correction yields a PWR with both a smaller slope and intercept, leading to a shallower relation than the uncorrected fit. Consistent with previous studies (ZC21; Lin22), we also performed fits by adopting a fixed slope ($a = -3.32$) and $\lambda$ = 1.9. Our analysis reveals that applying the zpt correction results in a shift of the fitted PWR intercept by approximately 0.1 mag compared to the case without the correction. This demonstrates that applying the zpt correction helps reduce systematic errors in Gaia parallaxes.

Additionally, in Figure~\ref{fig6:PWR}, one outlier J194806.54+260526.1 associated with FSR 0158 was marked by red open circle. Its RUWE of 0.934 indicates a reliable astrometric solution, while its \texttt{ipd\_frac\_multi\_peak}\footnote{\url{https://gea.esac.esa.int/archive/documentation/GDR3/Gaia_archive/chap_datamodel/sec_dm_main_source_catalogue/ssec_dm_gaia_source.html}} signals multiple peaks in the point-spread function, suggesting a potentially blended or complex source. Future Gaia DR4, with its expanded sample and higher precision, is expected to yield more reliable parallaxes for OC Cepheids and enable a more accurate calibration of the PWR.

\section{Summary}
\label{sect:summary}
In this study, we utilized Gaia DR3 data to search for OC Cepheids, focusing on identifying distant and long-period samples and updating OCs parameters to calibrate the PAR and PWR of CCs. By integrating multiple up-to-date OC and CC catalogs, we built a more comprehensive database. We compiled three OC catalogs and five CC catalogs including their spatial positions, kinematic parameters and photometry ($l$, $b$, $\varpi$, $\mu_{\alpha^{*}}$, $\mu_{\delta}$, $G$, $G_{BP}$, and $G_{RP}$).
Initial associations were established through positional cross-matching within 1 arcsecond using TOPCAT \citep{Taylor2005}. Membership was then confirmed by applying astrometric dispersions ($\delta_\varpi$, $\delta_\mu$) and angular distance relative to OCs and by examining CCs' distinct positions near the blue loop in the CMD.
Isochrone fitting was further used to refine key OC parameters such as age and extinction, providing the basis for subsequent PAR and PWR calibration.
As a result, we identified 110 CCs associated with 102 OCs, of which 41 CCs linked to 37 OCs were classified as OC Cepheids. Among these, 4 are newly discovered. Notably, BM Per linked to CWNU 3123 is the newly longest-period confirmed OC Cepheid with $\log P = 1.36$. Additionally, two newly discovered OC Cepheid candidates have distances exceeding 6 kpc.

Next, we performed fitting for the PAR and PWR using 41 OC Cepheids. Based on the ages derived from isochrone fitting, we calibrated the PAR as $\log \textit{Age}$ = (-0.595 $\pm$ 0.044) $\log P$ + (8.430 $\pm$ 0.042), $\sigma_{fit} = 0.057$. The scatter corresponds to an age uncertainty of about 14\%, demonstrating a precise and reliable relation for cluster age determination. Our results lie systematically below those of previous studies, indicating that the derived OC ages are slightly smaller. This discrepancy can be attributed to our use of more detailed isochrone fitting, including consideration of the red giant branch distribution and age determination, which may differ from the approaches adopted in previous studies. Moreover, when comparing our PAR calibration with other determinations of OC ages, additional differences are evident, suggesting that further optimization is required.
For the PWR, we systematically examined the effects of parallax selection and zero-point correction. Using cluster-averaged parallaxes effectively mitigates Gaia's systematics, while applying zpt corrections further improves the fit, yielding $W_G$ = (-3.615 $\pm$ 0.083) $\log P$ + (-2.379 $\pm$ 0.096), $\sigma_{fit} = 0.153$.

With the improved accuracy of observational data, we have expanded the sample of OC Cepheids. Nevertheless, even in the Gaia DR3 era, the census of OC Cepheids remains relatively small and incomplete, particularly for long-period and distant cases. Future releases, especially Gaia DR4, will enlarge the sample and enable more refined PAR and PWR calibrations, with broad applications to studies of the Milky Way's spiral structure and stellar evolution.

\section*{Acknowledgments}
We sincerely thank all those who provided valuable feedback and assistance for this work, especially the anonymous referee for the helpful comments and suggestions that improved the manuscript. This work was supported by National Natural Science Foundation of China (NSFC) through grants 12573023,12303024, the Natural Science Foundation of Sichuan Province (2024NSFSC0453), and the "Young Data Scientists" project of the National Astronomical Data Center (NADC2023YDS-07). A.R. is supported by the China Manned Space Program with grant no.CMS-CSST-2025-A13. Y.L. is supported by the NSFC12573035, and the CSST project: CMS-CSST-2021-A10. K.W. is supported by NSFC12373035,12173028. And this work has made use of data from the European Space Agency mission Gaia (\url{https://www.cosmos.esa.int/gaia}), processed by the Gaia Data Processing and Analysis Consortium (DPAC,\url{https://www.cosmos.esa.int/web/gaia/dpac/consortium}). Funding for the DPAC has been provided by national institutions, in particular the institutions participating in the Gaia Multilateral Agreement.


\appendix
\label{sect:addtions}
\setcounter{figure}{0}
\renewcommand{\thefigure}{A\arabic{figure}}

\setcounter{table}{0}
\renewcommand{\thetable}{A\arabic{table}}

\section{Data for OC Cepheid candidates and rejected samples}\label{appdendixa}
These tables list the OC Cepheid candidates and rejected samples identified in this work, continuing from Tables~\ref{table:table1}, \ref{table:table2}, and \ref{table:table3}. The complete table is available in machine-readable format in the online journal.
\newpage
\setlength{\tabcolsep}{1.5pt} 
\centering
\begin{longtable}{cccccccccccc} 
\caption{Astrometric parameters of the OC Cepheid Candidates and rejected samples.}\\
\hline\hline
Index & Cepheid&$\alpha (^{\circ})$ & $\delta (^{\circ})$ & $\varpi$ (mas) &$\mu_{\alpha^{*}}$ (mas/yr) & $\mu_{\delta}$ (mas/yr) &  $\delta_\varpi$&$\delta_{pm}$ & $R $ &Note &Cluster \\ 
\hline
\endfirsthead
\caption{Astrometric parameters of the OC Cepheid Candidates and rejected samples (continued).}\\
\hline\hline
Index & Cepheid&$\alpha (^{\circ})$ & $\delta (^{\circ})$ & $\varpi$ (mas) &$\mu_{\alpha^{*}}$ (mas/yr) & $\mu_{\delta}$ (mas/yr) &  $\delta_\varpi$&$\delta_{pm}$ & $R $ &Note &Cluster \\ 
\endhead
\multicolumn{12}{c}{OC Cepheid Candidates}\\
\hline
42 & J065.6485+45.5819 & 65.649 & 45.582 & 0.358 $\pm$ 0.033 & -0.18 $\pm$ 0.03 & -1.11 $\pm$ 0.02 & 1.70 & 1.58 & 12.20 & Y & CWNU 2580 \\
43 & OGLE-GD-CEP-0265 & 130.365 & -47.650 & 0.121 $\pm$ 0.030 & -2.16 $\pm$ 0.02 & 3.17 $\pm$ 0.02 & 0.22 & 0.98 & 0.53 & Y & CWNU 3981 \\
44 & V423 CMa & 110.813 & -29.722 & 0.145 $\pm$ 0.031 & -1.05 $\pm$ 0.02 & 1.83 $\pm$ 0.02 & 0.17 & 2.11 & 0.64 & Y & CWNU 3305 \\
45 & OGLE-GD-CEP-0519 & 156.999 & -57.177 & 0.264 $\pm$ 0.029 & -7.21 $\pm$ 0.02 & 3.98 $\pm$ 0.02 & 0.09 & 1.32 & 0.46 & Y & CWNU 3193 \\
46 & Gaia DR3 5254518760118884864 & 158.775 & -59.635 & 0.239 $\pm$ 0.029 & -5.56 $\pm$ 0.02 & 2.47 $\pm$ 0.02 & 0.08 & 1.90 & 0.65 & N6 & CWNU 3684 \\
47 & J233736.95+602243.8 & 354.404 & 60.379 & 0.383 $\pm$ 0.032 & -3.27 $\pm$ 0.02 & -1.67 $\pm$ 0.02 & 1.99 & 1.24 & 1.30 & Y & CWNU 3513 \\
48 & CM Sct & 280.612 & -5.341 & 0.444 $\pm$ 0.031 & -1.06 $\pm$ 0.02 & -1.41 $\pm$ 0.02 & 0.01 & 0.52 & 9.48 & N123456 & UBC 106 \\
49 & V0379 Cas & 6.650 & 60.798 & 0.524 $\pm$ 0.030 & -2.70 $\pm$ 0.02 & -1.31 $\pm$ 0.02 & 1.13 & 1.38 & 45.28 & N1345 & NGC 129 \\
50 & CG Cas & 0.247 & 60.959 & 0.296 $\pm$ 0.030 & -3.24 $\pm$ 0.02 & -1.67 $\pm$ 0.02 & 1.24 & 3.77 & 5.49 & N1234 & Berkeley 58 \\
51 & J015057.43+620110.6 & 27.739 & 62.019 & 0.214 $\pm$ 0.030 & -0.78 $\pm$ 0.02 & -0.18 $\pm$ 0.02 & 0.62 & 1.64 & 0.73 & Y & CWNU 3246 \\
52 & V824 Cas & 5.630 & 63.033 & 0.296 $\pm$ 0.029 & -2.87 $\pm$ 0.02 & -0.59 $\pm$ 0.02 & 0.99 & 1.41 & 31.24 & N3 & UBC409 \\
53 & AE Vel & 144.214 & -53.033 & 0.369 $\pm$ 0.029 & -5.14 $\pm$ 0.02 & 4.07 $\pm$ 0.02 & 0.42 & 0.92 & 1.39 & Y & CWNU 1903 \\
54 & MN Cam & 59.374 & 54.938 & 0.366 $\pm$ 0.034 & -0.26 $\pm$ 0.03 & -0.65 $\pm$ 0.02 & 0.50 & 2.25 & 26.71 & N5 & CWNU 2490 \\
55 & J053.9542+57.1479 & 53.954 & 57.148 & 0.165 $\pm$ 0.034 & -0.19 $\pm$ 0.03 & 0.28 $\pm$ 0.03 & 0.91 & 0.34 & 7.32 & Y & CWNU 3310 \\
56 & J192152.00+150346.9 & 290.467 & 15.063 & 0.253 $\pm$ 0.032 & -1.93 $\pm$ 0.02 & -4.12 $\pm$ 0.02 & 0.41 & 0.63 & 2.53 & N6 & CWNU 4287 \\
57 & WX Pup & 115.496 & -25.876 & 0.387 $\pm$ 0.030 & -2.16 $\pm$ 0.02 & 2.56 $\pm$ 0.02 & 0.96 & 3.78 & 23.68 & N145 & UBC 231 \\
58 & SX Vel & 131.223 & -46.343 & 0.501 $\pm$ 0.032 & -4.34 $\pm$ 0.02 & 4.92 $\pm$ 0.03 & 0.89 & 4.36 & 12.62 & N4 & Cl SX Vel \\
59 & Y Sct & 279.514 & -8.369 & 0.558 $\pm$ 0.033 & -0.74 $\pm$ 0.03 & -2.88 $\pm$ 0.02 & 0.25 & 0.15 & 37.69 & N5 & CWNU 337 \\
60 & KQ Sco & 252.911 & -45.427 & 0.472 $\pm$ 0.034 & -1.37 $\pm$ 0.03 & -2.50 $\pm$ 0.03 & 1.19 & 0.15 & 6.66 & N356 & UBC 1558 \\
\hline
\multicolumn{12}{c}{Rejected samples}\\
\hline
61 & Gaia DR3 5878427527969505024 & 217.659 & -60.903 & 0.322 $\pm$ 0.042 & -5.59 $\pm$ 0.04 & -3.30 $\pm$ 0.04 & 0.13 & 1.52 & 1.02 & N6 & Pismis 19 \\
62 & OGLE-GD-CEP-0095 & 108.397 & -9.836 & 1.640 $\pm$ 0.079 & -5.17 $\pm$ 0.07 & 2.88 $\pm$ 0.07 & 1.97 & 1.13 & 47.89 & N1 & Alessi 21 \\
63 & OGLE-GD-CEP-0039 & 97.233 & 11.125 & 0.257 $\pm$ 0.031 & 0.30 $\pm$ 0.02 & -0.34 $\pm$ 0.02 & 0.72 & 1.41 & 5.39 & Y & CWNU 4764 \\
64 & Gaia DR3 5614801095283190016 & 116.343 & -24.004 & 0.231 $\pm$ 0.034 & -1.22 $\pm$ 0.02 & 1.97 $\pm$ 0.03 & 0.60 & 0.91 & 0.64 & Y & CWNU 3272 \\
65 & OGLE-GD-CEP-0328 & 137.470 & -48.866 & 0.418 $\pm$ 0.058 & -5.58 $\pm$ 0.05 & 3.77 $\pm$ 0.06 & 0.61 & 1.56 & 2.73 & Y & SAI 104 \\
66 & Gaia DR3 5865126083064564736 & 200.824 & -63.656 & 0.183 $\pm$ 0.034 & -6.34 $\pm$ 0.02 & -1.17 $\pm$ 0.03 & 0.23 & 2.86 & 2.28 & Y & CWNU 3267 \\
67 & OGLE-GD-CEP-0270 & 130.887 & -47.750 & 1.412 $\pm$ 0.044 & -4.77 $\pm$ 0.05 & 3.36 $\pm$ 0.04 & 0.64 & 1.11 & 26.68 & N2 & IC 2395 \\
68 & J055122+2516.9 & 87.844 & 25.281 & 0.362 $\pm$ 0.040 & 0.61 $\pm$ 0.04 & -1.63 $\pm$ 0.03 & 0.12 & 3.54 & 28.98 & N5 & OC 0301 \\
69 & CU Ori & 92.855 & 11.800 & 0.188 $\pm$ 0.031 & 0.20 $\pm$ 0.02 & -0.25 $\pm$ 0.02 & 1.16 & 4.11 & 4.00 & Y & CWNU 3265 \\
70 & OGLE-GD-CEP-0055 & 100.231 & 9.623 & 1.340 $\pm$ 0.041 & -2.33 $\pm$ 0.04 & -3.89 $\pm$ 0.04 & 1.85 & 0.60 & 3.09 & N & Mon OB1-D \\
71 & SU Cas & 42.995 & 68.888 & 2.192 $\pm$ 0.068 & 3.10 $\pm$ 0.04 & -8.14 $\pm$ 0.06 & 1.75 & 3.44 & 2.34 & N2 & Theia 123 \\
72 & OGLE-GD-CEP-1175 & 250.196 & -48.702 & 0.816 $\pm$ 0.068 & 0.94 $\pm$ 0.07 & -3.92 $\pm$ 0.05 & 0.77 & 1.93 & 6.27 & N23 & NGC 6193 \\
73 & OGLE-GD-CEP-1609 & 116.367 & -36.580 & 0.112 $\pm$ 0.030 & -1.03 $\pm$ 0.02 & 2.29 $\pm$ 0.02 & 0.52 & 1.91 & 4.38 & Y & CWNU 3603 \\
74 & Gaia DR3 5599913639125075328 & 116.092 & -28.422 & 0.282 $\pm$ 0.035 & -1.80 $\pm$ 0.02 & 1.73 $\pm$ 0.03 & 0.08 & 0.35 & 5.12 & Y & Haffner 14 \\
75 & OGLE-GD-CEP-0270 & 156.746 & -57.676 & 0.384 $\pm$ 0.031 & -7.26 $\pm$ 0.02 & 3.68 $\pm$ 0.02 & 0.79 & 0.89 & 4.74 & Y & IC 2581 \\
76 & J093652.9-540325 & 144.221 & -54.057 & 0.173 $\pm$ 0.037 & -4.03 $\pm$ 0.03 & 3.44 $\pm$ 0.03 & 0.06 & 2.23 & 5.52 & Y & FSR 1500 \\
77 & J297.7863+25.3136 & 297.786 & 25.314 & 0.340 $\pm$ 0.032 & -3.02 $\pm$ 0.02 & -6.29 $\pm$ 0.02 & 2.08 & 1.55 & 2.29 & N236 & Czernik 41 \\
78 & OGLE-GD-CEP-0329 & 137.497 & -48.670 & 1.423 $\pm$ 0.048 & -7.52 $\pm$ 0.04 & 3.68 $\pm$ 0.04 & 0.54 & 0.57 & 3.84 & Y & HSC 2193 \\
79 & J233736.95+602243.8 & 354.404 & 60.379 & 0.383 $\pm$ 0.032 & -3.27 $\pm$ 0.02 & -1.67 $\pm$ 0.02 & 2.38 & 7.39 & 11.36 & N23 & SAI 149 \\
80 & J075840-3330.2 & 119.679 & -33.590 & 0.292 $\pm$ 0.029 & -2.27 $\pm$ 0.02 & 3.30 $\pm$ 0.02 & 0.63 & 1.10 & 2.30 & N236 & UBC 1424 \\
81 & ST Tau & 86.263 & 13.576 & 0.916 $\pm$ 0.043 & 0.19 $\pm$ 0.04 & -2.32 $\pm$ 0.03 & 1.65 & 8.04 & 25.49 & N4 & Cl ST Tau \\
82 & V554 Cyg & 301.838 & 40.178 & 0.091 $\pm$ 0.029 & -1.97 $\pm$ 0.02 & -3.24 $\pm$ 0.02 & 0.73 & 3.89 & 1.47 & Y & CWNU 2959 \\
83 & V733 Cyg & 323.514 & 53.310 & 0.238 $\pm$ 0.029 & -2.71 $\pm$ 0.02 & -3.34 $\pm$ 0.02 & 0.76 & 6.93 & 20.03 & N123 & Kronberger 84 \\
84 & DY Car & 163.137 & -60.527 & 0.220 $\pm$ 0.029 & -7.14 $\pm$ 0.02 & 3.12 $\pm$ 0.02 & 0.80 & 4.75 & 16.18 & N23 & UBC 1490 \\
85 & OGLE-GD-CEP-1216 & 280.248 & -5.767 & 0.201 $\pm$ 0.048 & -2.80 $\pm$ 0.05 & -5.71 $\pm$ 0.04 & 0.10 & 3.22 & 1.06 & Y & CWNU 3968 \\\hline
86 & V Lac & 342.158 & 56.322 & 0.496 $\pm$ 0.031 & -3.24 $\pm$ 0.02 & -1.44 $\pm$ 0.02 & 0.90 & 3.60 & 28.03 & N14 & FSR 0384 \\
87 & V824 Cas & 5.630 & 63.033 & 0.296 $\pm$ 0.029 & -2.87 $\pm$ 0.02 & -0.59 $\pm$ 0.02 & 1.48 & 3.04 & 19.64 & N5 & SAI 4 \\
88 & X Lac & 342.263 & 56.428 & 0.520 $\pm$ 0.032 & -3.30 $\pm$ 0.02 & -1.44 $\pm$ 0.02 & 0.13 & 3.28 & 28.34 & N14 & FSR 0384 \\
89 & DP Vel & 142.566 & -53.059 & 0.335 $\pm$ 0.029 & -4.31 $\pm$ 0.02 & 3.35 $\pm$ 0.02 & 3.13 & 0.88 & 12.01 & N1236 & UBC491 \\
90 & AU Cas & 13.804 & 64.261 & 0.206 $\pm$ 0.029 & -1.61 $\pm$ 0.02 & 0.03 $\pm$ 0.02 & 1.48 & 5.23 & 33.06 & N23 & UBC 412 \\
91 & V0378 Cen & 199.864 & -62.541 & 0.536 $\pm$ 0.080 & -3.54 $\pm$ 0.07 & -2.12 $\pm$ 0.08 & 0.08 & 9.21 & 9.87 & N4 & Cl V0378 Cen \\
92 & SV Cru & 179.474 & -62.646 & 0.238 $\pm$ 0.030 & -6.60 $\pm$ 0.02 & 1.54 $\pm$ 0.02 & 1.89 & 14.93 & 4.66 & N16 & Ruprecht 97 \\
93 & GM Cas & 36.736 & 60.571 & 0.417 $\pm$ 0.032 & -0.90 $\pm$ 0.02 & -0.38 $\pm$ 0.02 & 1.89 & 1.73 & 39.44 & N5 & COIN-Gaia 36 \\
94 & V2340 Cyg & 322.187 & 48.978 & 0.300 $\pm$ 0.030 & -3.67 $\pm$ 0.02 & -4.10 $\pm$ 0.02 & 1.62 & 2.20 & 9.04 & Y & Platais 1 \\
95 & GH Lup & 231.160 & -52.854 & 0.864 $\pm$ 0.034 & -1.34 $\pm$ 0.03 & -2.20 $\pm$ 0.03 & 0.20 & 5.37 & 35.21 & N4 & UBC 533 \\
96 & TY Sct & 280.533 & -4.293 & 0.371 $\pm$ 0.031 & -1.11 $\pm$ 0.02 & -2.47 $\pm$ 0.02 & 0.01 & 3.06 & 12.99 & N56 & Trumpler 35 \\
97 & V438 Cyg & 304.726 & 40.064 & 0.530 $\pm$ 0.031 & -3.32 $\pm$ 0.02 & -4.56 $\pm$ 0.02 & 2.22 & 5.27 & 2.49 & N14 & UBC 375 \\
98 & Z Sct & 280.739 & -5.821 & 0.357 $\pm$ 0.032 & -0.38 $\pm$ 0.03 & -2.21 $\pm$ 0.02 & 0.91 & 1.82 & 26.87 & N5 & NGC 6683 \\
99 & OGLE-GD-CEP-1231 & 281.094 & -2.113 & 0.297 $\pm$ 0.060 & -0.68 $\pm$ 0.06 & -2.50 $\pm$ 0.05 & 1.04 & 1.47 & 1.65 & Y & CWNU 3845 \\
100 & VY Sgr & 273.019 & -20.704 & 0.412 $\pm$ 0.036 & 0.31 $\pm$ 0.03 & -1.55 $\pm$ 0.03 & 0.13 & 2.48 & 11.18 & N5 & CWNU 1841 \\
101 & SZ Cas & 36.807 & 59.461 & 0.407 $\pm$ 0.032 & 0.27 $\pm$ 0.02 & -0.65 $\pm$ 0.02 & 0.71 & 4.35 & 28.28 & N23 & UBC 190 \\
102 & V1788 Cyg & 310.655 & 38.457 & 0.376 $\pm$ 0.034 & -2.71 $\pm$ 0.03 & -4.45 $\pm$ 0.03 & 0.30 & 1.73 & 6.35 & N23 & OC 0125 \\
103 & J280.6497-05.2660 & 280.650 & -5.266 & 0.117 $\pm$ 0.060 & -1.58 $\pm$ 0.06 & -3.21 $\pm$ 0.05 & 2.47 & 4.05 & 2.31 & N3 & Teutsch 145 \\
104 & OGLE-GD-CEP-1012 & 223.597 & -60.458 & 0.176 $\pm$ 0.038 & -5.23 $\pm$ 0.03 & -3.53 $\pm$ 0.03 & 5.47 & 4.31 & 7.02 & N3 & Teutsch 80 \\
105 & XZ Car & 166.056 & -60.980 & 0.473 $\pm$ 0.032 & -7.28 $\pm$ 0.02 & 2.62 $\pm$ 0.02 & 0.35 & 9.28 & 23.89 & N4 & Ruprecht 93 \\
106 & J300.0102+29.1869 & 300.010 & 29.187 & 0.360 $\pm$ 0.045 & -2.50 $\pm$ 0.04 & -6.17 $\pm$ 0.04 & 0.70 & 1.82 & 1.98 & N26 & FSR 0172 \\
107 & RY Sco & 267.718 & -33.706 & 0.764 $\pm$ 0.042 & 1.49 $\pm$ 0.03 & -1.39 $\pm$ 0.02 & 1.55 & 3.48 & 49.86 & N5 & HSC 2961 \\
108 & WZ Car & 163.828 & -60.940 & 0.284 $\pm$ 0.032 & -5.89 $\pm$ 0.03 & 2.25 $\pm$ 0.03 & 2.03 & 3.58 & 25.22 & N3 & UBC1496 \\
109 & 2MASS J17184740-3817292 & 259.698 & -38.291 & 0.230 $\pm$ 0.045 & -1.94 $\pm$ 0.04 & -3.42 $\pm$ 0.03 & 5.18 & 25.79 & 8.31 & N123 & BH 222 \\
110 & Gaia DR3 4041579687419836672 & 267.721 & -34.586 & 0.104 $\pm$ 0.034 & -0.06 $\pm$ 0.03 & -5.87 $\pm$ 0.02 & 0.25 & 5.78 & 0.88 & Y & NGC 6453 \\
\hline
\end{longtable}
\tablecomments{Positions, parallaxes, and proper motions are listed for each OC Cepheid. The astrometric dispersions ($\delta_\varpi$, $\delta_{pm}$) denote the deviations of Cepheids from their host OCs in parallax and proper motion, respectively. $R$ denotes angular distance of CC from cluster center. Y: newly identified OC Cepheids; N: previously identified OC Cepheids. References—(1) \citet{zc2021}; (2) \citet{Hao2022}; (3) \citet{lin2022}; (4) \citet{mauricio2023}; (5) \citet{wang2024}; (6) \citet{Majaess2024}.}

\newpage
\setlength{\tabcolsep}{2.5pt} 
\centering
\begin{longtable}{ccccclllcc}
\caption{Pulsation mode, period, Gaia-band magnitudes, and metallicity of CCs.}\\
\hline\hline
Index &Cepheid & Gaia ID& Mode& P (days) & $G$ (mag)	&$G_{BP}$ (mag)	& $G_{RP}$ (mag)&	[M/H]&	Reference (CCs)\\
\endfirsthead
\caption{Pulsation mode, period, Gaia-band magnitudes, and metallicity of CCs. (continued)} \\ 
\hline\hline
Index &Cepheid & Gaia ID& Mode& P (days) & $G$ (mag)	&$G_{BP}$ (mag)	& $G_{RP}$ (mag)&	[M/H]&	Reference (CCs)\\
\hline
\endhead
\hline
\multicolumn{10}{c}{OC Cepheid Candidates}\\
\hline
42 & J065.6485+45.5819 & 232773865804463232 & 1O & 2.78 & 12.27 & 13.47 & 11.20 &   & abcde \\
43 & OGLE-GD-CEP-0265 & 5329323277776517120 & F & 3.18 & 13.99 & 14.81 & 13.09 &   & acde \\
44 & V423 CMa & 5605736961940589568 & F & 3.34 & 12.85 & 13.31 & 12.22 &   & ace \\
45 & OGLE-GD-CEP-0519 & 5351749294895015168 & 1O & 3.72 & 11.87 & 13.00 & 10.82 &   & cde \\
46 & Gaia DR3 5254518760118884864$^{\mathrm{*}}$ & 5254518760118884864 & 1O & 3.77 & 11.63 & 12.50 & 10.70 &   & e \\
47 & J233736.95+602243.8 & 2012181324799904512 & F & 3.90 & 12.92 & 14.38 & 11.74 &   & abe \\
48 & CM Sct & 4253603501158148736 & F & 3.92 & 10.52 & 11.39 & 9.59 & 0.12 & acdef \\
49 & V0379 Cas & 428839329030983040 & 1O & 4.31 & 8.69 & 9.35 & 7.89 & 0.12 & cdef \\
50 & CG Cas & 429385923752386944 & F & 4.37 & 10.84 & 11.61 & 9.97 & 0.06 & acf \\
51 & J015057.43+620110.6 & 511362109099936384 & 1O & 4.69 & 12.54 & 13.70 & 11.48 &   & de \\
52 & V824 Cas & 430629986799994880 & 1O & 5.35 & 10.59 & 11.45 & 9.67 &   & cde \\
53 & AE Vel & 5309174967720762496 & F & 7.13 & 9.69 & 10.49 & 8.82 & 0.11 & acdef \\
54 & MN Cam & 444443941646255232 & F & 8.17 & 11.13 & 12.39 & 10.04 & -0.02 & acdef \\
55 & J053.9542+57.1479 & 448977846560597248 & F & 8.58 & 14.10 & 15.93 & 12.82 &   & abcde \\
56 & J192152.00+150346.9 & 4320029877589560448 & F & 8.78 & 13.35 & 15.34 & 12.03 &   & be \\
57 & WX Pup & 5613972681993587200 & F & 8.94 & 8.73 & 9.28 & 8.03 & -0.01 & cdef \\
58 & SX Vel & 5329838158460391296 & F & 9.55 & 8.00 & 8.50 & 7.33 & -0.18 & cdef \\
59 & Y Sct & 4156450099578283776 & F & 10.34 & 8.87 & 9.82 & 7.87 & 0.2 & acdef \\
60 & KQ Sco & 5964193485048327808 & F & 28.69 & 8.85 & 10.08 & 7.76 & 0.52 & acdef \\
\hline
\multicolumn{10}{c}{Rejected samples}\\
\hline
61 & Gaia DR3 5878427527969505024 & 5878427527969505024 & 1O & 0.29 & 15.68 & 17.02 & 14.51 &   & e \\
62 & OGLE-GD-CEP-0095$^{\mathrm{*}}$ & 3047933166636348800 & 1O & 0.33 & 17.01 & 18.04 & 16.02 &   & c \\
63 & OGLE-GD-CEP-0039 & 3330987346981211392 & 1O2O & 1.10 & 12.57 & 13.21 & 11.77 &   & bce \\
64 & Gaia DR3 5614801095283190016$^{\mathrm{*}}$ & 5614801095283190016 & F & 1.12 & 11.65 & 11.82 & 11.32 &   & ab \\
65 & OGLE-GD-CEP-0328$^{\mathrm{*}}$ & 5326773441577337472 & 1O & 1.18 & 16.72 & 17.99 & 15.58 &   & ce \\
66 & Gaia DR3 5865126083064564736 & 5865126083064564736 & 1O & 1.60 & 13.77 & 14.69 & 12.80 &   & e \\
67 & OGLE-GD-CEP-0270$^{\mathrm{*}}$ & 5329151646581429248 & 1O & 1.62 & 16.16 & 17.07 & 15.11 &   & ce \\
68 & J055122+2516.9 & 3428975209070530176 & 1O & 1.72 & 11.19 & 11.86 & 10.38 &   & ade \\
69 & CU Ori & 3330404094719897984 & 1O & 1.86 & 13.10 & 13.79 & 12.27 &   & acde \\
70 & OGLE-GD-CEP-0055$^{\mathrm{*}}$ & 3326697194343548416 & 1O & 1.90 & 15.05 & 15.95 & 14.09 &   & bc \\
71 & SU Cas & 541716257882226560 & 1O & 1.95 & 5.74 & 6.14 & 5.14 & 0.12 & acef \\
72 & OGLE-GD-CEP-1175$^{\mathrm{*}}$ & 5940956994758917760 & 1O & 1.98 & 16.72 & 17.83 & 15.67 &   & ce \\
73 & OGLE-GD-CEP-1609 & 5587355360904762624 & F1O & 2.56 & 14.08 & 14.78 & 13.27 &   & ade \\
74 & Gaia DR3 5599913639125075328$^{\mathrm{*}}$ & 5599913639125075328 & F & 2.68 & 15.03 & 15.32 & 14.57 &   & abc \\
75 & OGLE-GD-CEP-0270$^{\mathrm{*}}$ & 5351710915065605120 & F & 2.78 & 8.65 & 8.79 & 8.32 &   & abcc \\
76 & J093652.9-540325 & 5309048249001936000 & F1O & 2.93 & 14.95 & 16.39 & 13.77 &   & ace \\
77 & J297.7863+25.3136 & 2026716726778925568 & 1O & 2.94 & 12.10 & 13.82 & 10.85 &   & abcde \\
78 & OGLE-GD-CEP-0329$^{\mathrm{*}}$ & 5326780347884791936 & 1O & 3.30 & 16.20 & 17.50 & 14.96 &   & c \\
79 & J233736.95+602243.8$^{\mathrm{*}}$ & 2012181324799904512 & F & 3.90 & 12.92 & 14.41 & 11.74 &   & e \\
80 & J075840-3330.2$^{\mathrm{*}}$ & 5594087803735560704 & F & 4.03 & 11.87 & 12.18 & 11.34 &   & ab \\
81 & ST Tau & 3346584610911570176 & F & 4.03 & 7.91 & 8.43 & 7.22 & -0.14 & acdef \\
82 & V554 Cyg & 2074102933481447808 & F & 4.33 & 13.76 & 14.45 & 12.89 &   & acde \\
83 & V733 Cyg & 2172686554763264768 & F & 4.56 & 11.96 & 13.00 & 10.95 &   & abcde \\
84 & DY Car & 5338207327417387520 & F & 4.67 & 10.94 & 11.54 & 10.20 & 0.04 & acef \\
85 & OGLE-GD-CEP-1216 & 4253472311292427520 & F1O & 4.81 & 14.01 & 15.92 & 12.65 &   & bcde \\\hline\\
86 & V Lac & 2003938801532007808 & F & 4.98 & 8.61 & 9.14 & 7.93 & 0.04 & acdef \\
87 & V824 Cas$^{\mathrm{*}}$ & 430629986799994880 & 1O & 5.35 & 10.64 & 11.56 & 9.71 & -0.08 & f \\
88 & X Lac & 2004036486267748352 & 1O & 5.44 & 8.10 & 8.62 & 7.40 & 0.05 & cdef \\
89 & DP Vel & 5312211543964567040 & F & 5.48 & 11.18 & 12.08 & 10.24 &   & acde \\
90 & AU Cas & 524192787020949376 & F & 5.62 & 12.19 & 13.33 & 11.13 &   & abcd \\
91 & V0378 Cen$^{\mathrm{*}}$ & 5868286904025306624 & F & 6.46 & 17.48 & 18.19 & 16.63 &   & c \\
92 & SV Cru & 5334563996911649408 & F & 7.00 & 11.43 & 12.35 & 10.48 &   & ace \\
93 & GM Cas & 465365517618881536 & F & 7.47 & 10.84 & 12.11 & 9.73 & -0.13 & abcdef \\
94 & V2340 Cyg & 2170785842762144512 & F & 7.97 & 10.10 & 10.88 & 9.24 & 0.28 & acde \\
95 & GH Lup & 5888819699572539264 & F & 9.29 & 7.22 & 7.88 & 6.43 & 0.13 & aef \\
96 & TY Sct & 4256744187345732224 & F & 11.05 & 9.94 & 11.08 & 8.90 & 0.34 & acdef \\
97 & V438 Cyg & 2062267687221984128 & F & 11.21 & 9.85 & 11.18 & 8.73 & 0.3 & acdef \\
98 & Z Sct & 4253435825622331392 & F & 12.90 & 9.08 & 9.87 & 8.21 & 0.12 & acdef \\
99 & OGLE-GD-CEP-1231 & 4259861783820291072 & F & 12.93 & 12.45 & 15.22 & 11.03 &   & bce \\
100 & VY Sgr & 4094007532908816896 & F & 13.56 & 10.35 & 11.71 & 9.20 & 0.33 & abcdef \\
101 & SZ Cas & 459263743483634304 & F & 13.64 & 9.14 & 10.09 & 8.17 &   & acef \\
102 & V1788 Cyg & 2063349022550766720 & F & 14.09 & 11.19 & 12.88 & 9.98 & 0.37 & bcef \\
103 & J280.6497-05.2660 & 4256607740489330304 & F & 14.36 & 12.83 & 15.36 & 11.42 &   & abce \\
104 & OGLE-GD-CEP-1012 & 5877982466278766208 & F & 15.97 & 12.09 & 13.97 & 10.80 &   & e \\
105 & XZ Car & 5338036117182452096 & F & 16.65 & 8.15 & 8.85 & 7.35 & 0.16 & aef \\
106 & J300.0102+29.1869 & 2030036736506607616 & F & 18.42 & 12.40 & 15.10 & 11.00 &   & abce \\
107 & RY Sco & 4041690364529590144 & F & 20.32 & 7.31 & 8.28 & 6.36 & 0.01 & acef \\
108 & WZ Car & 5338182833266967680 & F & 23.02 & 8.83 & 9.49 & 8.03 & 0.06 & acdef \\
109 & 2MASS J17184740-3817292 & 5972735423411223552 & F & 23.30 & 13.01 & 15.35 & 11.63 &   & bce \\
110 & Gaia DR3 4041579687419836672$^{\mathrm{*}}$ & 4041579687419836672 & F & 27.18 & 13.57 & 14.34 & 12.62 &   & a \\
\hline
\end{longtable}
\tablecomments{Metallicity values are taken from the CC catalog of \citet{Ripepi2022}. For CCs' photometry, the intensity-averaged magnitudes ($G$, ${G_{\rm RP}}$, and $G_{\rm BP}$) were extracted from \texttt{gaiadr3.vari\_cepheid} \citep{Ripepi2023}. In other cases, indicated with an asterisk next to the Cepheid, Gaia DR3 photometry was adopted. References—a: \citet{skowron2019}; b: \citet{chen2019}; c: \citet{chen2020}; d: \citet{inno2021}; e: \citet{pietrukowicz2021}; f: \citet{Ripepi2022}.}

\newpage
\centering
\setlength{\tabcolsep}{12pt} 
\begin{longtable}{cccccccccccc}
\caption{Recalibrated parallax, extinction, distance modulus, and age of OCs from isochrone fitting.}\\
\hline\hline
Index & Cluster &$\varpi$ (mas) &$A_v$ (mag)& $\mu$ (mag)& log ${Age}$&Referrence (OCs)\\
\hline
\endfirsthead
\caption{Recalibrated parallax, extinction, distance modulus, and age of OCs from isochrone fitting. (Continued)} \\ 
\hline\hline
Index & Cluster &$\varpi$ (mas) &$A_v$ (mag)& $\mu$ (mag)& log ${Age}$&Referrence (OCs)\\
\hline
\endhead
\multicolumn{12}{c}{OC Cepheid Candidates}\\
\hline
42 & CWNU 2580 & 0.416 $\pm$ 0.008 & 2.30 & 11.91 & 8.90 $\pm$ 0.04 & C \\
43 & CWNU 3981 & 0.113 $\pm$ 0.015 & 2.60 & 14.73 & 8.20 $\pm$ 0.05 & A \\
44 & CWNU 3305 & 0.139 $\pm$ 0.010 & 0.05 & 14.29 & 9.15 $\pm$ 0.05 & A \\
45 & CWNU 3193 & 0.267 $\pm$ 0.010 & 3.46 & 12.87 & 7.96 $\pm$ 0.04 & A \\
46 & CWNU 3684 & 0.242 $\pm$ 0.010 & 2.50 & 13.08 & 8.10 $\pm$ 0.05 & A \\
47 & CWNU 3513 & 0.312 $\pm$ 0.015 & 4.90 & 12.53 & 8.07 $\pm$ 0.03 & A \\
48 & UBC 106 & 0.444 $\pm$ 0.002 & 2.40 & 11.76 & 7.95 $\pm$ 0.10 & B \\
49 & NGC 129 & 0.558 $\pm$ 0.002 & 1.55 & 11.27 & 7.95 $\pm$ 0.05 & B \\
50 & Berkeley 58 & 0.333 $\pm$ 0.003 & 2.10 & 12.39 & 8.15 $\pm$ 0.05 & B \\
51 & CWNU 3246 & 0.194 $\pm$ 0.012 & 3.48 & 13.56 & 7.99 $\pm$ 0.03 & A \\
52 & UBC409 & 0.267 $\pm$ 0.004 & 1.80 & 12.87 & 7.90 $\pm$ 0.05 & D \\
53 & CWNU 1903 & 0.382 $\pm$ 0.009 & 2.45 & 12.09 & 7.86 $\pm$ 0.04 & C \\
54 & CWNU 2490 & 0.348 $\pm$ 0.012 & 2.60 & 12.29 & 8.10 $\pm$ 0.05 & C \\
55 & CWNU 3310 & 0.197 $\pm$ 0.012 & 5.40 & 13.52 & 8.05 $\pm$ 0.05 & A \\
56 & CWNU 4287 & 0.238 $\pm$ 0.017 & 6.58 & 13.12 & 7.75 $\pm$ 0.05 & A \\
57 & UBC 231 & 0.357 $\pm$ 0.005 & 1.10 & 12.24 & 7.81 $\pm$ 0.06 & B \\
58 & Cl SX Vel & 0.531 $\pm$ 0.010 & 1.45 & 11.38 & 7.84 $\pm$ 0.04 & E \\
59 & CWNU 337 & 0.567 $\pm$ 0.009 & 3.80 & 11.23 & 7.60 $\pm$ 0.05 & B \\
60 & UBC 1558 & 0.431 $\pm$ 0.006 & 2.85 & 11.83 & 7.57 $\pm$ 0.04 & B \\
\hline
\multicolumn{12}{c}{Rejected samples}\\
\hline
61 & Pismis 19 & 0.328 $\pm$ 0.004 & 4.65 & 12.42 & 8.60 $\pm$ 0.10 & B \\
62 & Alessi 21 & 1.796 $\pm$ 0.004 & 0.30 & 8.73 & 7.70 $\pm$ 0.15 & B \\
63 & CWNU 4764 & 0.233 $\pm$ 0.014 & 1.80 & 13.16 & 8.75 $\pm$ 0.10 & A \\
64 & CWNU 3272 & 0.209 $\pm$ 0.011 & 1.55 & 13.40 & 7.60 $\pm$ 0.10 & A \\
65 & SAI 104 & 0.382 $\pm$ 0.011 & 6.00 & 12.09 & 7.35 $\pm$ 0.10 & B \\
66 & CWNU 3267 & 0.175 $\pm$ 0.006 & 2.70 & 13.79 & 8.85 $\pm$ 0.05 & A \\
67 & IC 2395 & 1.440 $\pm$ 0.003 & 0.40 & 9.21 & 7.00 $\pm$ 0.10 & B \\
68 & OC 0301 & 0.367 $\pm$ 0.009 & 1.81 & 12.18 & 8.80 $\pm$ 0.05 & B \\
69 & CWNU 3265 & 0.227 $\pm$ 0.011 & 2.05 & 13.22 & 8.30 $\pm$ 0.10 & A \\
70 & Mon OB1-D & 1.417 $\pm$ 0.006 & 1.50 & 9.24 & 7.30 $\pm$ 0.10 & B \\
71 & Theia 123 & 2.311 $\pm$ 0.005 & 0.80 & 8.18 & 8.48 $\pm$ 0.07 & B \\
72 & NGC 6193 & 0.869 $\pm$ 0.003 & 1.50 & 10.31 & 7.00 $\pm$ 0.10 & B \\
73 & CWNU 3603 & 0.133 $\pm$ 0.025 & 0.95 & 14.38 & 8.95 $\pm$ 0.05 & A \\
74 & Haffner 14 & 0.280 $\pm$ 0.003 & 1.95 & 12.77 & 8.30 $\pm$ 0.10 & B \\
75 & IC 2581 & 0.409 $\pm$ 0.003 & 1.20 & 11.94 & 7.50 $\pm$ 0.10 & B \\
76 & FSR 1500 & 0.175 $\pm$ 0.007 & 4.50 & 13.78 & 7.50 $\pm$ 0.05 & B \\
77 & Czernik 41 & 0.407 $\pm$ 0.003 & 4.55 & 11.95 & 7.15 $\pm$ 0.05 & B \\
78 & HSC 2193 & 1.450 $\pm$ 0.013 & 4.60 & 9.19 & 7.00 $\pm$ 0.10 & B \\
79 & SAI 149 & 0.304 $\pm$ 0.009 & 4.00 & 12.59 & 8.15 $\pm$ 0.10 & B \\
80 & UBC 1424 & 0.274 $\pm$ 0.007 & 1.60 & 12.81 & 8.21 $\pm$ 0.06 & B \\
81 & Cl ST Tau & 0.989 $\pm$ 0.008 & 1.00 & 10.03 & 8.20 $\pm$ 0.03 & E \\
82 & CWNU 2959 & 0.114 $\pm$ 0.012 & 1.80 & 14.72 & 9.10 $\pm$ 0.05 & A \\
83 & Kronberger 84 & 0.215 $\pm$ 0.006 & 2.59 & 13.34 & 8.00 $\pm$ 0.05 & B \\
84 & UBC 1490 & 0.244 $\pm$ 0.005 & 1.20 & 13.06 & 8.25 $\pm$ 0.10 & B \\
85 & CWNU 3968 & 0.206 $\pm$ 0.012 & 5.65 & 13.43 & 8.25 $\pm$ 0.05 & A \\\hline
86 & FSR 0384 & 0.524 $\pm$ 0.006 & 1.15 & 11.40 & 8.20 $\pm$ 0.10 & B \\
87 & SAI 4 & 0.340 $\pm$ 0.006 & 1.95 & 12.34 & 7.80 $\pm$ 0.05 & B \\
88 & FSR 0384 & 0.524 $\pm$ 0.006 & 1.15 & 11.40 & 8.20 $\pm$ 0.10 & B \\
89 & UBC491 & 0.242 $\pm$ 0.004 & 2.20 & 13.08 & 8.30 $\pm$ 0.10 & D \\
90 & UBC 412 & 0.252 $\pm$ 0.008 & 2.48 & 13.00 & 8.15 $\pm$ 0.05 & B \\
91 & Cl V0378 Cen & 0.542 $\pm$ 0.011 & 1.45 & 11.33 & 8.50 $\pm$ 0.05 & E \\
92 & Ruprecht 97 & 0.295 $\pm$ 0.005 & 1.56 & 12.65 & 8.15 $\pm$ 0.10 & B \\
93 & COIN-Gaia 36 & 0.478 $\pm$ 0.004 & 2.55 & 11.60 & 7.65 $\pm$ 0.05 & B \\
94 & Platais 1 & 0.350 $\pm$ 0.007 & 1.40 & 12.28 & 8.25 $\pm$ 0.10 & A \\
95 & UBC 533 & 0.857 $\pm$ 0.005 & 0.60 & 10.33 & 8.41 $\pm$ 0.06 & B \\
96 & Trumpler 35 & 0.371 $\pm$ 0.005 & 3.50 & 12.15 & 7.80 $\pm$ 0.05 & B \\
97 & UBC 375 & 0.599 $\pm$ 0.005 & 3.35 & 11.11 & 7.85 $\pm$ 0.05 & B \\
98 & NGC 6683 & 0.327 $\pm$ 0.005 & 2.10 & 12.43 & 7.50 $\pm$ 0.05 & B \\
99 & CWNU 3845 & 0.360 $\pm$ 0.008 & 3.50 & 12.22 & 8.30 $\pm$ 0.10 & A \\
100 & CWNU 1841 & 0.407 $\pm$ 0.009 & 3.80 & 11.95 & 7.25 $\pm$ 0.05 & C \\
101 & UBC 190 & 0.384 $\pm$ 0.006 & 3.09 & 12.08 & 7.10 $\pm$ 0.10 & B \\
102 & OC 0125 & 0.365 $\pm$ 0.006 & 2.45 & 12.19 & 8.10 $\pm$ 0.05 & B \\
103 & Teutsch 145 & 0.268 $\pm$ 0.008 & 5.22 & 12.86 & 8.55 $\pm$ 0.10 & B \\
104 & Teutsch 80 & 0.383 $\pm$ 0.005 & 5.05 & 12.08 & 8.05 $\pm$ 0.10 & B \\
105 & Ruprecht 93 & 0.484 $\pm$ 0.003 & 0.52 & 11.57 & 8.65 $\pm$ 0.05 & B \\
106 & FSR 0172 & 0.328 $\pm$ 0.005 & 4.80 & 12.42 & 8.00 $\pm$ 0.10 & B \\
107 & HSC 2961 & 0.698 $\pm$ 0.007 & 2.95 & 10.78 & 7.30 $\pm$ 0.05 & B \\
108 & UBC1496 & 0.217 $\pm$ 0.009 & 1.25 & 13.32 & 7.90 $\pm$ 0.05 & D \\
109 & BH 222 & 0.466 $\pm$ 0.006 & 3.75 & 11.66 & 7.20 $\pm$ 0.10 & B \\
110 & NGC 6453 & 0.095 $\pm$ 0.006 & 4.70 & 15.11 & 7.45 $\pm$ 0.05 & B \\
\hline  
\end{longtable} 
\tablecomments{References—A: \citet{he2023}; B: \citet{huntemily2024}; C: \citet{he2022}; D:~\citet{CG2020}; E:~\citet{mauricio2023}.}


\section{CMDs of remaining samples}
\label{appendixc}   
Analogous to Figure~\ref{fig2:newOCCs}, this figure presents the CMDs of the remaining 106 OCs, including 37 OC Cepheids, 19 OC Cepheid candidates, and 50 rejected samples.
\begin{figure*}[htbp]
\centering
\includegraphics[width=0.23\linewidth]{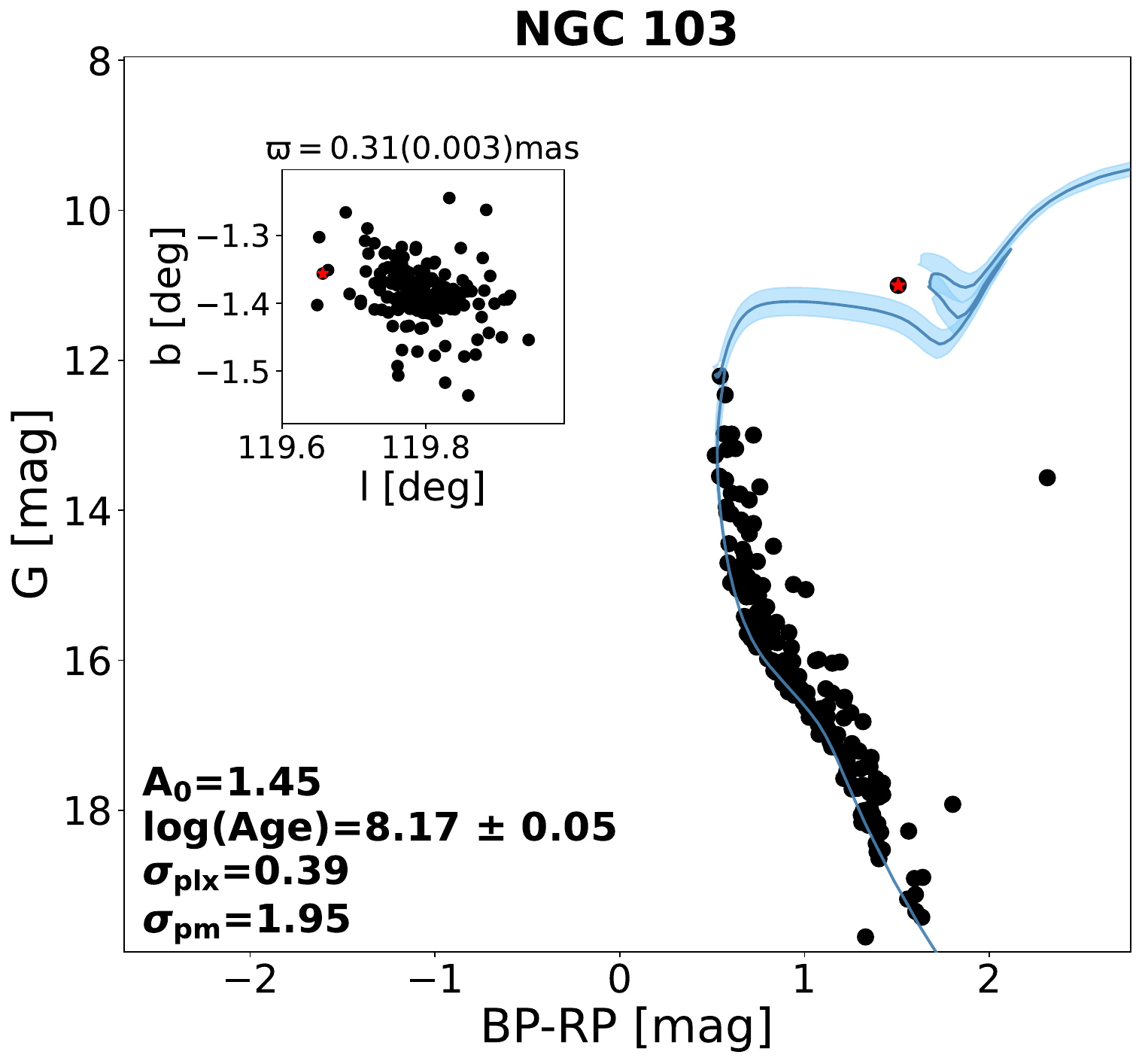}\hspace{0.02\linewidth}
\includegraphics[width=0.23\linewidth]{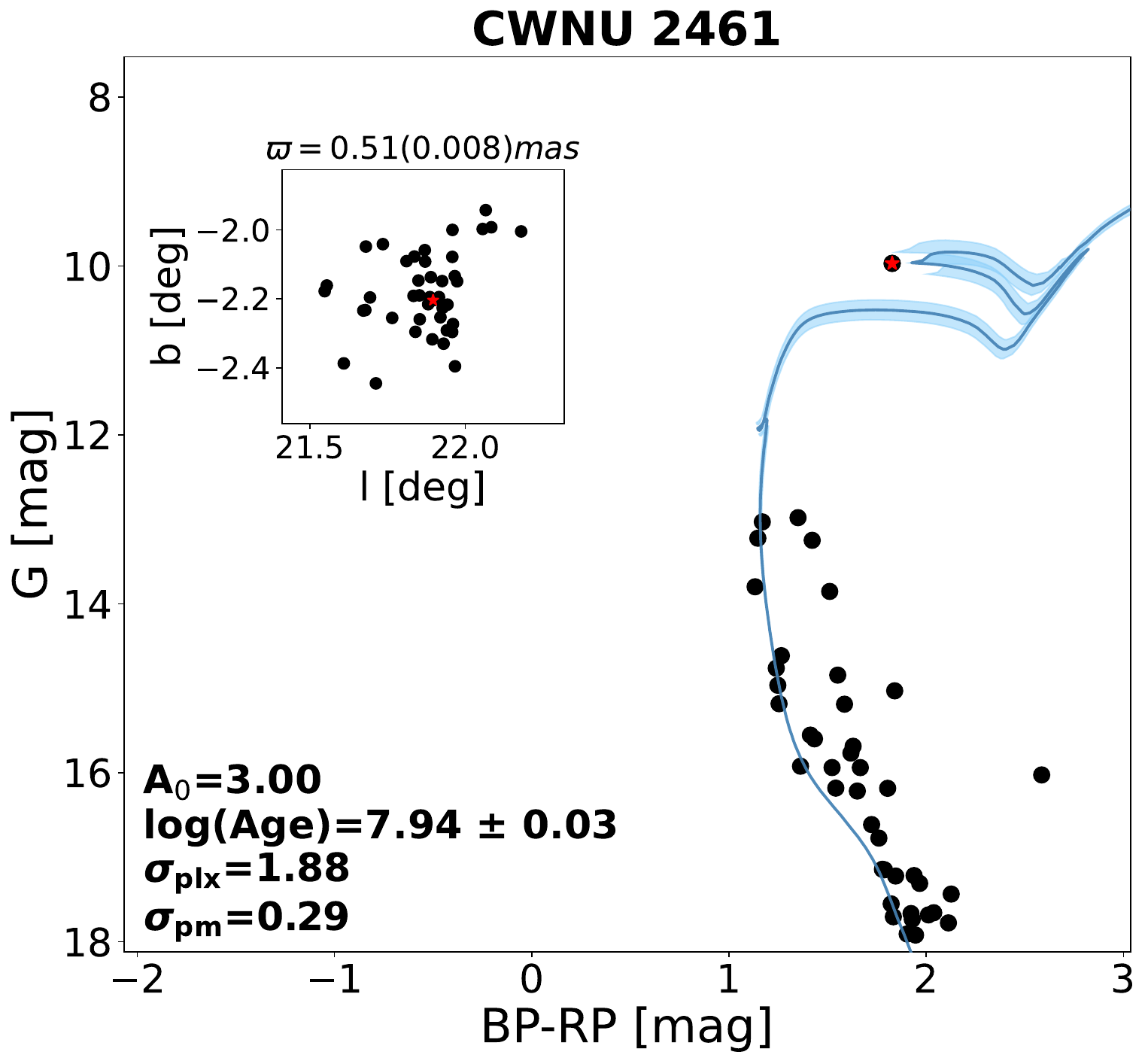}\hspace{0.02\linewidth}
\includegraphics[width=0.23\linewidth]{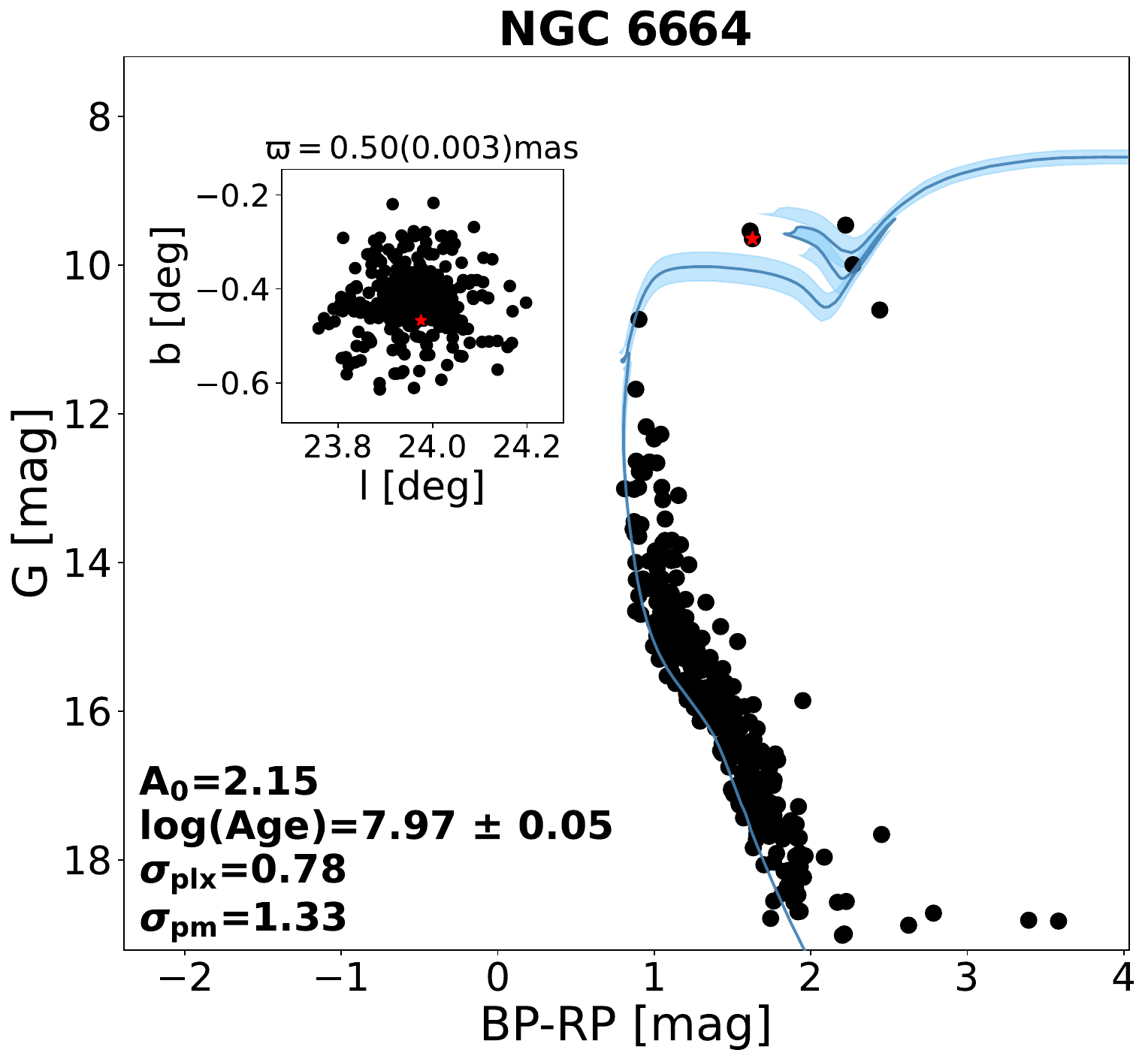}\hspace{0.02\linewidth}
\includegraphics[width=0.23\linewidth]{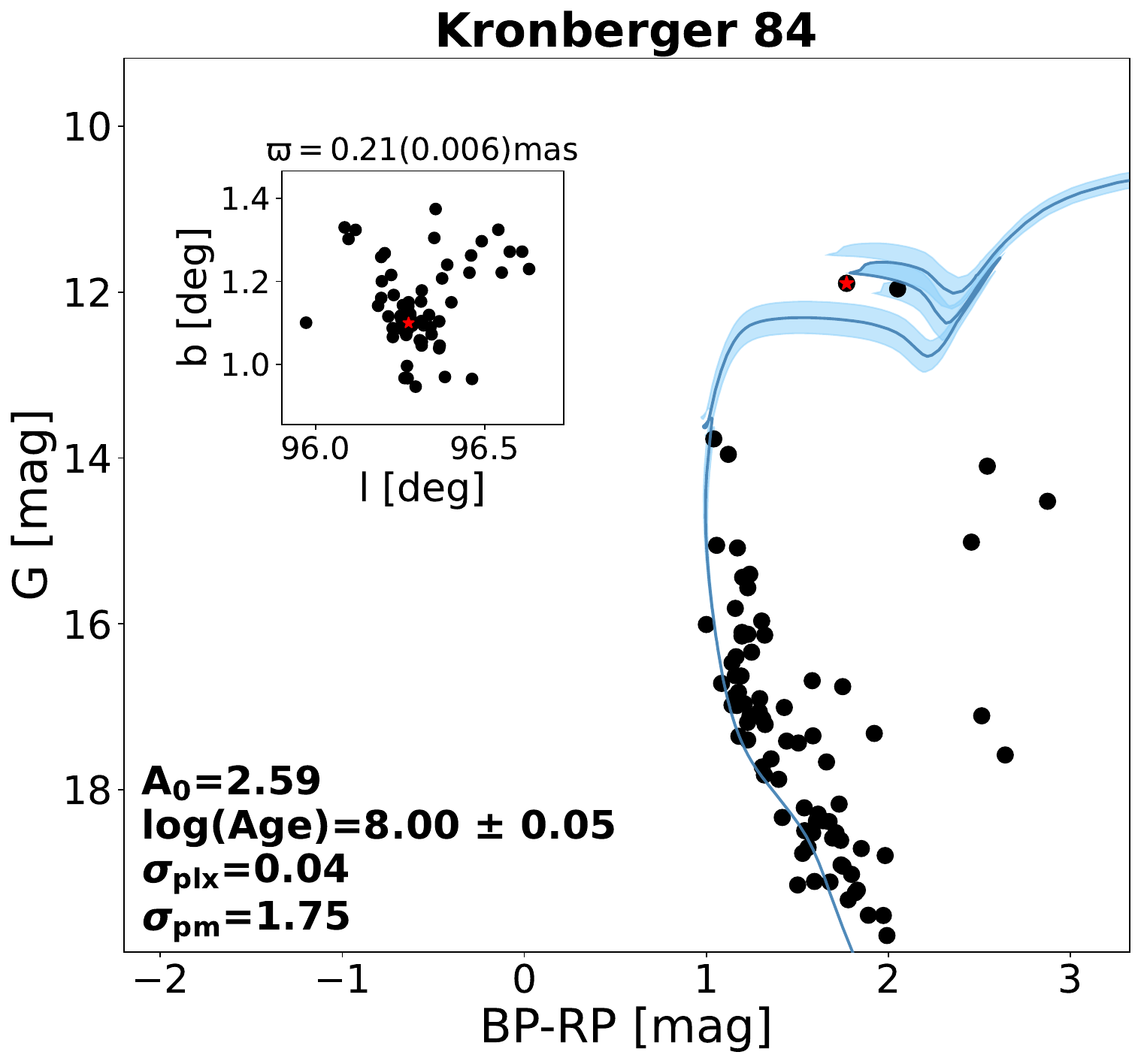}\\[0.4cm]

\includegraphics[width=0.23\linewidth]{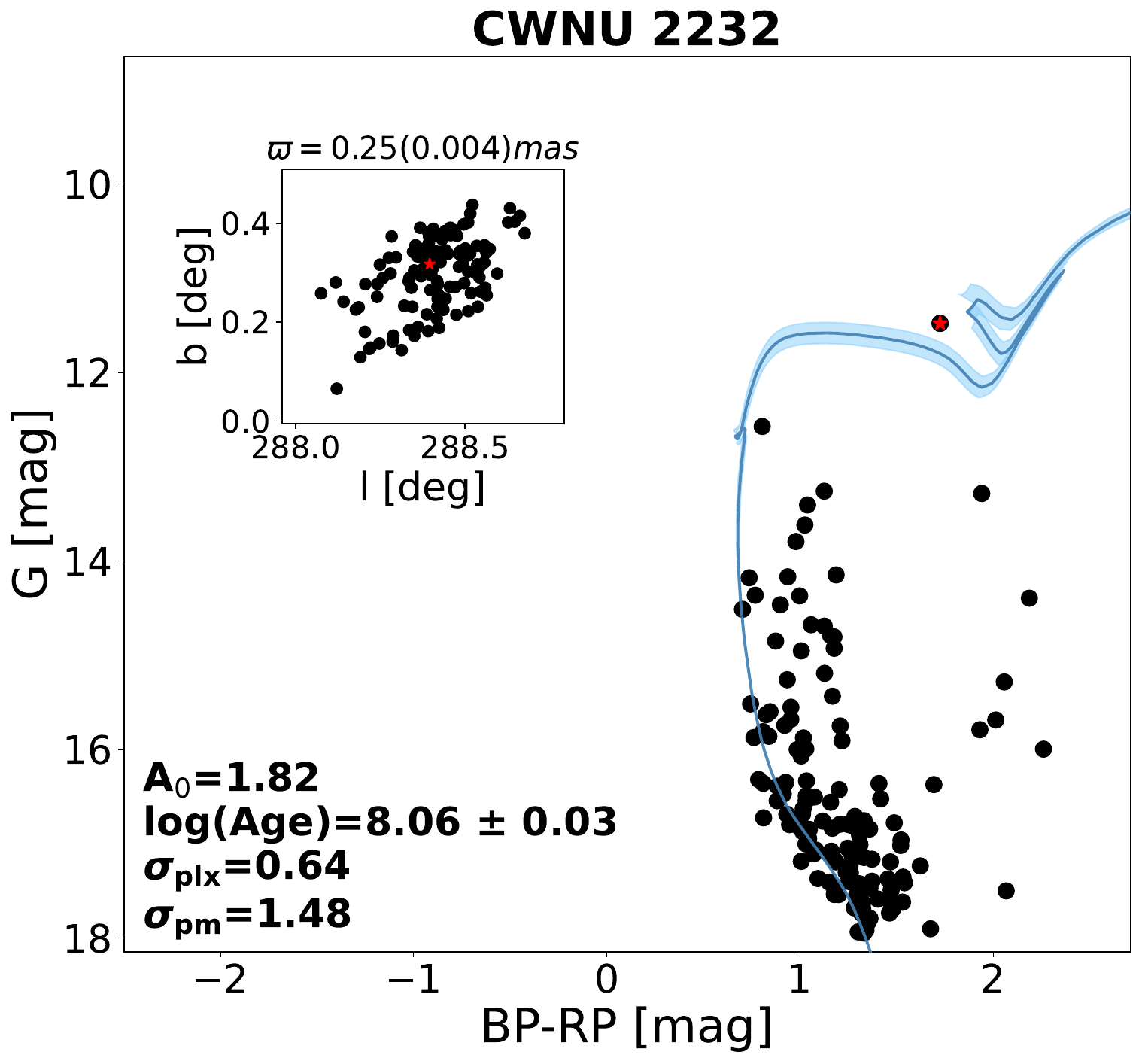}\hspace{0.02\linewidth}
\includegraphics[width=0.23\linewidth]{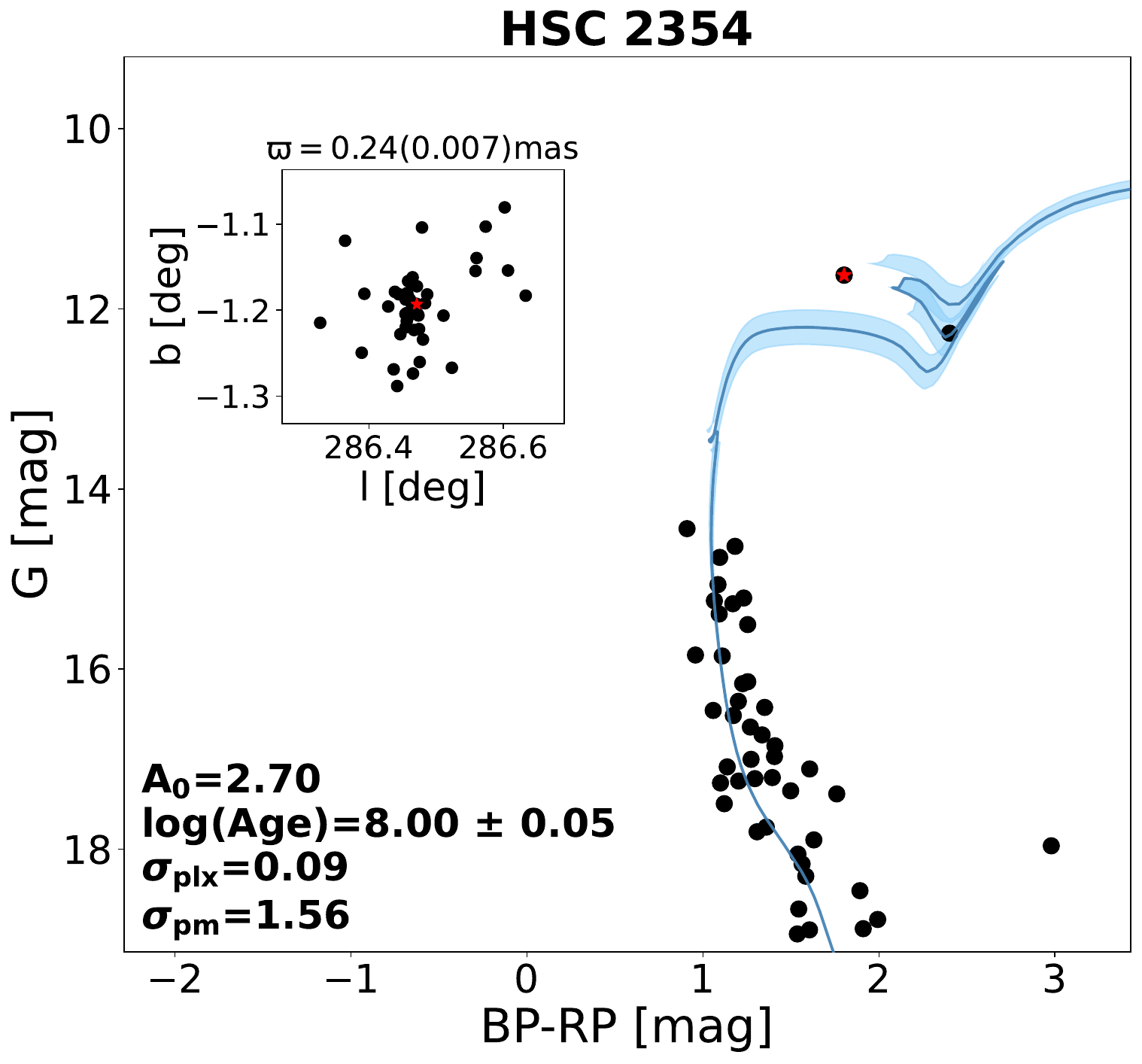}\hspace{0.02\linewidth}
\includegraphics[width=0.23\linewidth]{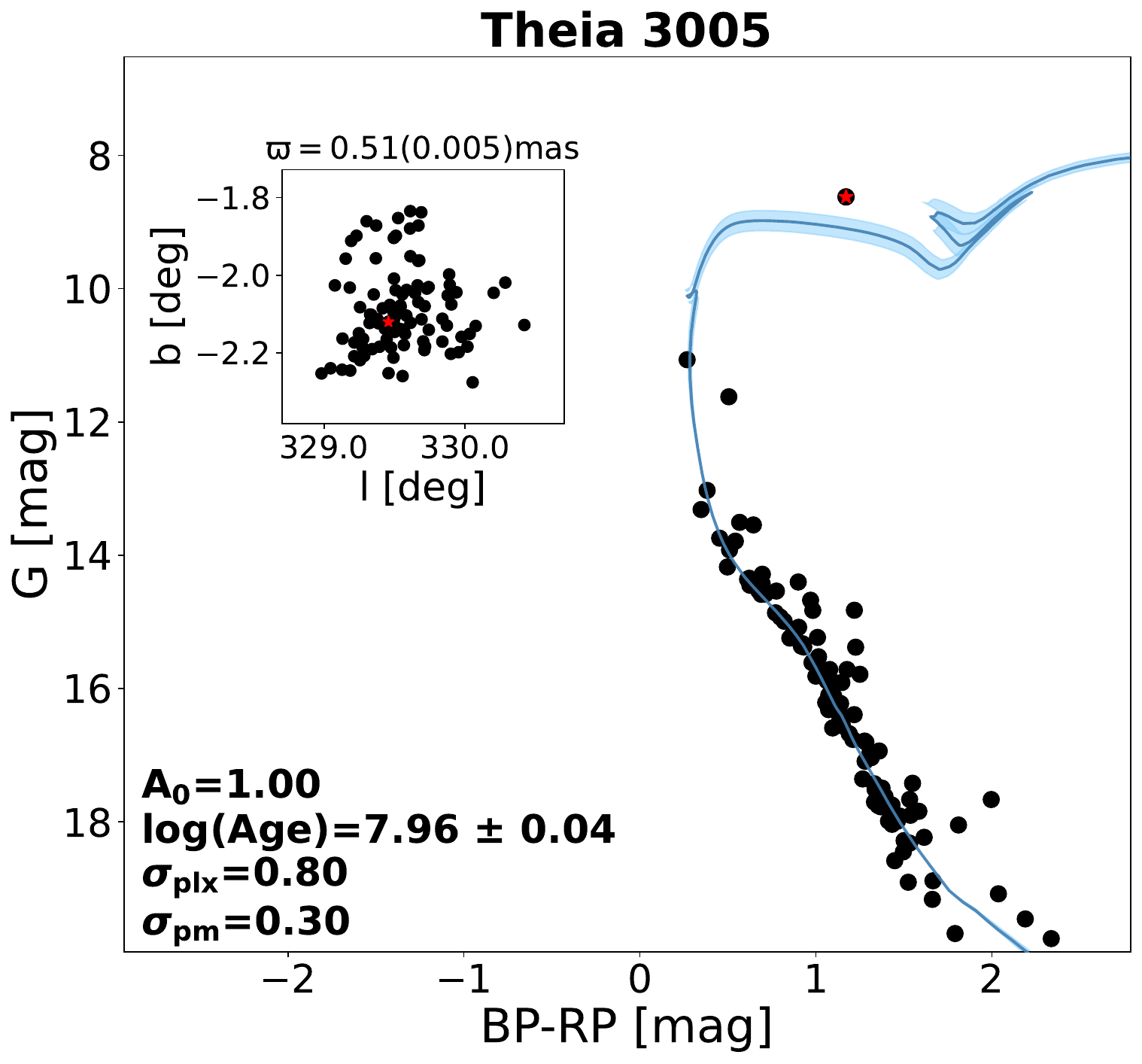}\hspace{0.02\linewidth}
\includegraphics[width=0.23\linewidth]{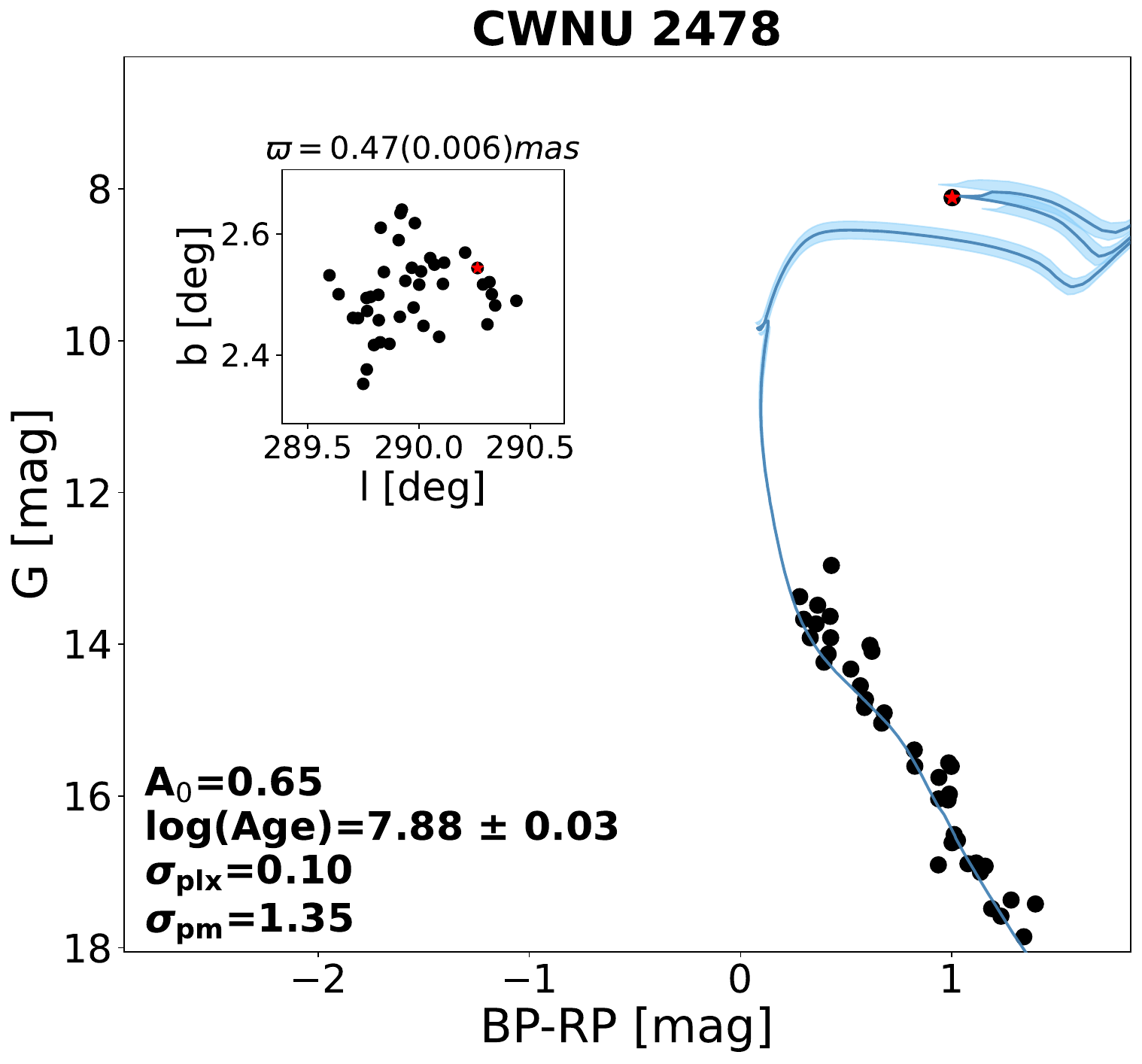}\\[0.4cm]

\includegraphics[width=0.23\linewidth]{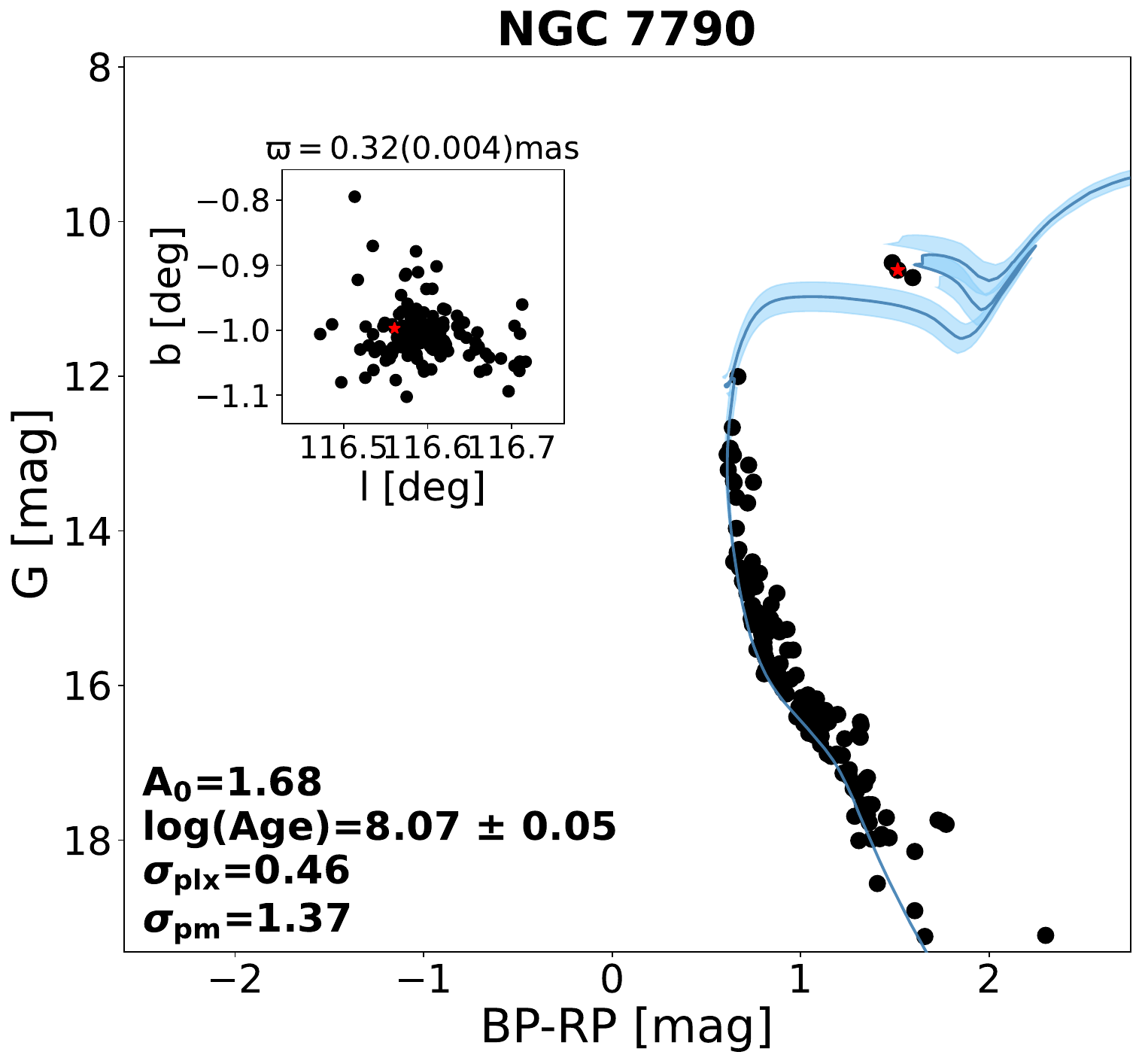}\hspace{0.02\linewidth}
\includegraphics[width=0.23\linewidth]{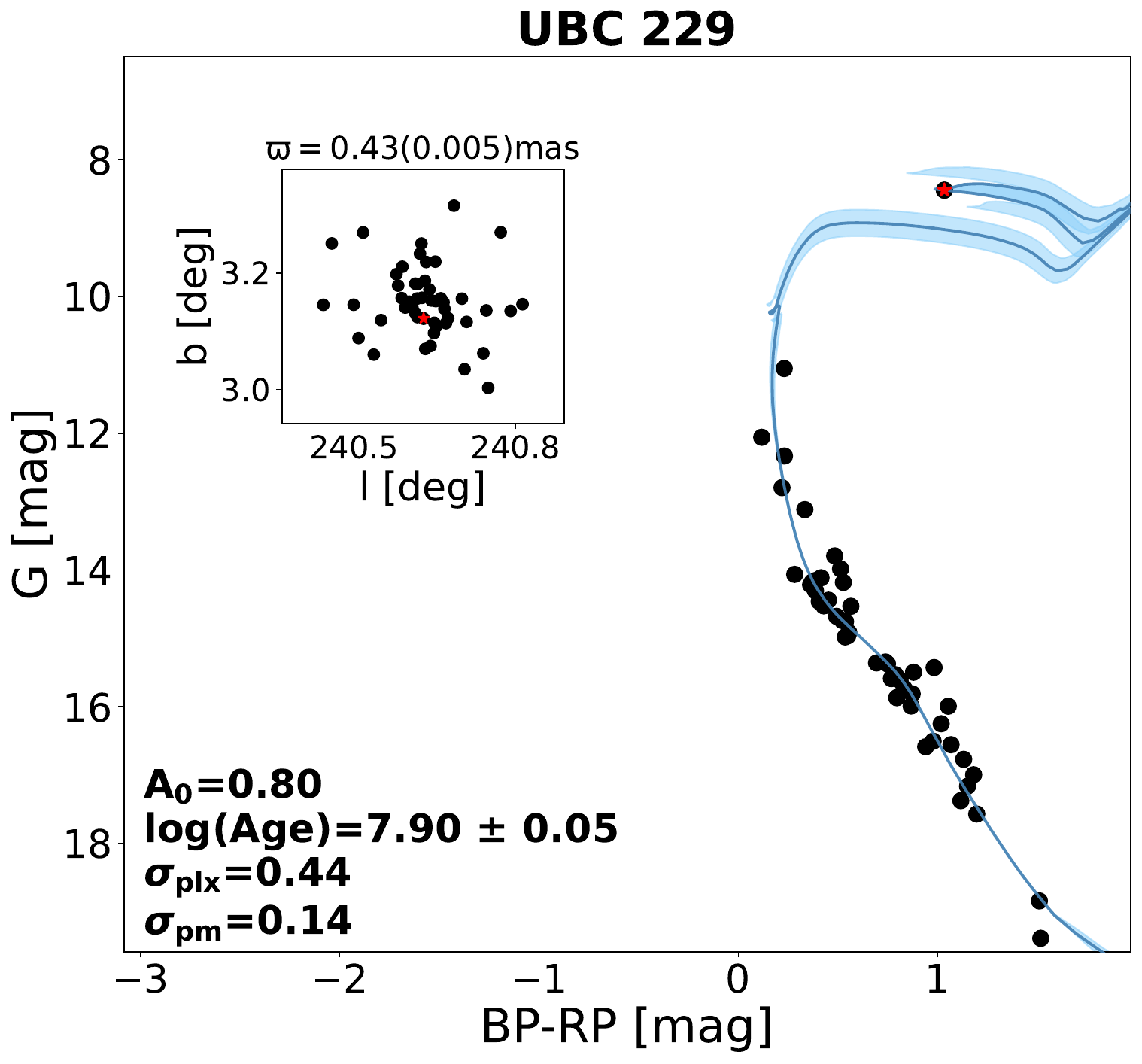}\hspace{0.02\linewidth}
\includegraphics[width=0.23\linewidth]{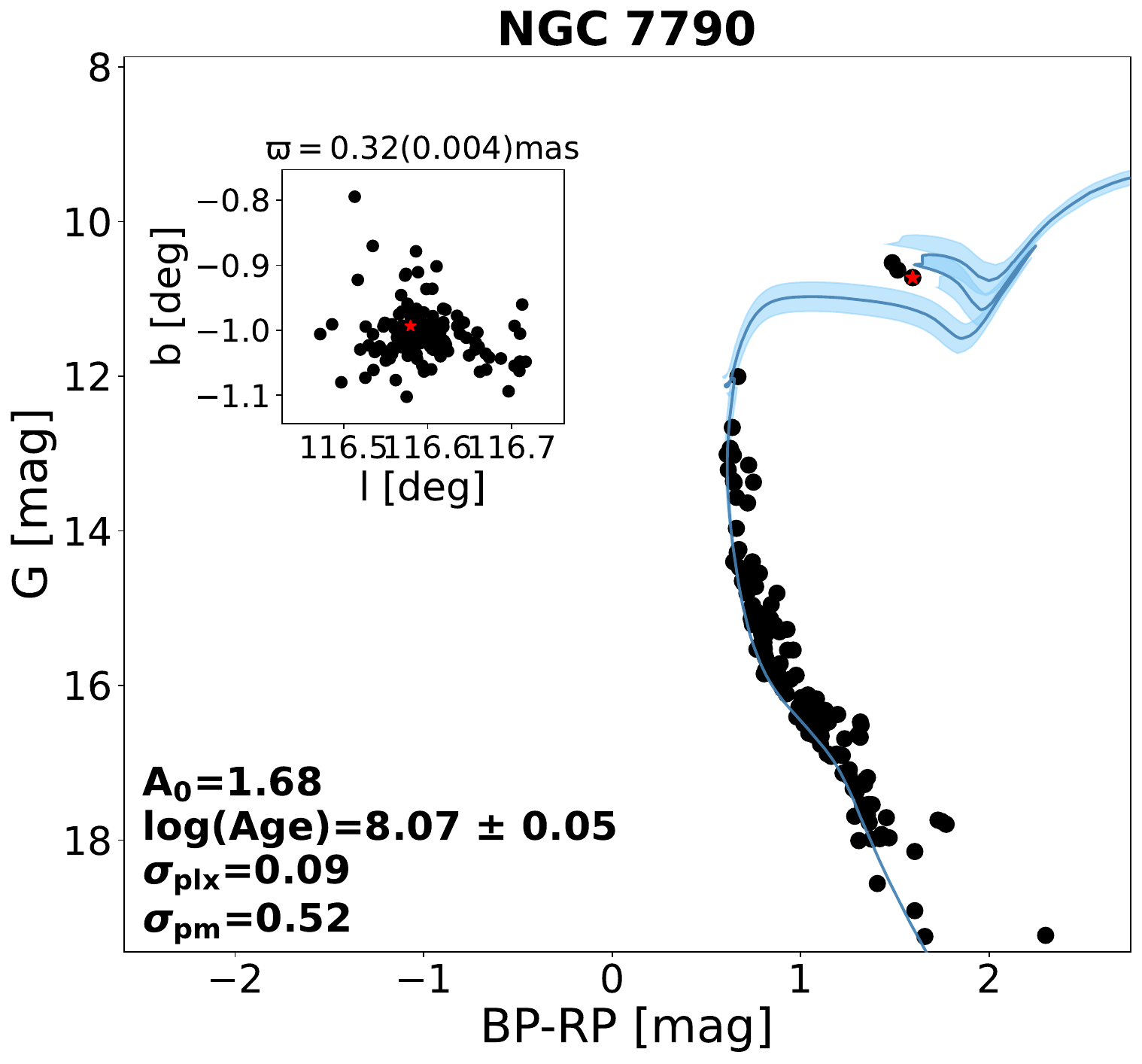}\hspace{0.02\linewidth}
\includegraphics[width=0.23\linewidth]{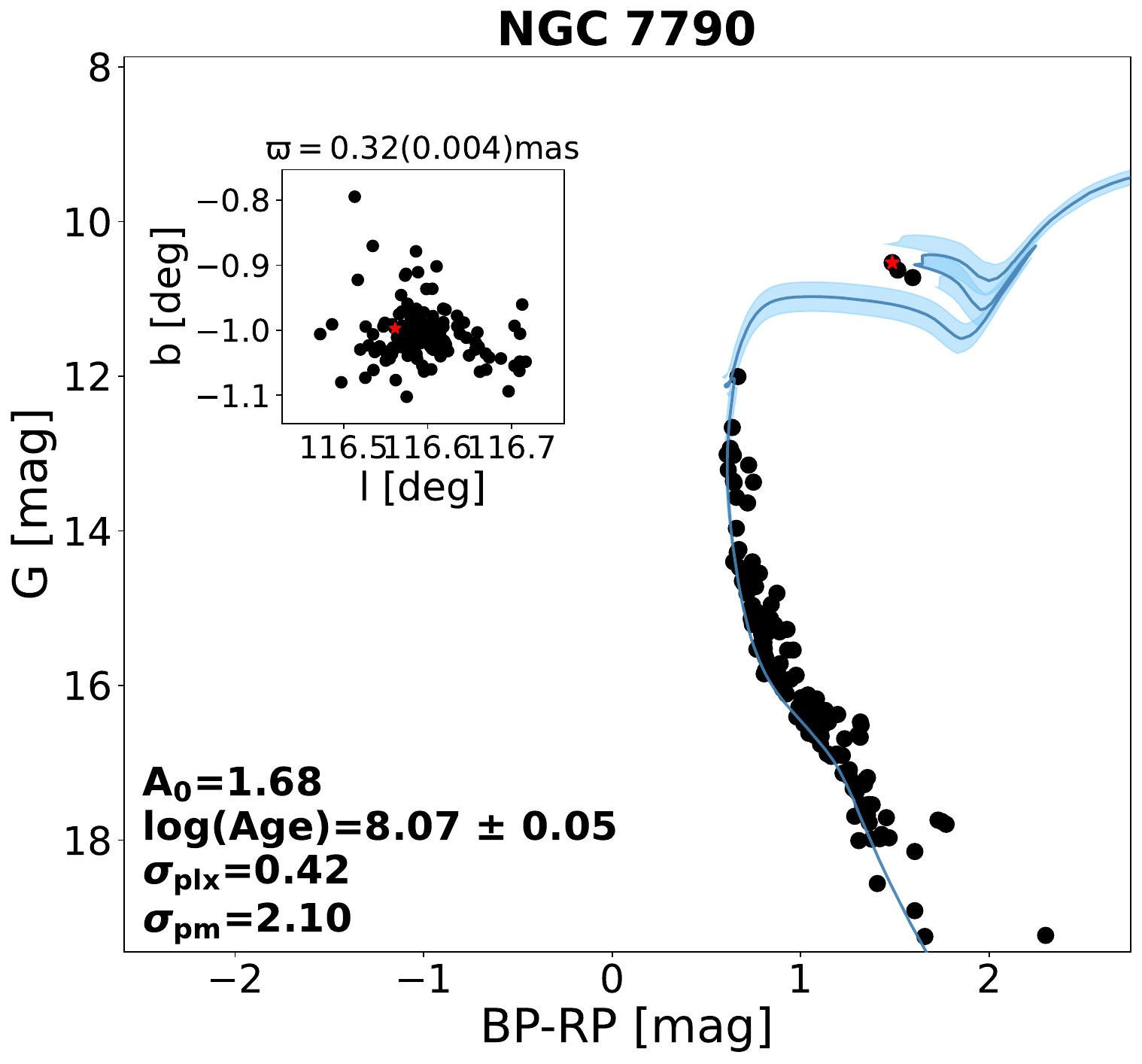}\\[0.4cm]

\includegraphics[width=0.23\linewidth]{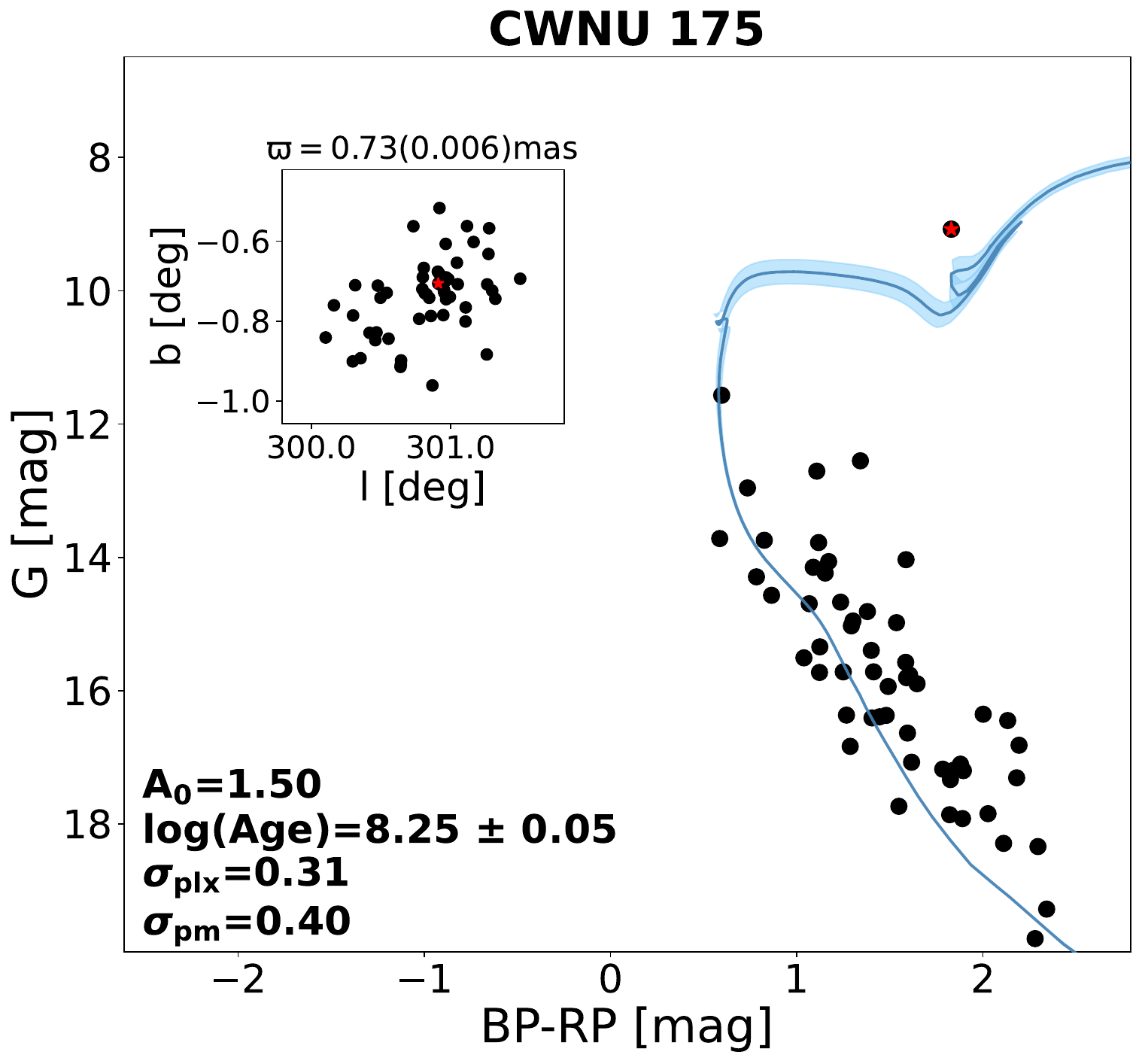}\hspace{0.02\linewidth}
\includegraphics[width=0.23\linewidth]{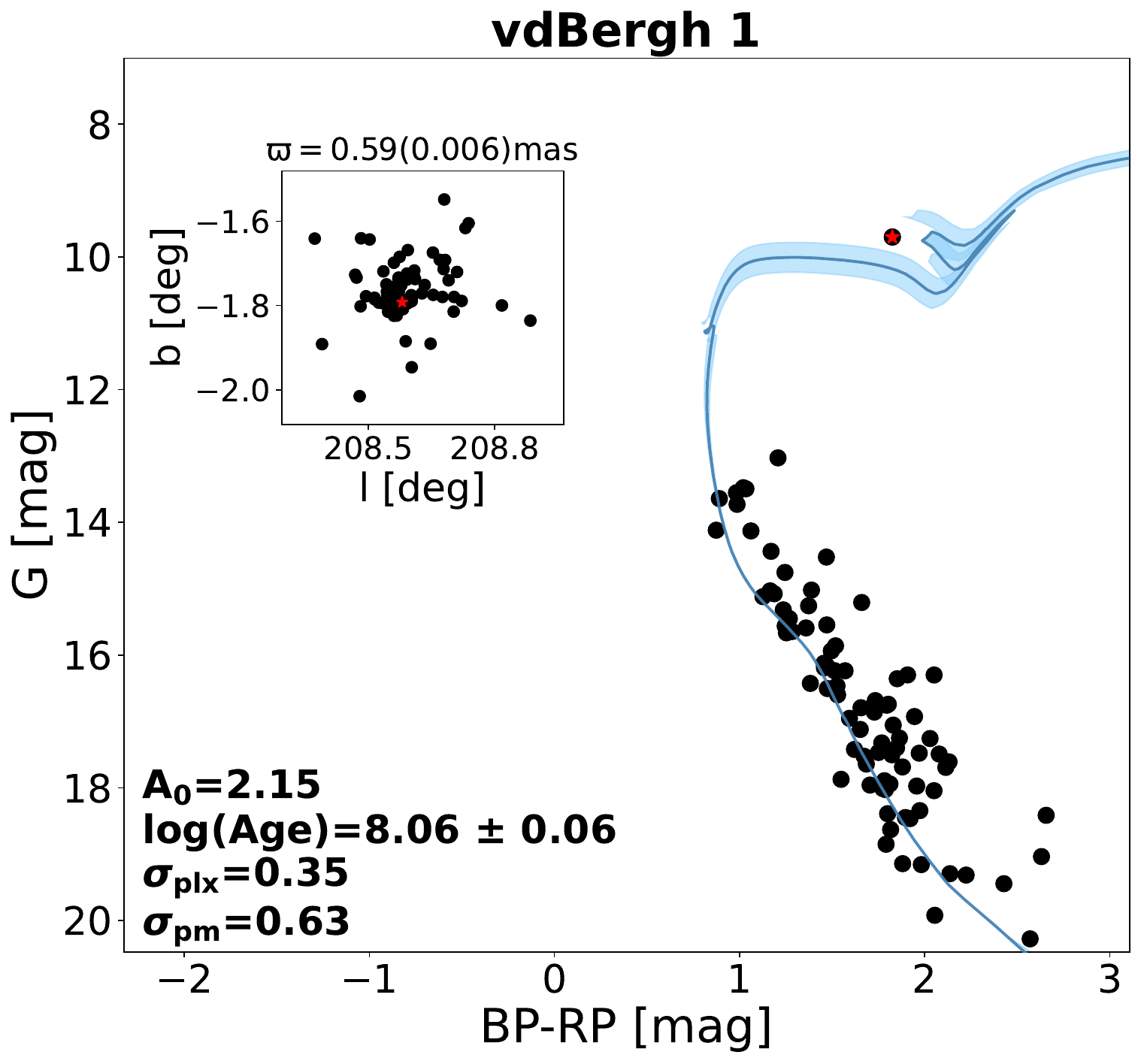}\hspace{0.02\linewidth}
\includegraphics[width=0.23\linewidth]{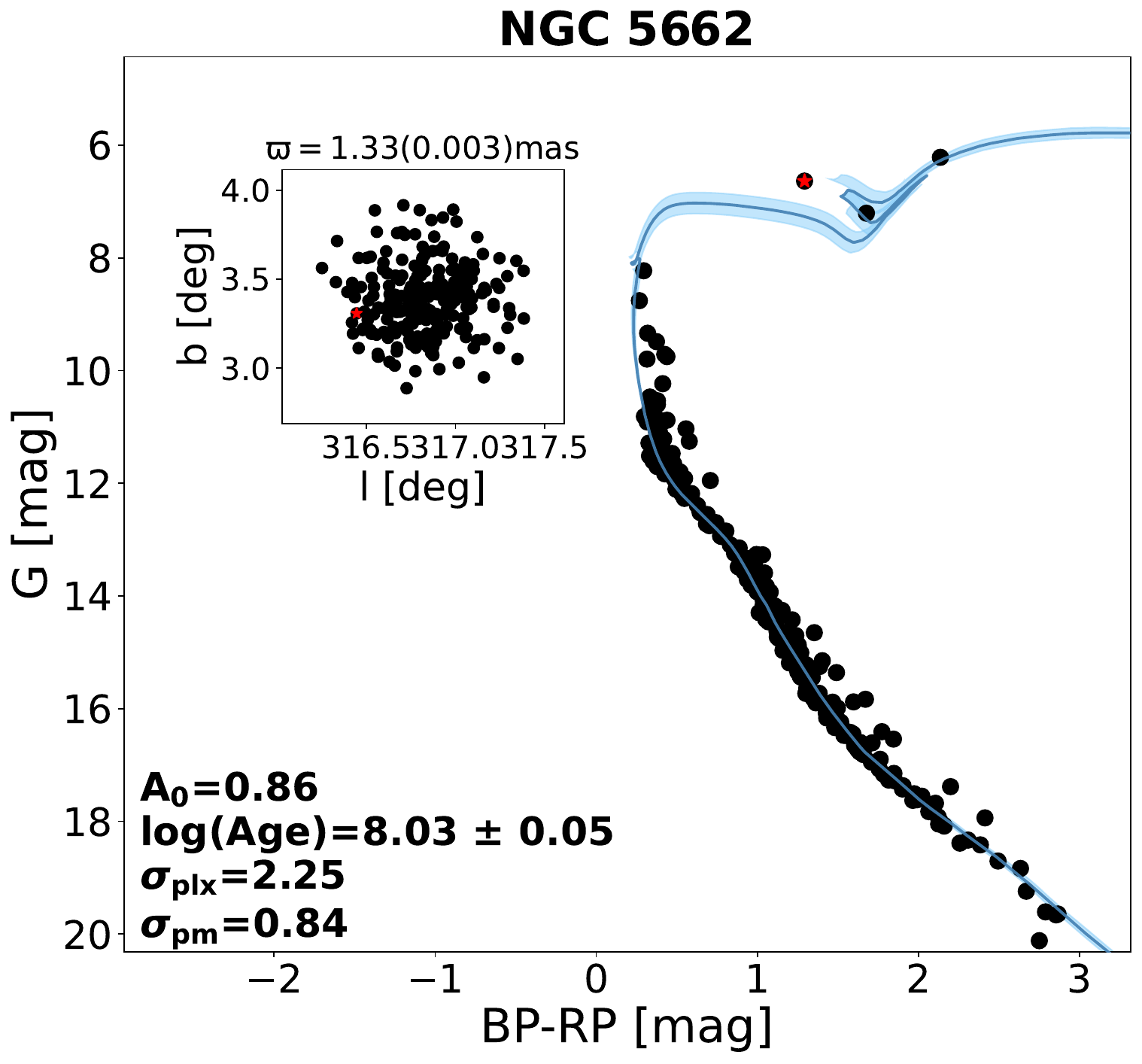}\hspace{0.02\linewidth}
\includegraphics[width=0.23\linewidth]{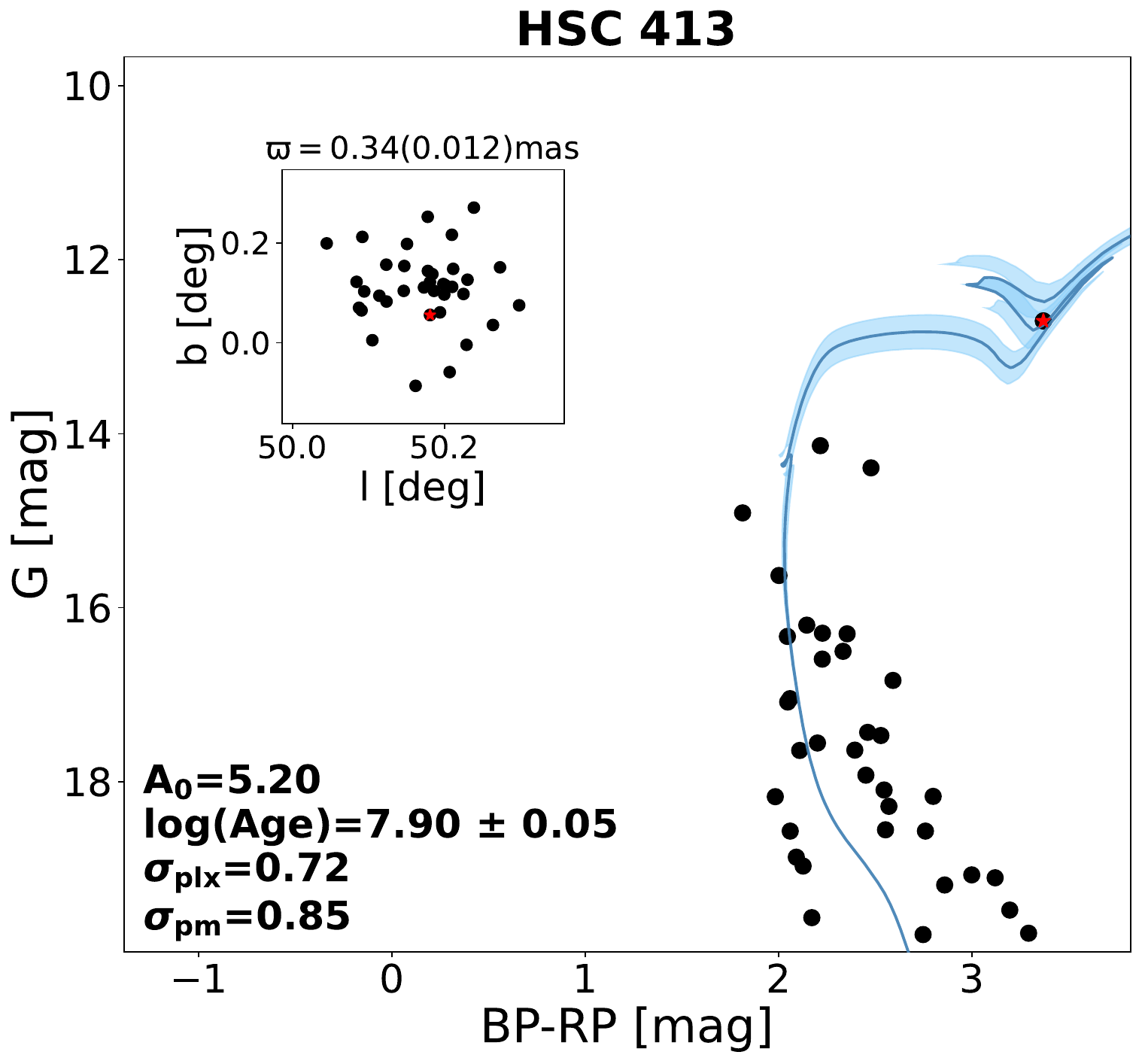}\\[0.4cm]

\includegraphics[width=0.23\linewidth]{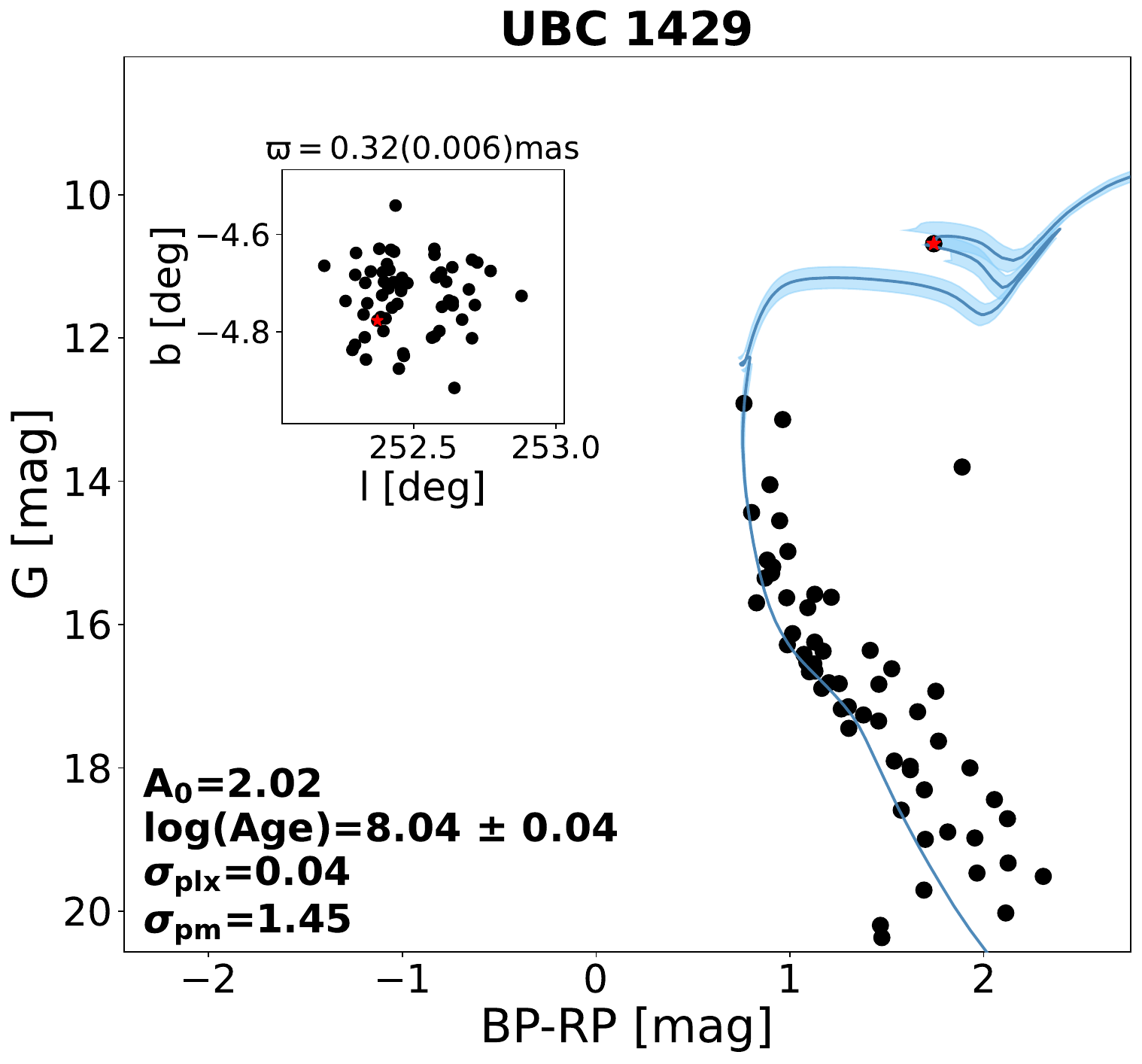}\hspace{0.02\linewidth}
\includegraphics[width=0.23\linewidth]{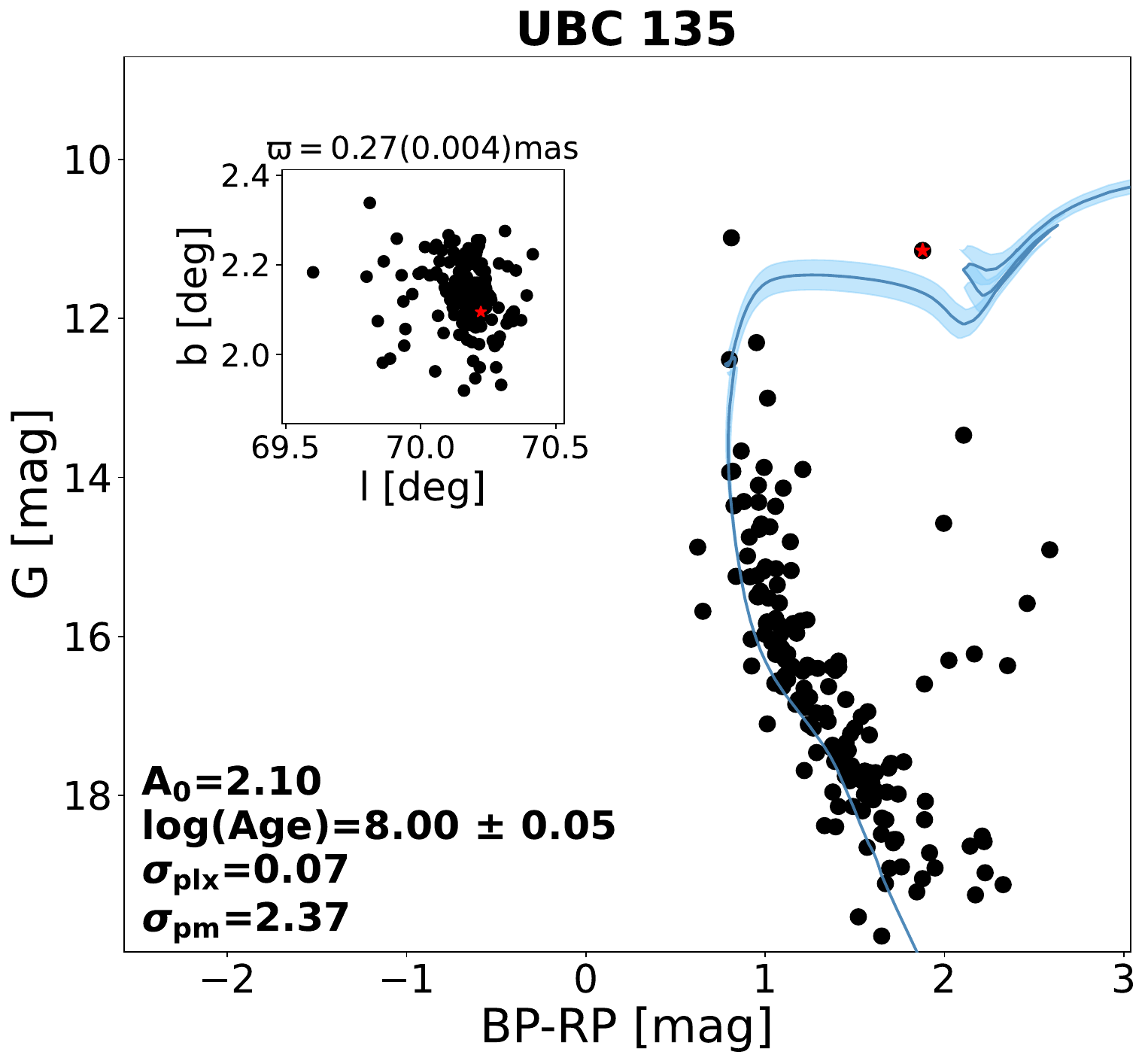}\hspace{0.02\linewidth}
\includegraphics[width=0.23\linewidth]{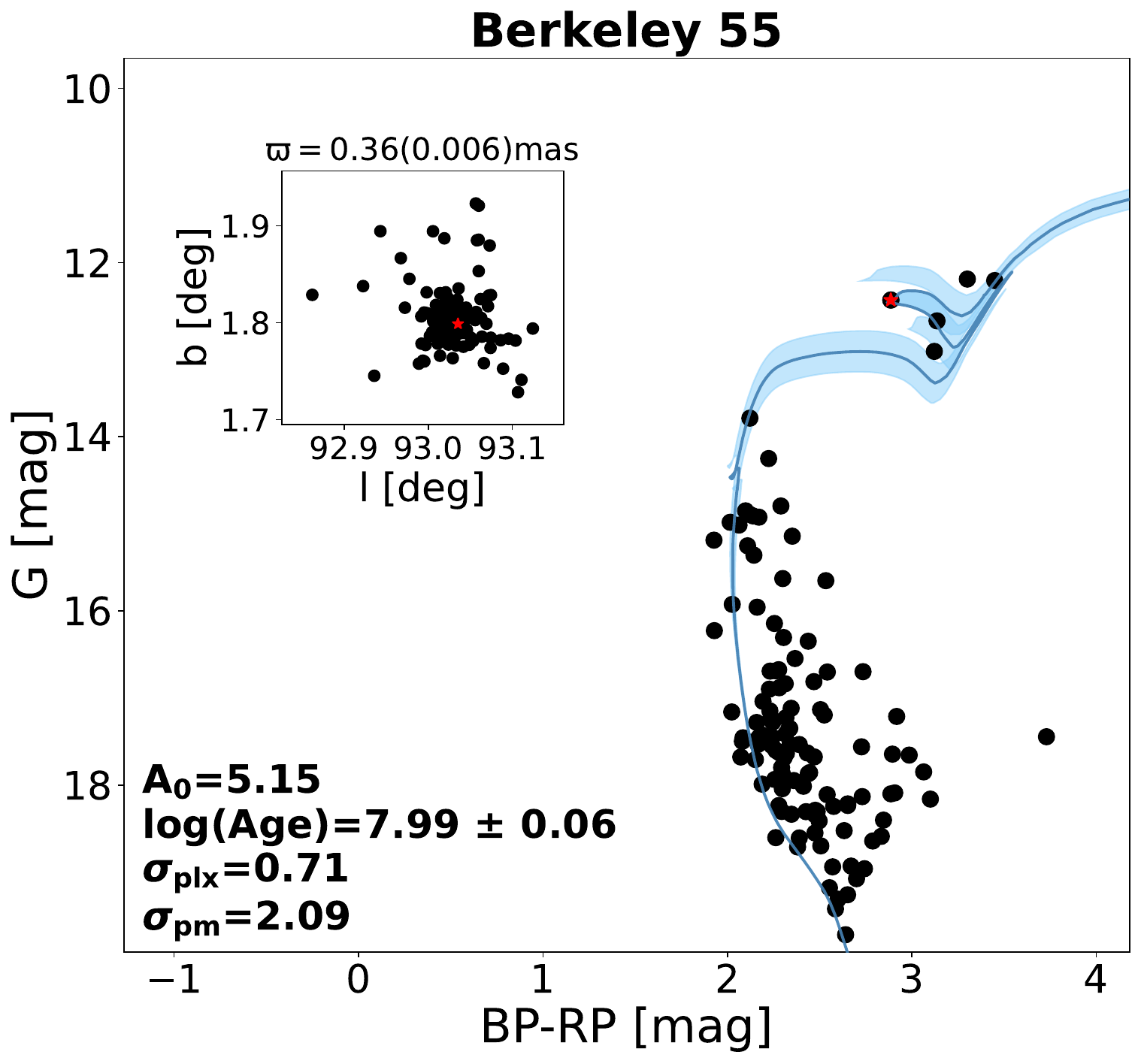}\hspace{0.02\linewidth}
\includegraphics[width=0.23\linewidth]{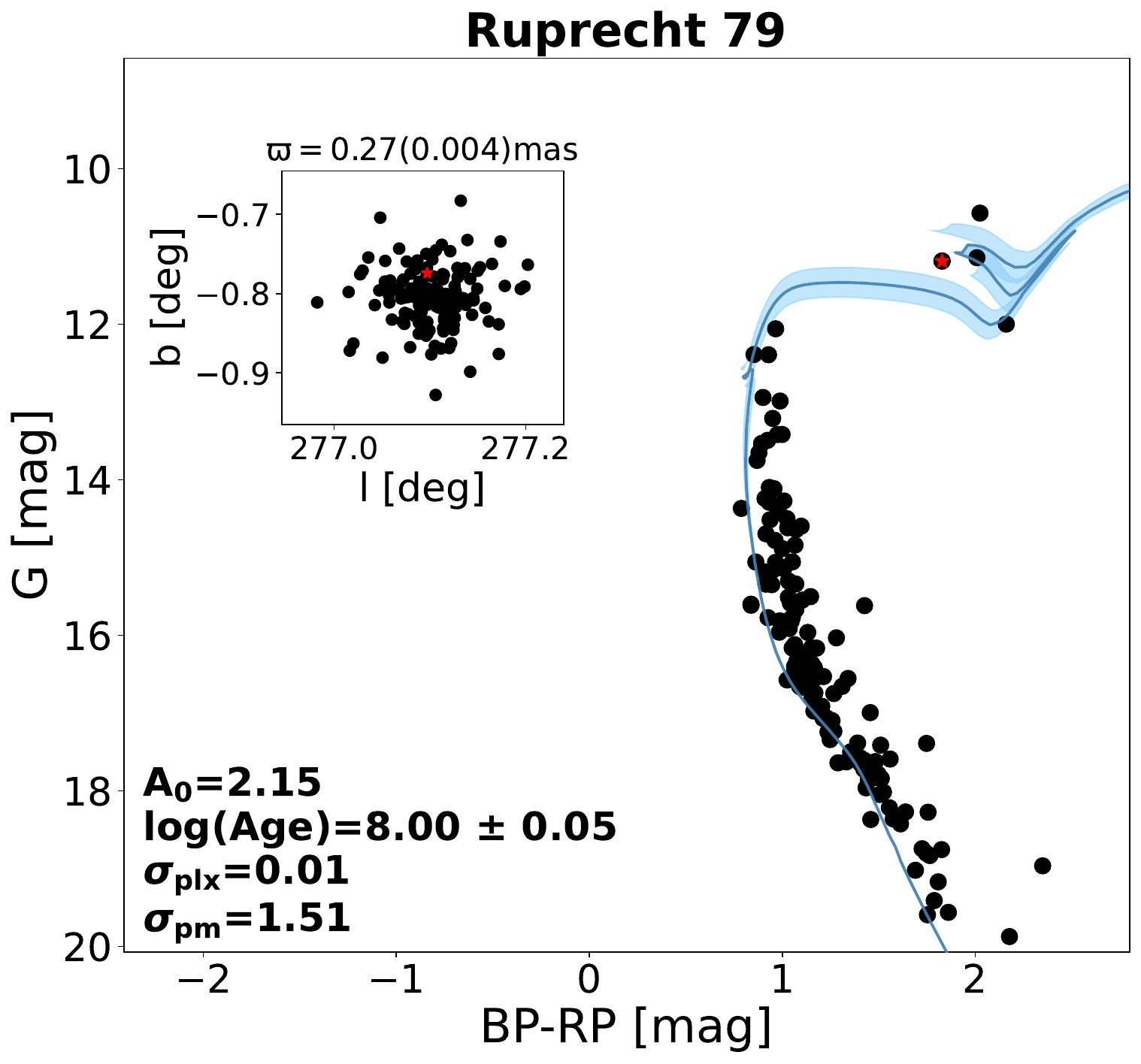}\\[0.4cm]

\caption{CMDs of the remaining part of OC Cepheids, which also contain the extinction, age and distance parameters of the OCs, namely, 
NGC 103, CWNU 2461, NGC 6664, Kronberger 84 (J213533.70+533049.3), CWNU 2232, HSC 2354, Theia 3005, CWNU 2478, 
NGC 7790 (CE Cas B), UBC 229, NGC 7790 (CF Cas), NGC 7790 (CE Cas A), CWNU 175, vdBergh 1, NGC 5662, HSC 413, 
UBC 1429, UBC 135, Berkeley 55, and Ruprecht 79.}
\label{fig:cmd_new}
\end{figure*}

\begin{figure*}[htbp]
\centering
\includegraphics[width=0.23\linewidth]{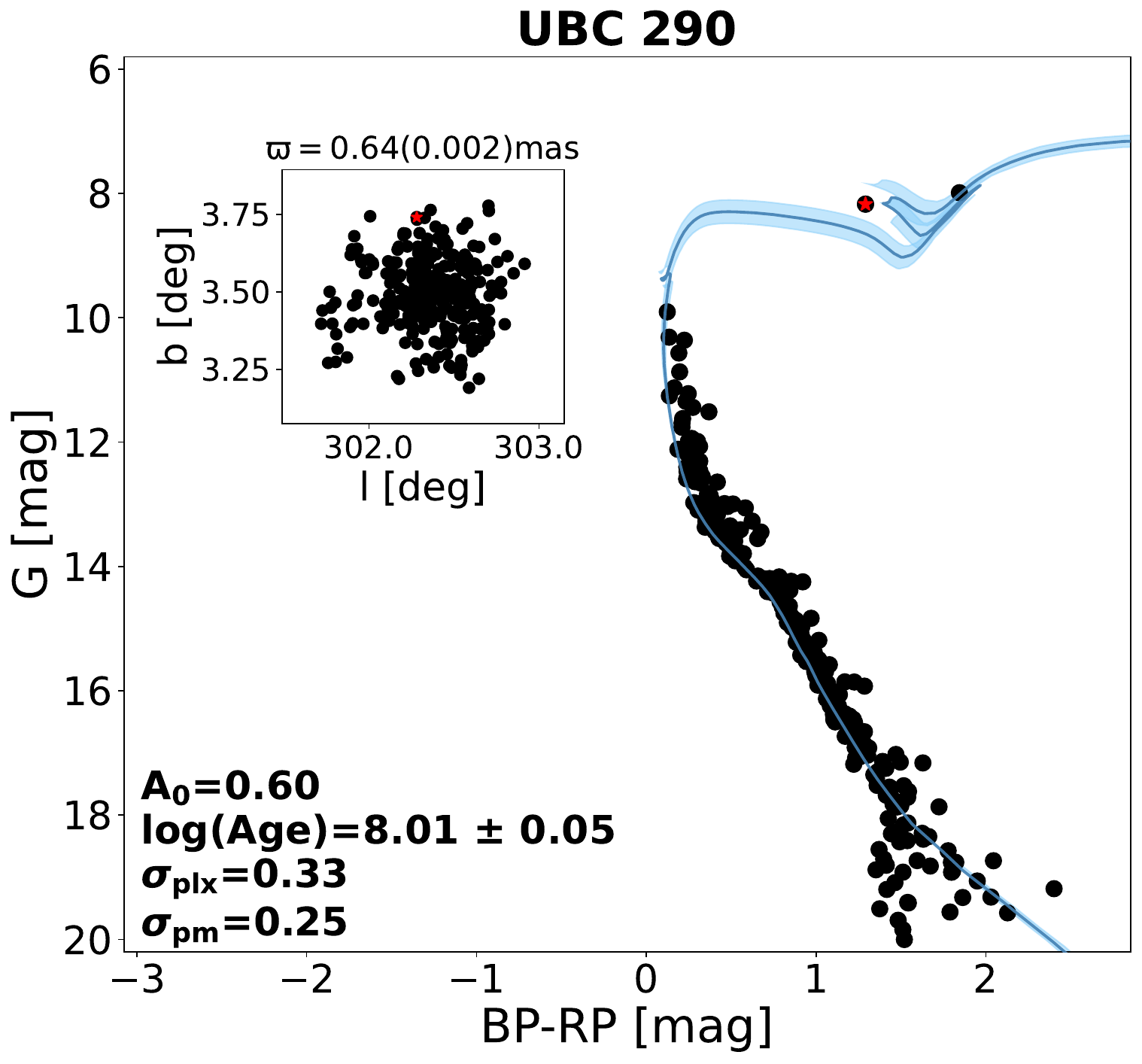}\hspace{0.02\linewidth}
\includegraphics[width=0.23\linewidth]{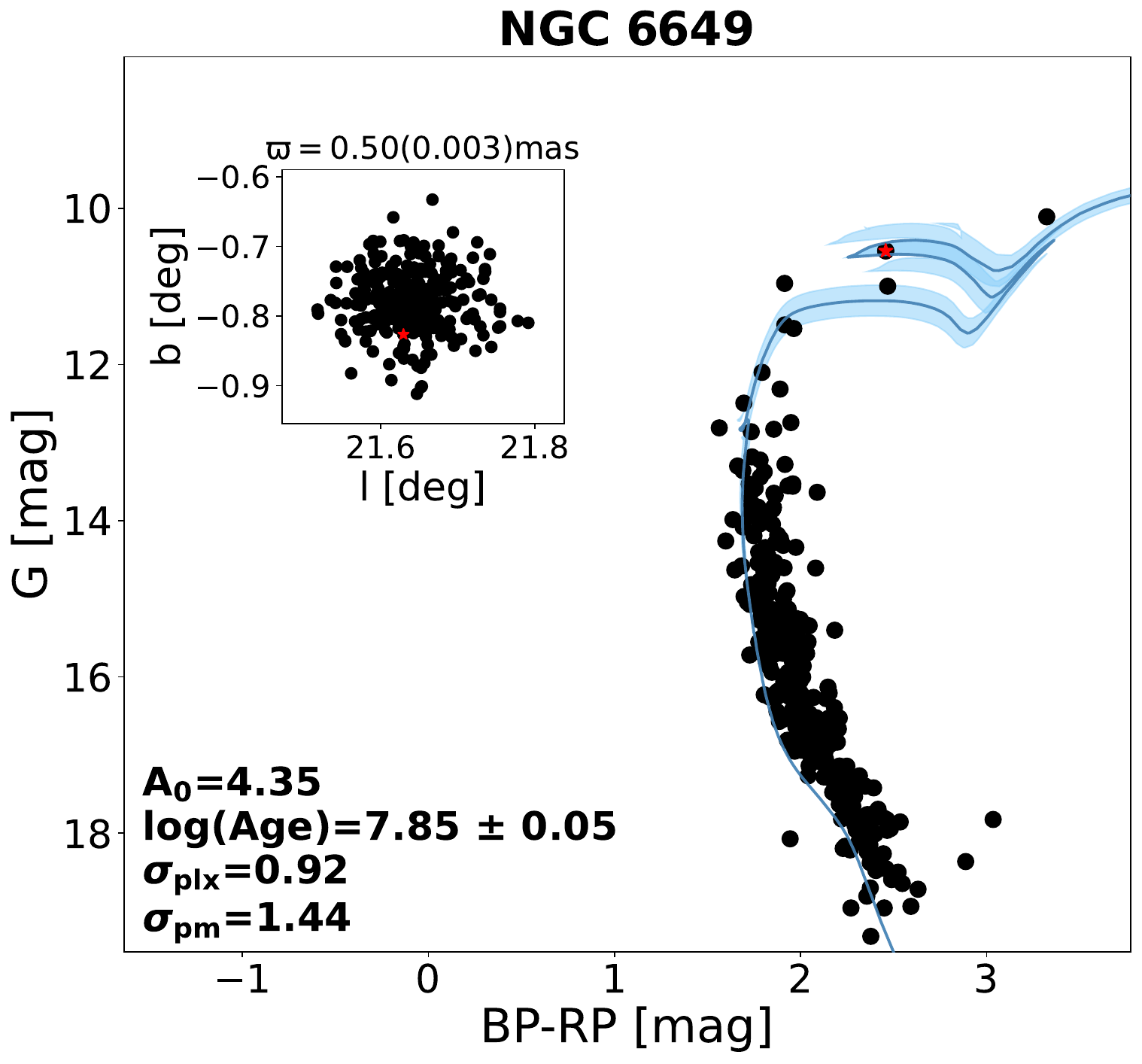}\hspace{0.02\linewidth}
\includegraphics[width=0.23\linewidth]{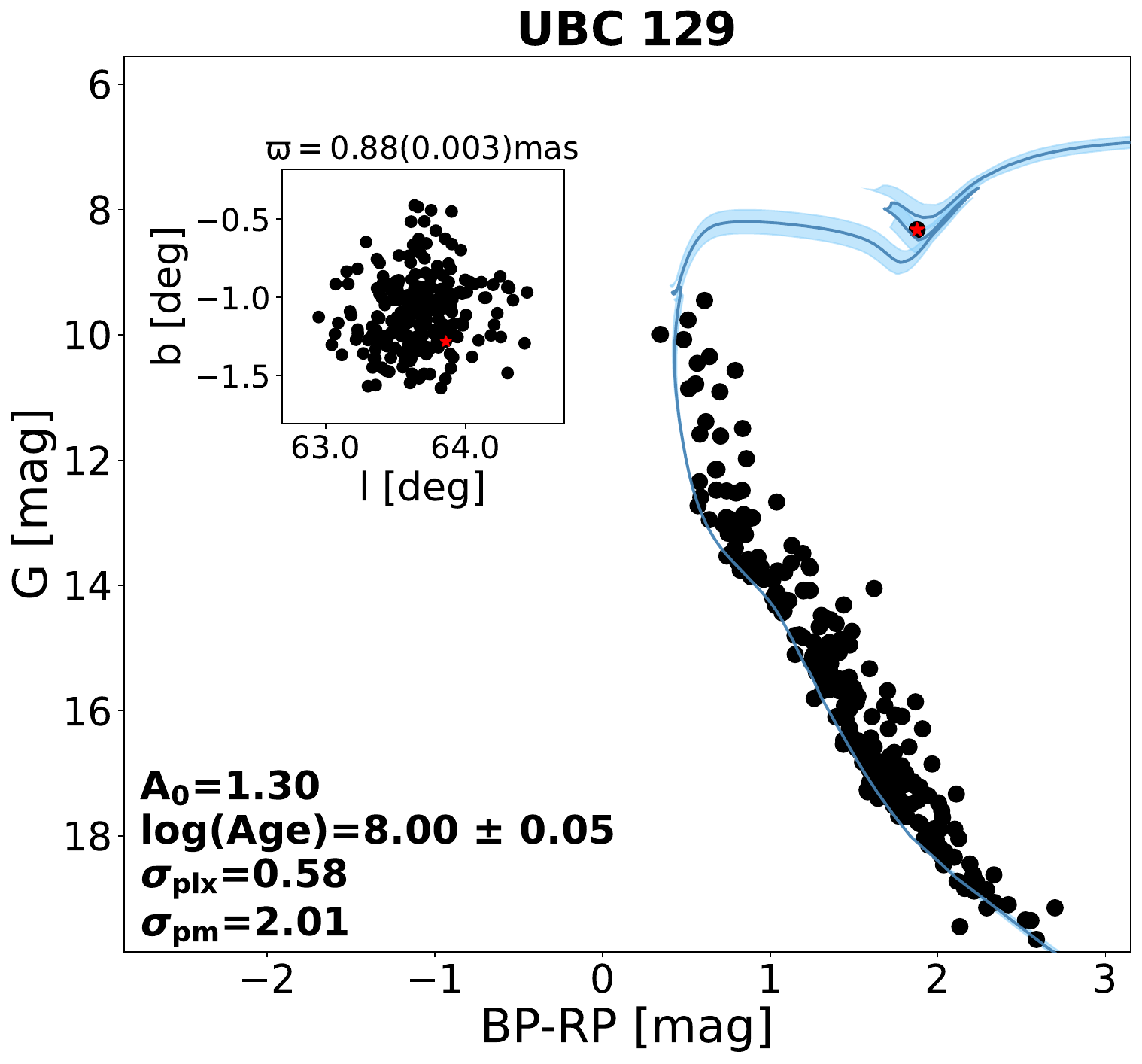}\hspace{0.02\linewidth}
\includegraphics[width=0.23\linewidth]{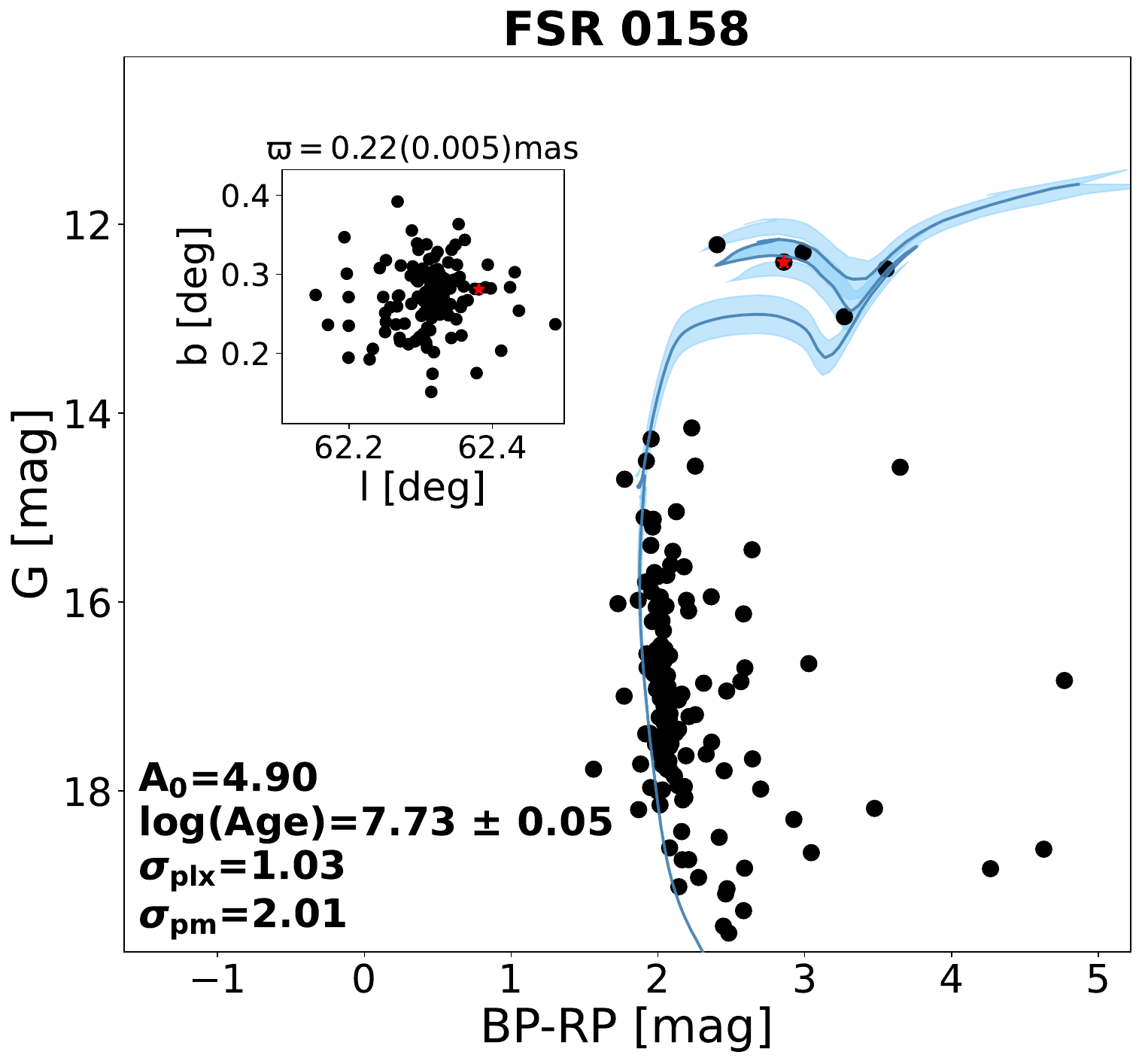}\\[0.4cm]

\includegraphics[width=0.23\linewidth]{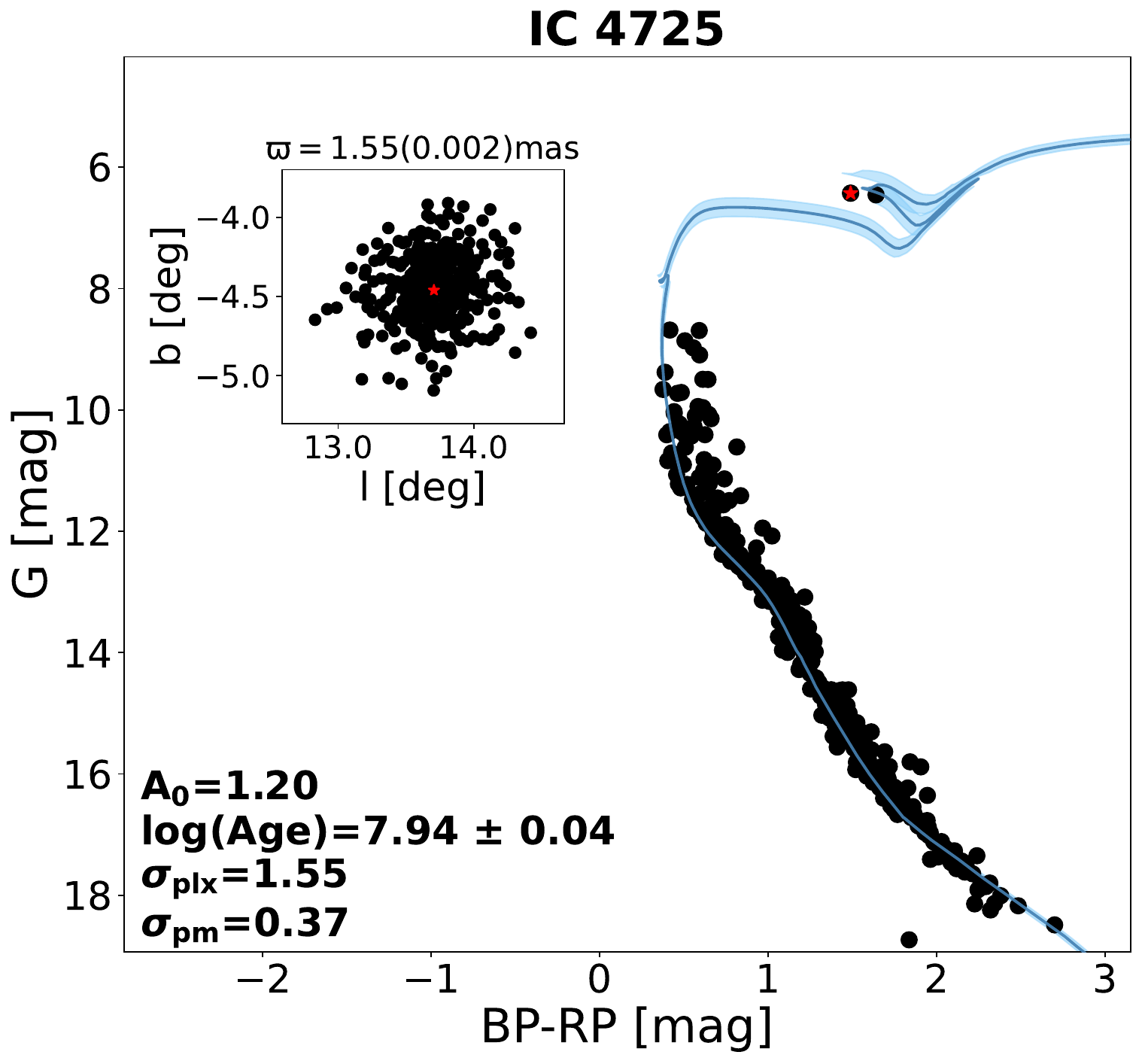}\hspace{0.02\linewidth}
\includegraphics[width=0.23\linewidth]{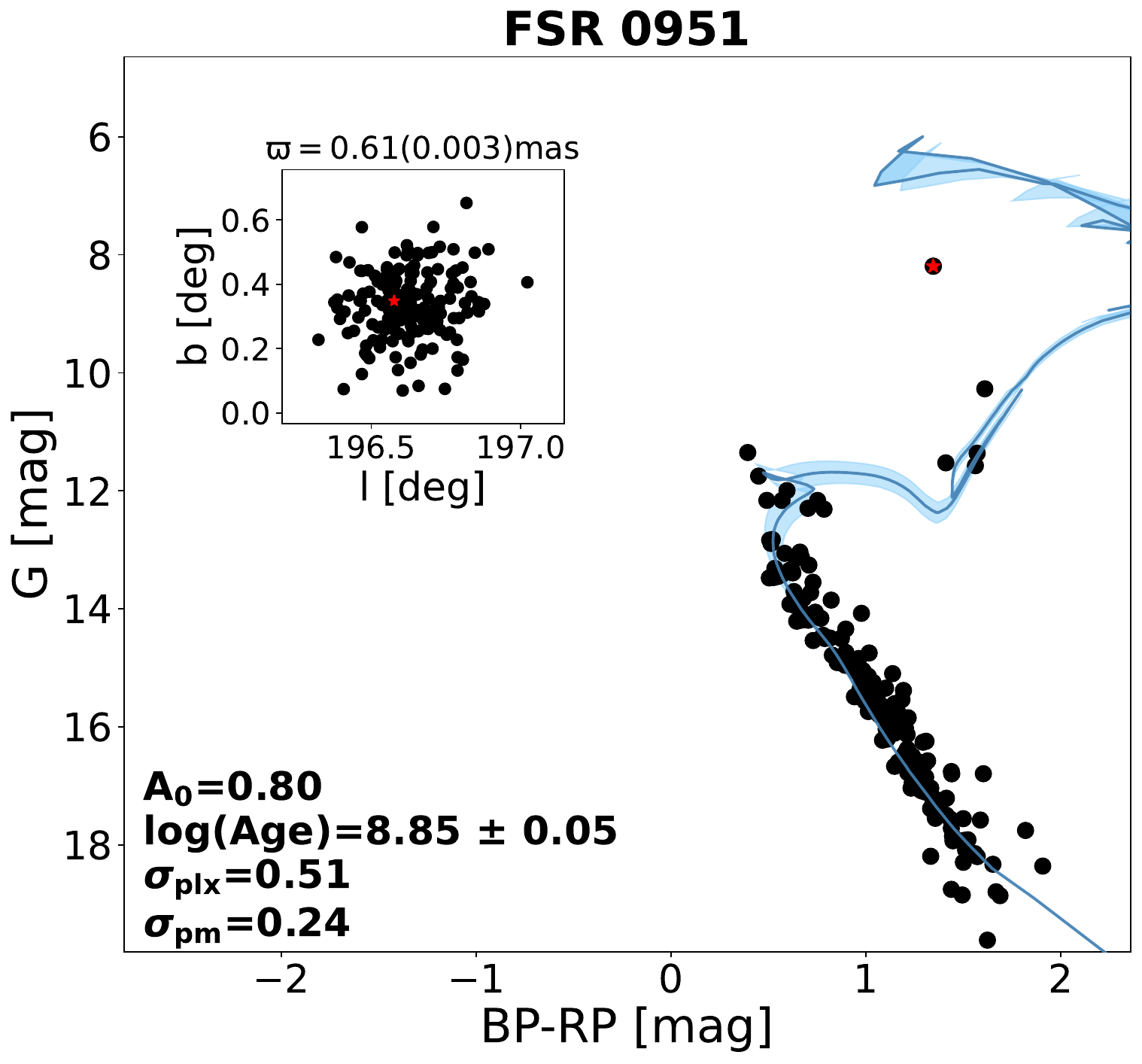}\hspace{0.02\linewidth}
\includegraphics[width=0.23\linewidth]{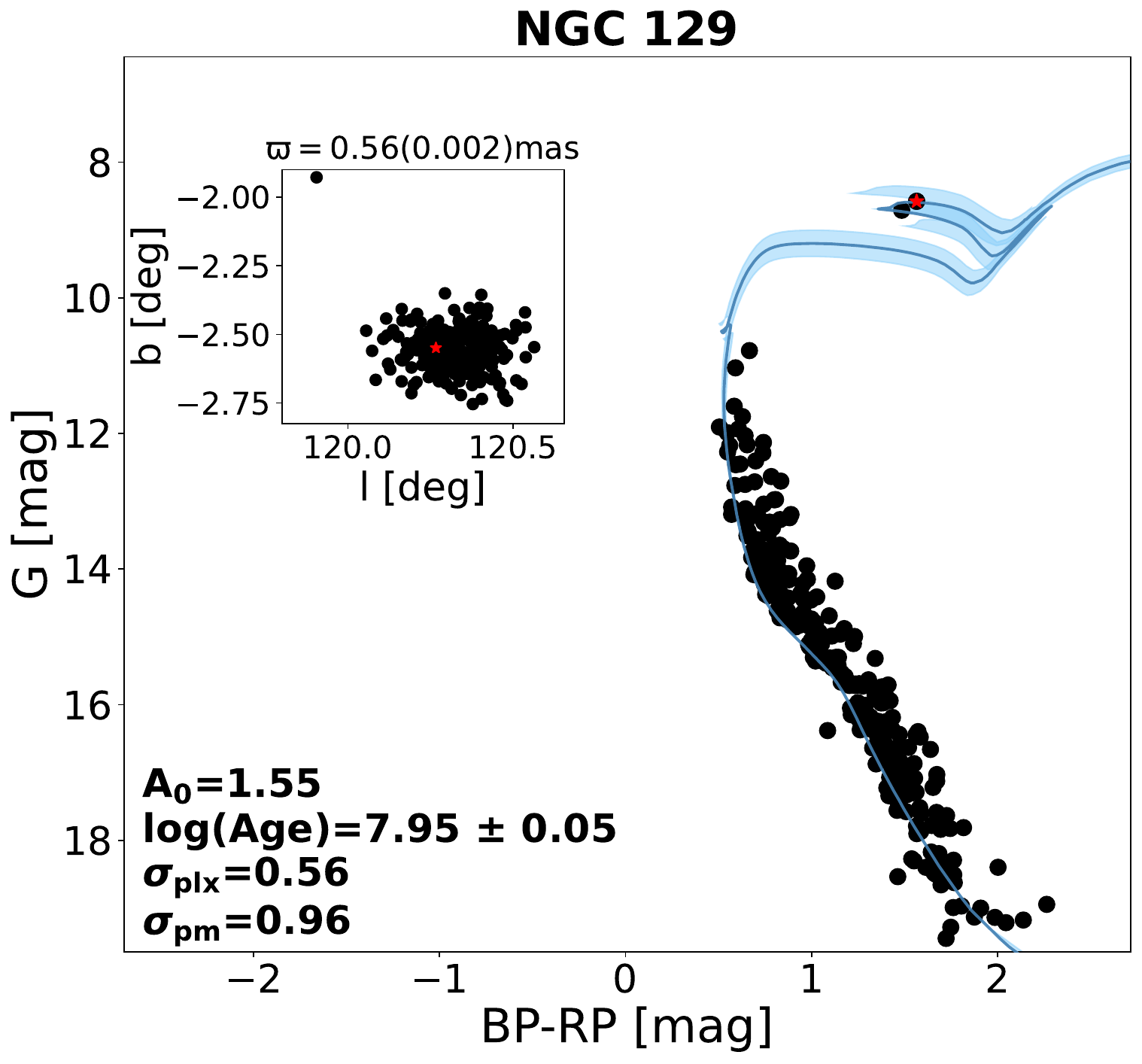}\hspace{0.02\linewidth}
\includegraphics[width=0.23\linewidth]{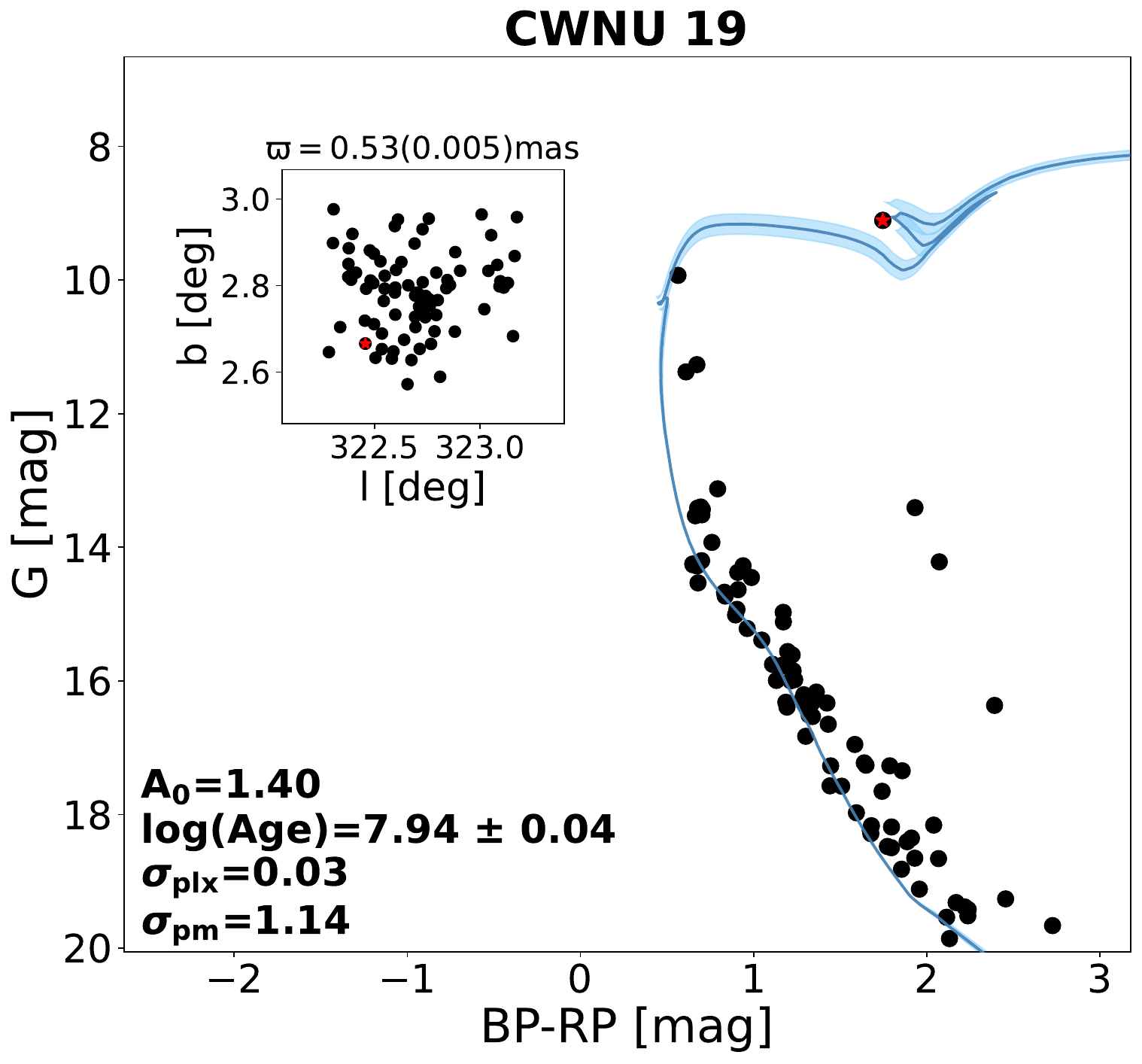}\\[0.4cm]

\includegraphics[width=0.23\linewidth]{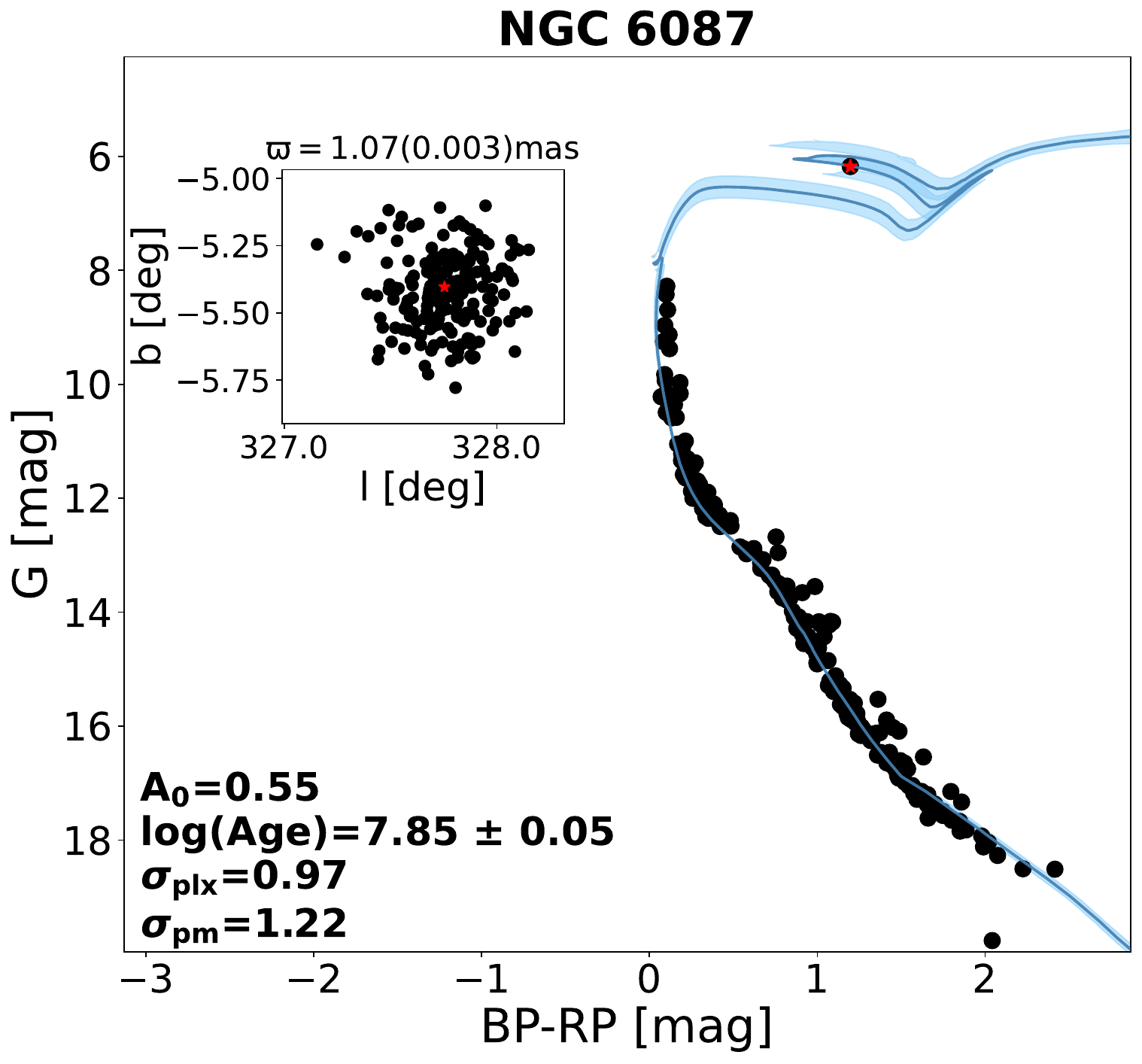}\hspace{0.02\linewidth}
\includegraphics[width=0.23\linewidth]{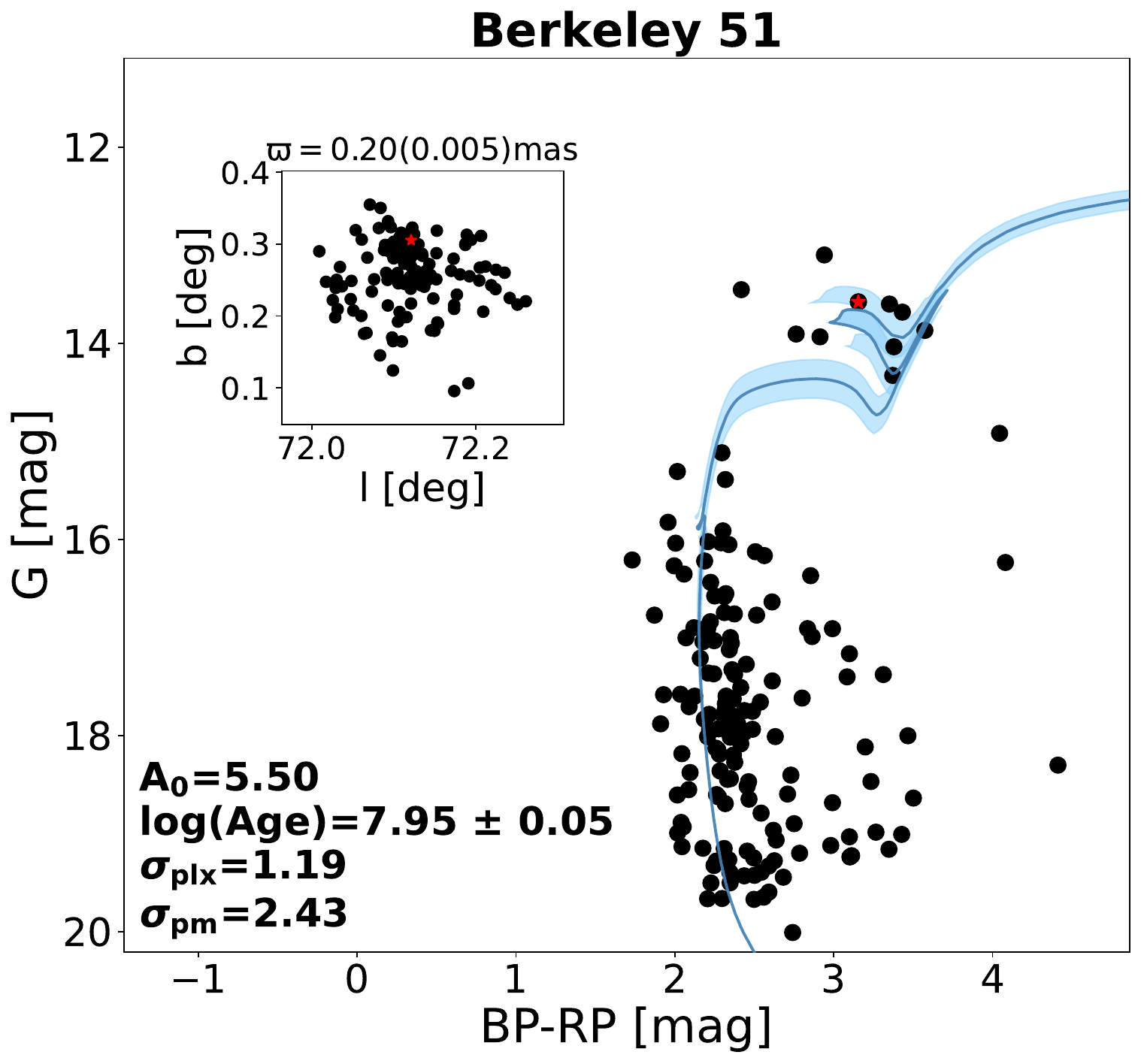}\hspace{0.02\linewidth}
\includegraphics[width=0.23\linewidth]{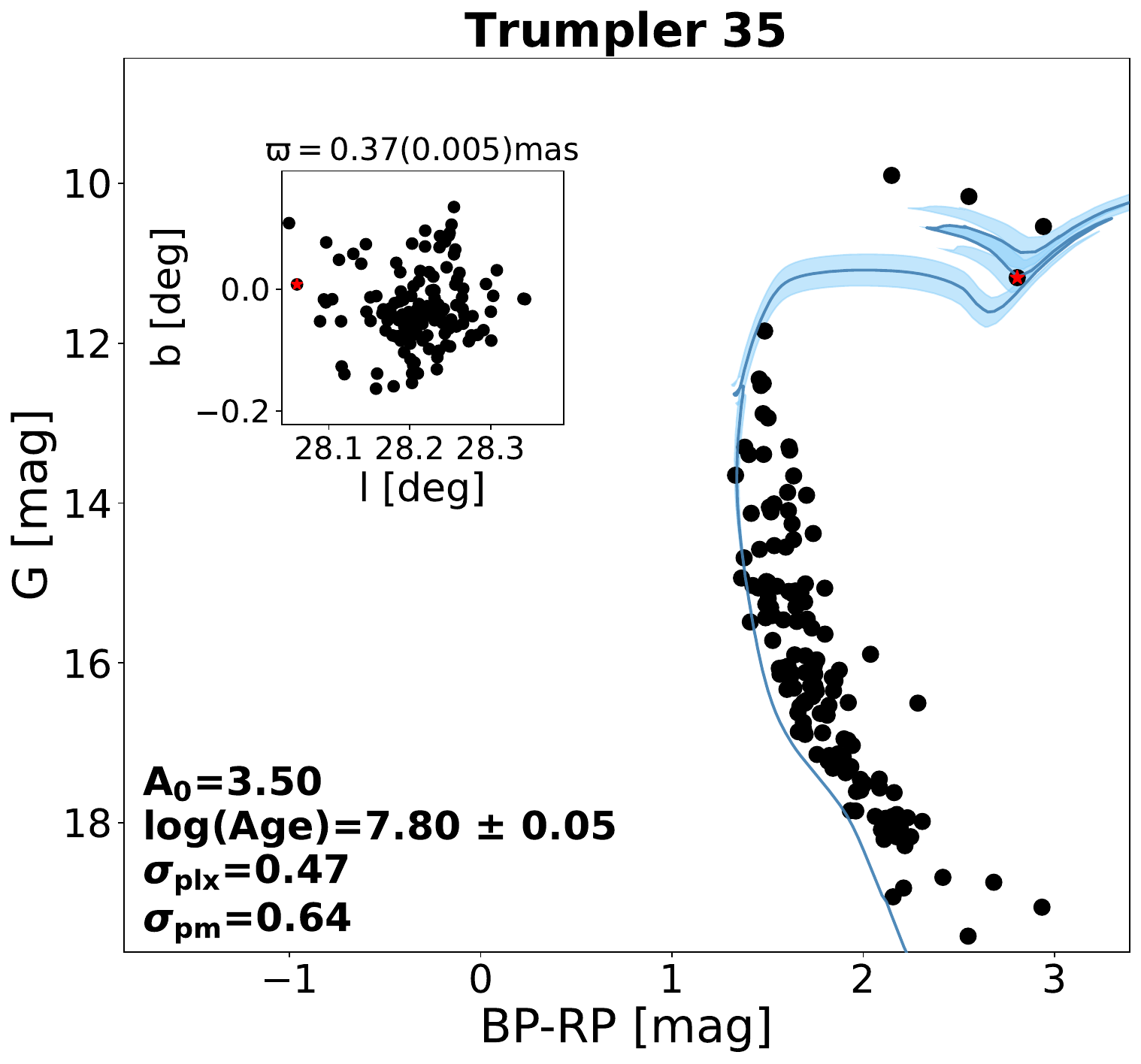}\hspace{0.02\linewidth}
\includegraphics[width=0.23\linewidth]{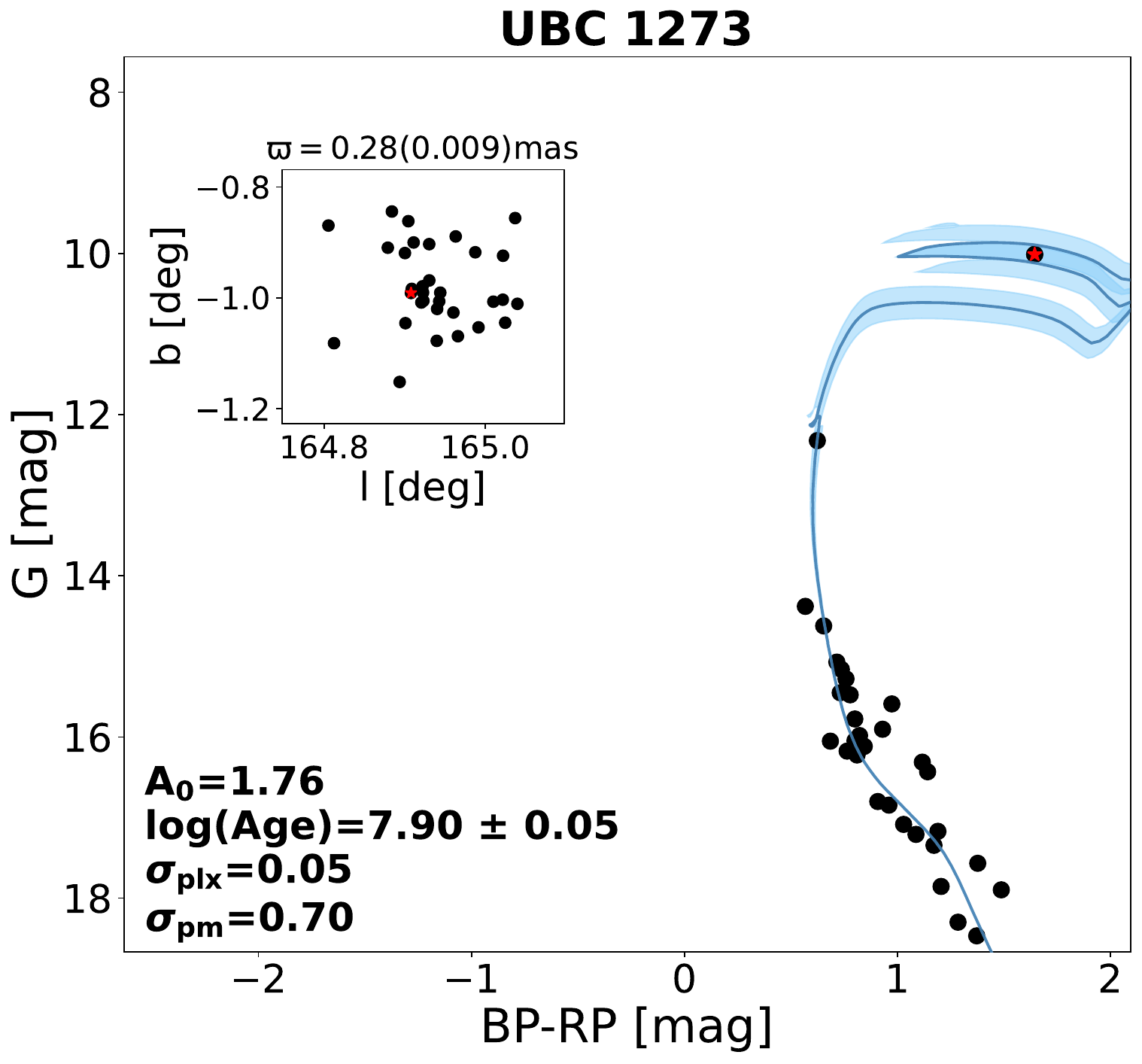}\\[0.4cm]

\includegraphics[width=0.23\linewidth]{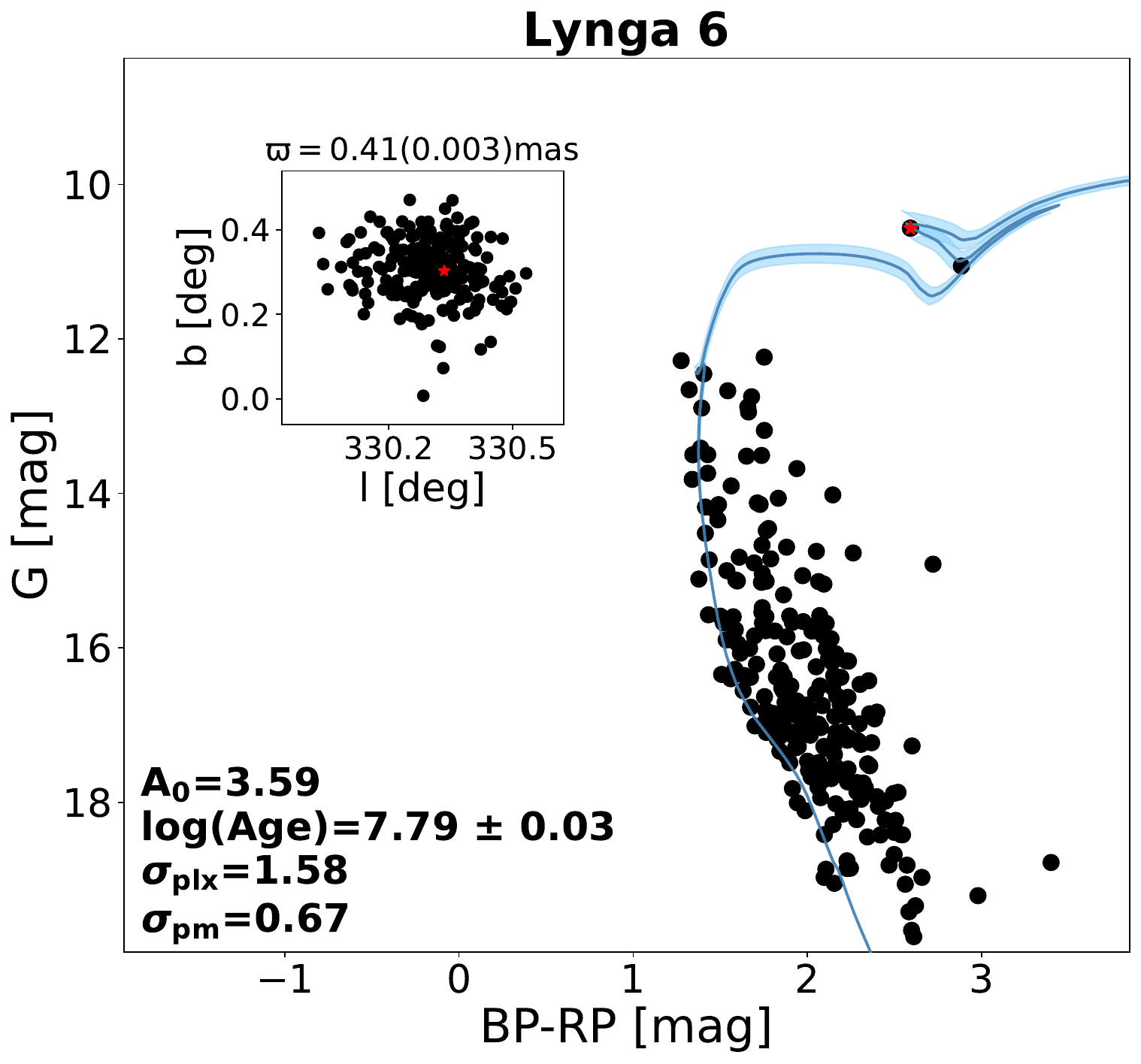}\hspace{0.02\linewidth}
\includegraphics[width=0.23\linewidth]{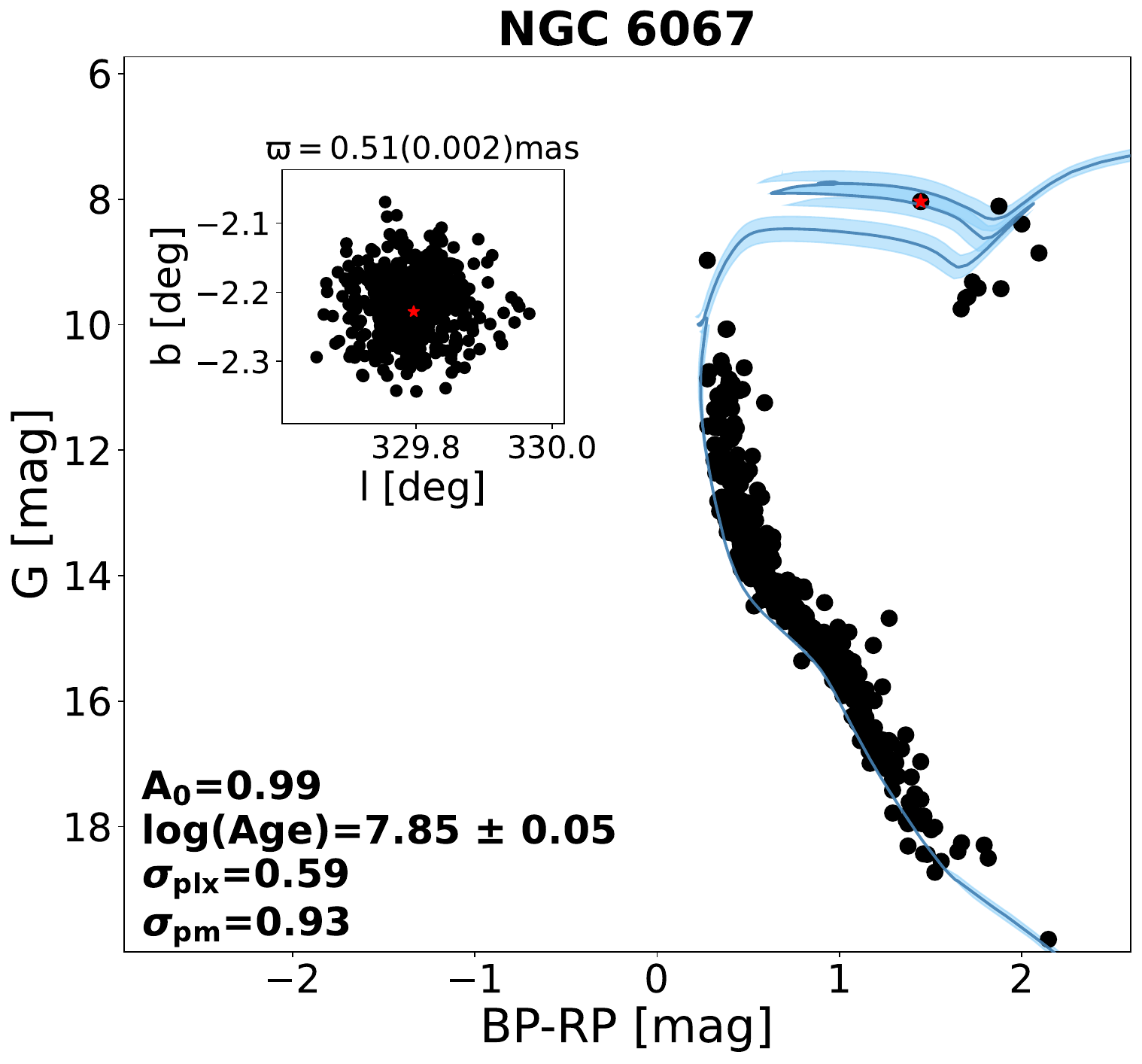}\hspace{0.02\linewidth}
\includegraphics[width=0.23\linewidth]{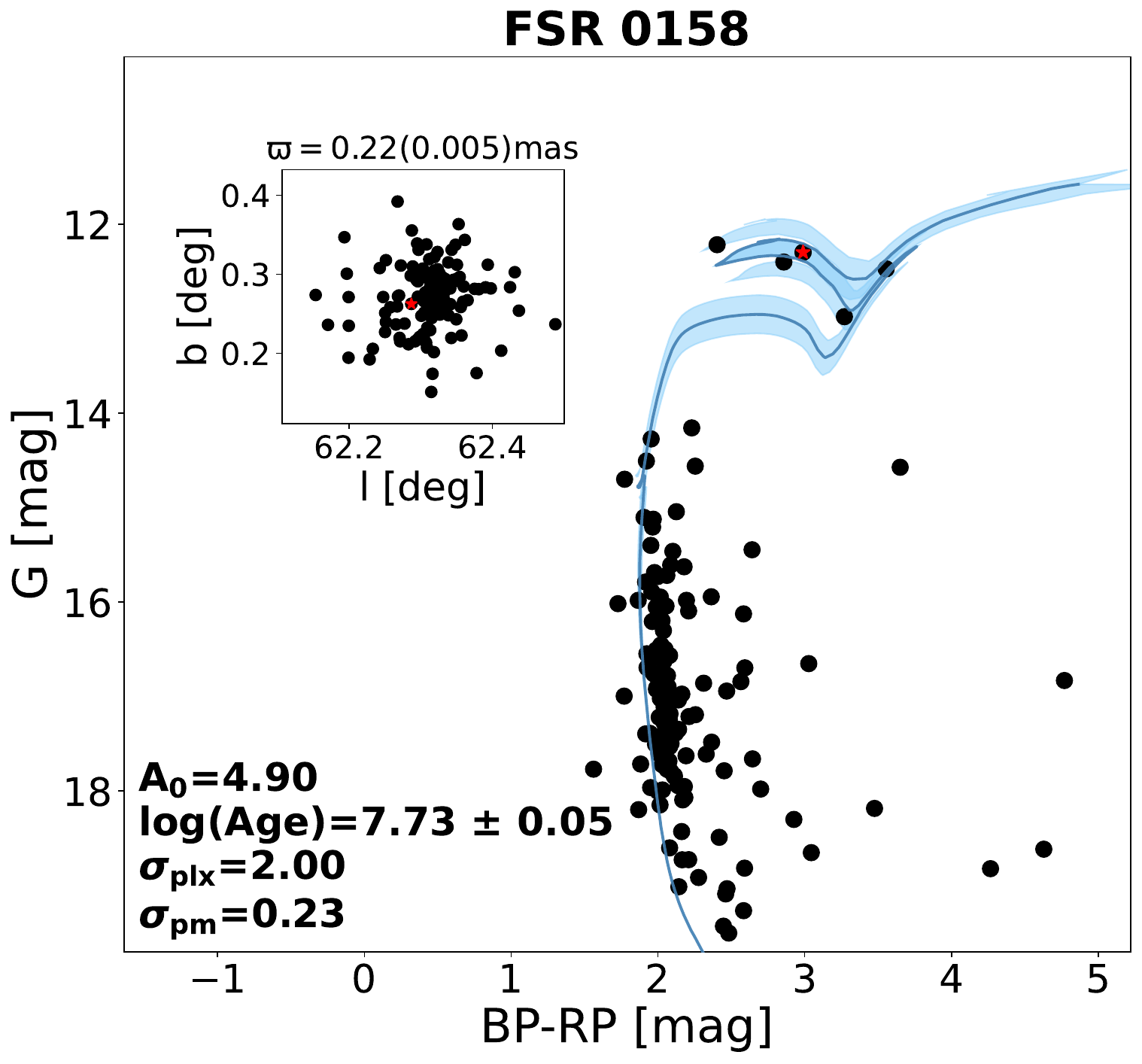}\hspace{0.02\linewidth}
\includegraphics[width=0.23\linewidth]{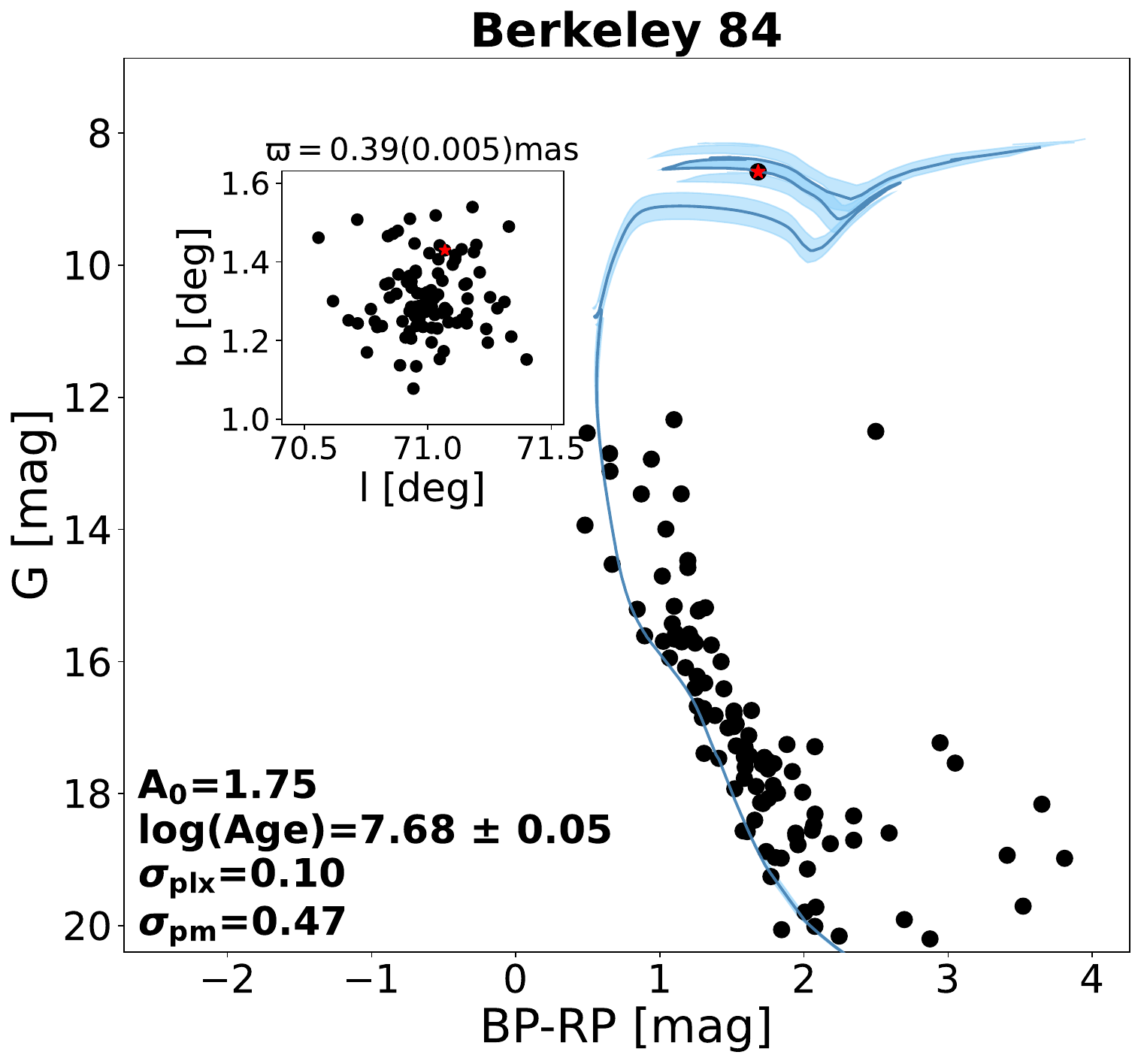}\\[0.4cm]

\includegraphics[width=0.23\linewidth]{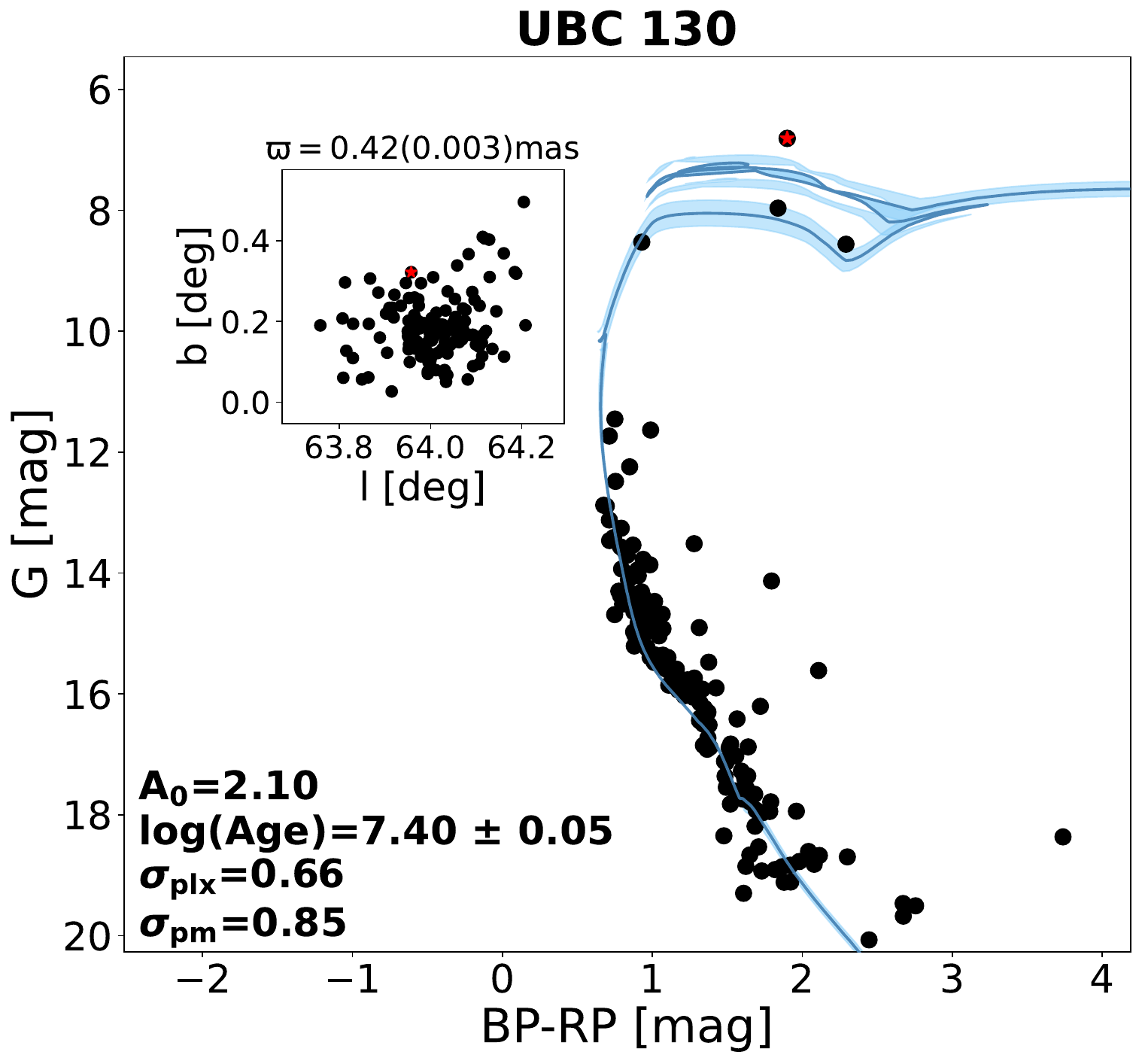}\hspace{0.02\linewidth}
\includegraphics[width=0.23\linewidth]{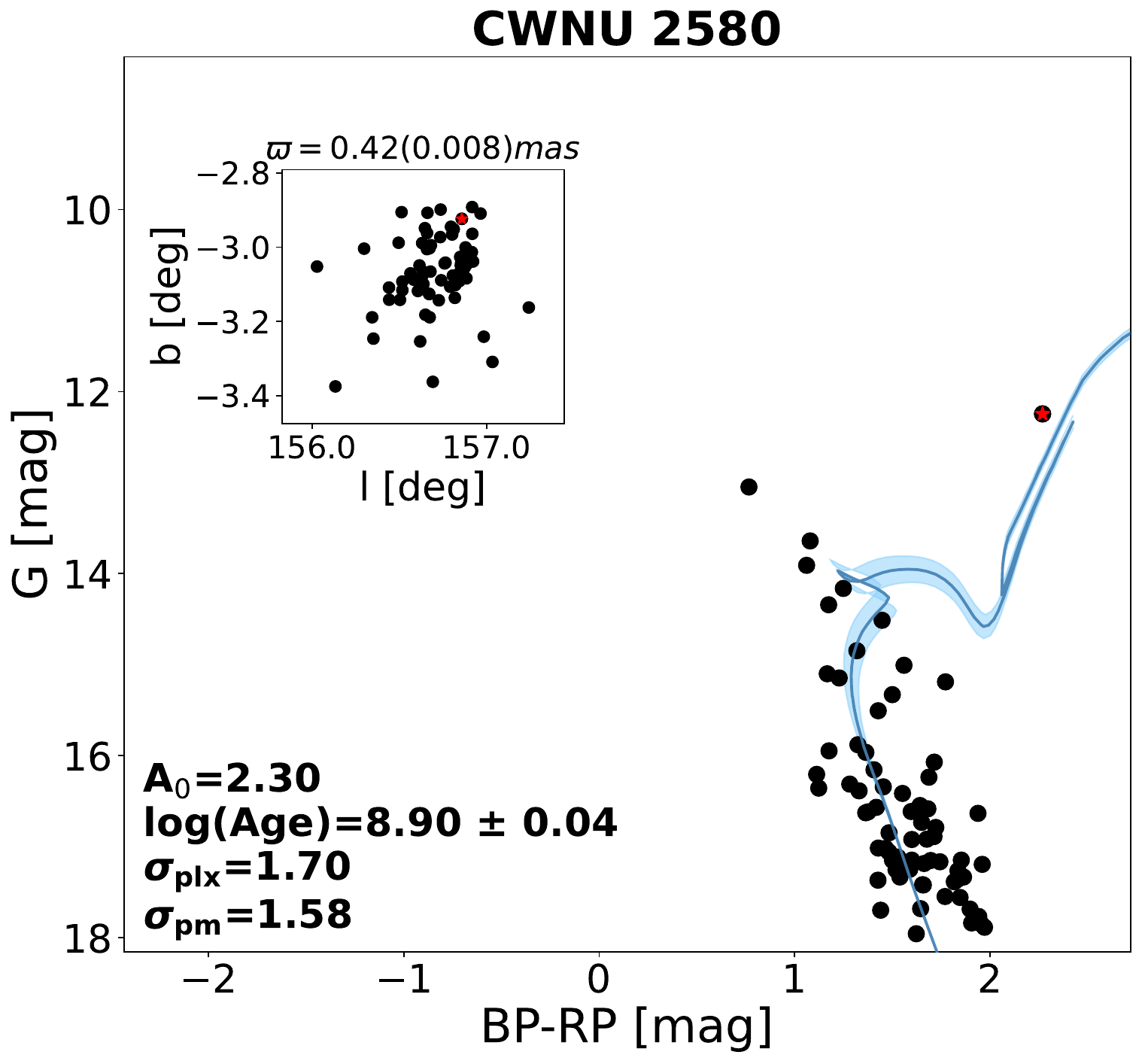}\hspace{0.02\linewidth}
\includegraphics[width=0.23\linewidth]{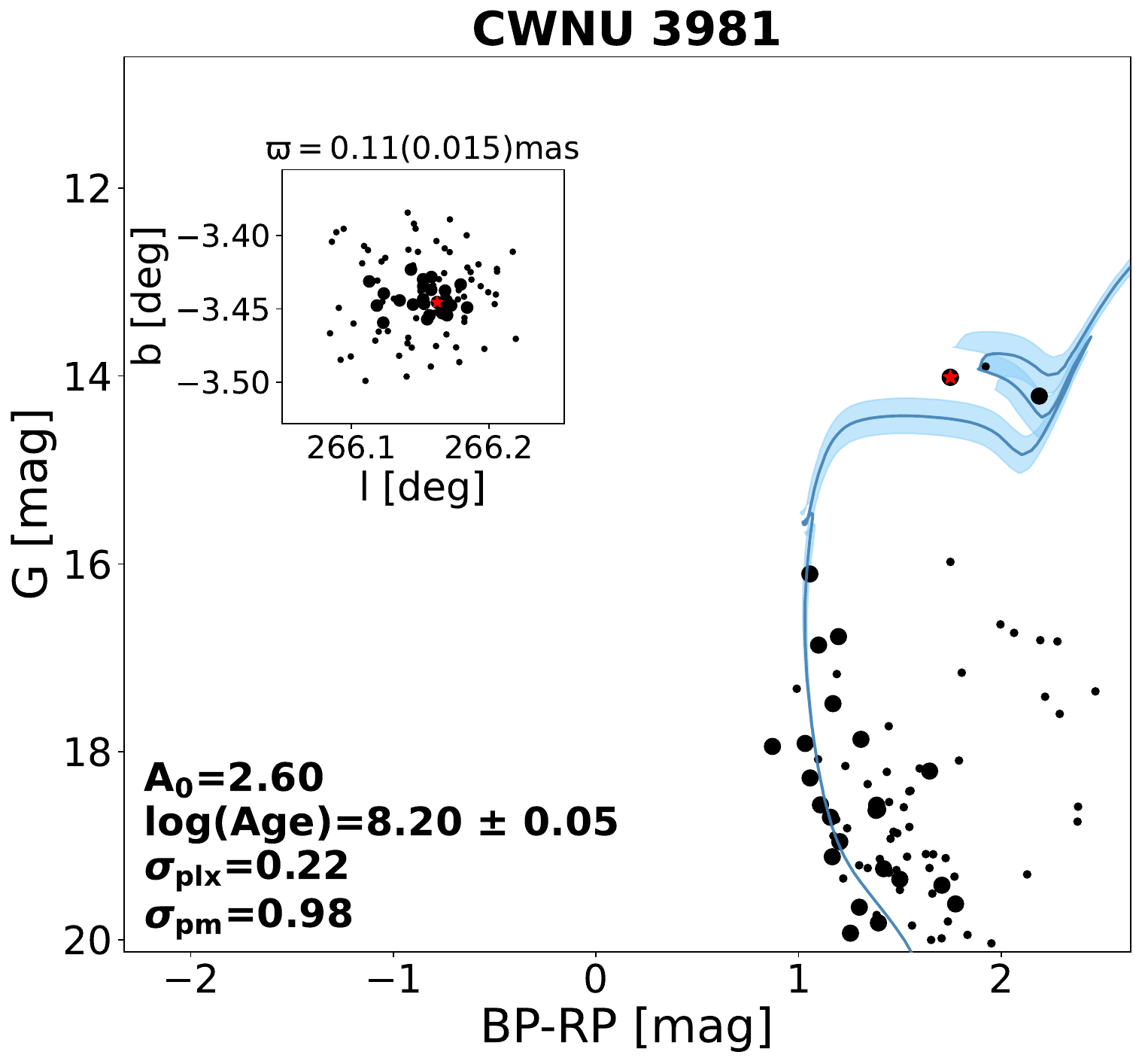}\hspace{0.02\linewidth}
\includegraphics[width=0.23\linewidth]{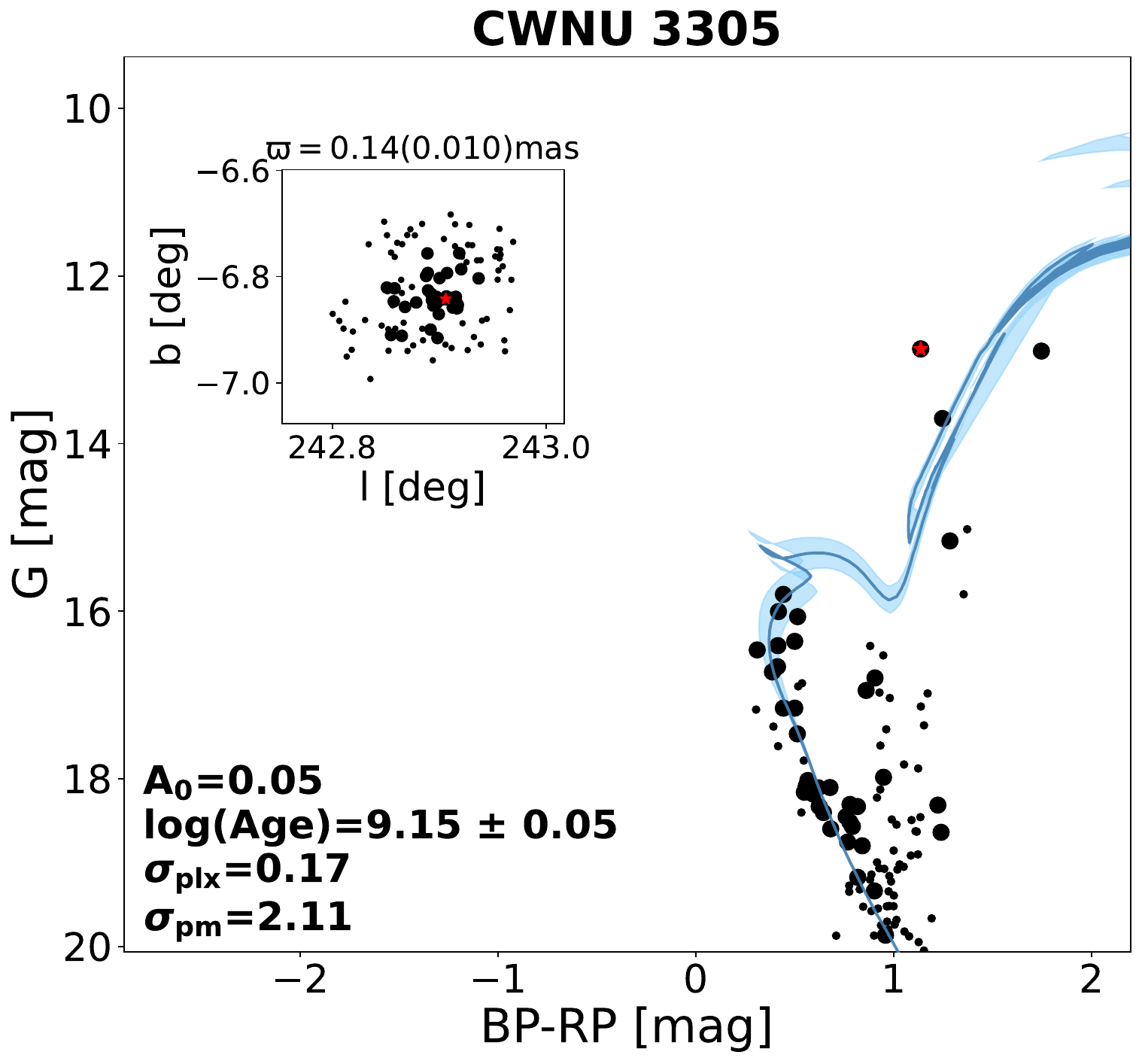}\\[0.4cm]

\caption{CMDs of the remaining part of OC Cepheids, which also contain the extinction, age and distance parameters of the OCs, namely, 
UBC 290, NGC 6649, UBC 129, FSR 0158 (J194806.54+260526.1), IC 4725, FSR 0951, NGC 129 (DL Cas), CWNU 19, NGC 6087, 
Berkeley 51, Trumpler 35 (CN Sct), UBC 1273, Lynga 6, NGC 6067, FSR 0158 (GQ Vul), Berkeley 84,and UBC 130. 
CMDs of OC Cepheid candidates, namely, CWNU 2580, CWNU 3981, and CWNU 3305.}
\label{fig:cmd_set2}
\end{figure*}
\begin{figure*}[htbp]
\centering
\includegraphics[width=0.23\linewidth]{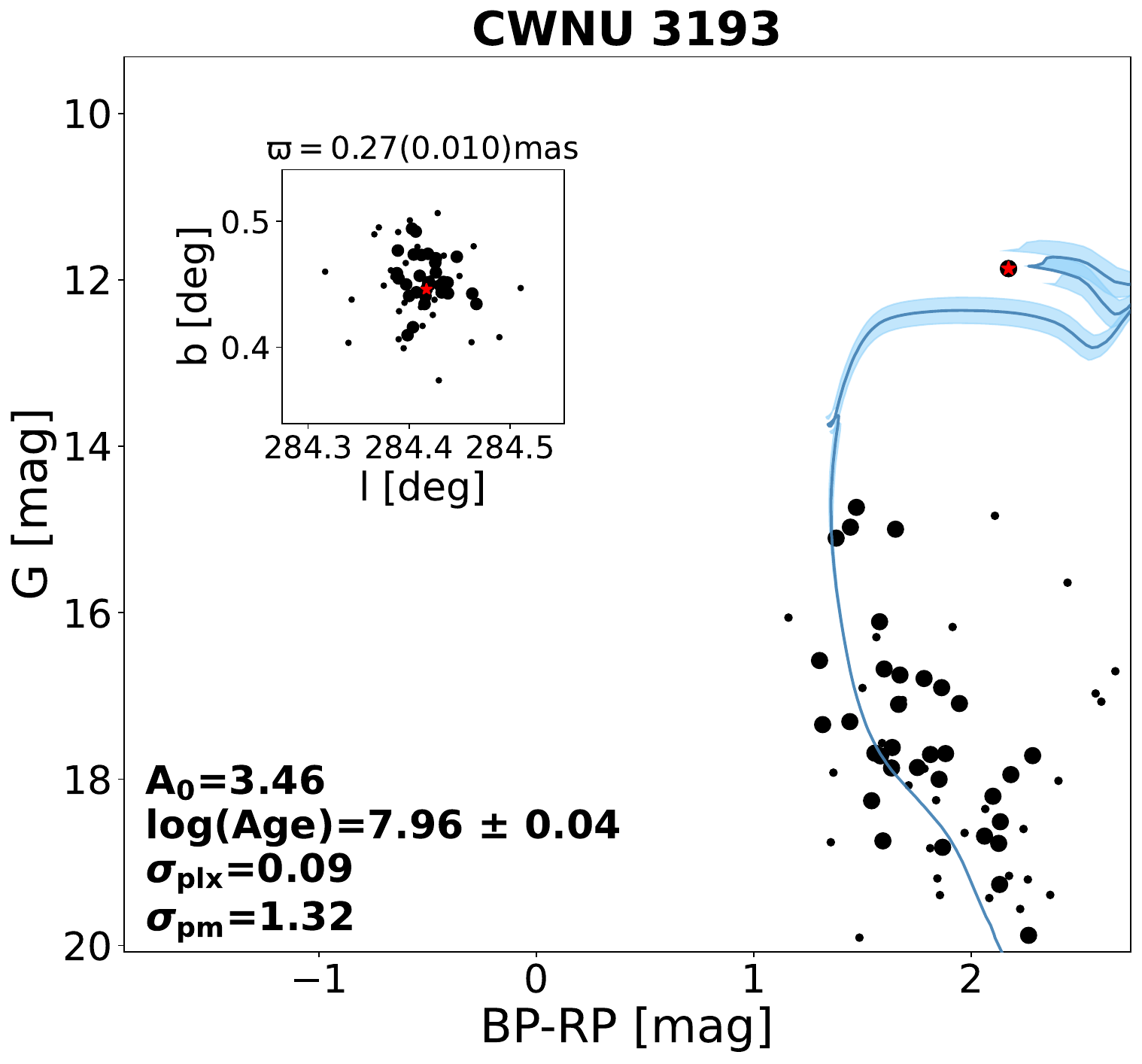}\hspace{0.02\linewidth}
\includegraphics[width=0.23\linewidth]{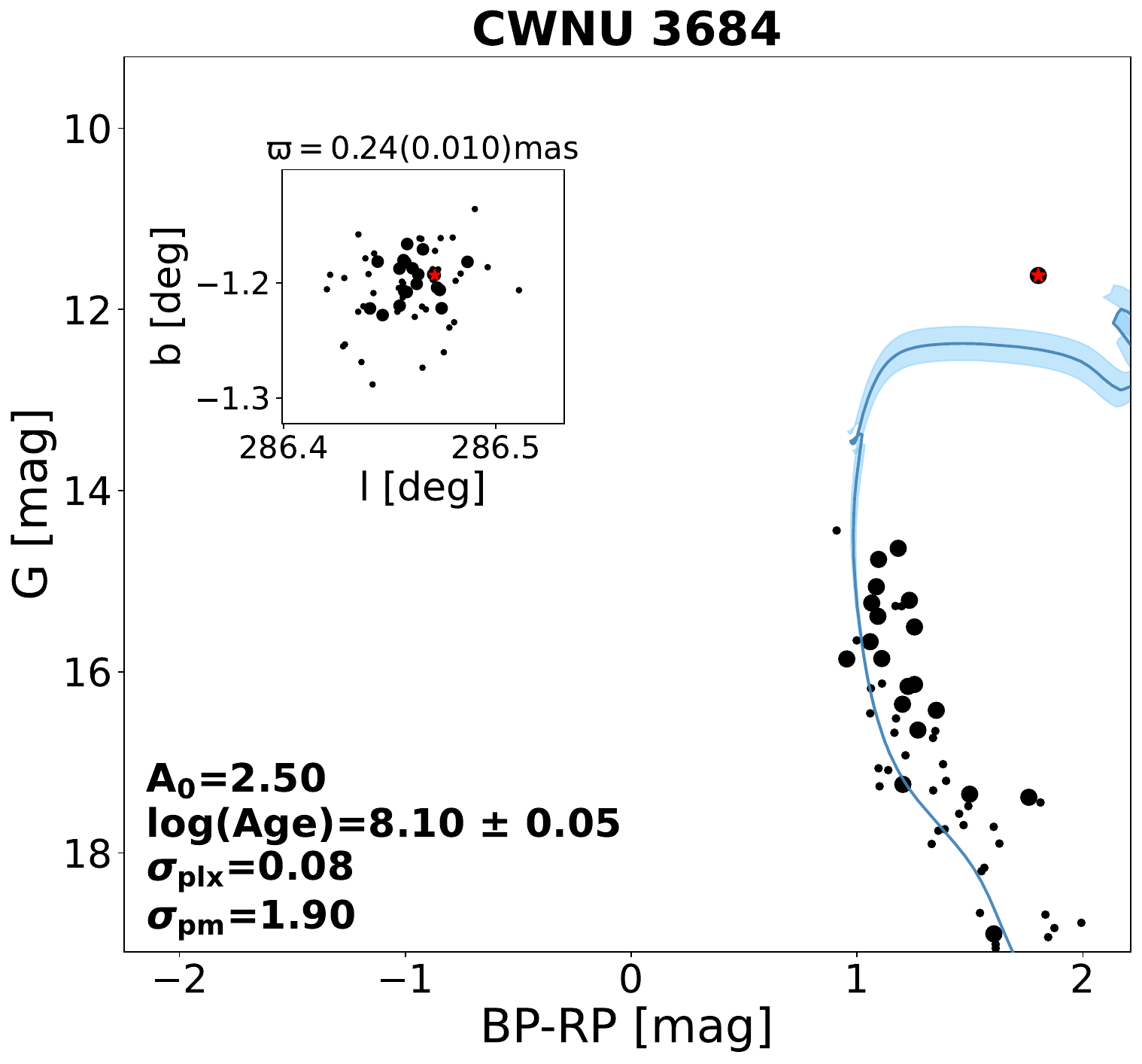}\hspace{0.02\linewidth}
\includegraphics[width=0.23\linewidth]{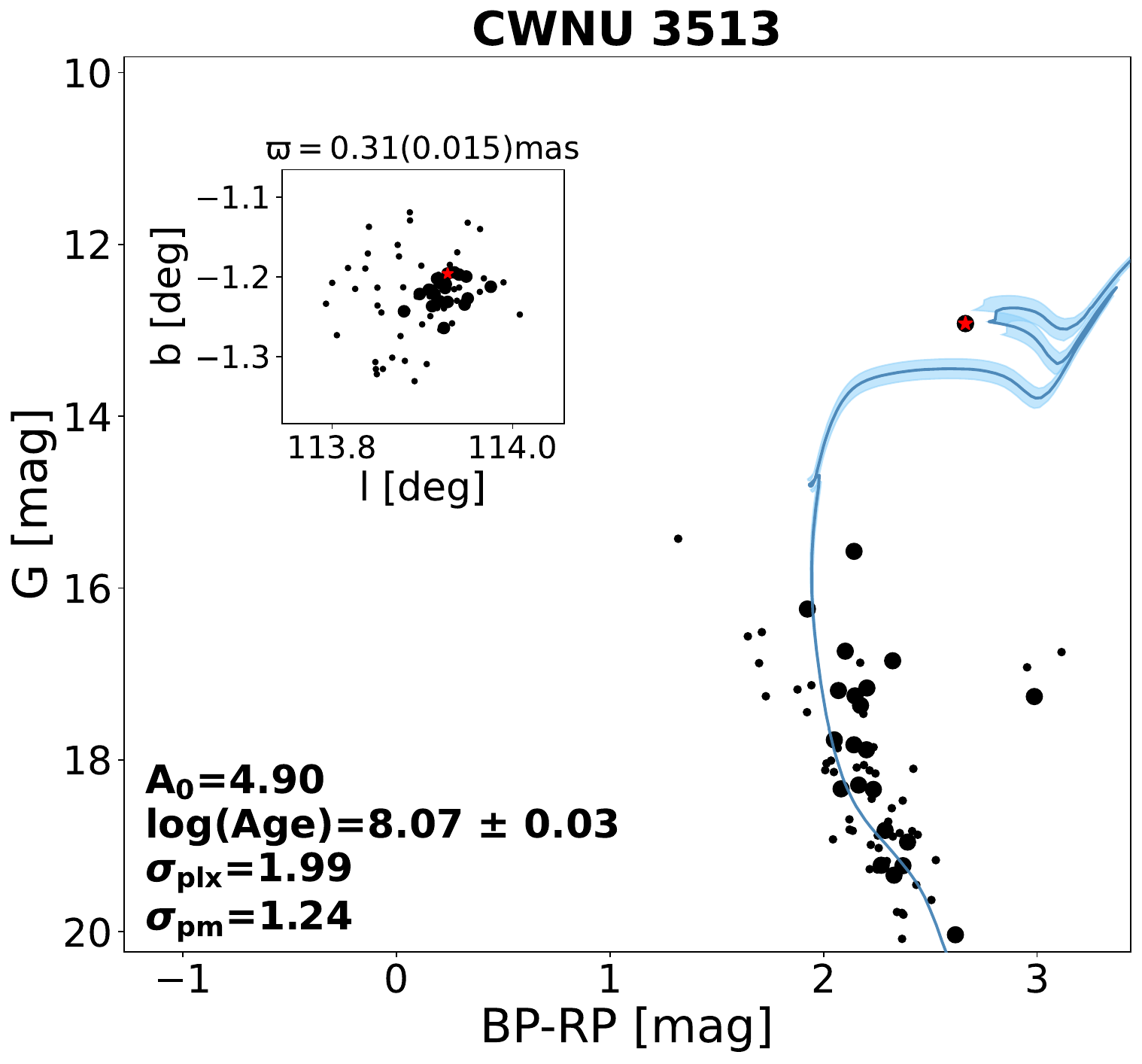}\hspace{0.02\linewidth}
\includegraphics[width=0.23\linewidth]{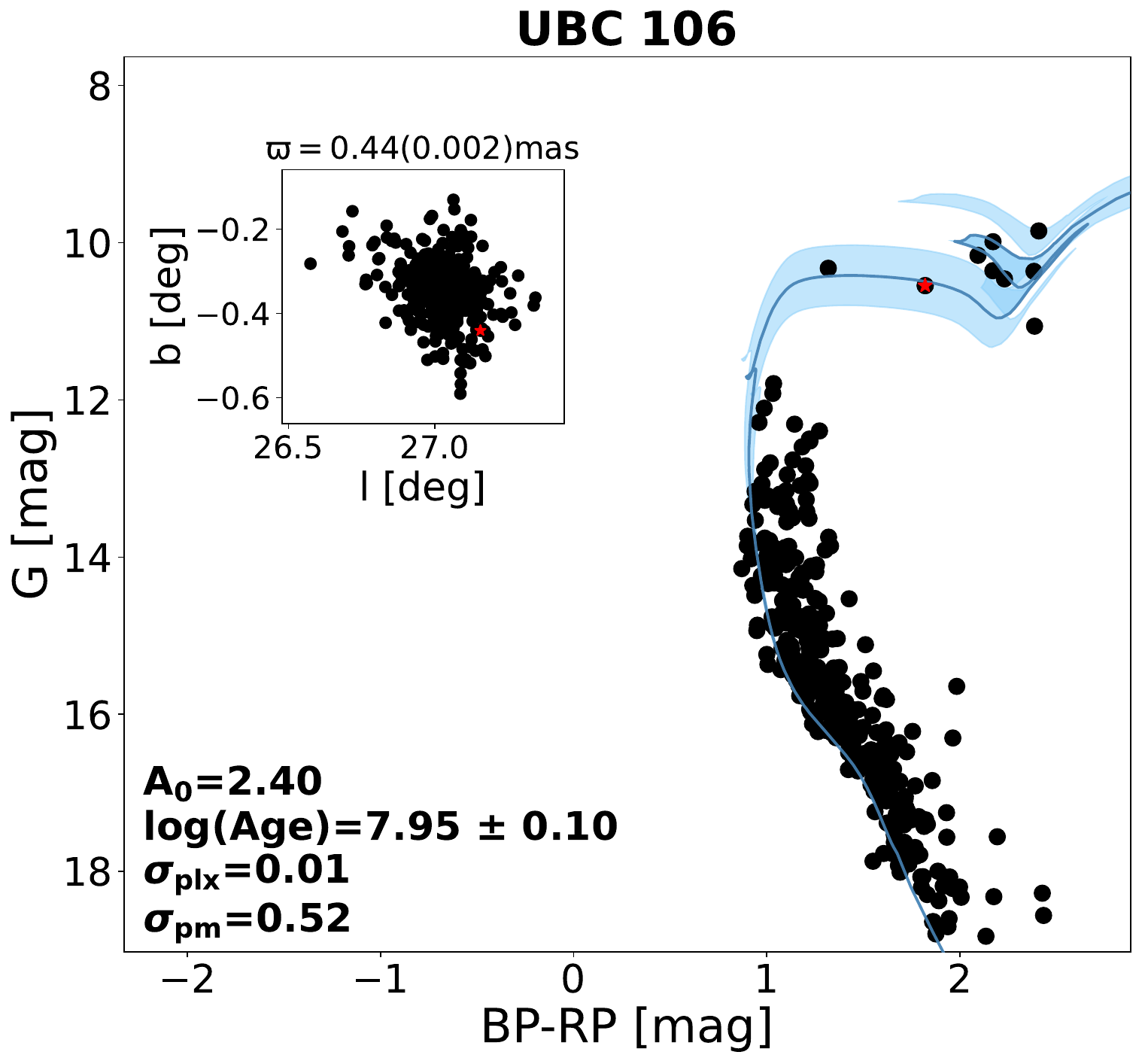}\\[0.4cm]

\includegraphics[width=0.23\linewidth]{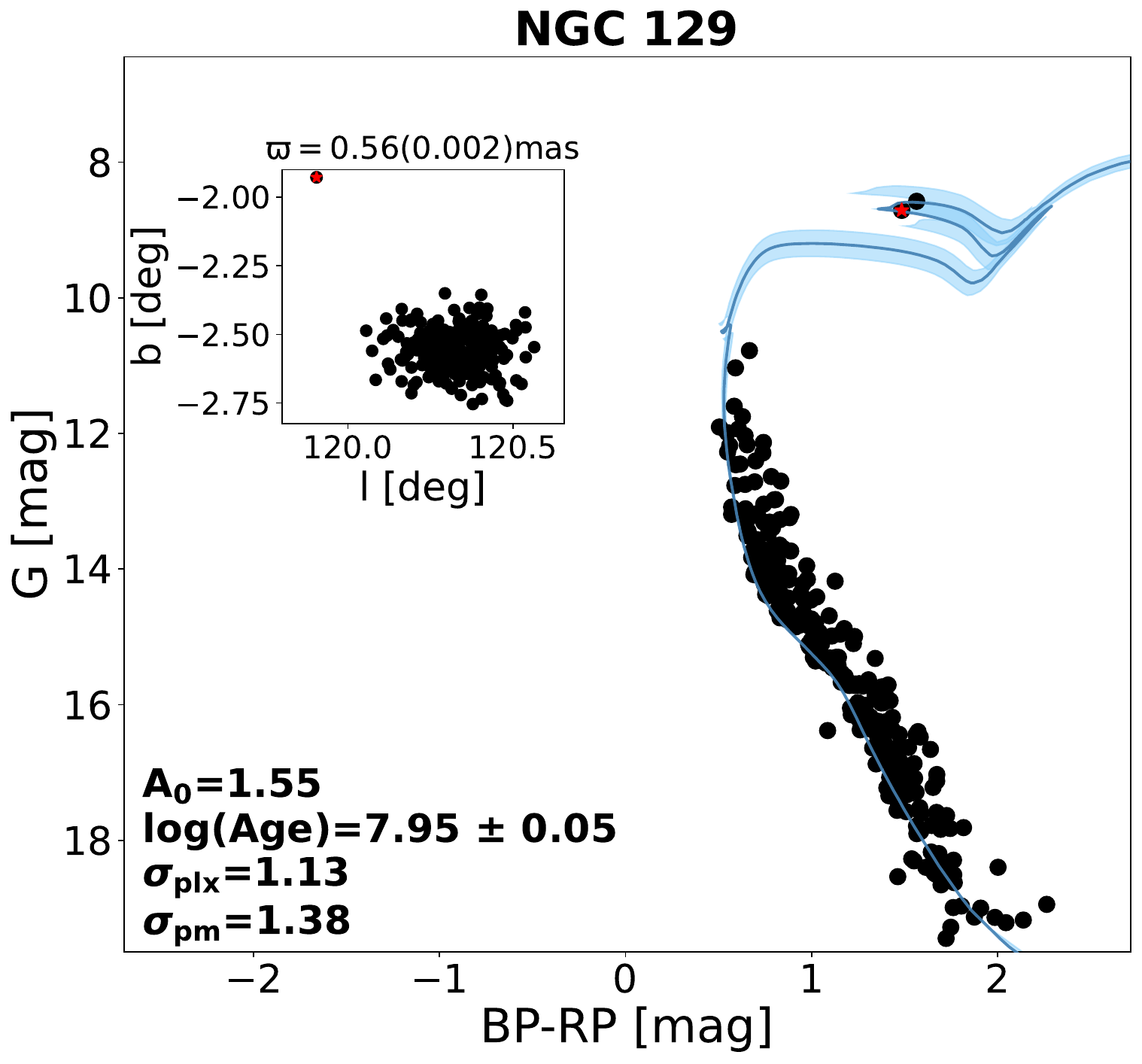}\hspace{0.02\linewidth}
\includegraphics[width=0.23\linewidth]{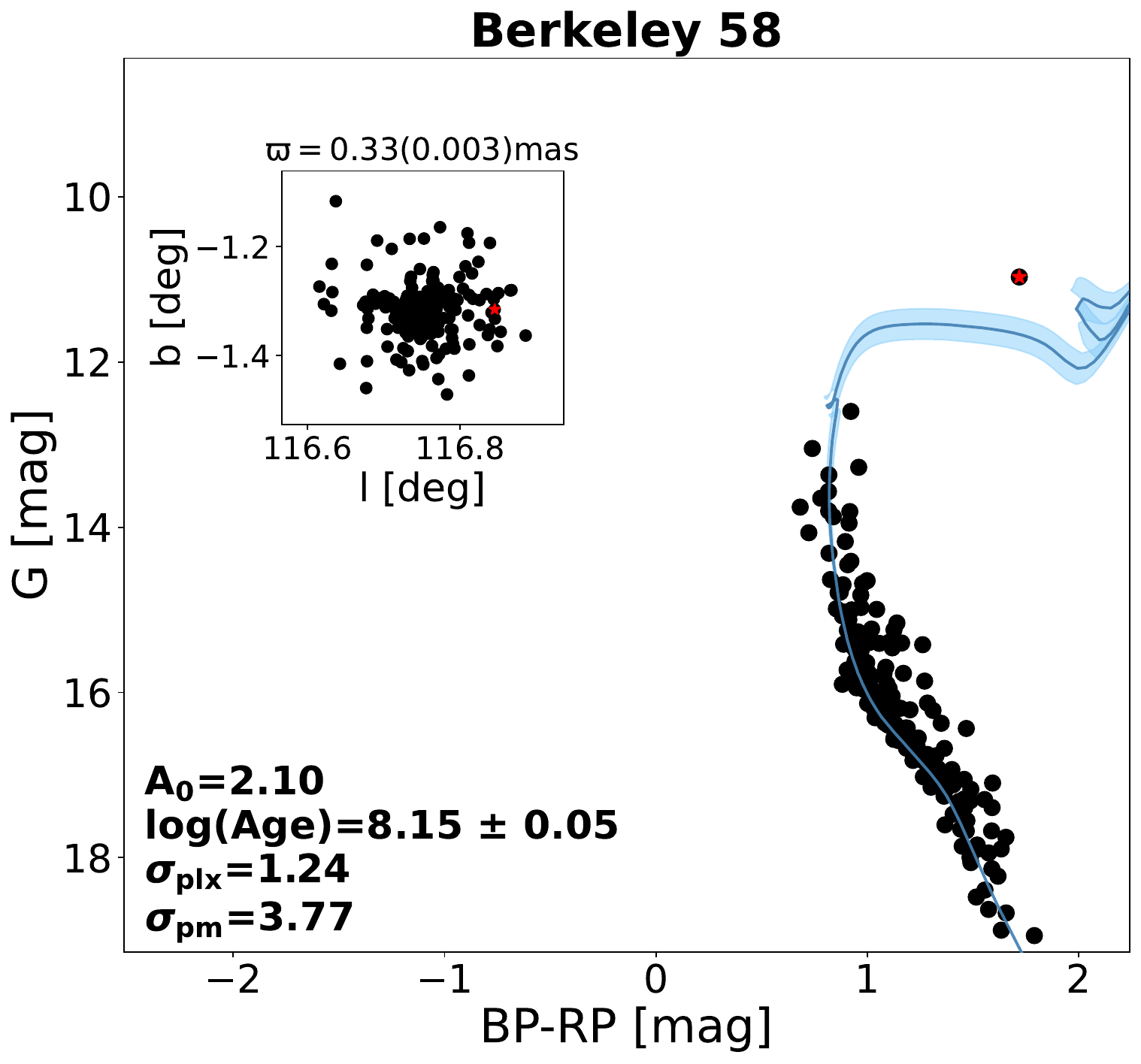}\hspace{0.02\linewidth}
\includegraphics[width=0.23\linewidth]{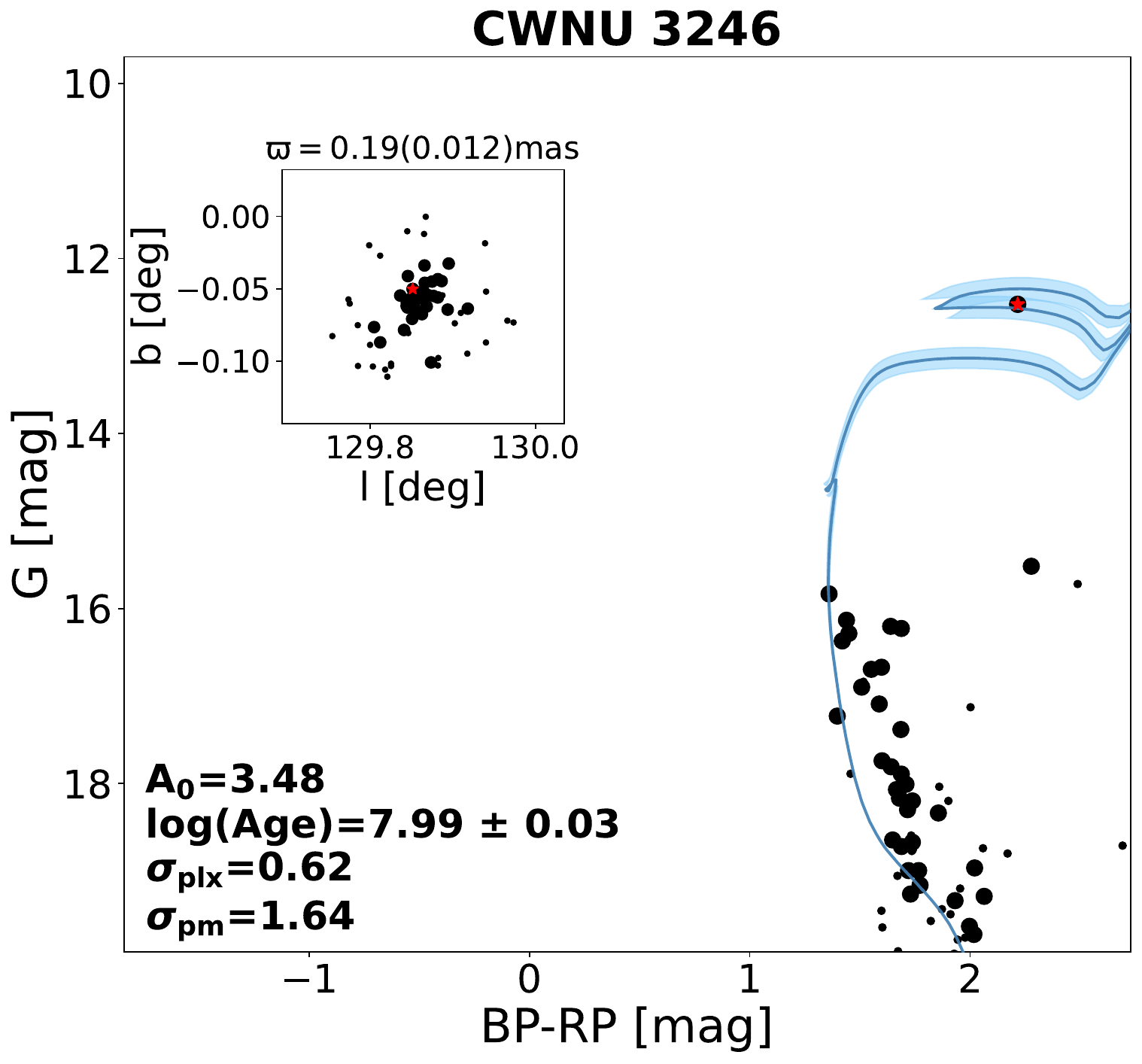}\hspace{0.02\linewidth}
\includegraphics[width=0.23\linewidth]{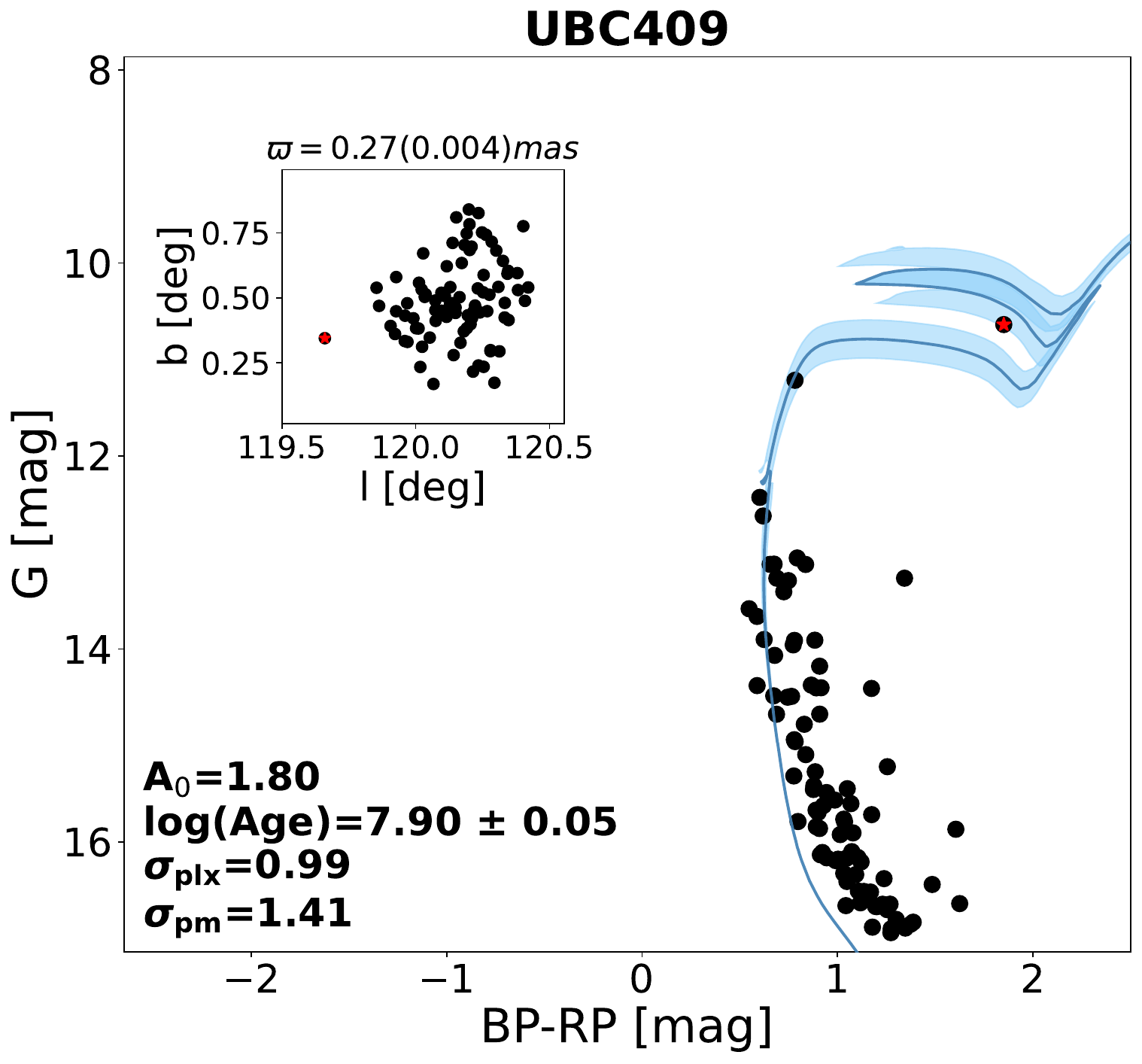}\\[0.4cm]

\includegraphics[width=0.23\linewidth]{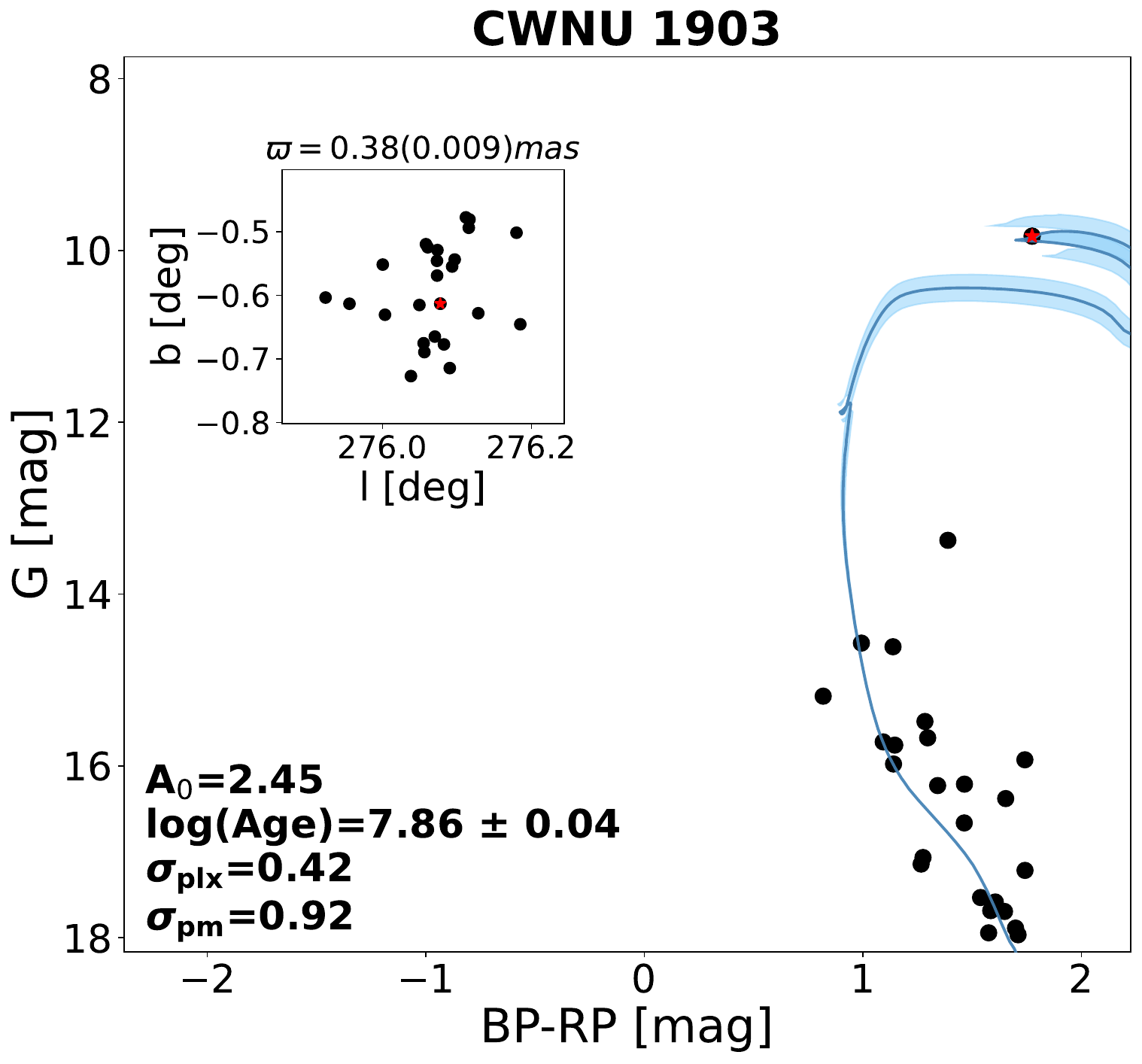}\hspace{0.02\linewidth}
\includegraphics[width=0.23\linewidth]{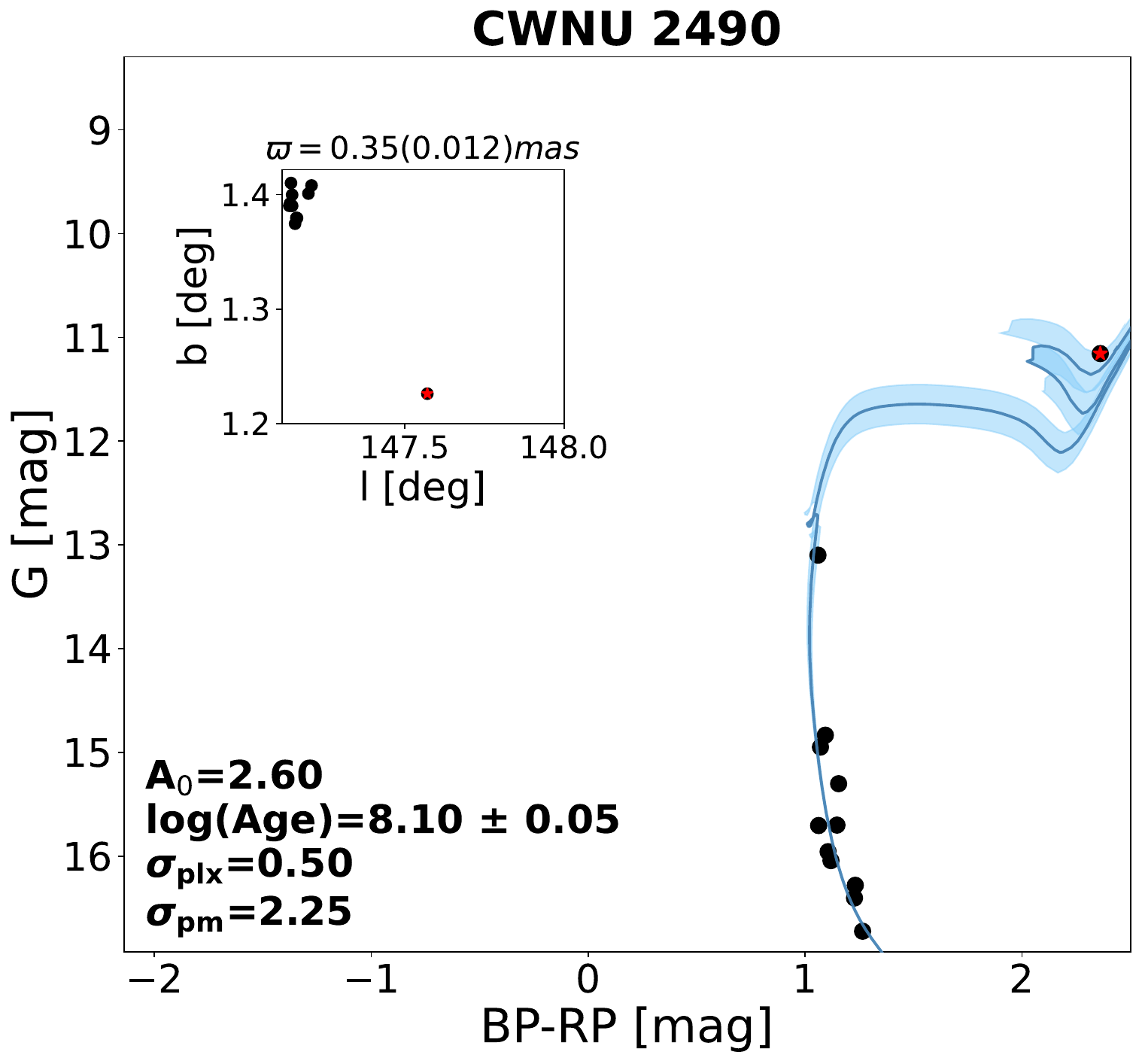}\hspace{0.02\linewidth}
\includegraphics[width=0.23\linewidth]{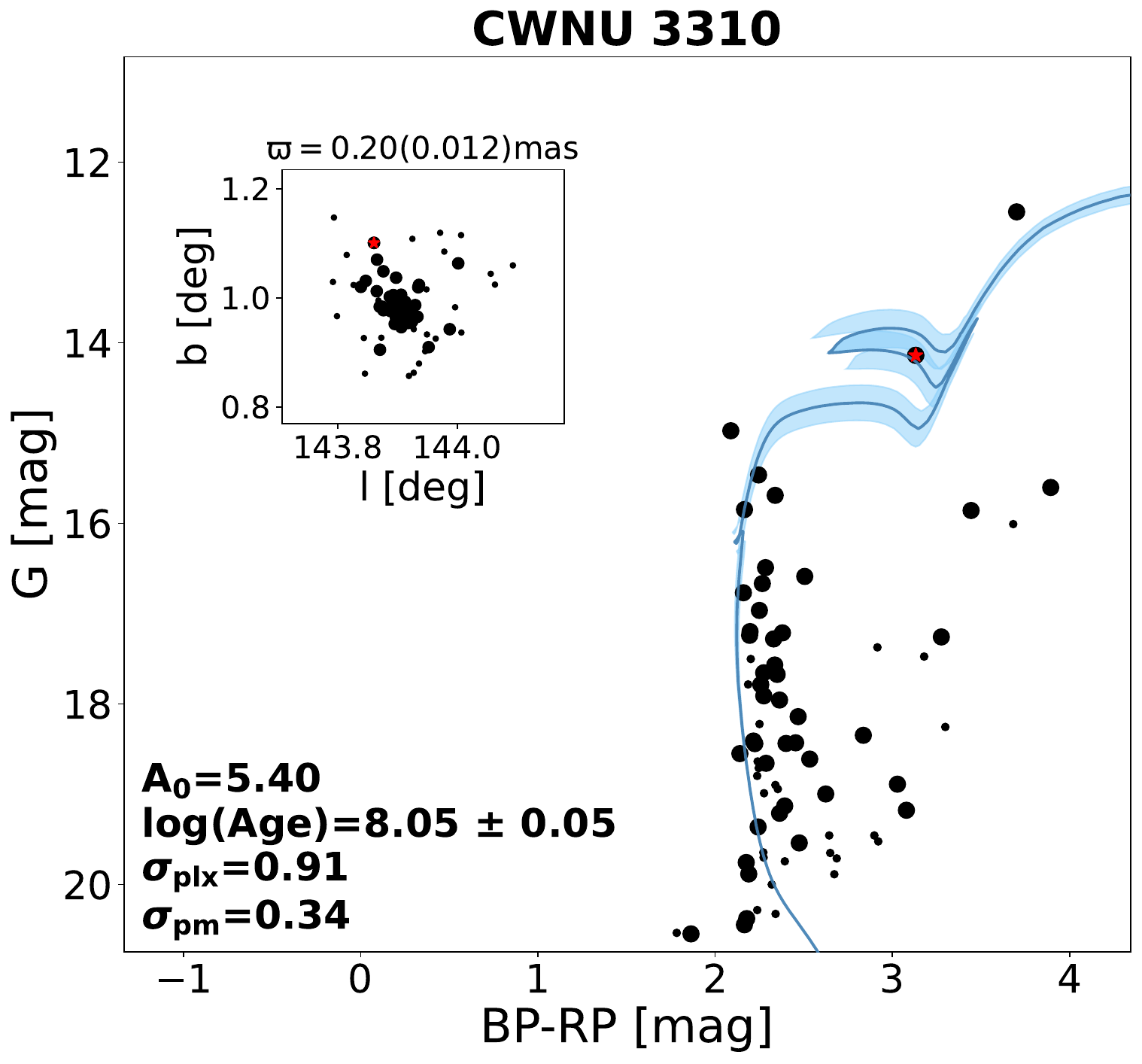}\hspace{0.02\linewidth}
\includegraphics[width=0.23\linewidth]{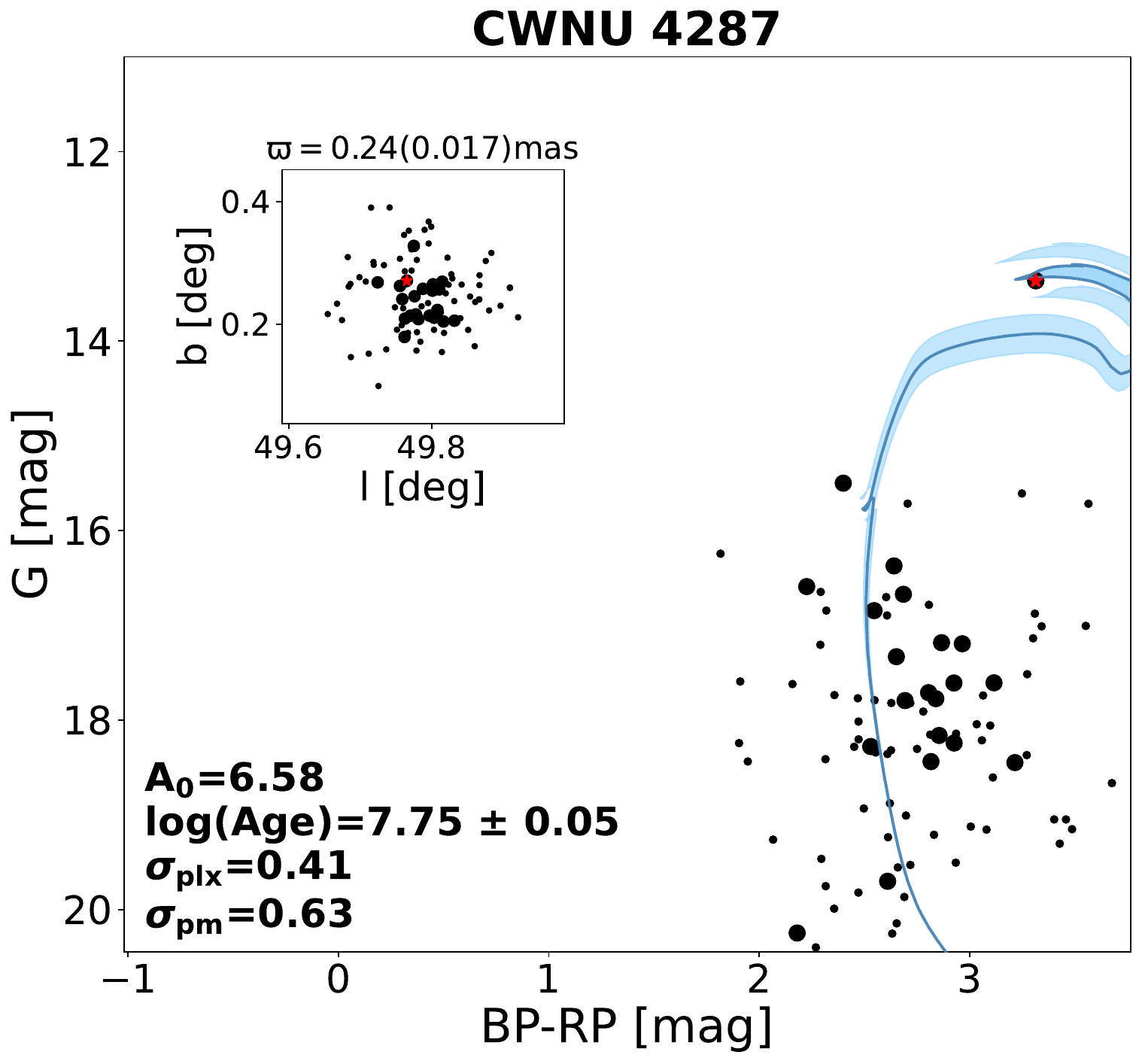}\\[0.4cm]

\includegraphics[width=0.23\linewidth]{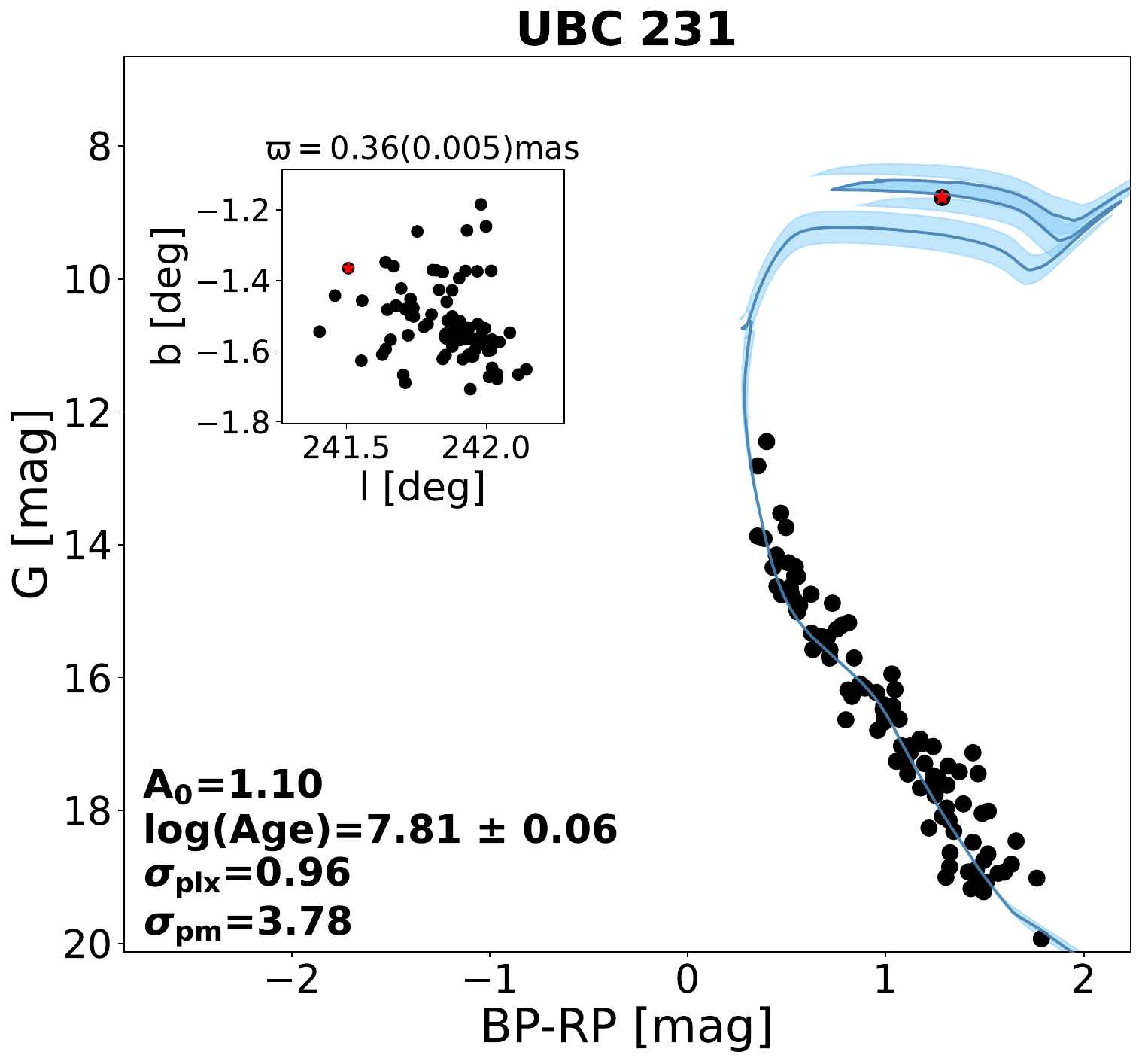}\hspace{0.02\linewidth}
\includegraphics[width=0.23\linewidth]{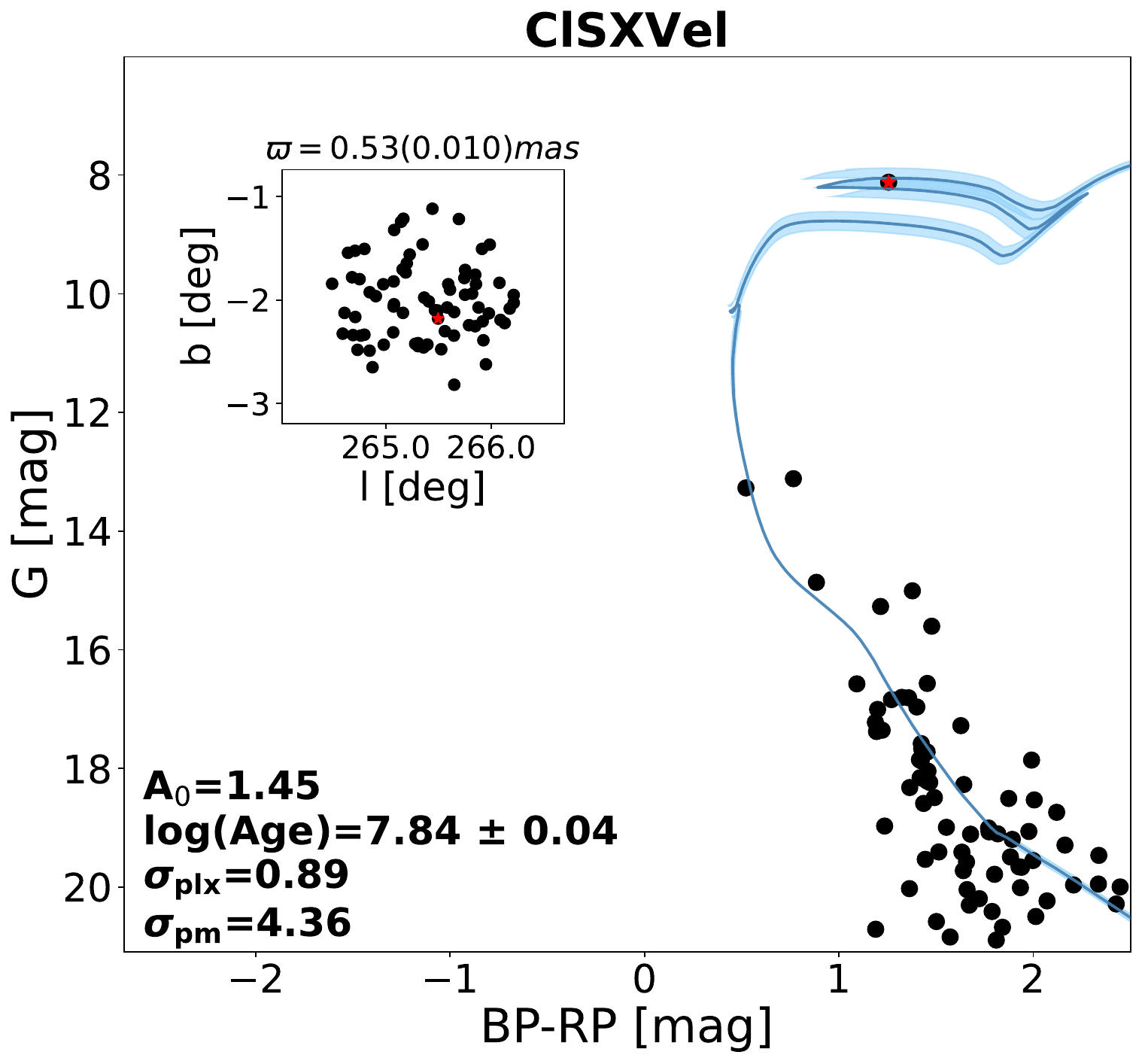}\hspace{0.02\linewidth}
\includegraphics[width=0.23\linewidth]{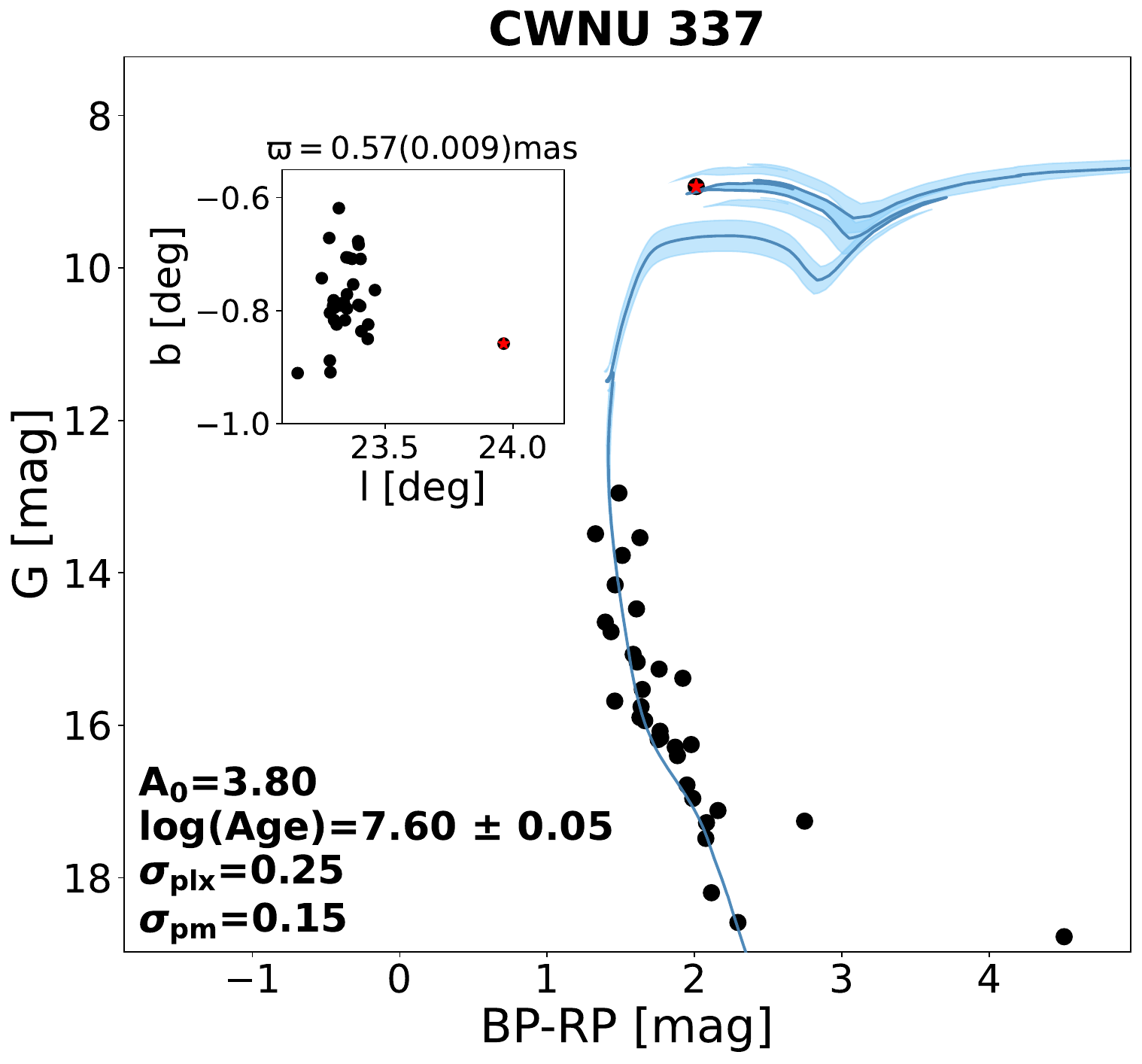}\hspace{0.02\linewidth}
\includegraphics[width=0.23\linewidth]{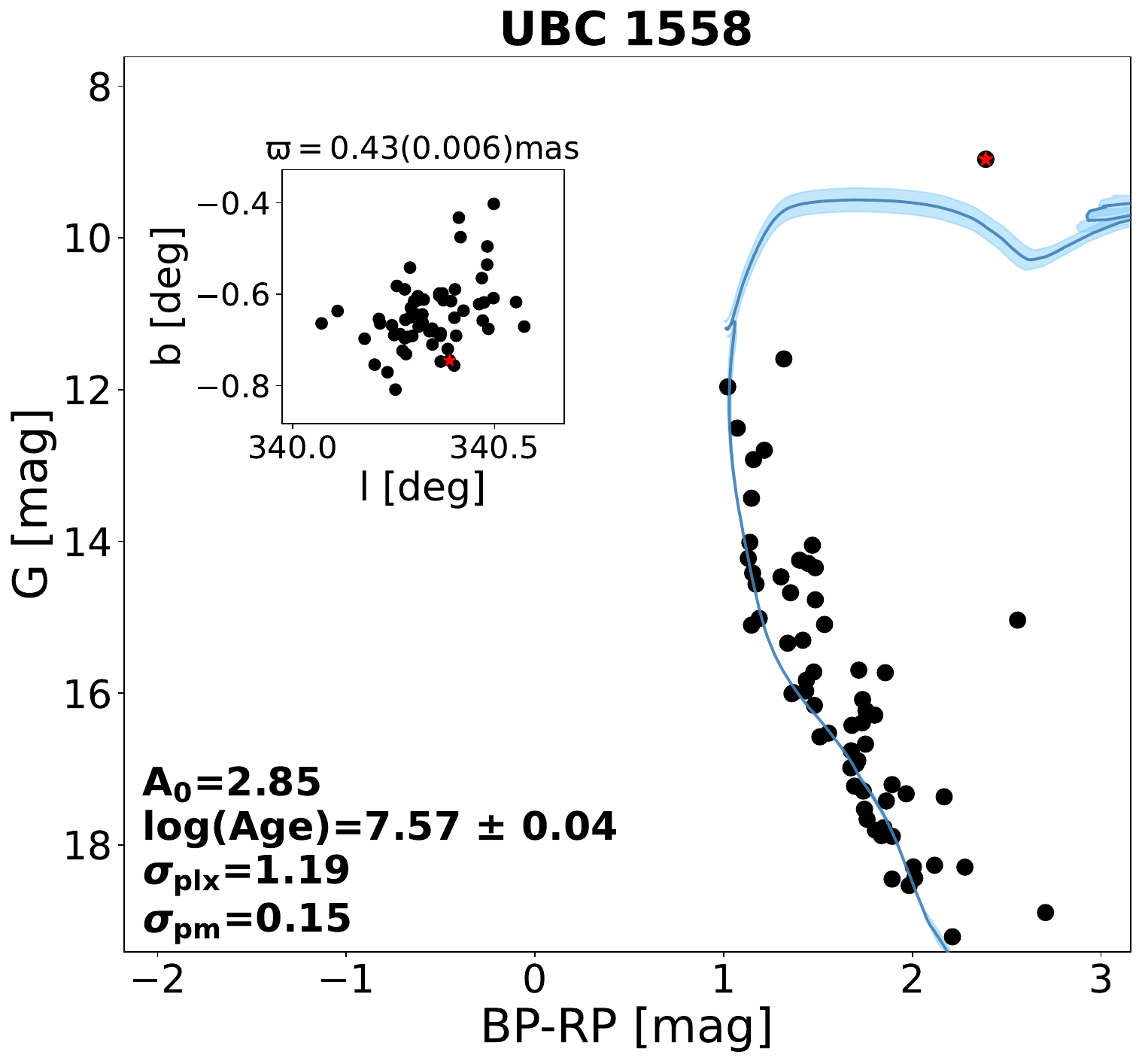}\\[0.4cm]

\includegraphics[width=0.23\linewidth]{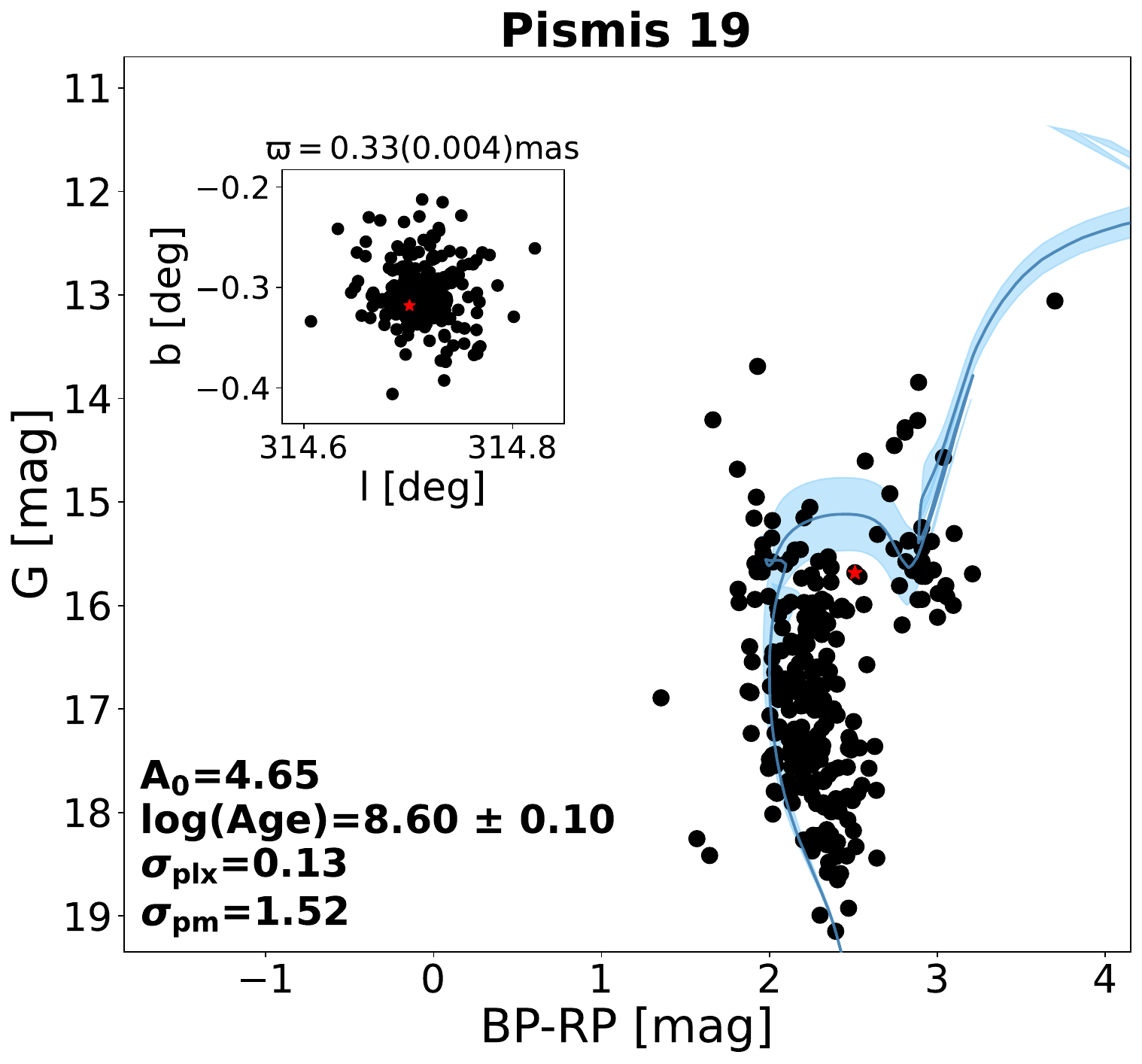}\hspace{0.02\linewidth}
\includegraphics[width=0.23\linewidth]{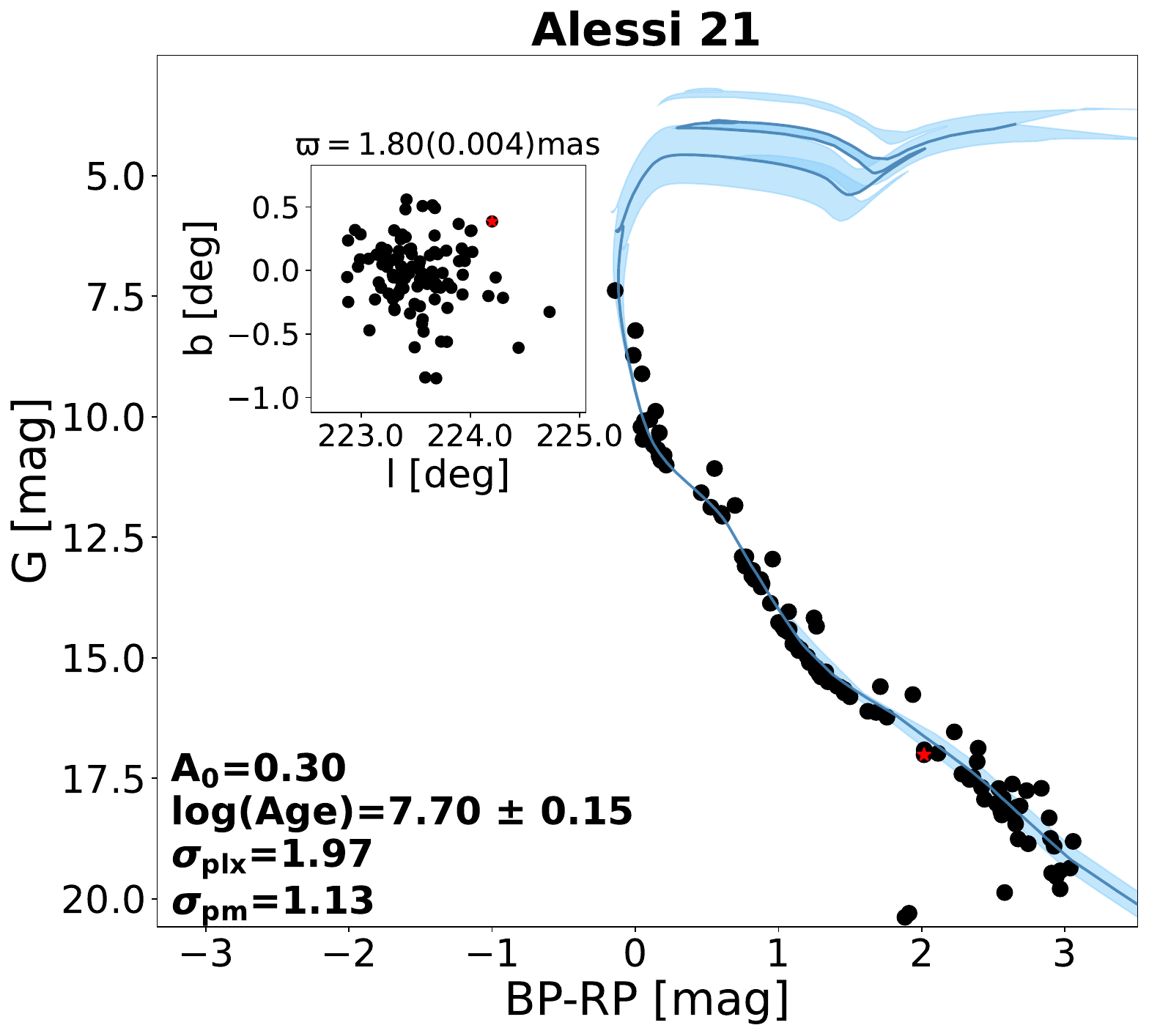}\hspace{0.02\linewidth}
\includegraphics[width=0.23\linewidth]{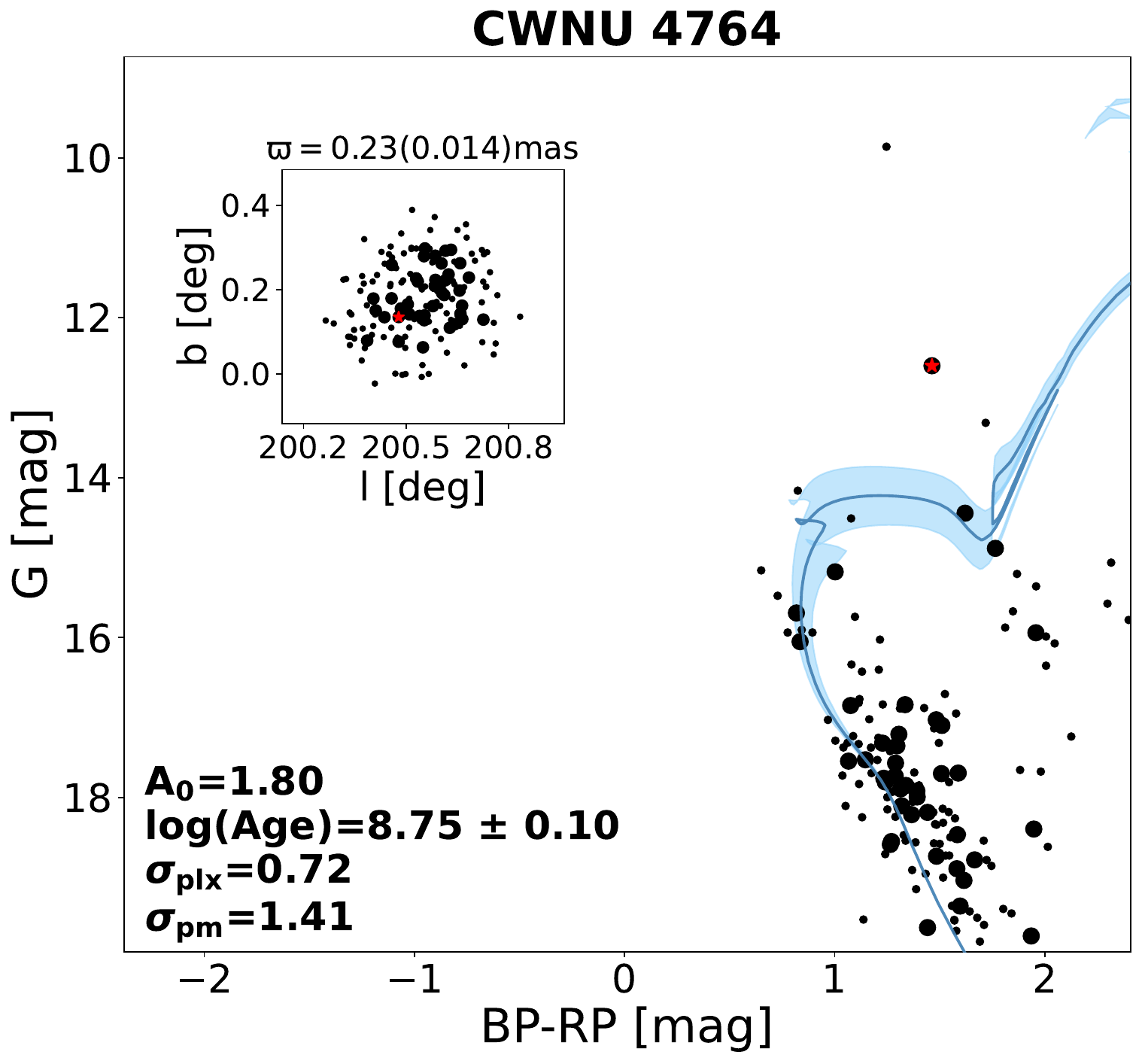}\hspace{0.02\linewidth}
\includegraphics[width=0.23\linewidth]{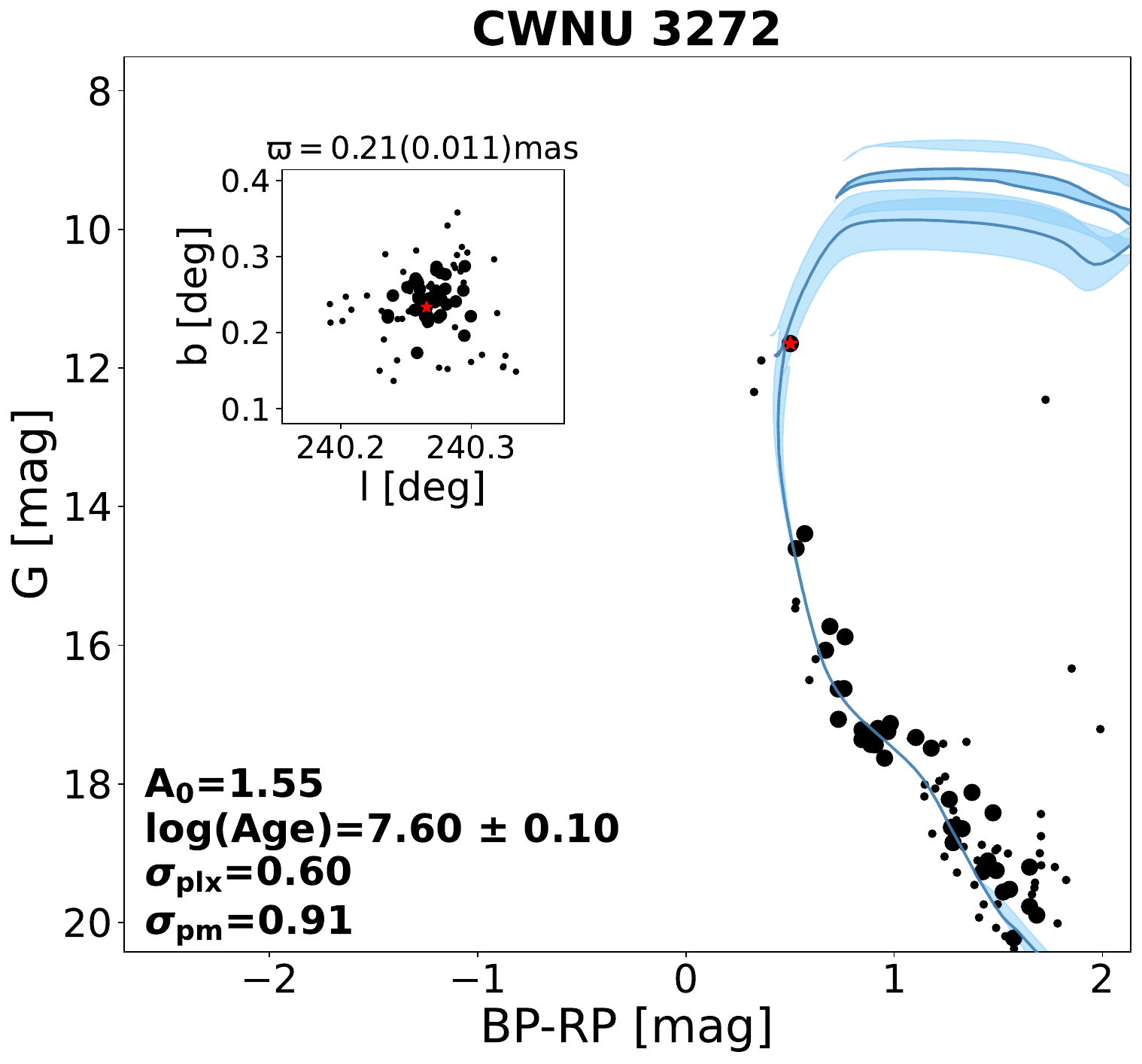}\\[0.4cm]

\caption{CMDs of the remaining part of OC Cepheid candidates, which also contain the extinction, age and distance parameters of the OCs, namely,
CWNU 3193, CWNU 3684, CWNU 3513, UBC 106, NGC 129 (V0379 Cas), Berkeley 58, CWNU 3246, UBC 409, CWNU 1903, CWNU 2490, 
CWNU 3310, CWNU 4287, UBC 231, Cl SX Vel, CWNU 337,and UBC 1558. CMDs of the remaining rejected OC Cepheid samples, namely, Pismis 19, Alessi 21, CWNU 4764, and CWNU 3272.}
\label{fig:cmd_set3}
\end{figure*}
\begin{figure*}[htbp]
\centering
\includegraphics[width=0.23\linewidth]{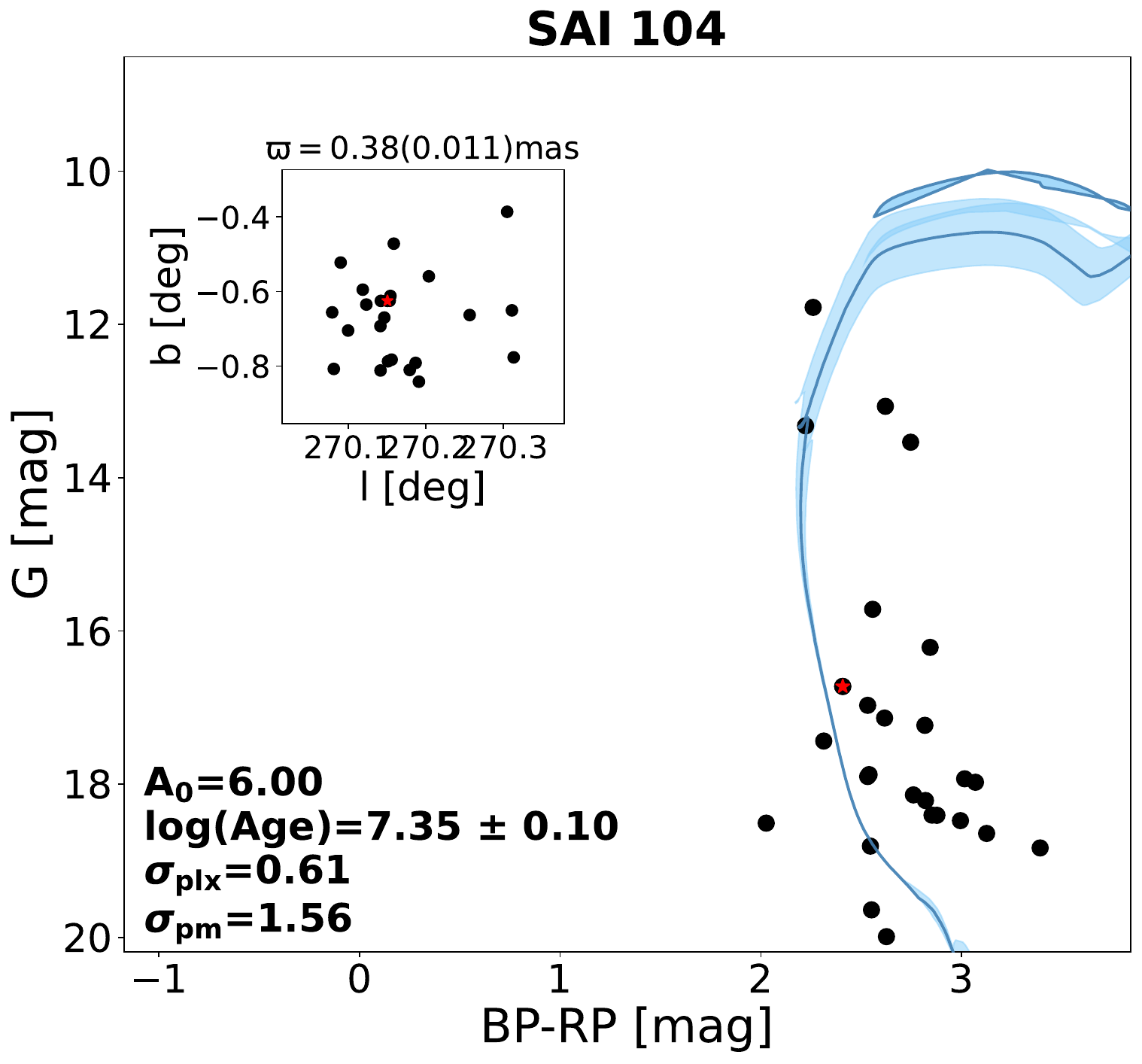}\hspace{0.02\linewidth}
\includegraphics[width=0.23\linewidth]{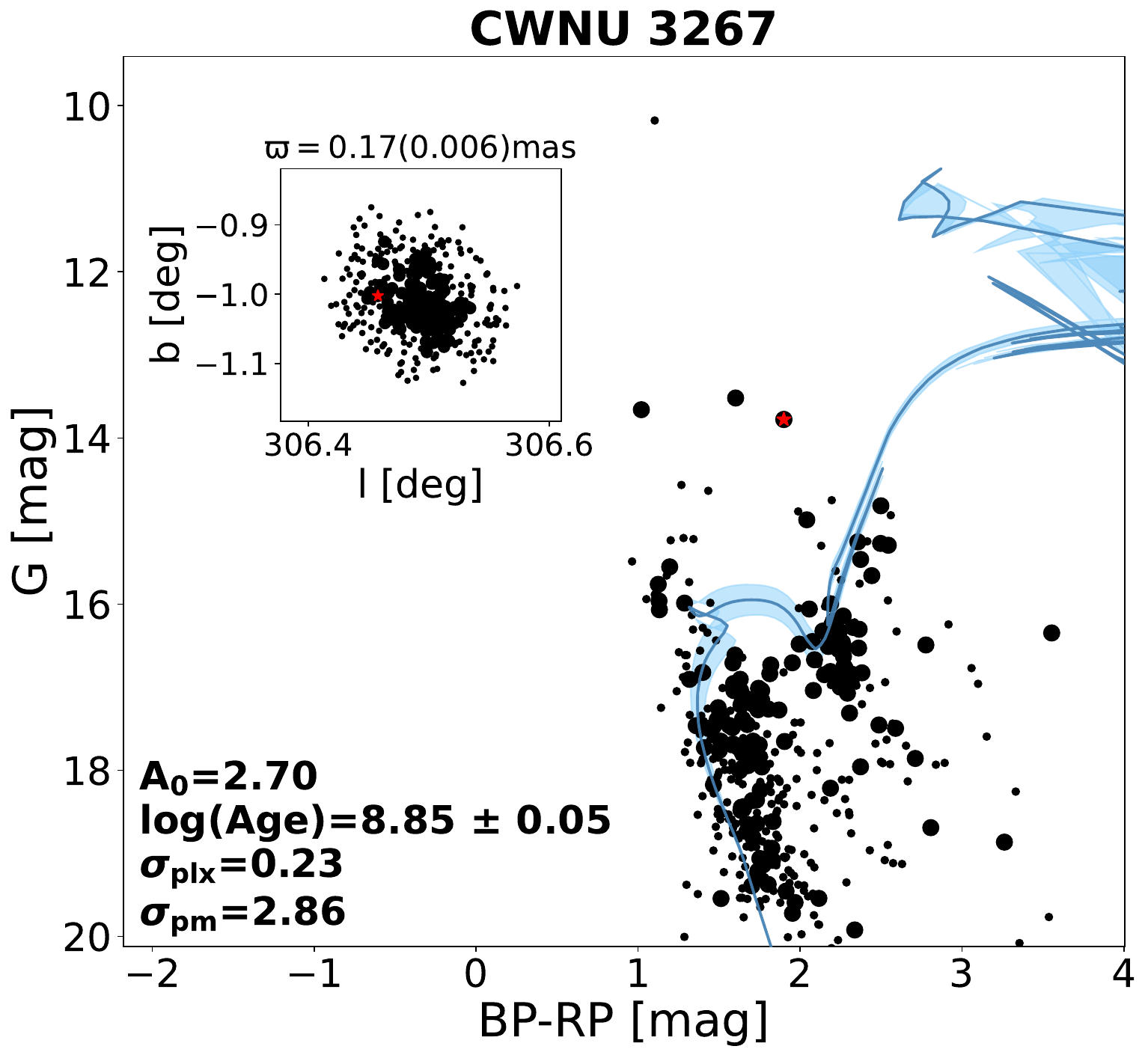}\hspace{0.02\linewidth}
\includegraphics[width=0.23\linewidth]{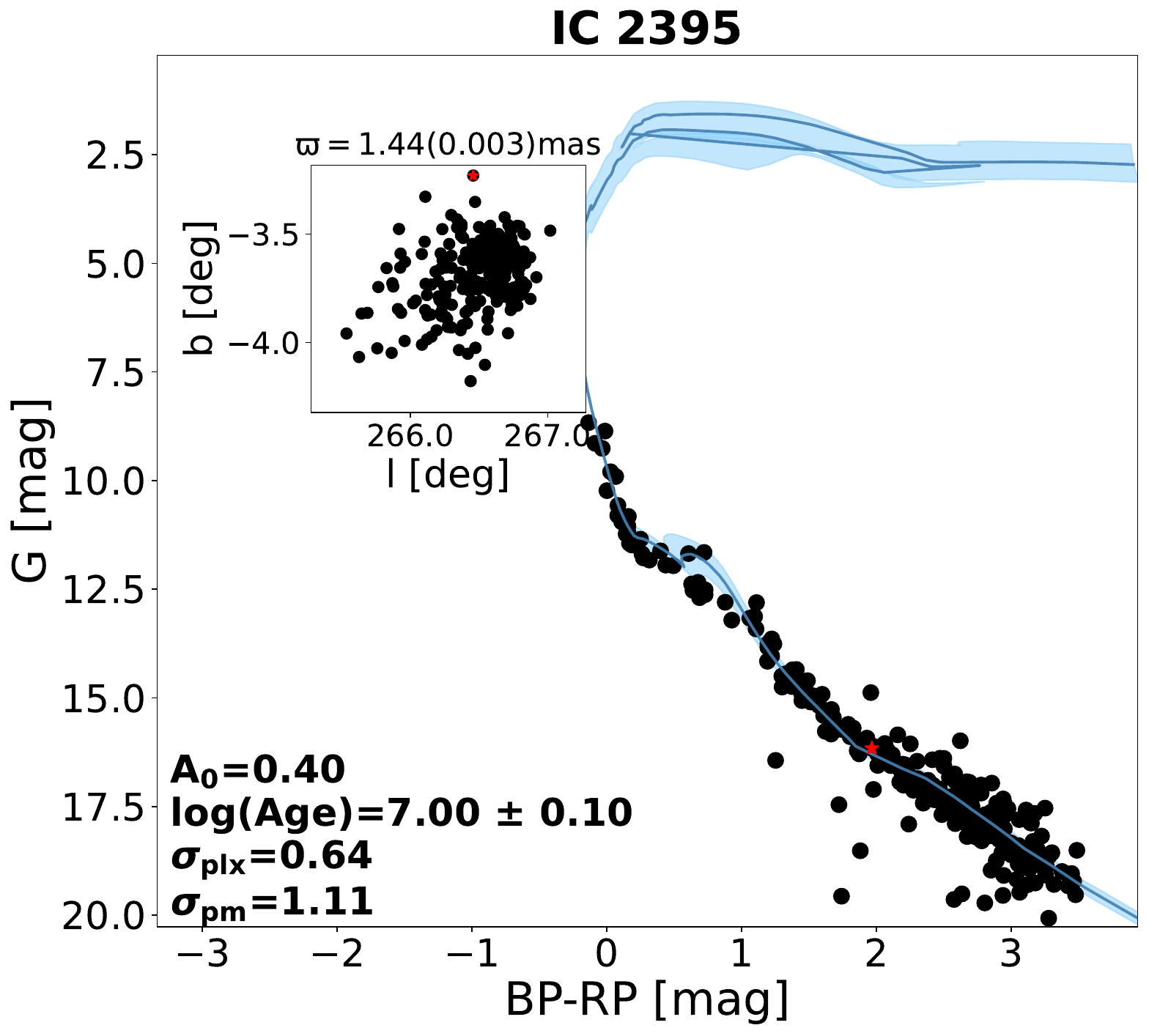}\hspace{0.02\linewidth}
\includegraphics[width=0.23\linewidth]{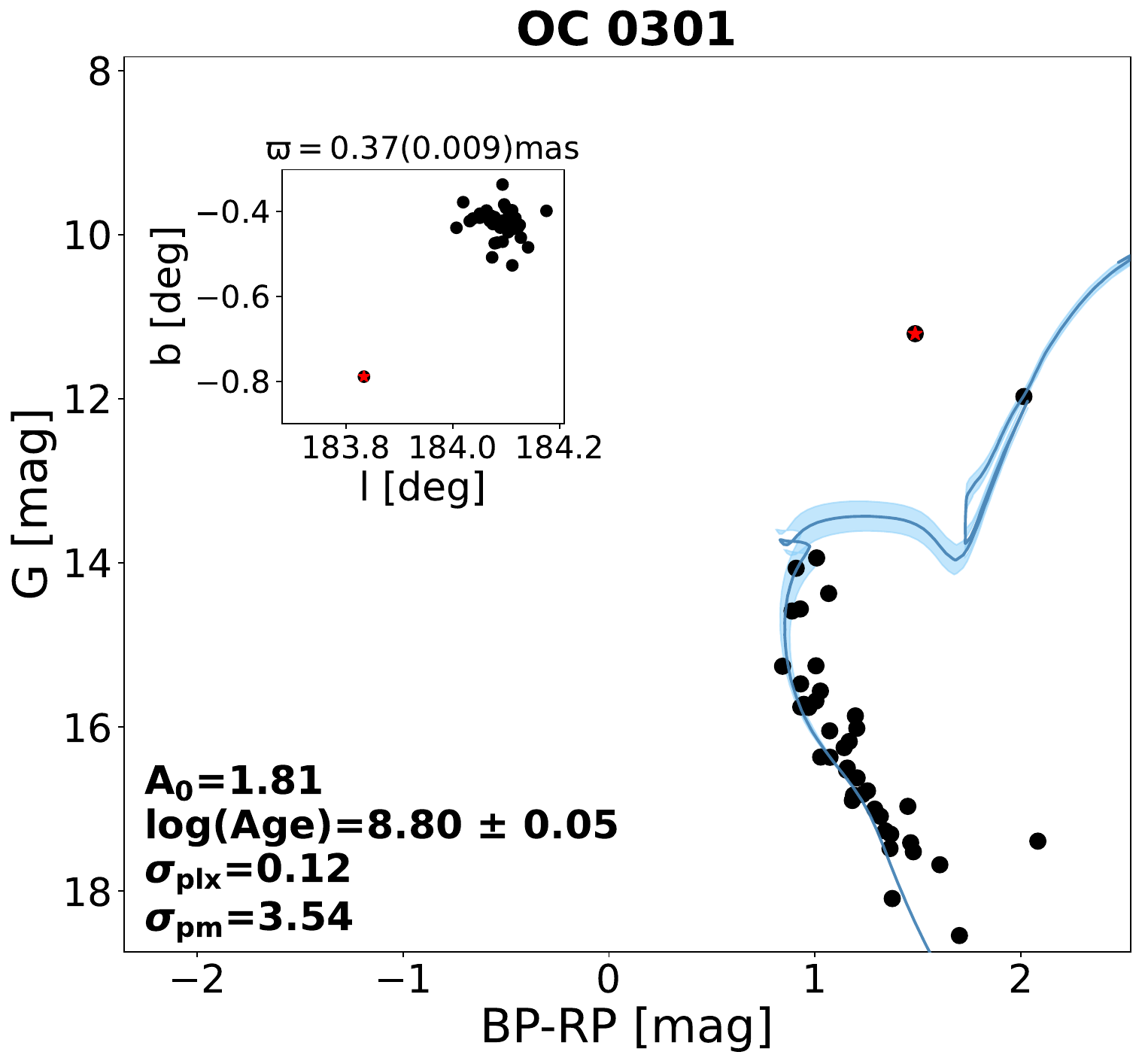}\\[0.4cm]

\includegraphics[width=0.23\linewidth]{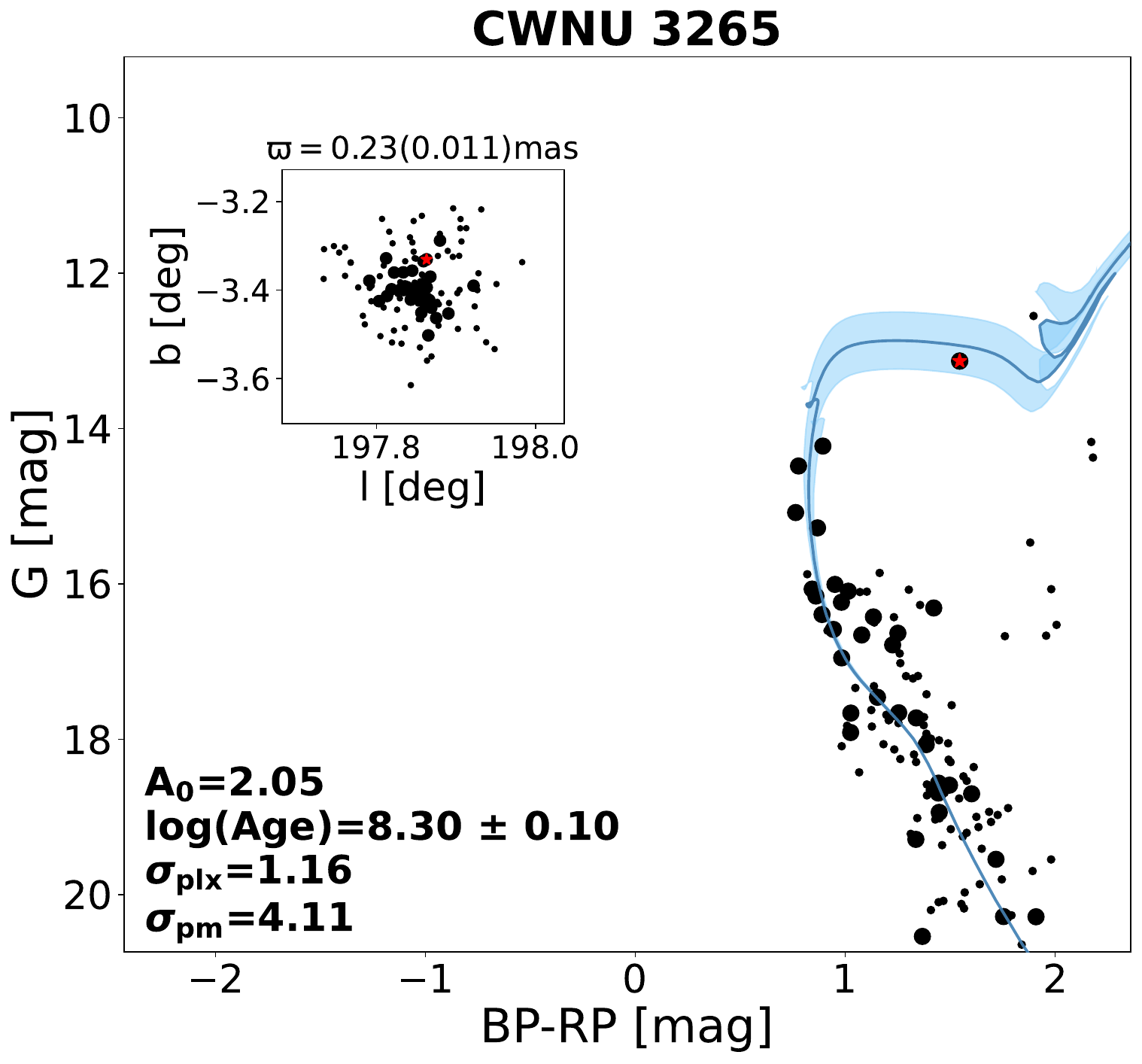}\hspace{0.02\linewidth}
\includegraphics[width=0.23\linewidth]{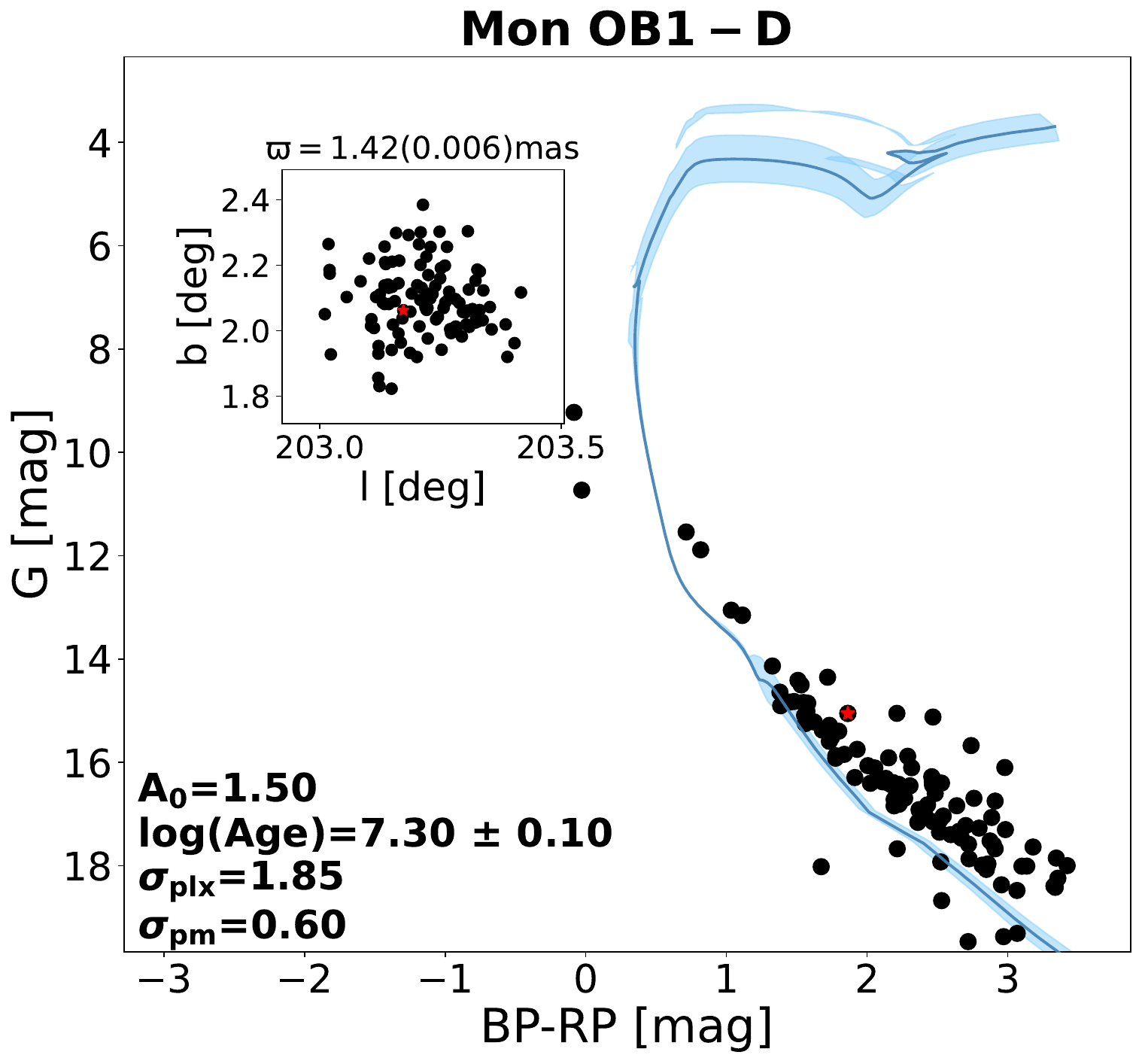}\hspace{0.02\linewidth}
\includegraphics[width=0.23\linewidth]{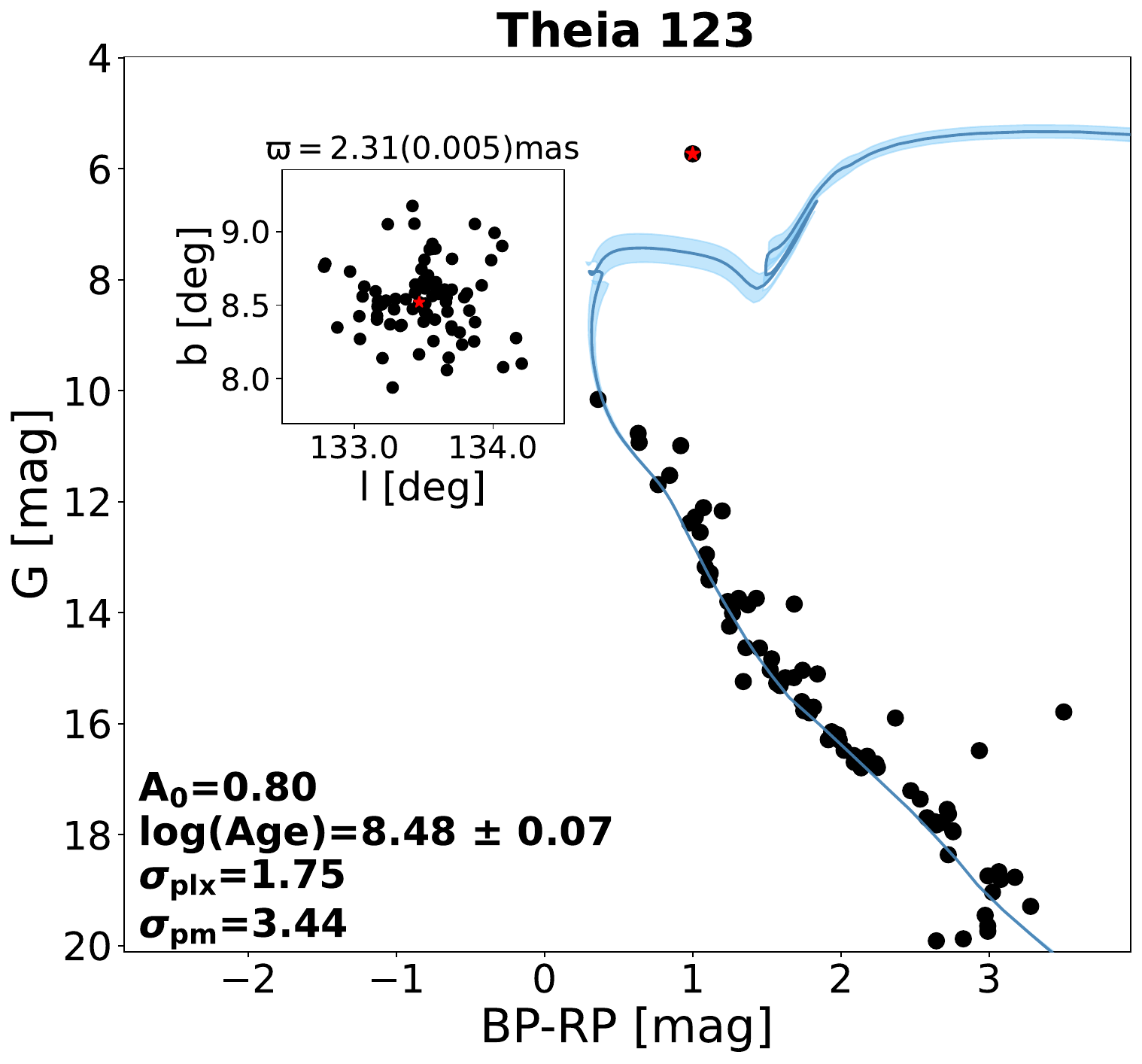}\hspace{0.02\linewidth}
\includegraphics[width=0.23\linewidth]{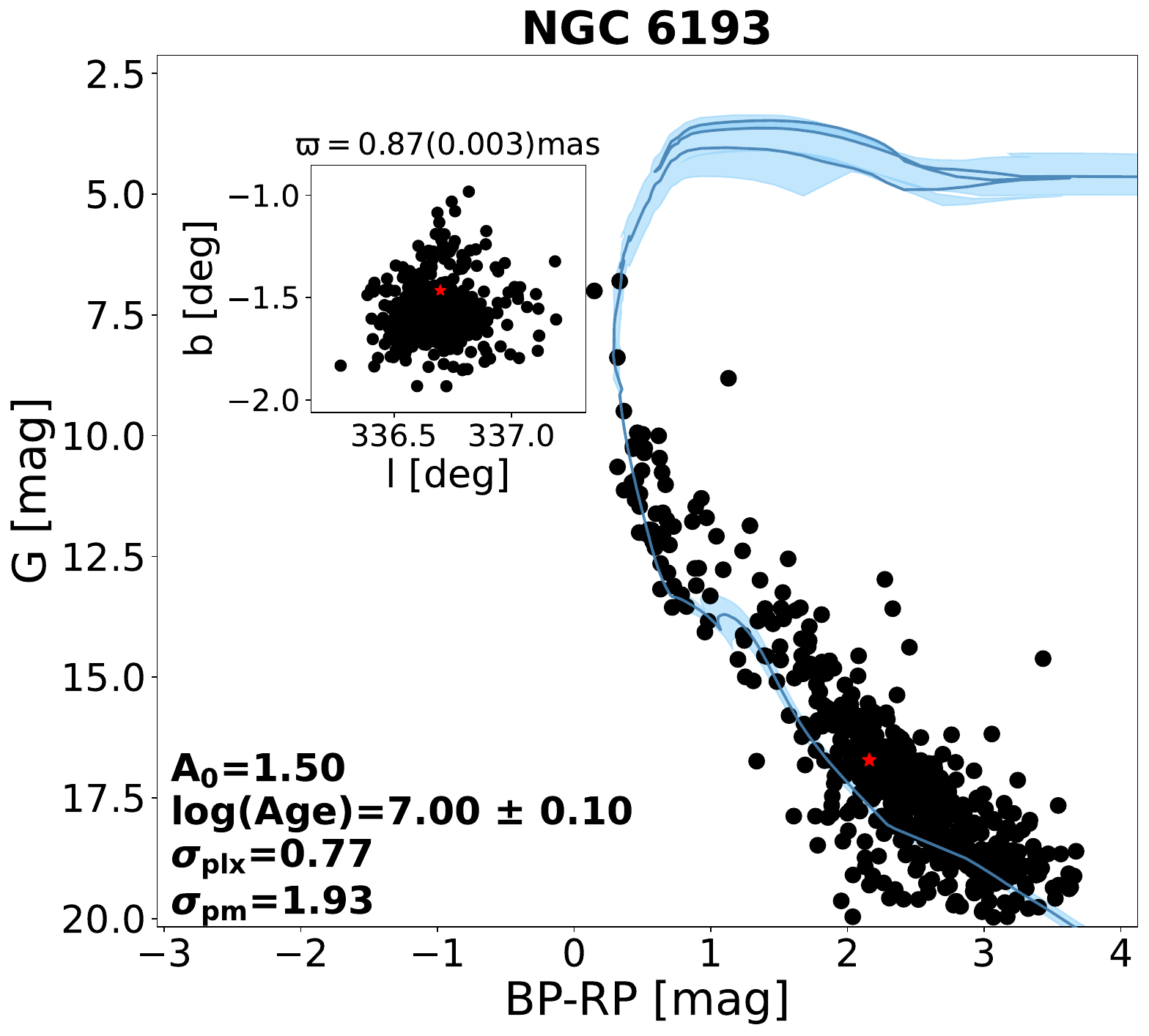}\\[0.4cm]

\includegraphics[width=0.23\linewidth]{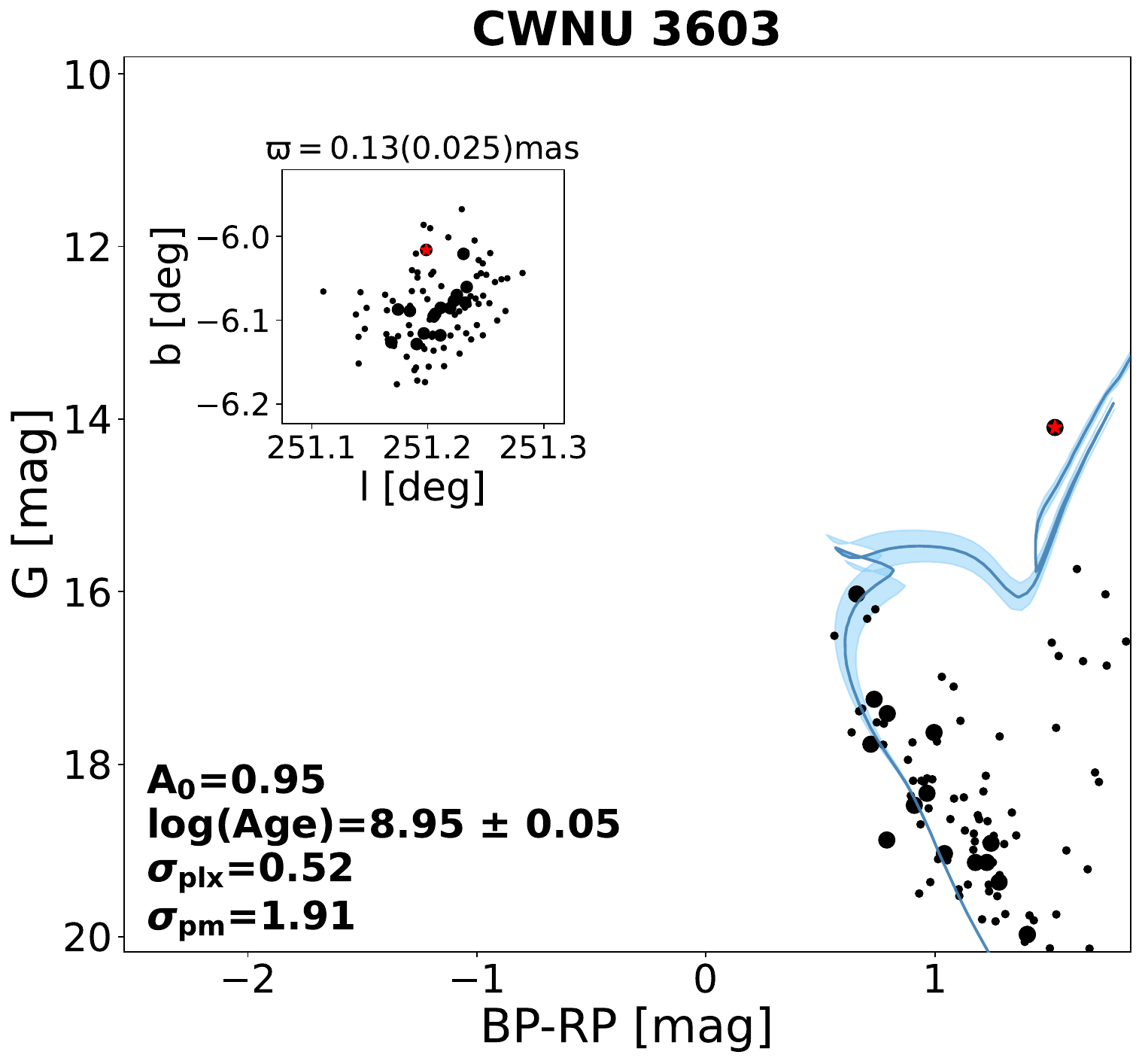}\hspace{0.02\linewidth}
\includegraphics[width=0.23\linewidth]{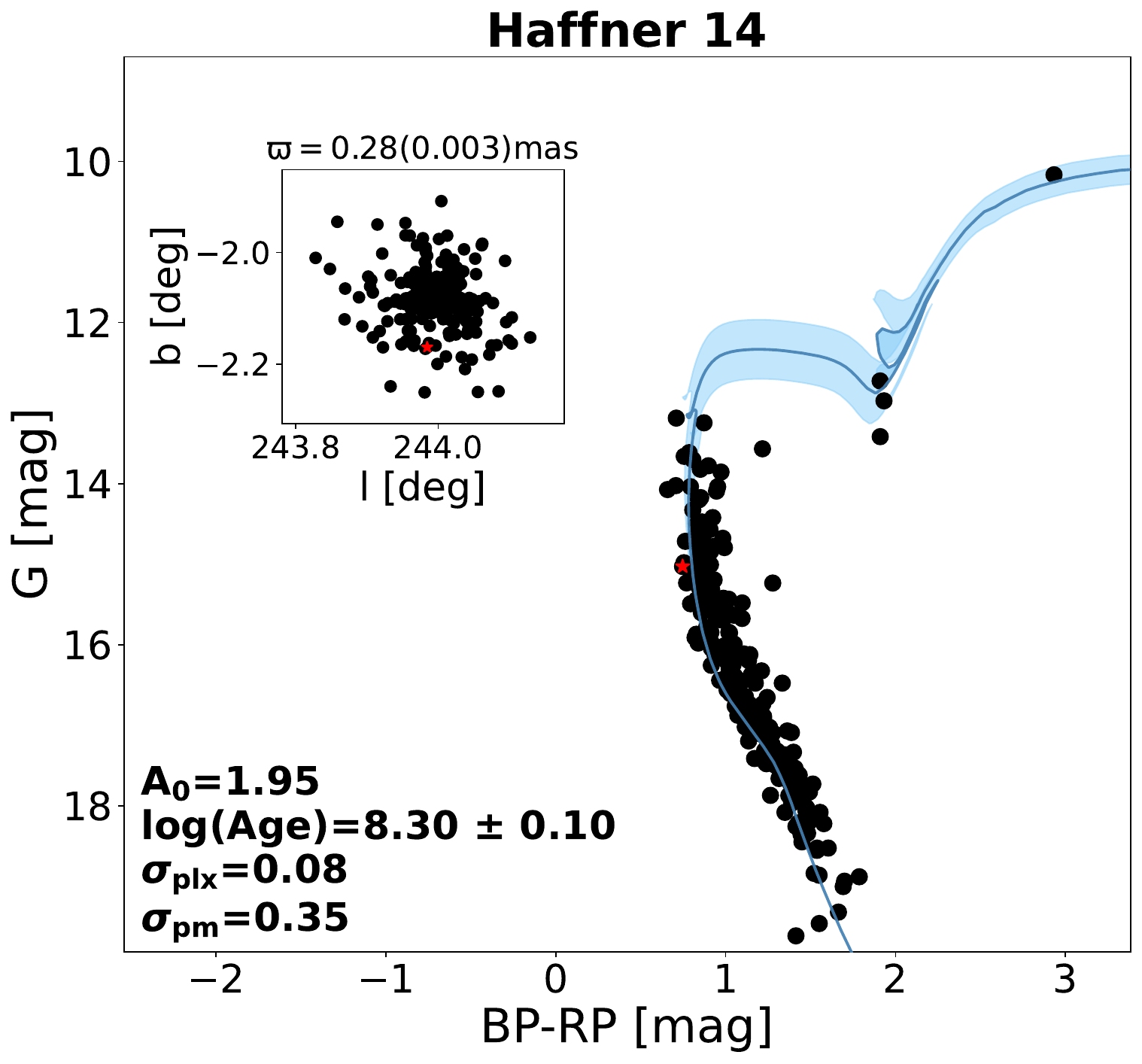}\hspace{0.02\linewidth}
\includegraphics[width=0.23\linewidth]{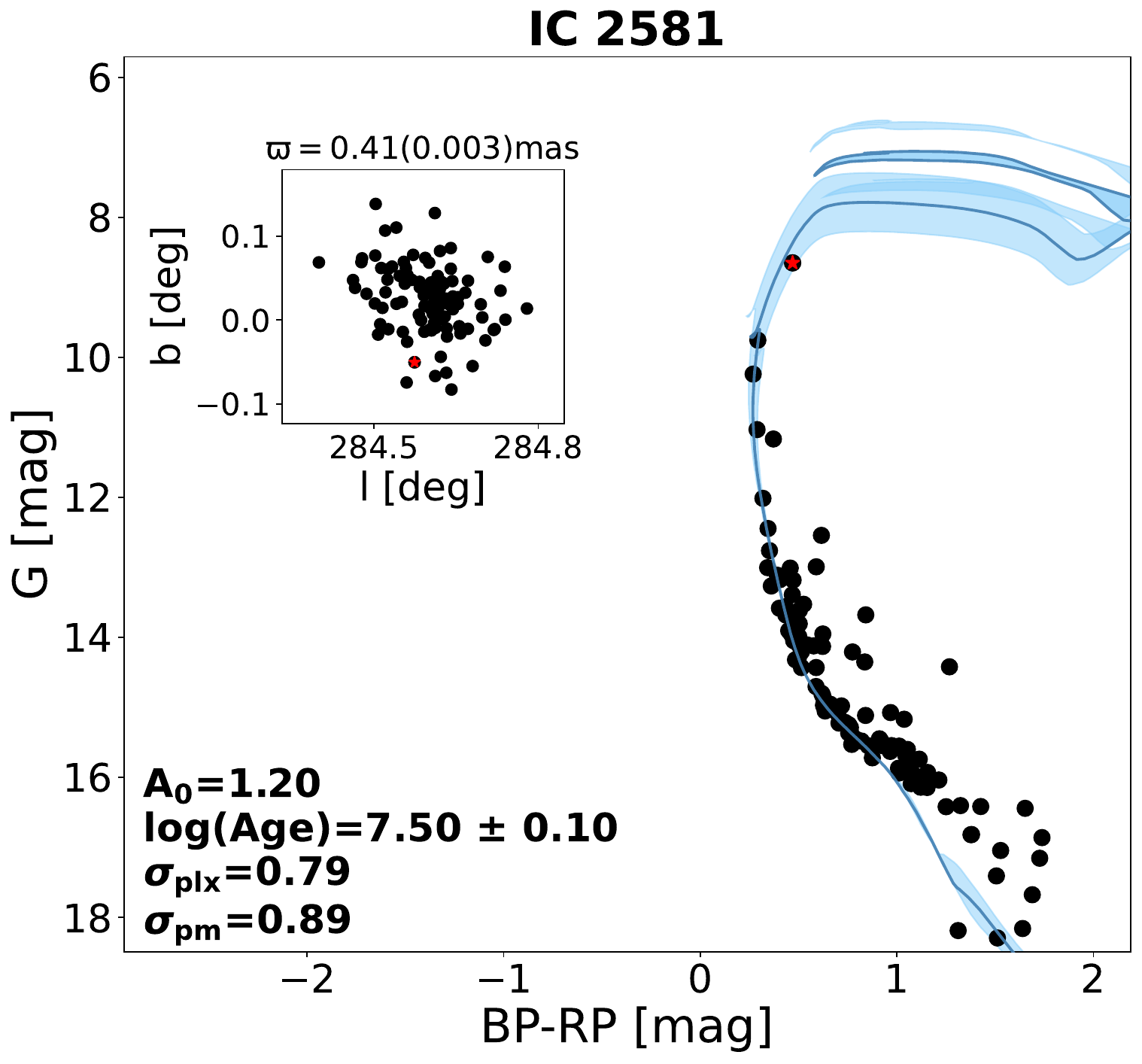}\hspace{0.02\linewidth}
\includegraphics[width=0.23\linewidth]{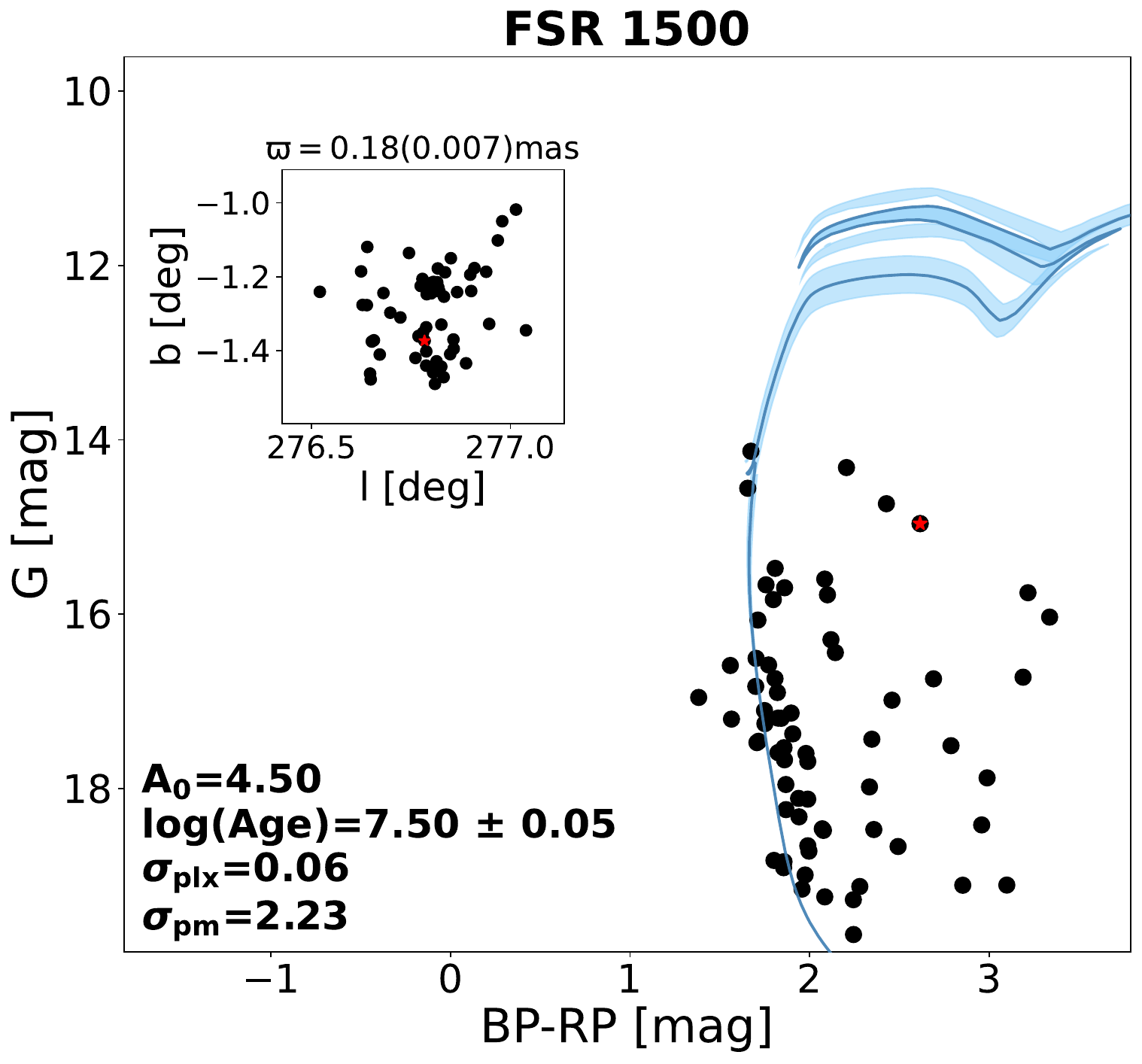}\\[0.4cm]

\includegraphics[width=0.23\linewidth]{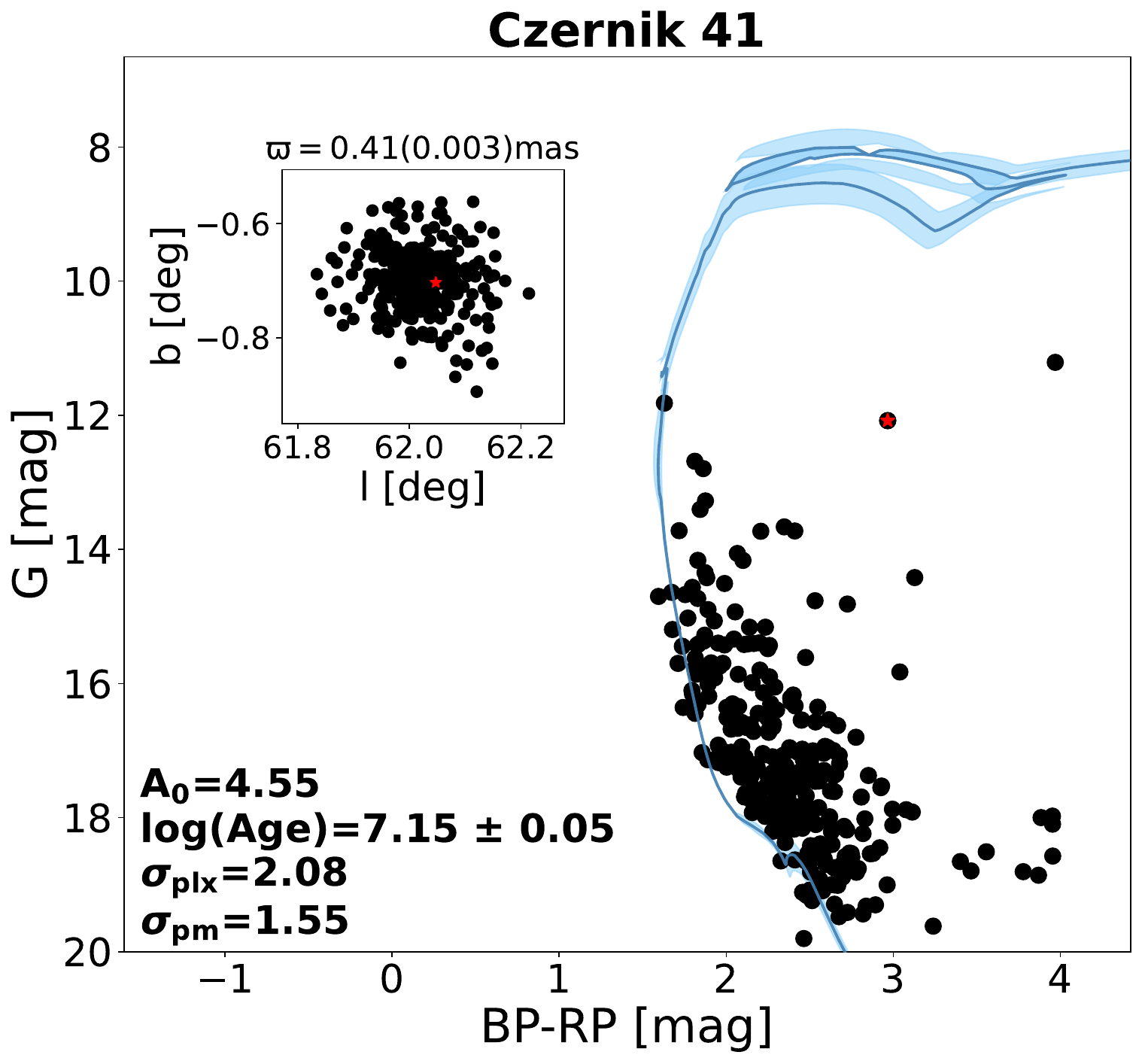}\hspace{0.02\linewidth}
\includegraphics[width=0.23\linewidth]{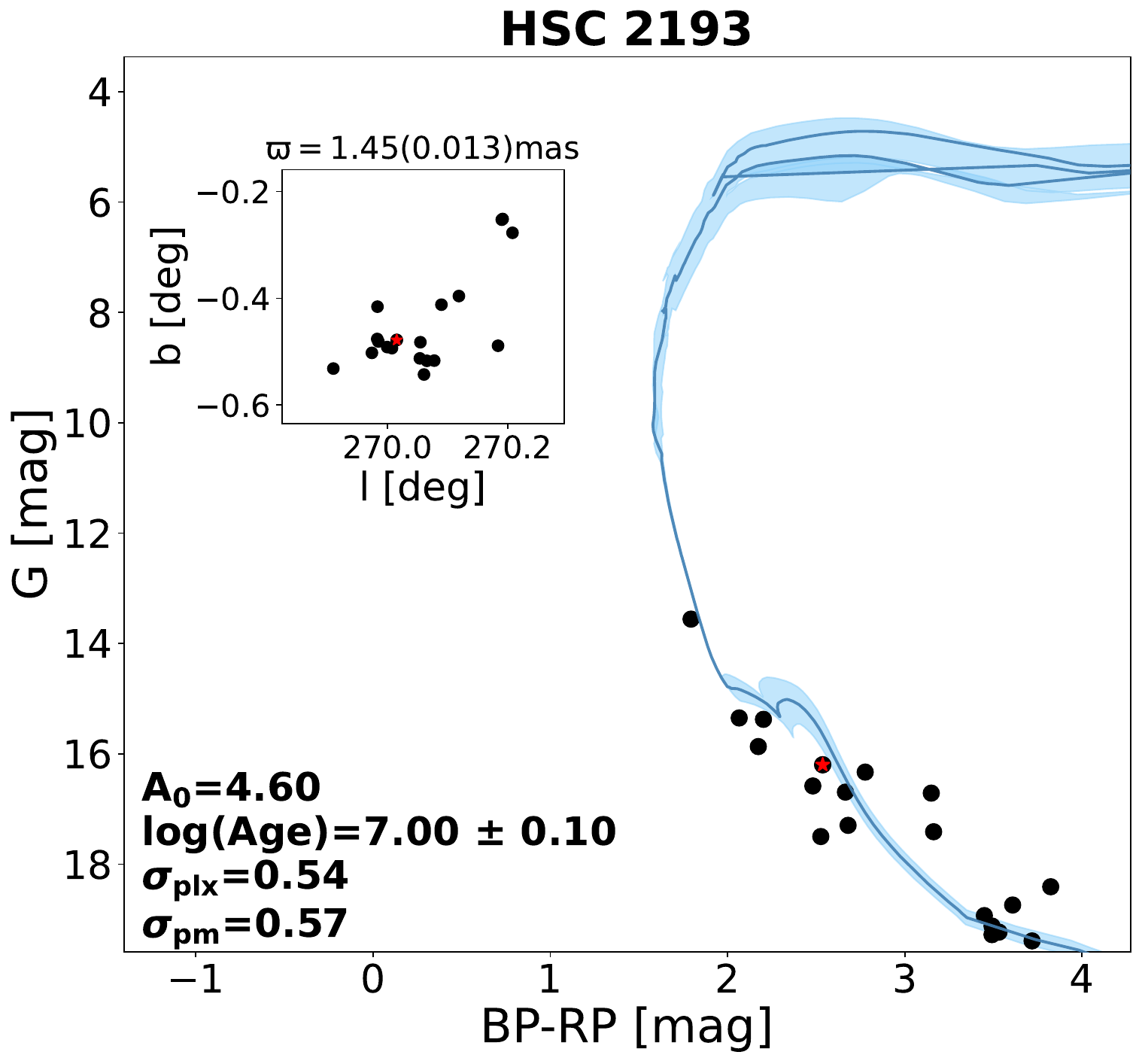}\hspace{0.02\linewidth}
\includegraphics[width=0.23\linewidth]{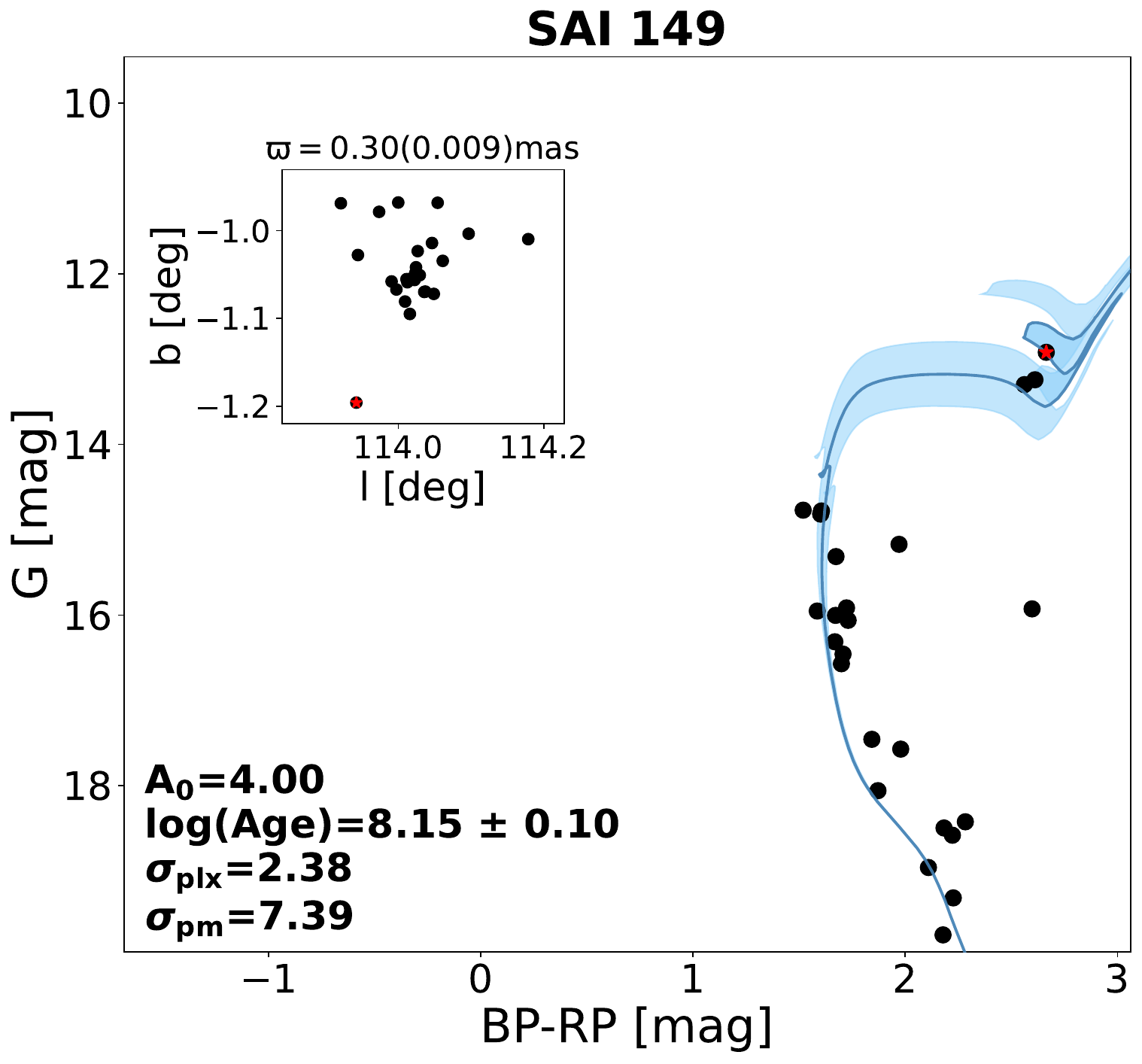}\hspace{0.02\linewidth}
\includegraphics[width=0.23\linewidth]{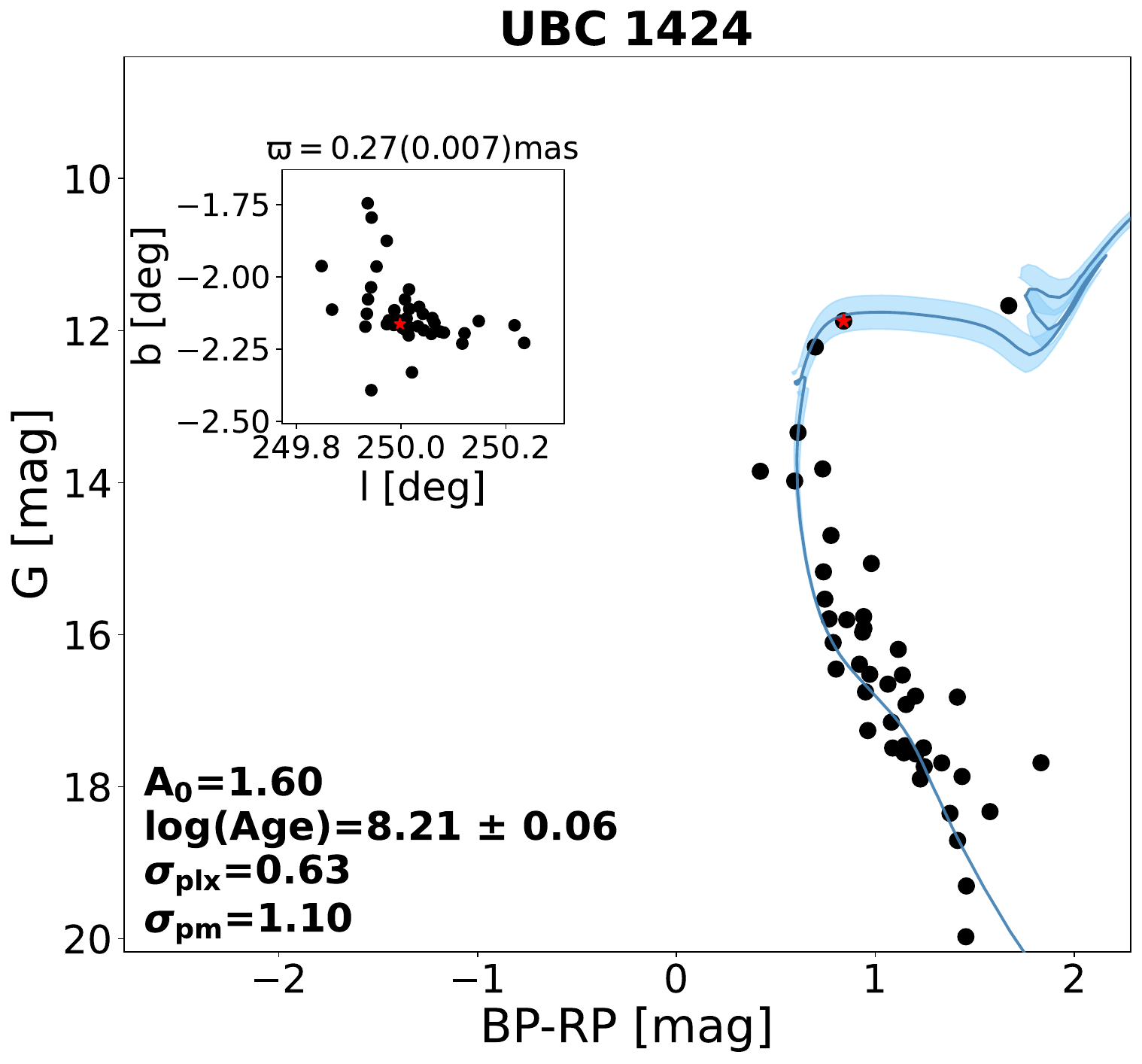}\\[0.4cm]

\includegraphics[width=0.23\linewidth]{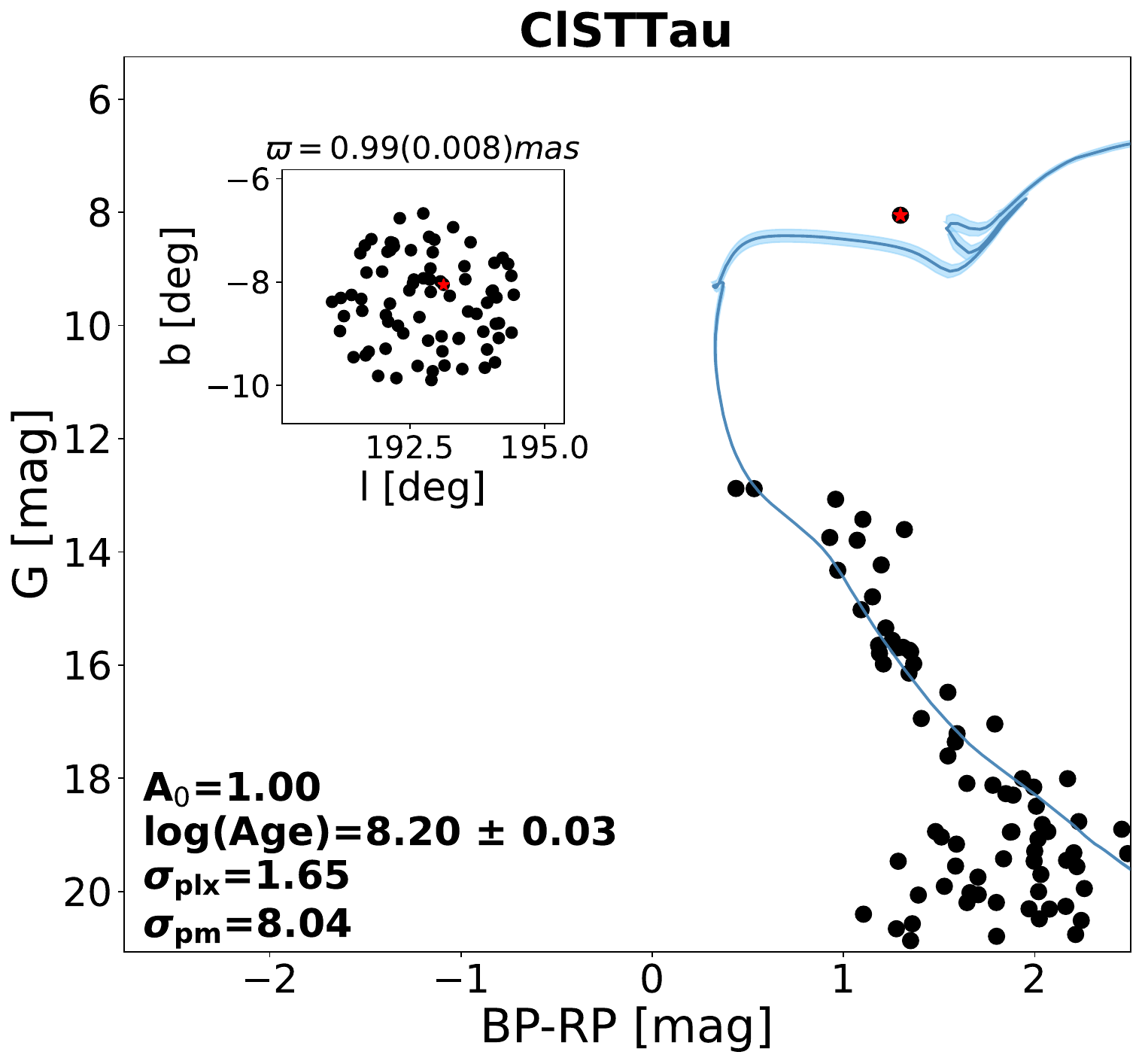}\hspace{0.02\linewidth}
\includegraphics[width=0.23\linewidth]{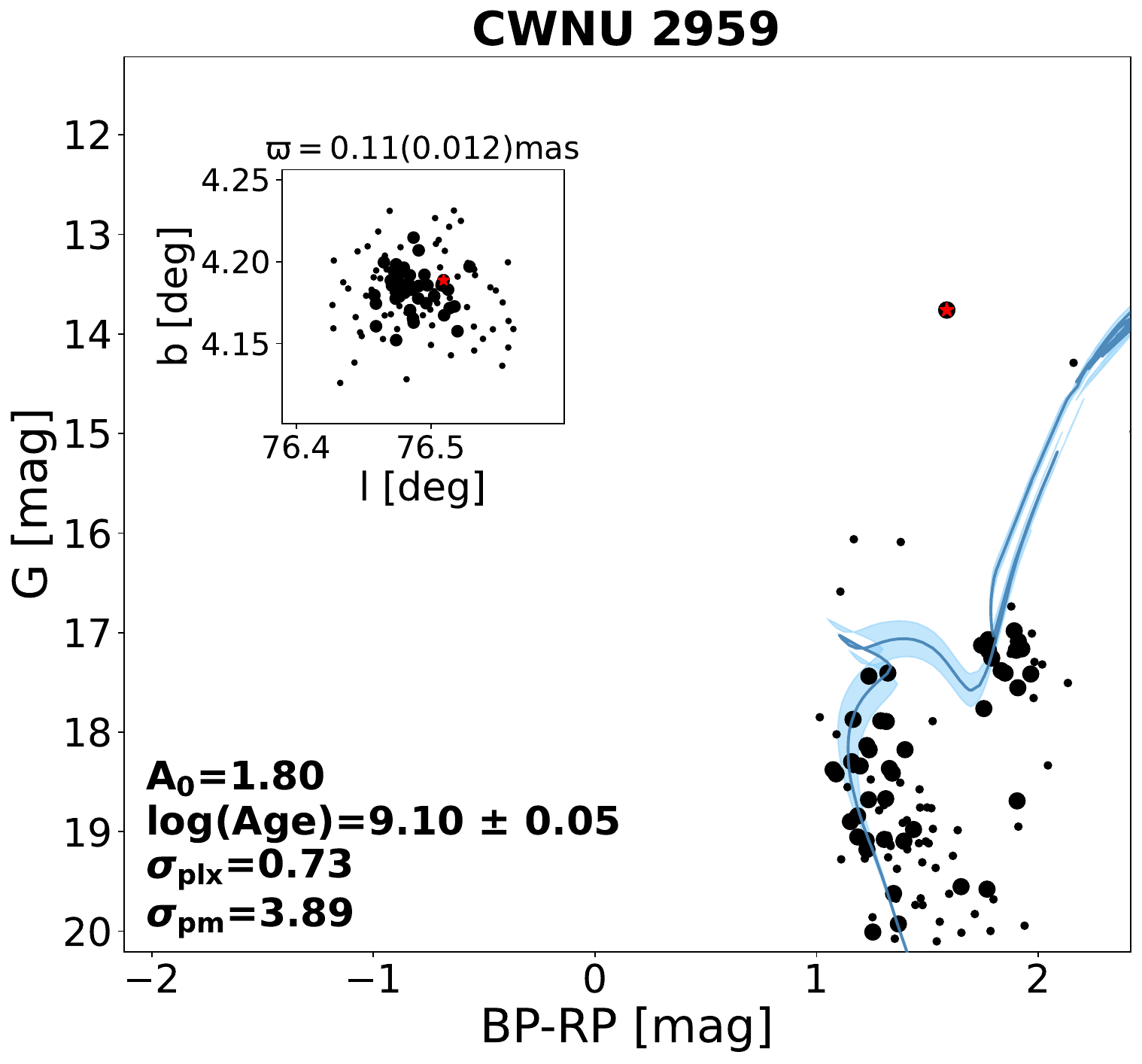}\hspace{0.02\linewidth}
\includegraphics[width=0.23\linewidth]{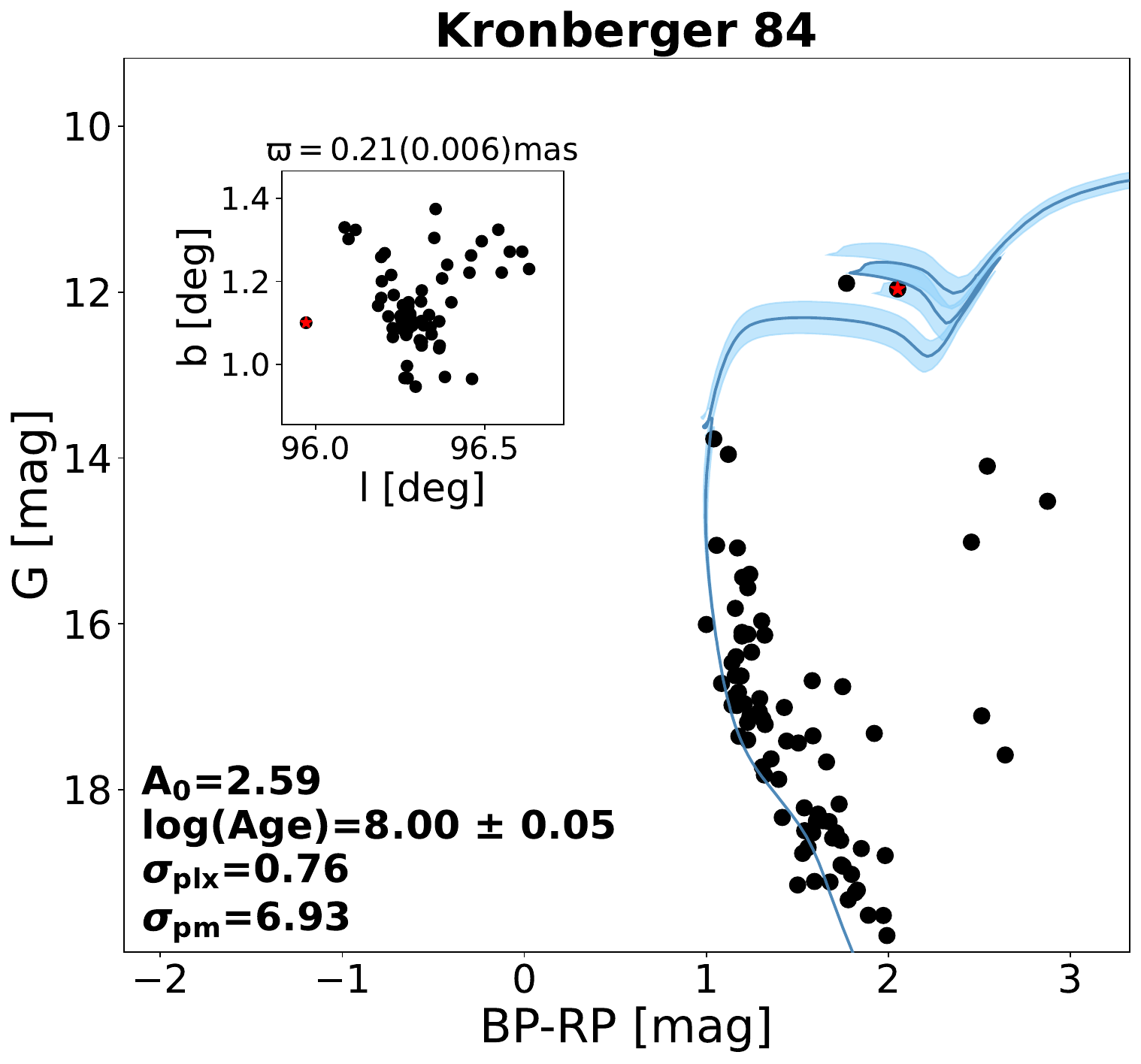}\hspace{0.02\linewidth}
\includegraphics[width=0.23\linewidth]{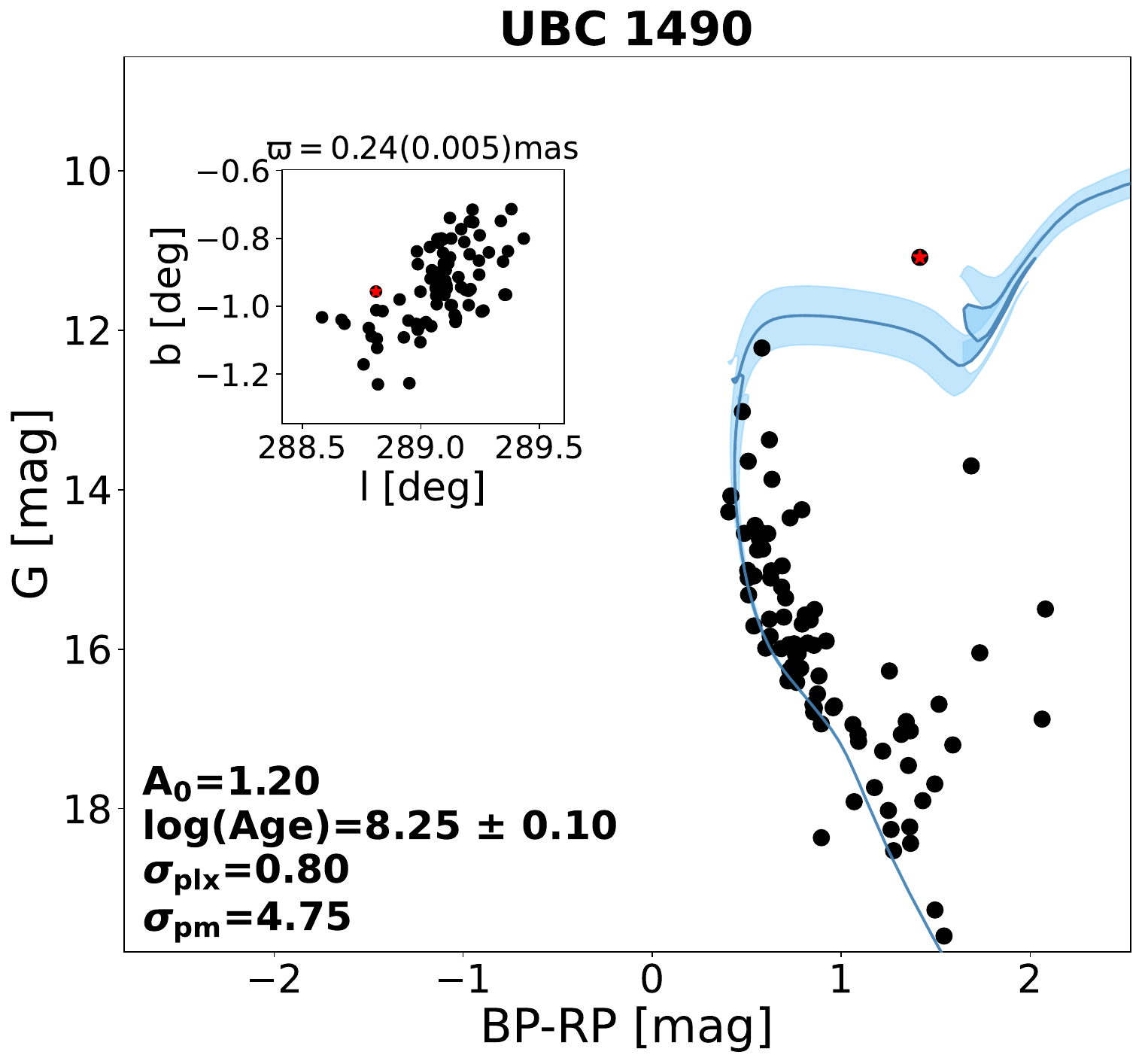}\\[0.4cm]

\caption{CMDs of the remaining rejected OC Cepheid samples, which also contain the extinction, age and distance parameters of the OCs, namely,
SAI 104, CWNU 3267, IC 2395, OC 0301, CWNU 3265, Mon OB1-D, Theia 123, NGC 6193, CWNU 3603, Haffner 14, IC 2581, 
FSR 1500, Czernik 41, HSC 2193, SAI 149, UBC 1424, Cl ST Tau, CWNU 2959, Kronberger 84 (V733 Cyg), and UBC 1490.}
\label{fig:cmd_set4}
\end{figure*}

\begin{figure*}[htbp]
\centering
\includegraphics[width=0.23\linewidth]{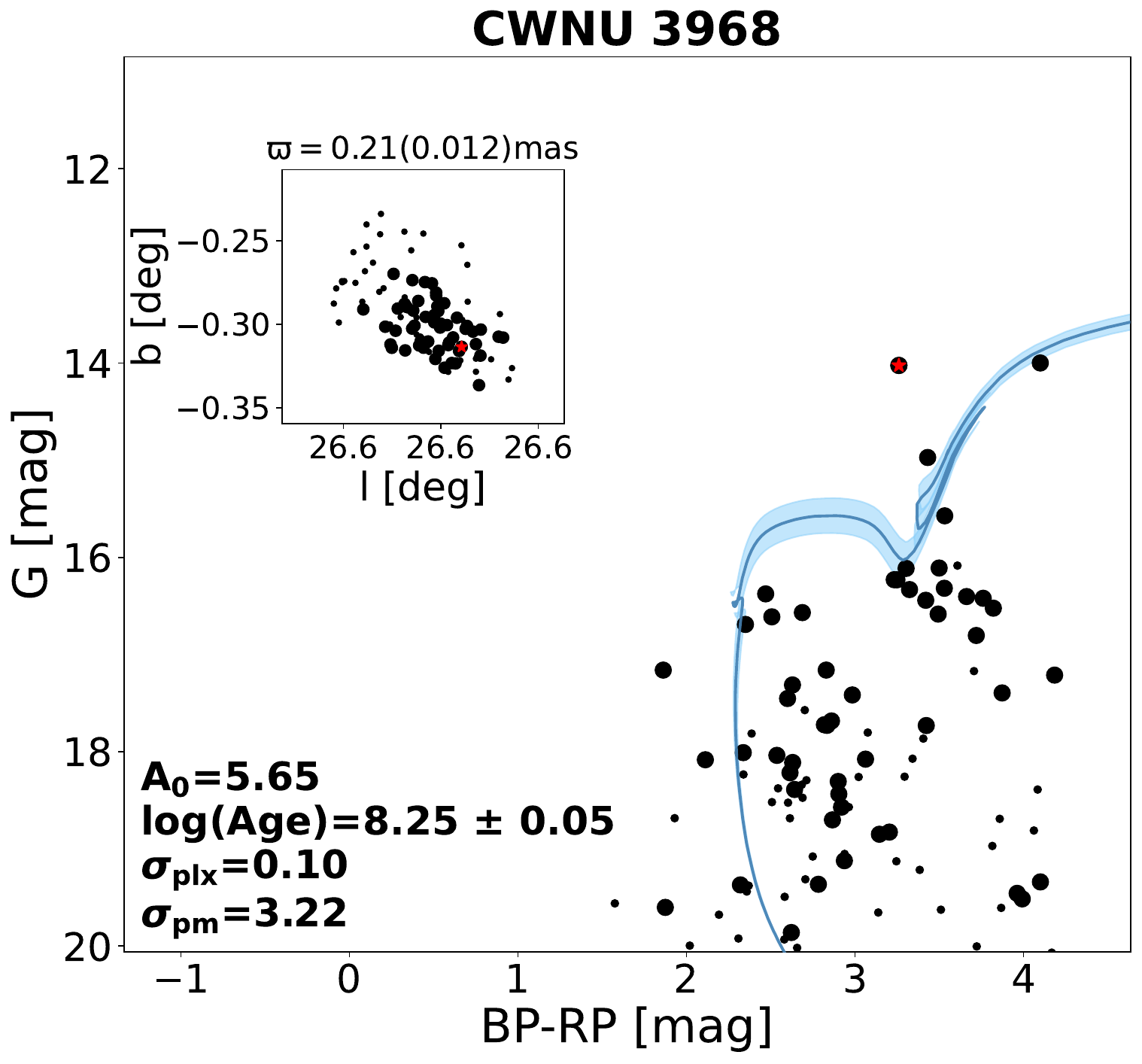}\hspace{0.02\linewidth}
\includegraphics[width=0.23\linewidth]{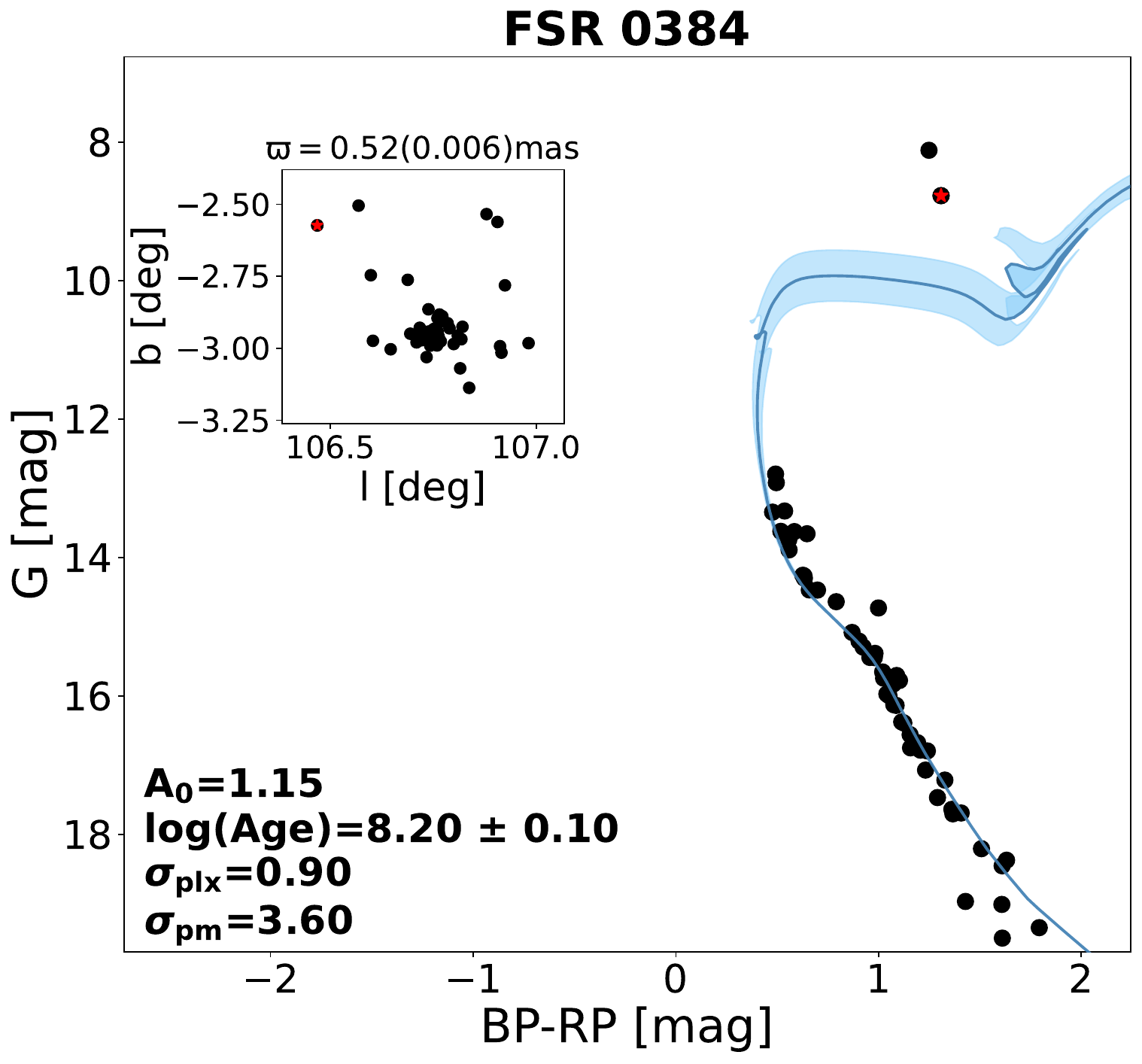}\hspace{0.02\linewidth}
\includegraphics[width=0.23\linewidth]{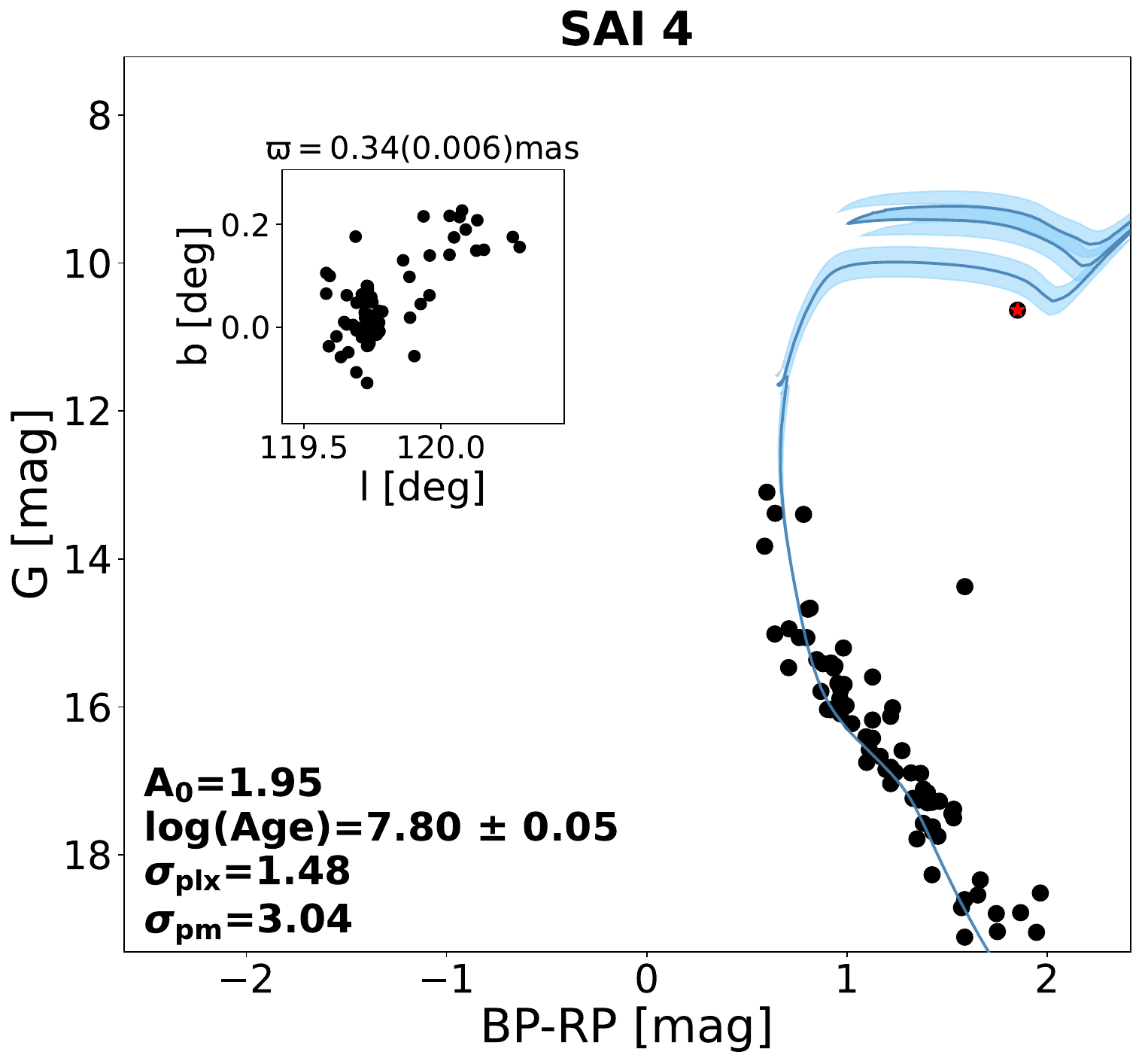}\hspace{0.02\linewidth}
\includegraphics[width=0.23\linewidth]{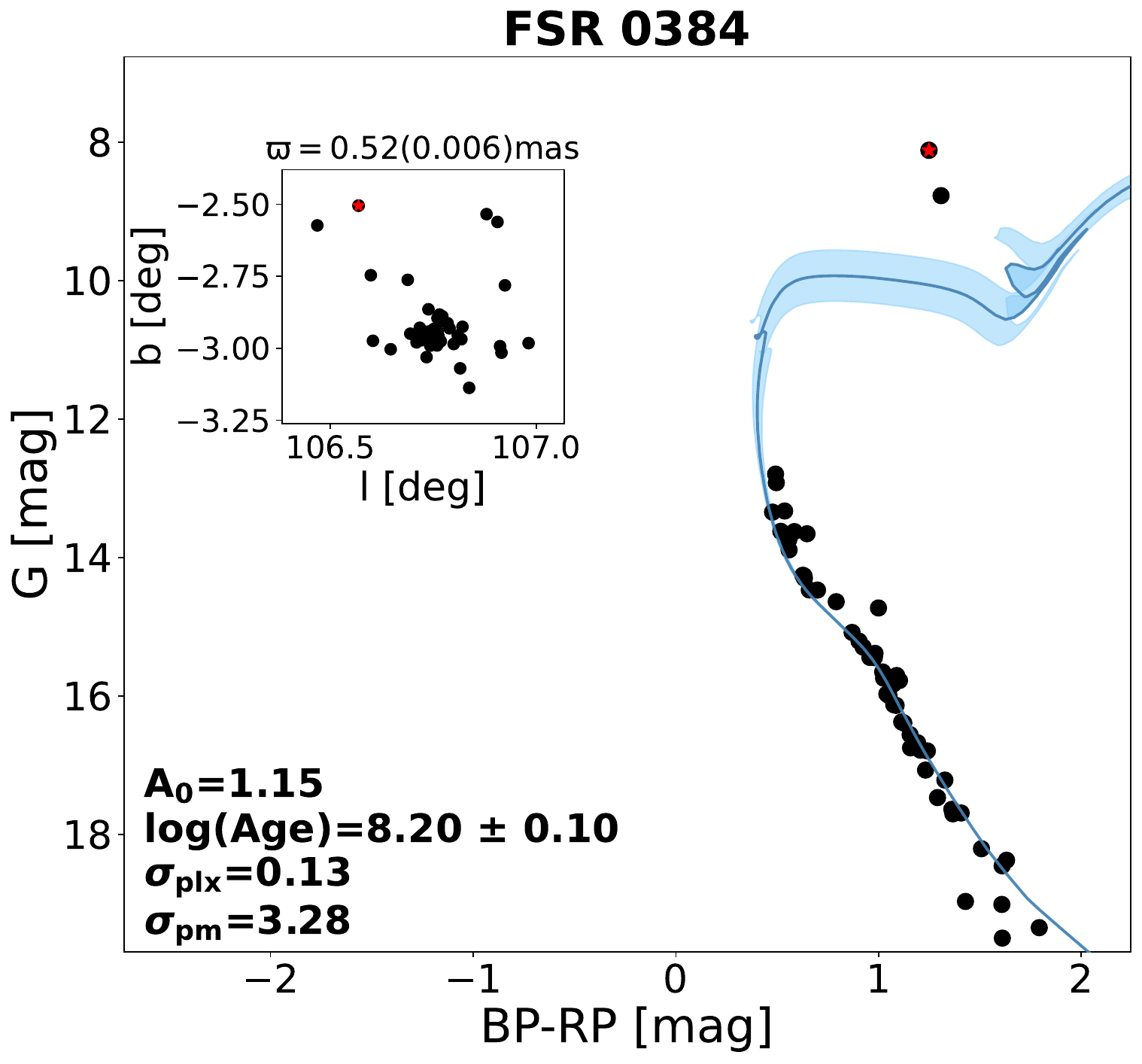}\\[0.4cm]

\includegraphics[width=0.23\linewidth]{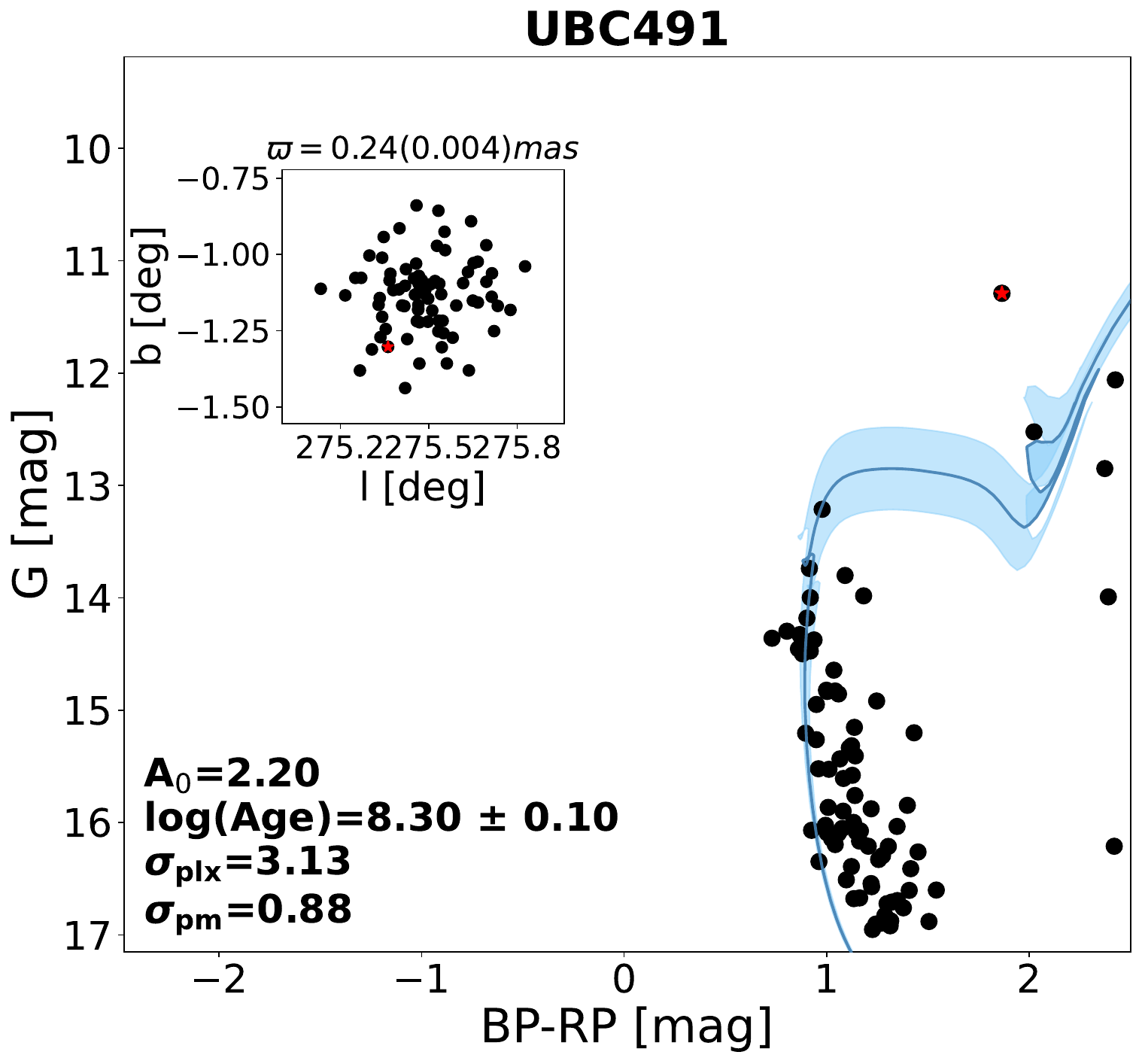}\hspace{0.02\linewidth}
\includegraphics[width=0.23\linewidth]{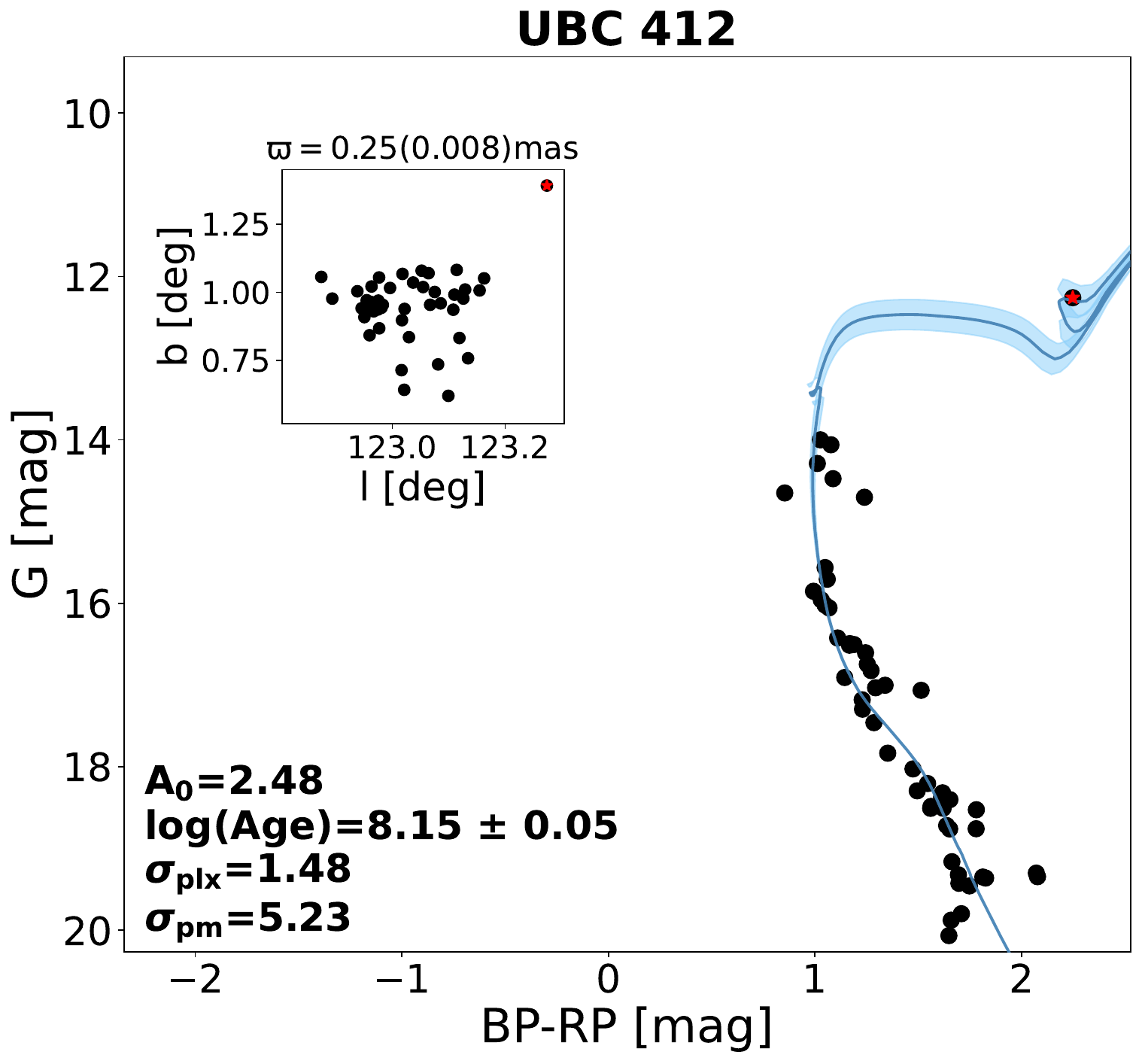}\hspace{0.02\linewidth}
\includegraphics[width=0.23\linewidth]{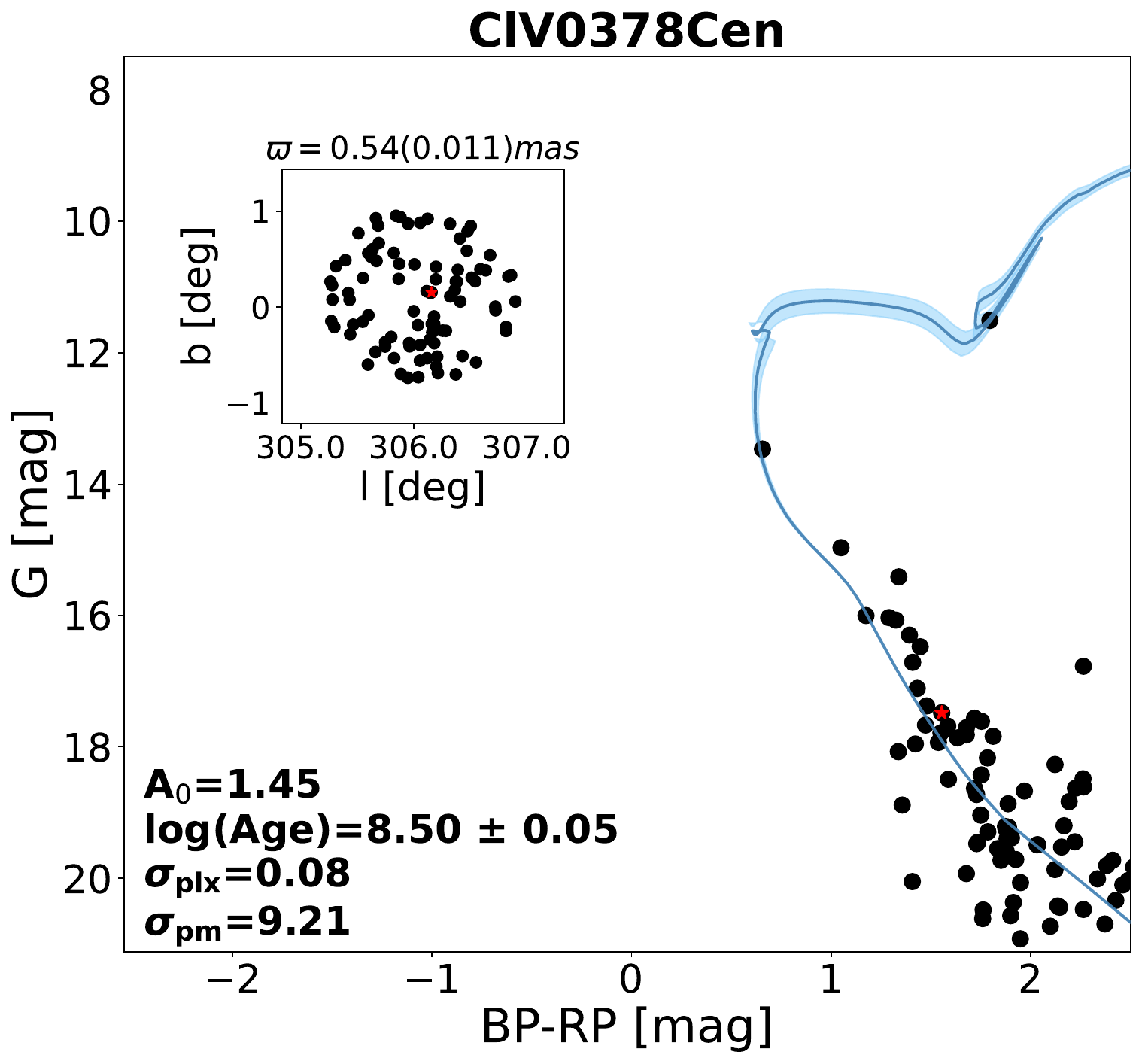}\hspace{0.02\linewidth}
\includegraphics[width=0.23\linewidth]{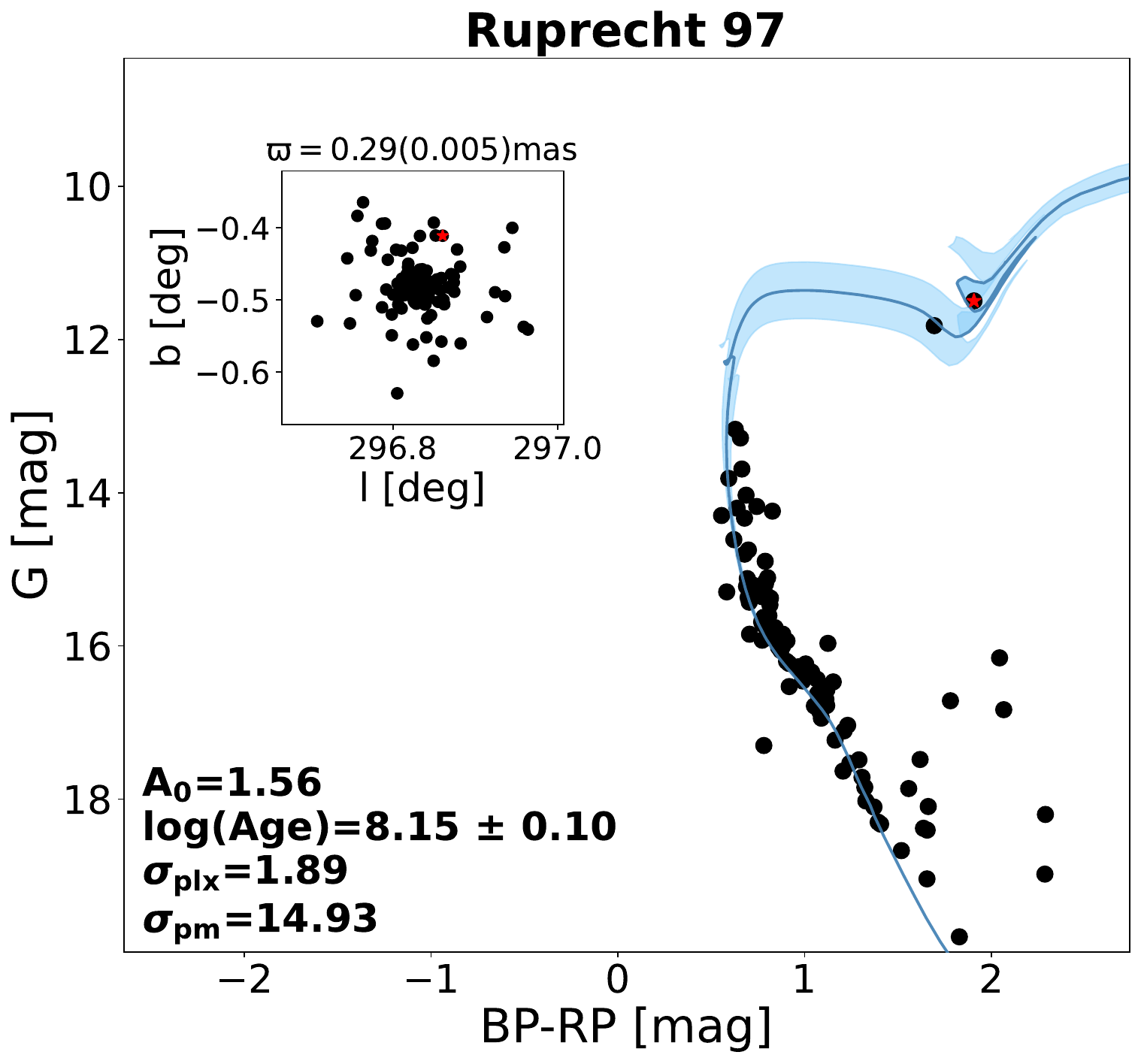}\\[0.4cm]

\includegraphics[width=0.23\linewidth]{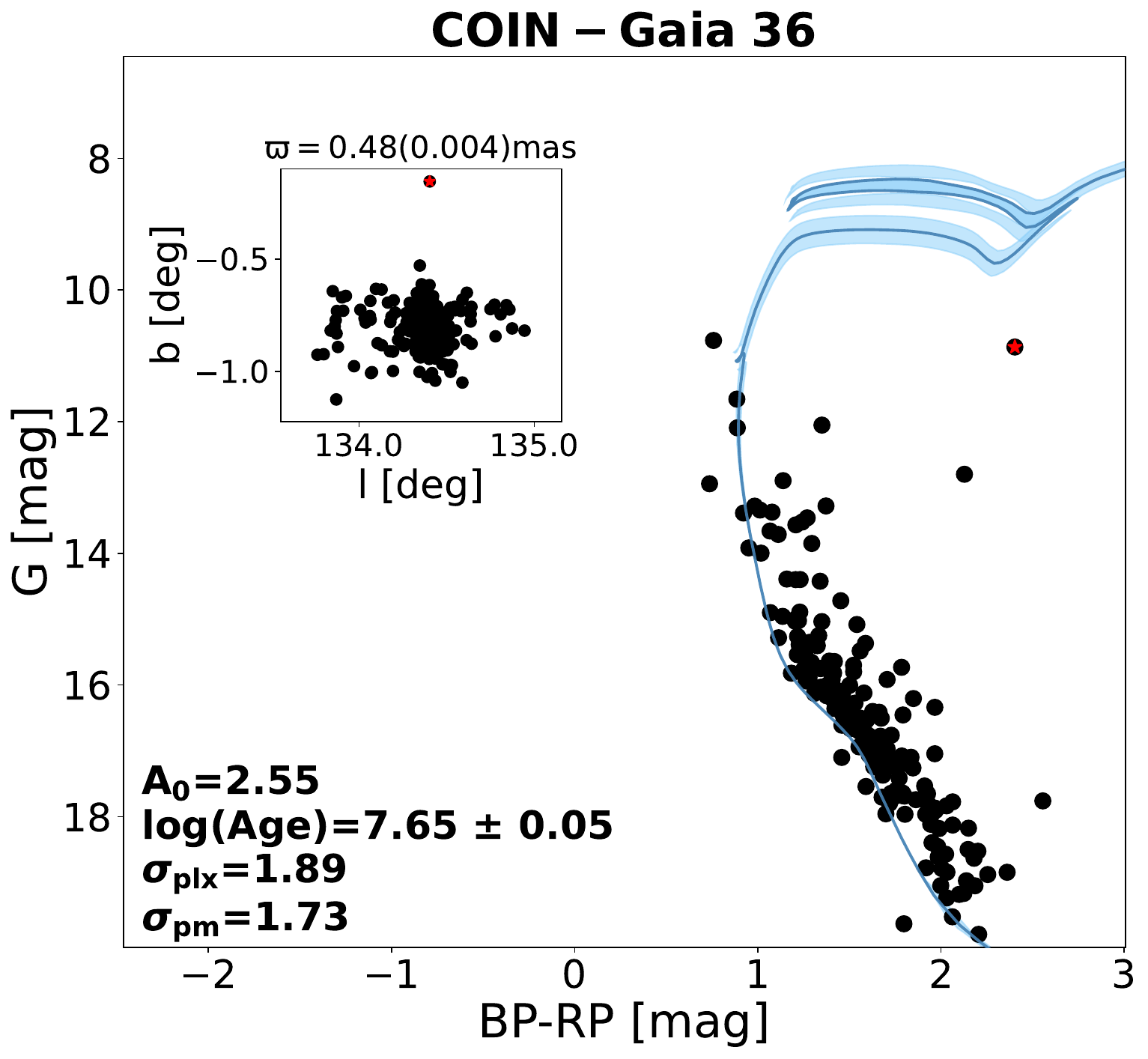}\hspace{0.02\linewidth}
\includegraphics[width=0.23\linewidth]{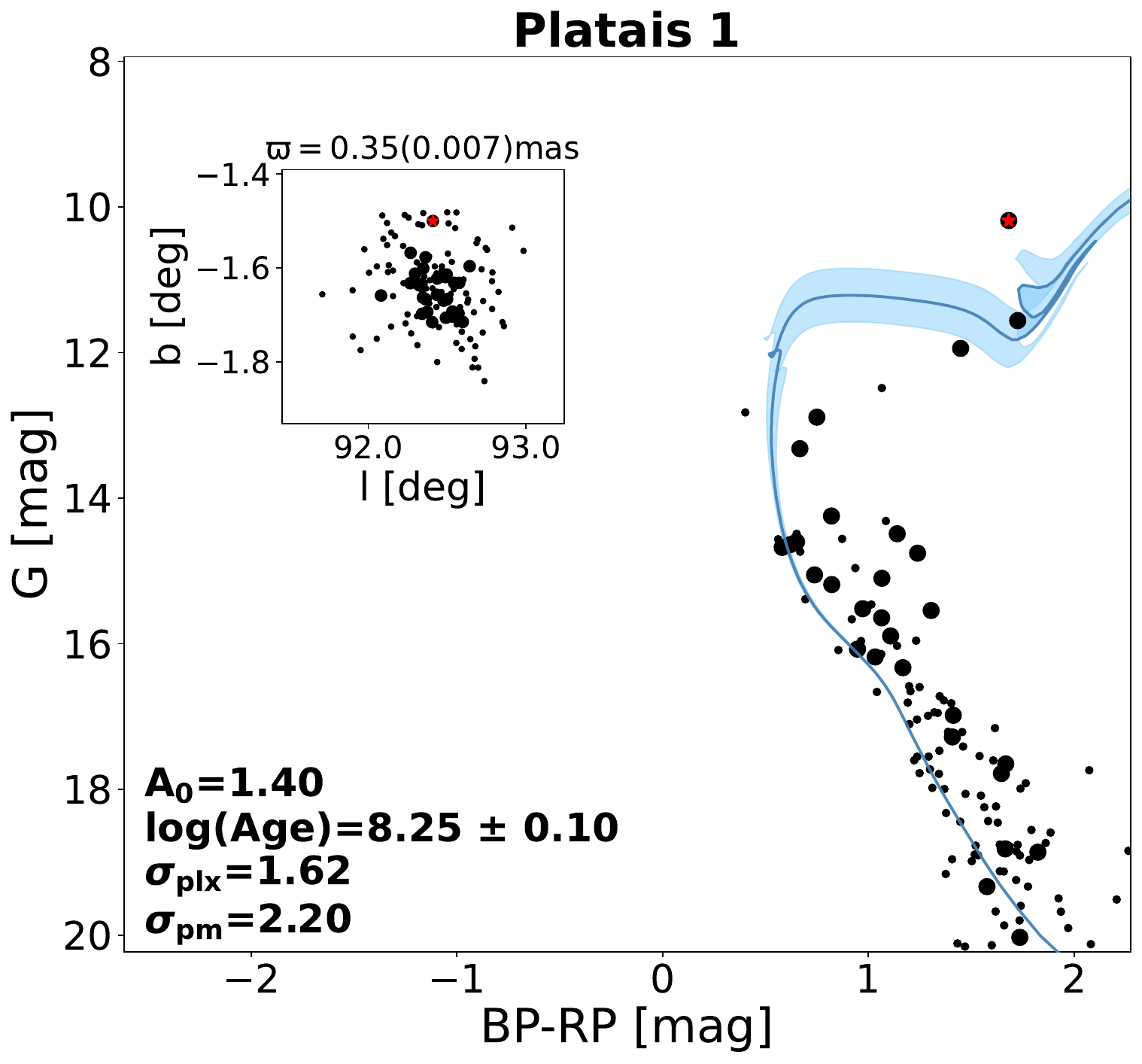}\hspace{0.02\linewidth}
\includegraphics[width=0.23\linewidth]{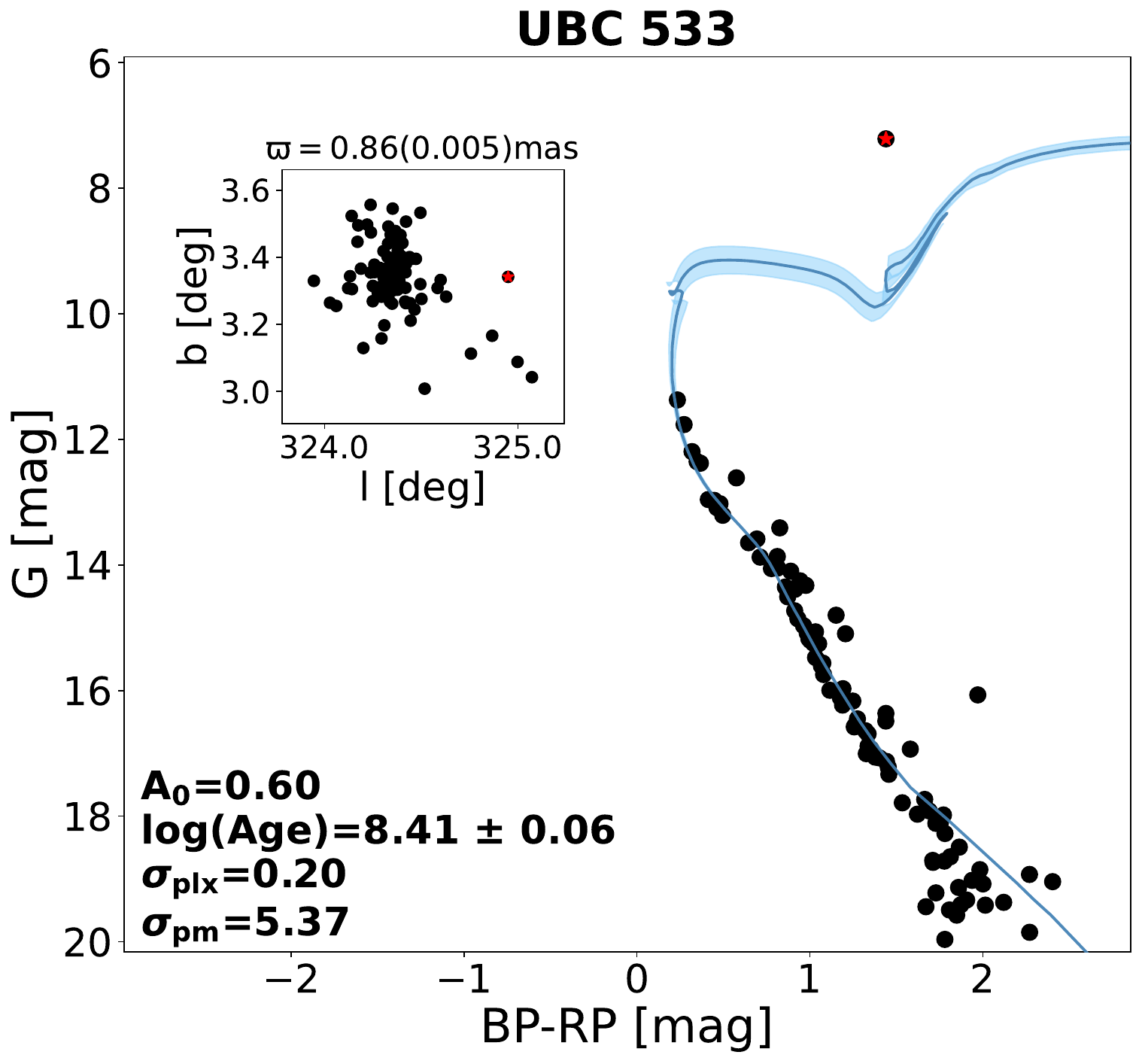}\hspace{0.02\linewidth}
\includegraphics[width=0.23\linewidth]{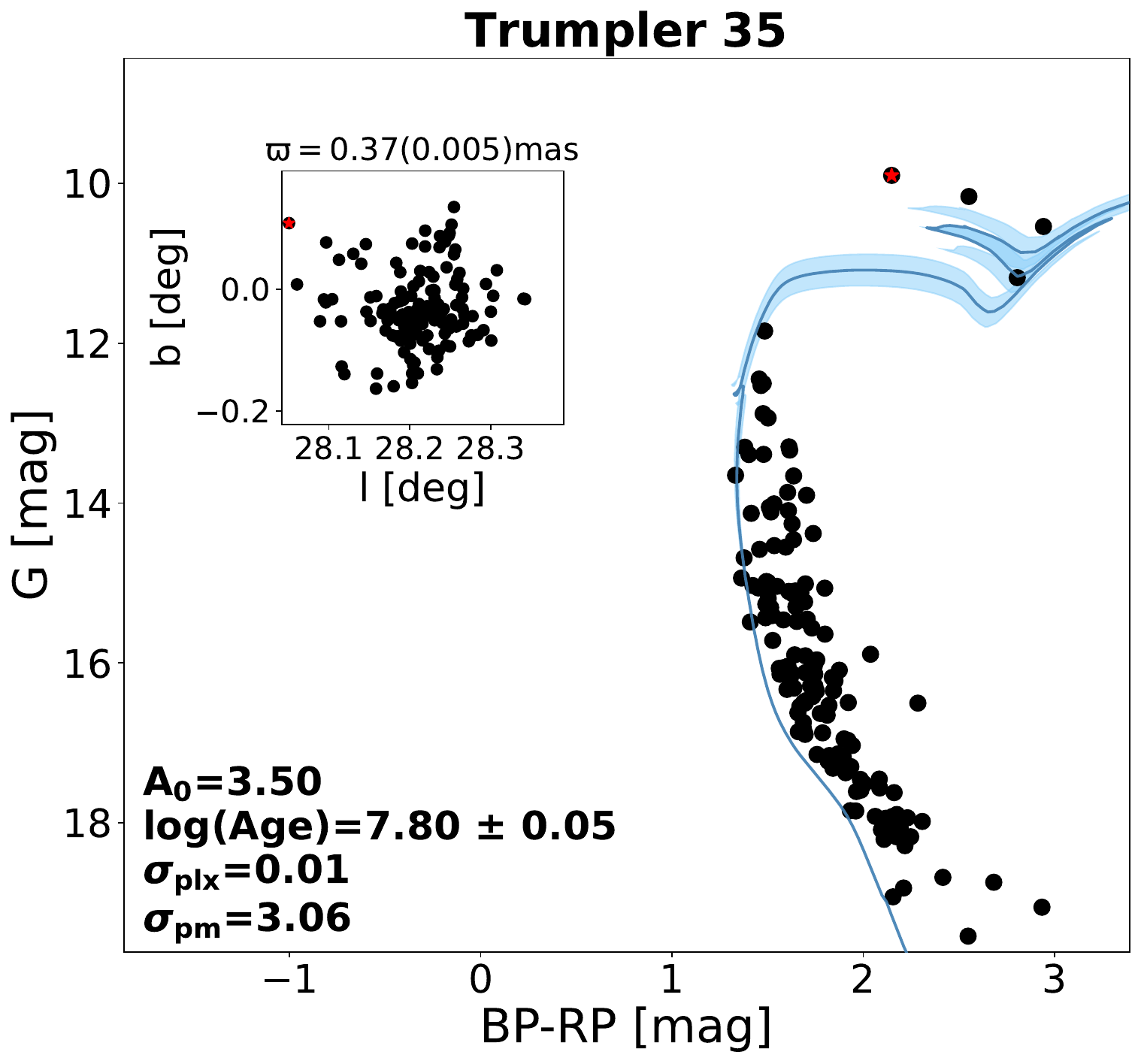}\\[0.4cm]

\includegraphics[width=0.23\linewidth]{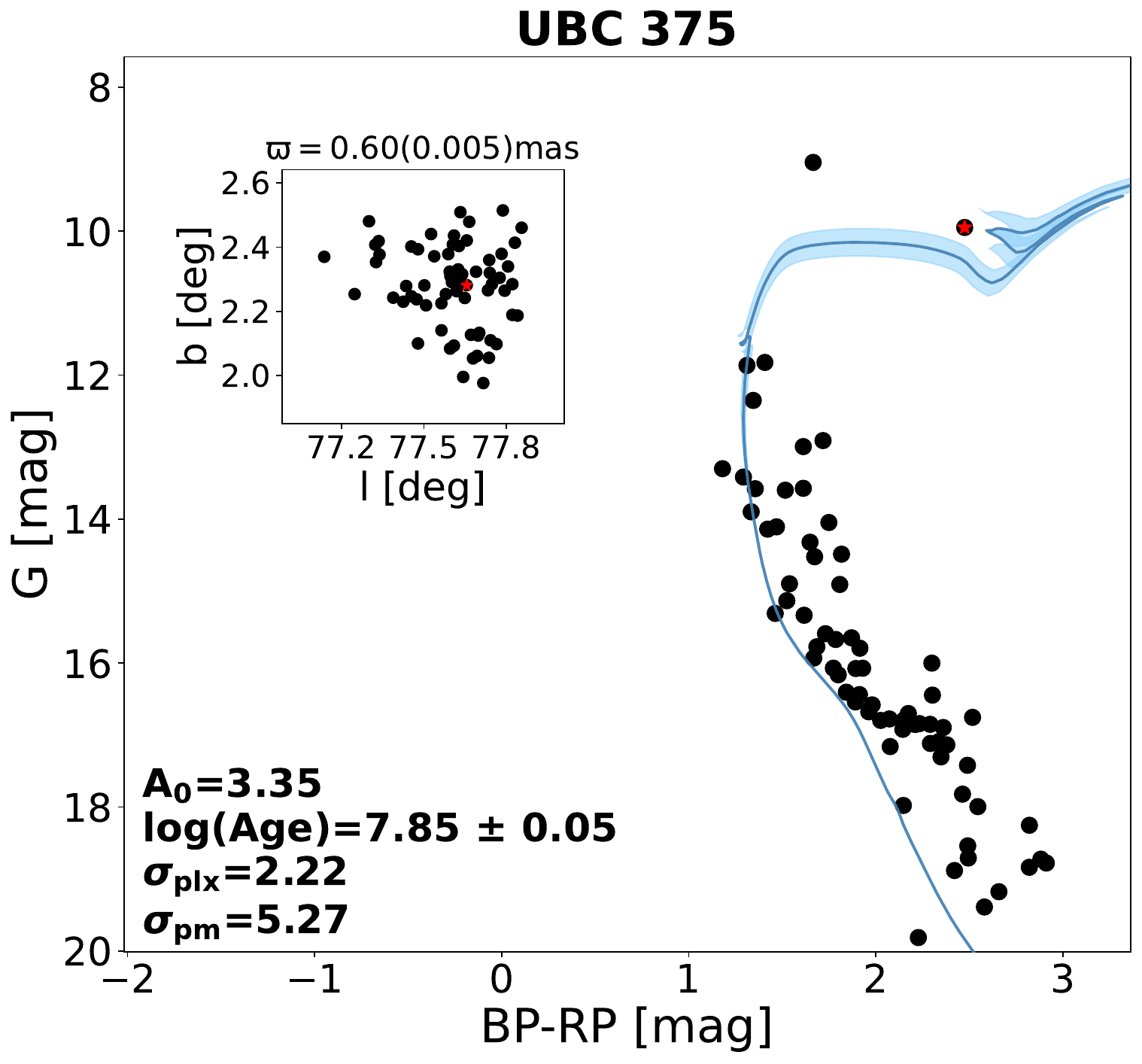}\hspace{0.02\linewidth}
\includegraphics[width=0.23\linewidth]{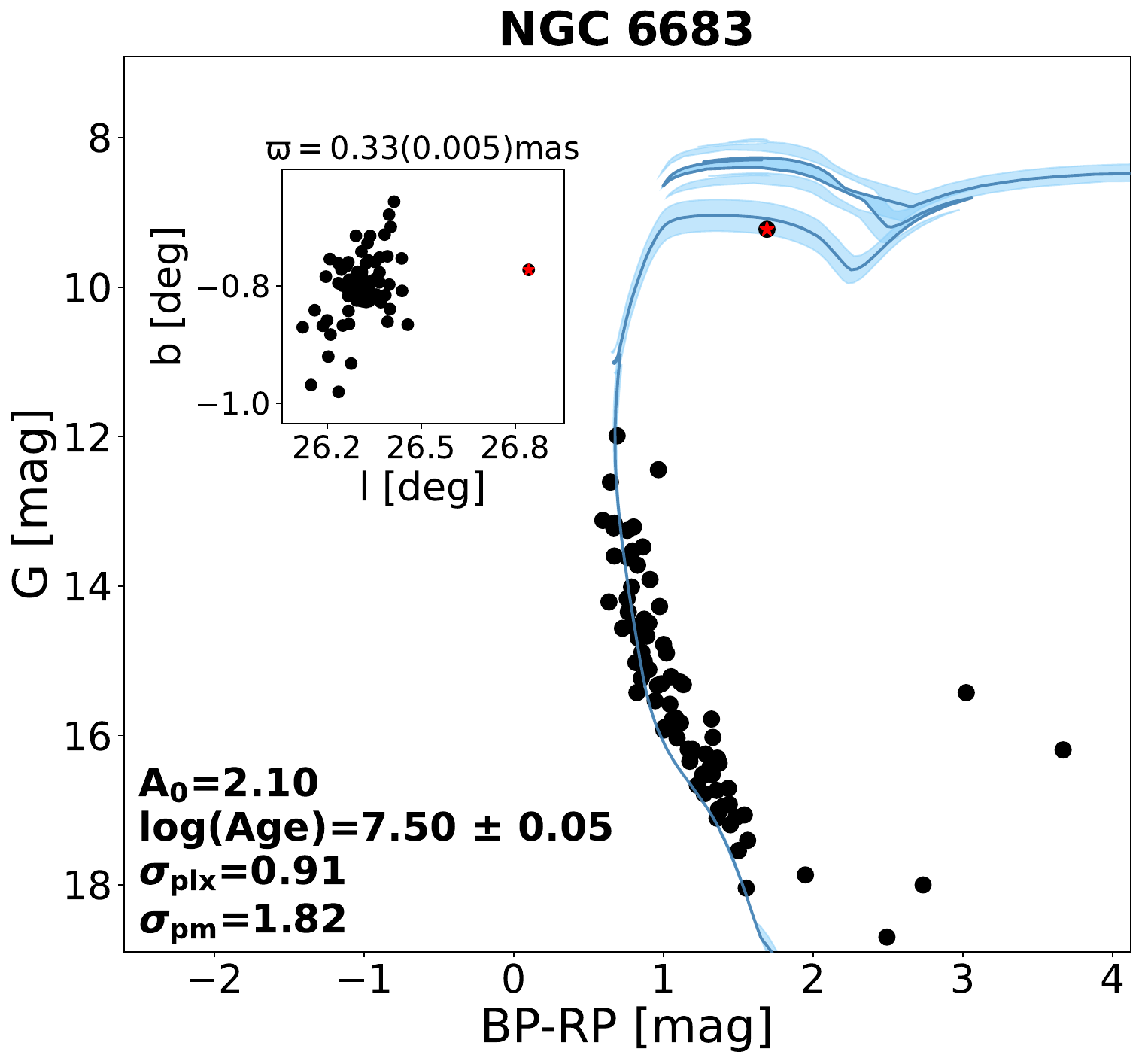}\hspace{0.02\linewidth}
\includegraphics[width=0.23\linewidth]{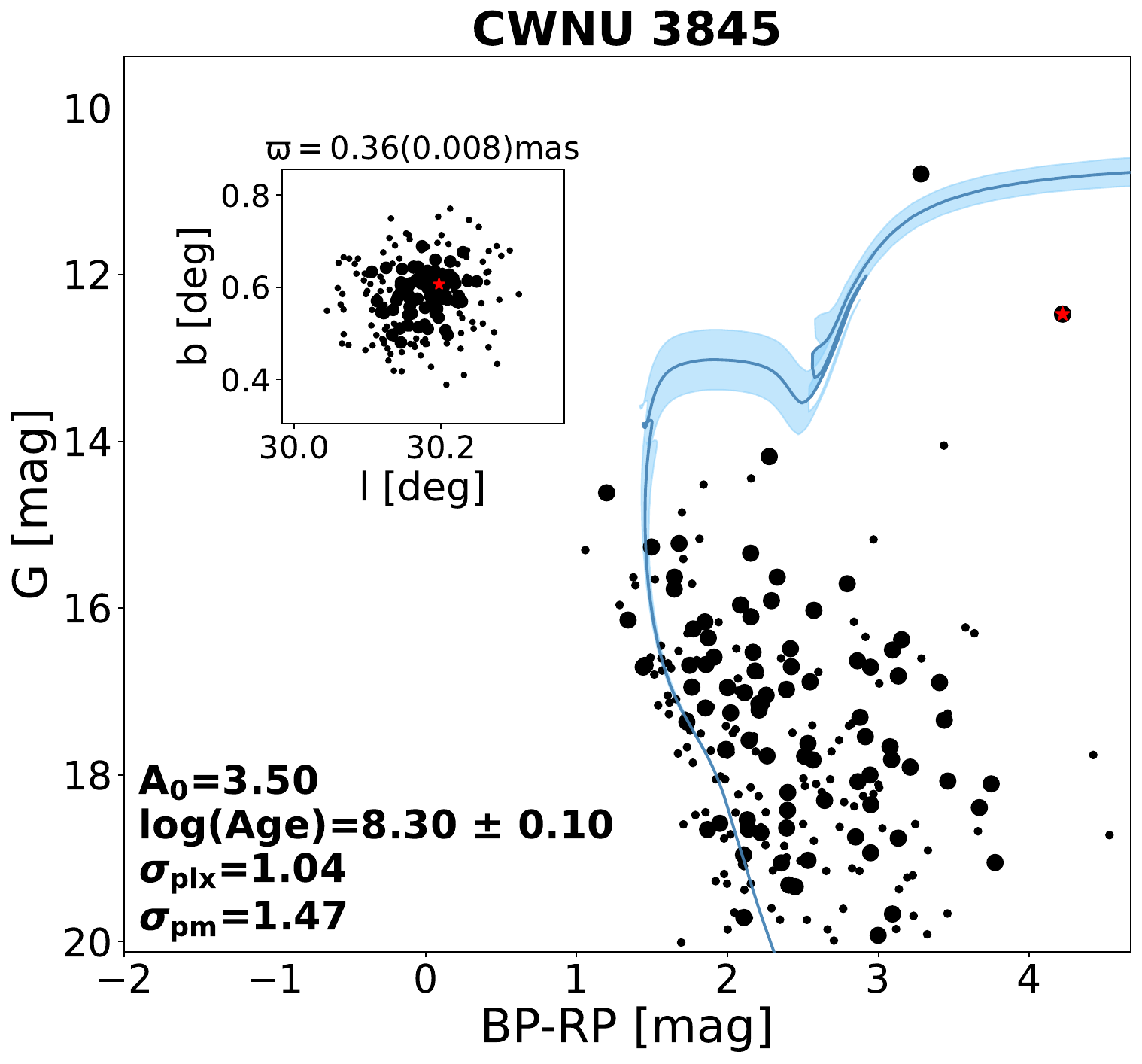}\hspace{0.02\linewidth}
\includegraphics[width=0.23\linewidth]{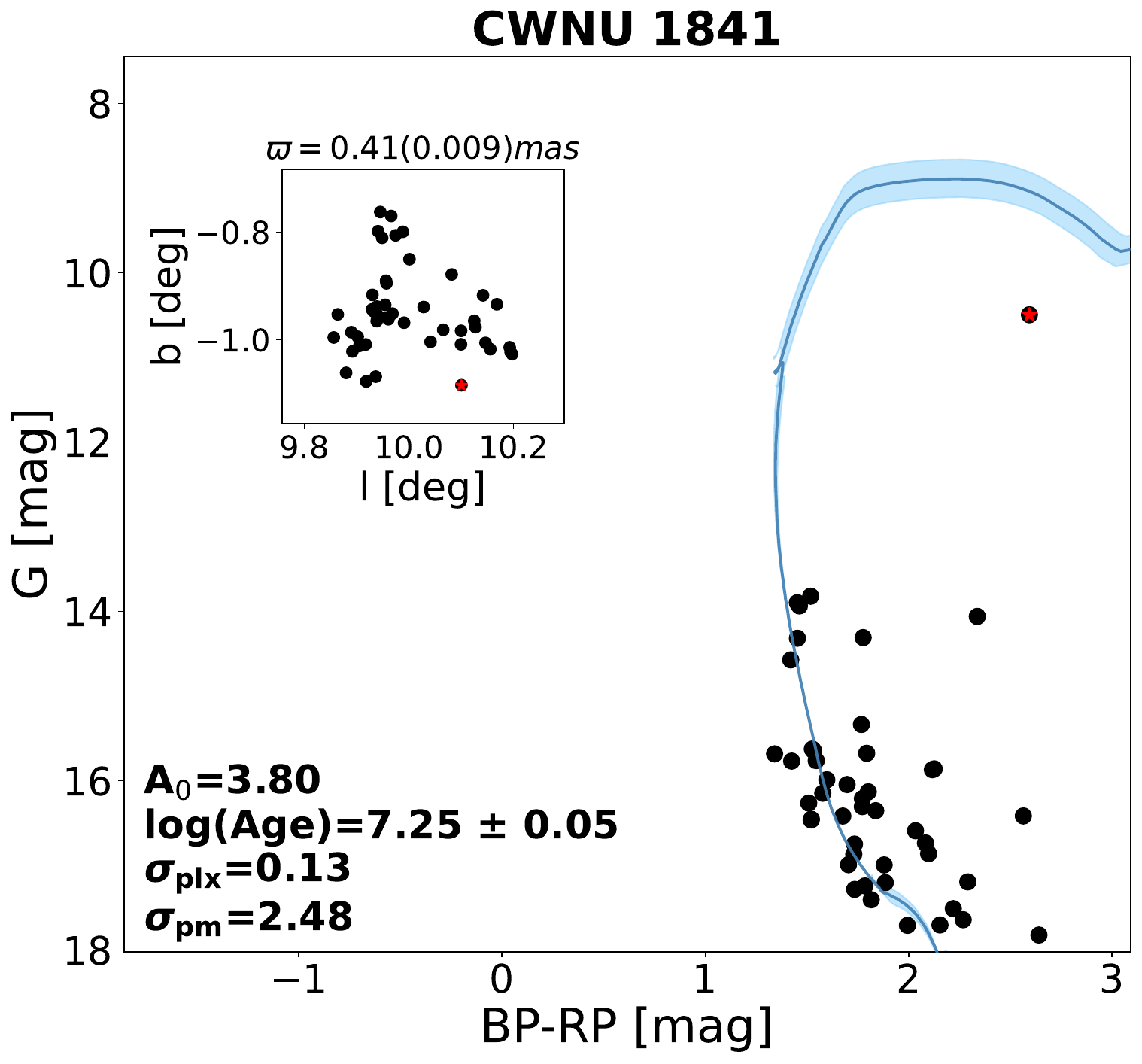}\\[0.4cm]

\includegraphics[width=0.23\linewidth]{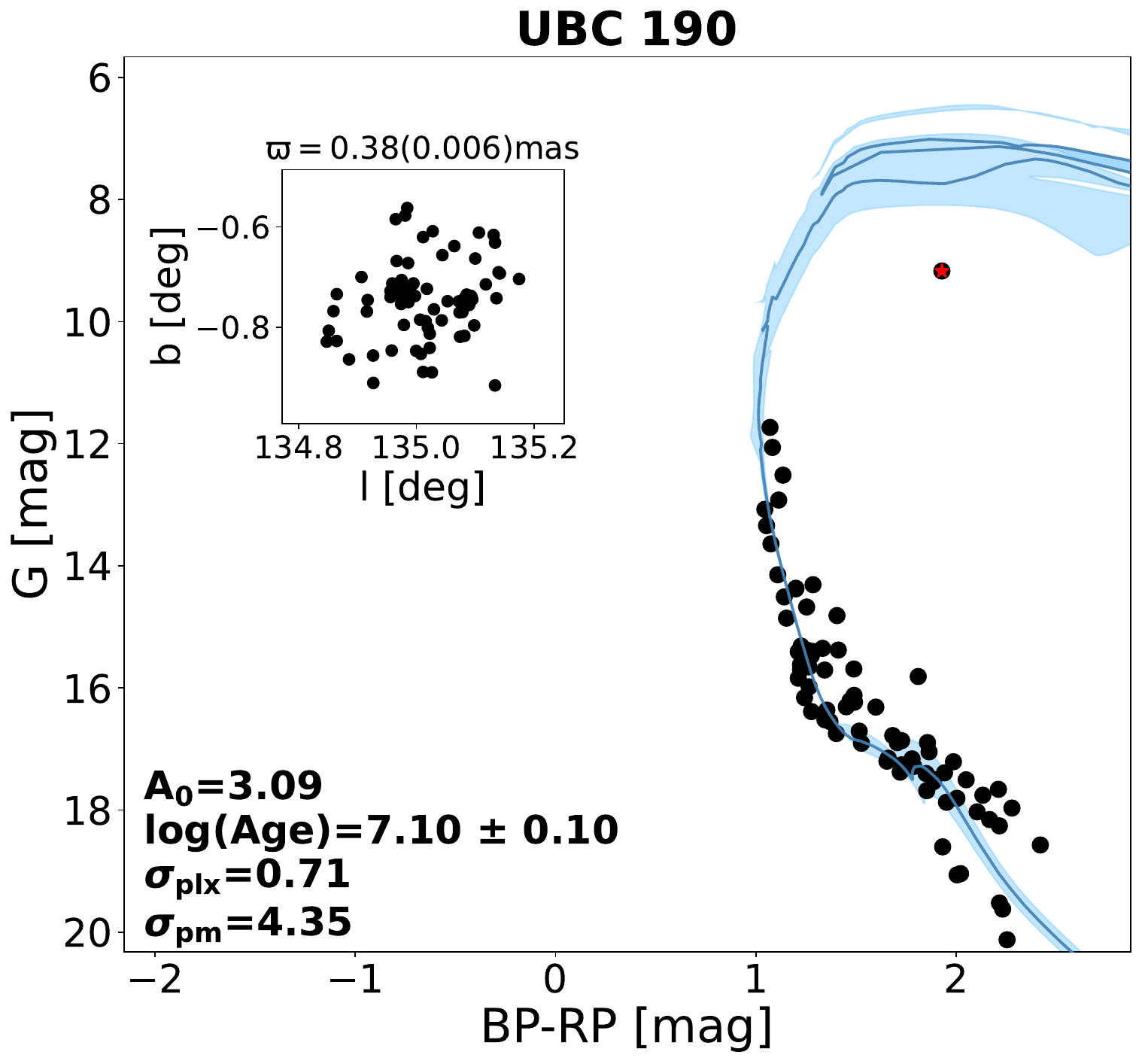}\hspace{0.02\linewidth}
\includegraphics[width=0.23\linewidth]{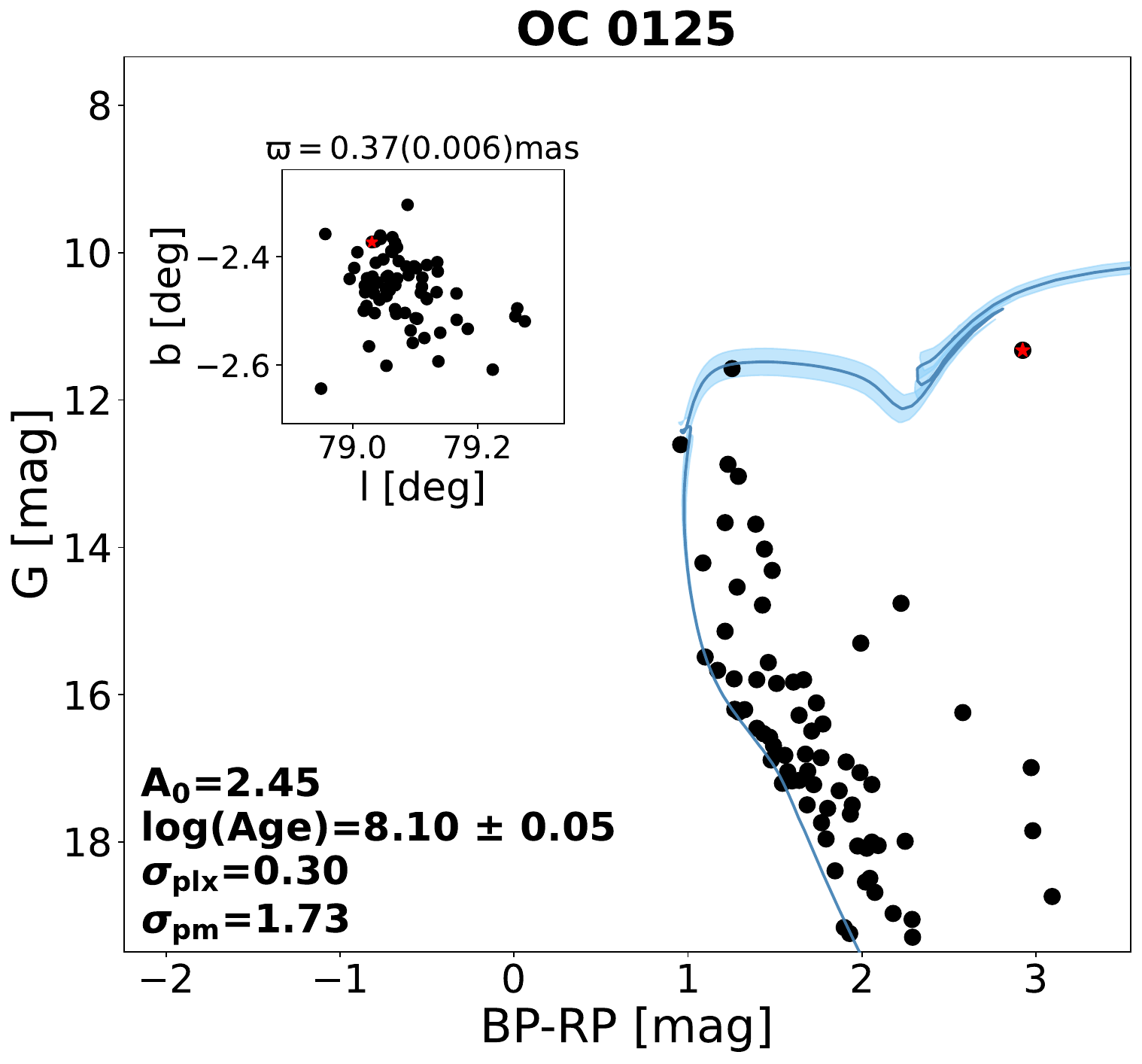}\hspace{0.02\linewidth}
\includegraphics[width=0.23\linewidth]{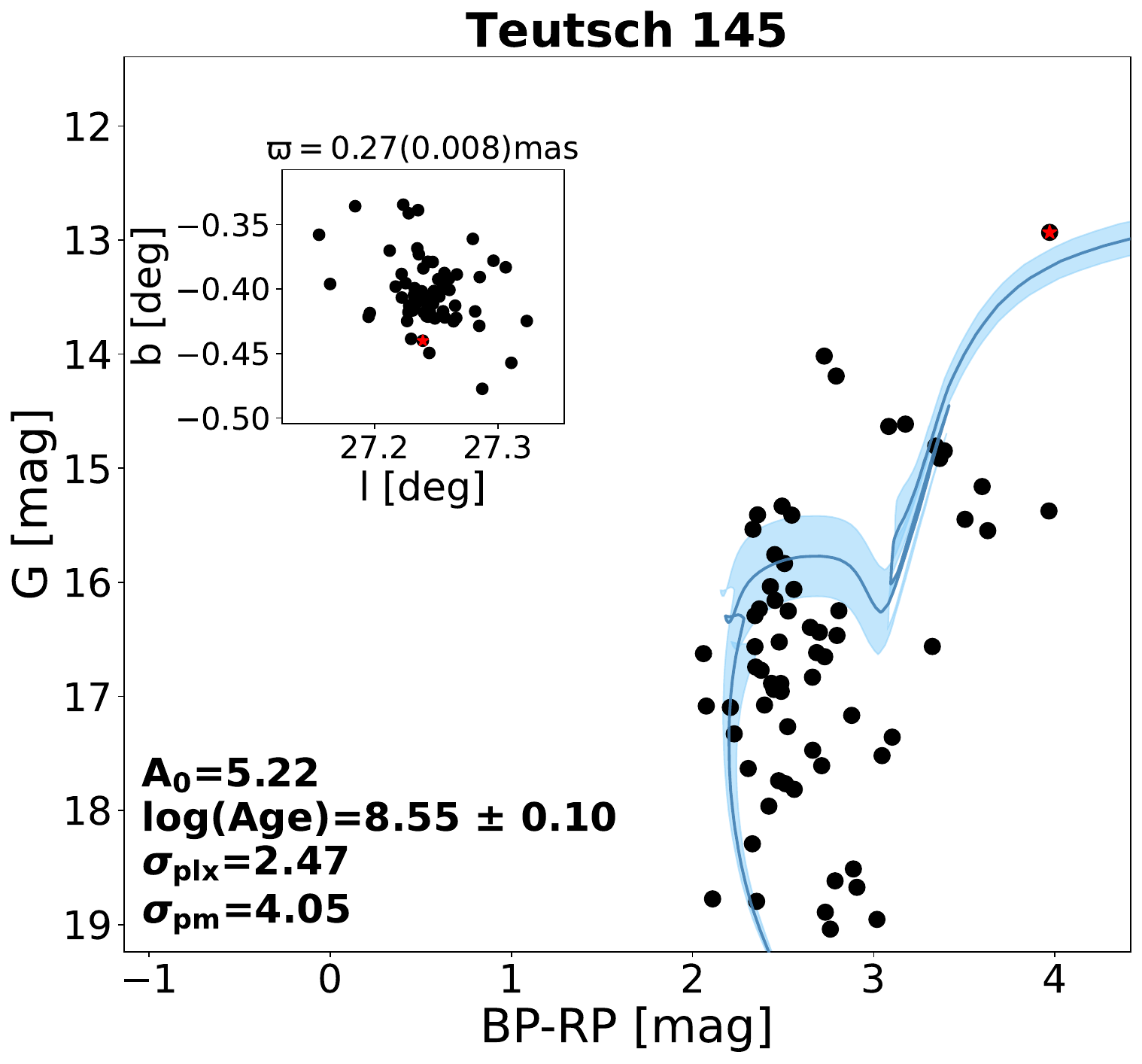}\hspace{0.02\linewidth}
\includegraphics[width=0.23\linewidth]{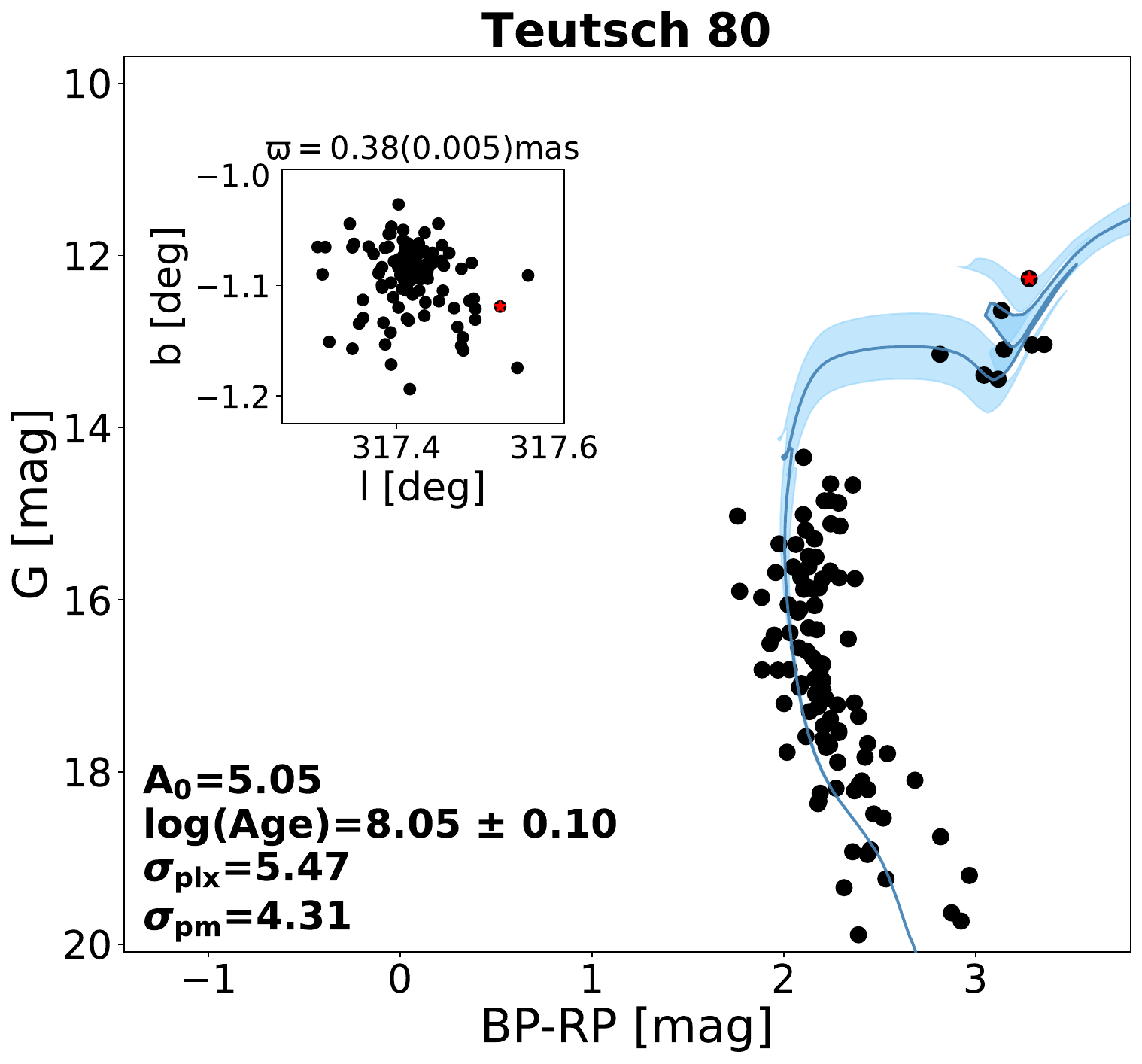}\\[0.4cm]

\caption{CMDs of the remaining rejected OC Cepheid samples, which also contain the extinction, age and distance parameters of the OCs, namely,
CWNU 3968, FSR 0384 (V Lac), SAI 4, FSR 0384 (X Lac), UBC 491, UBC 412, Cl V0378 Cen, Ruprecht 97, COIN-Gaia 36, Platais 1, 
UBC 533, Trumpler 35 (TY Sct), UBC 375, NGC 6683, CWNU 3845, CWNU 1841, UBC 190, OC 0125, Teutsch 145, and Teutsch 80.}
\label{fig:cmd_set5}
\end{figure*}
\begin{figure*}
\centering
\includegraphics[width=0.23\linewidth]{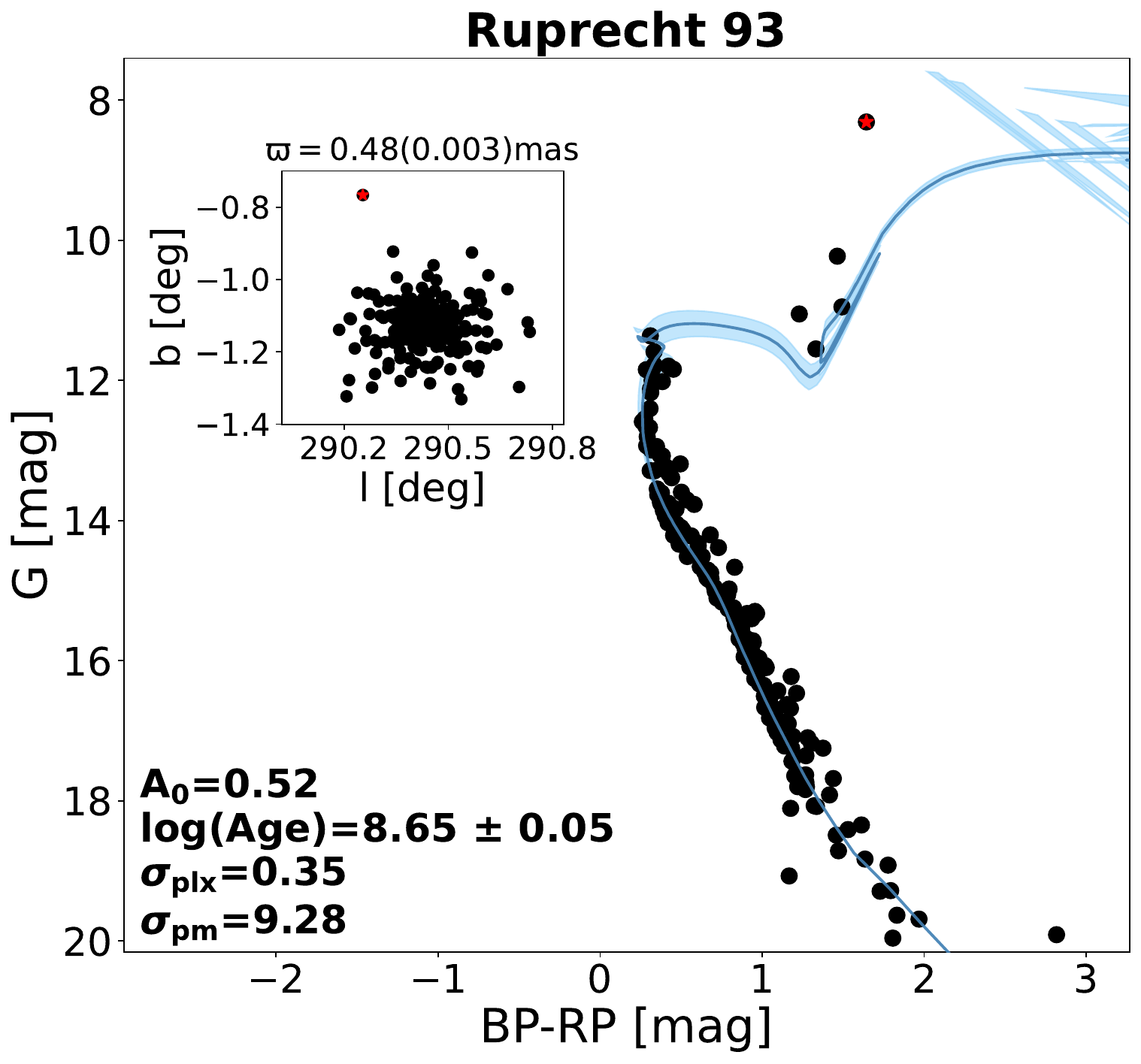}\hspace{0.02\linewidth}
\includegraphics[width=0.23\linewidth]{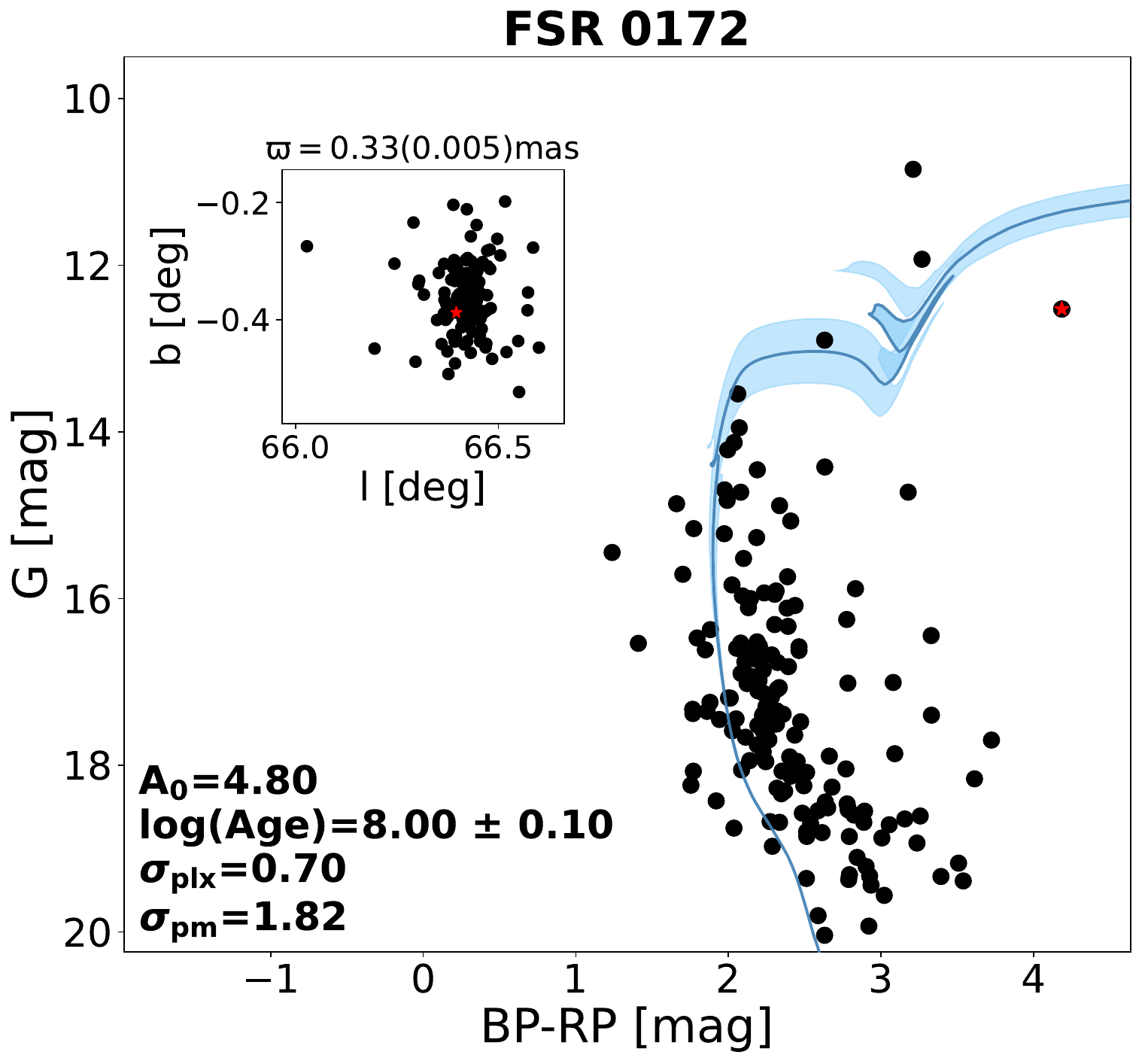}\hspace{0.02\linewidth}
\includegraphics[width=0.23\linewidth]{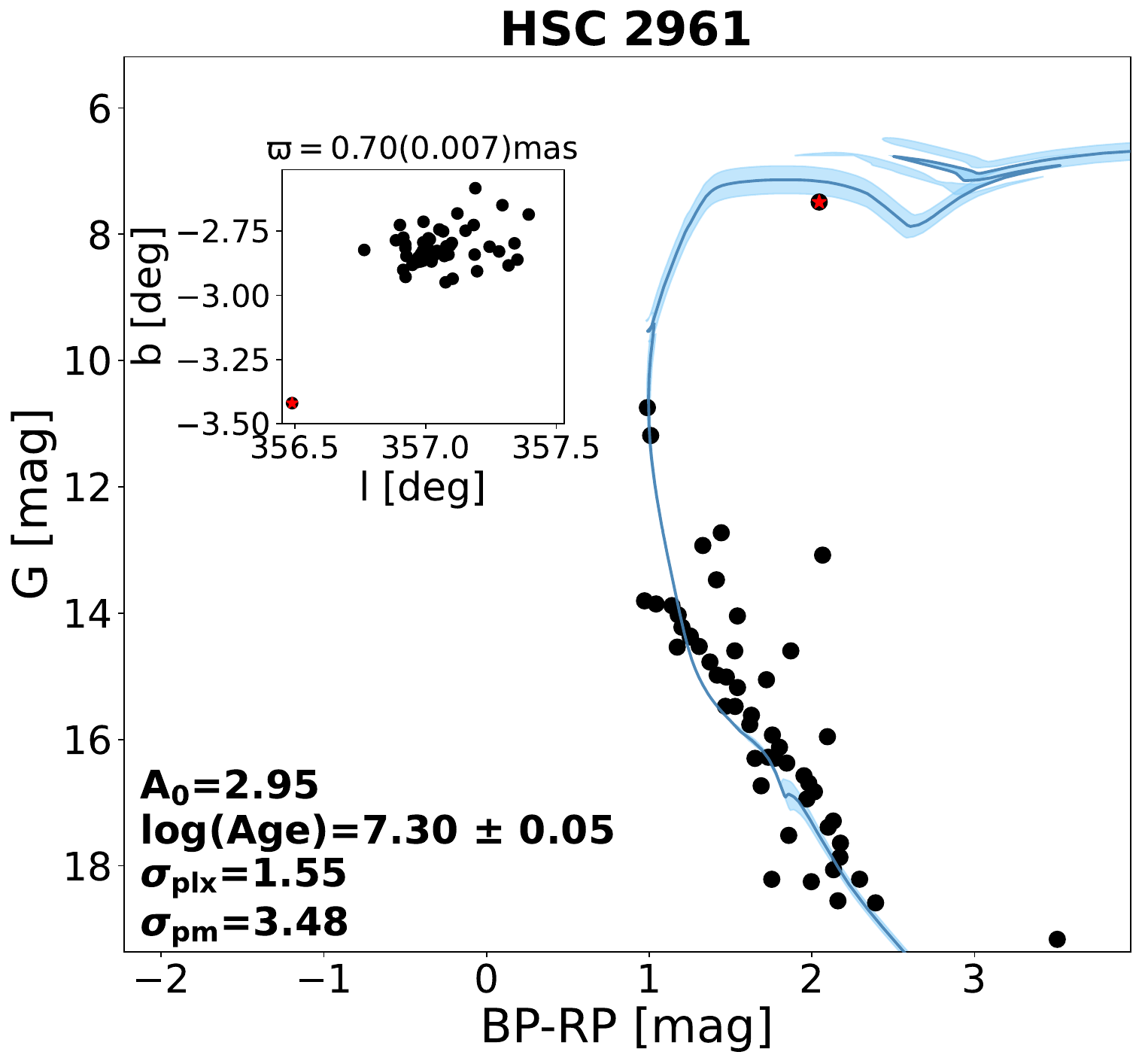}\hspace{0.02\linewidth}
\includegraphics[width=0.23\linewidth]{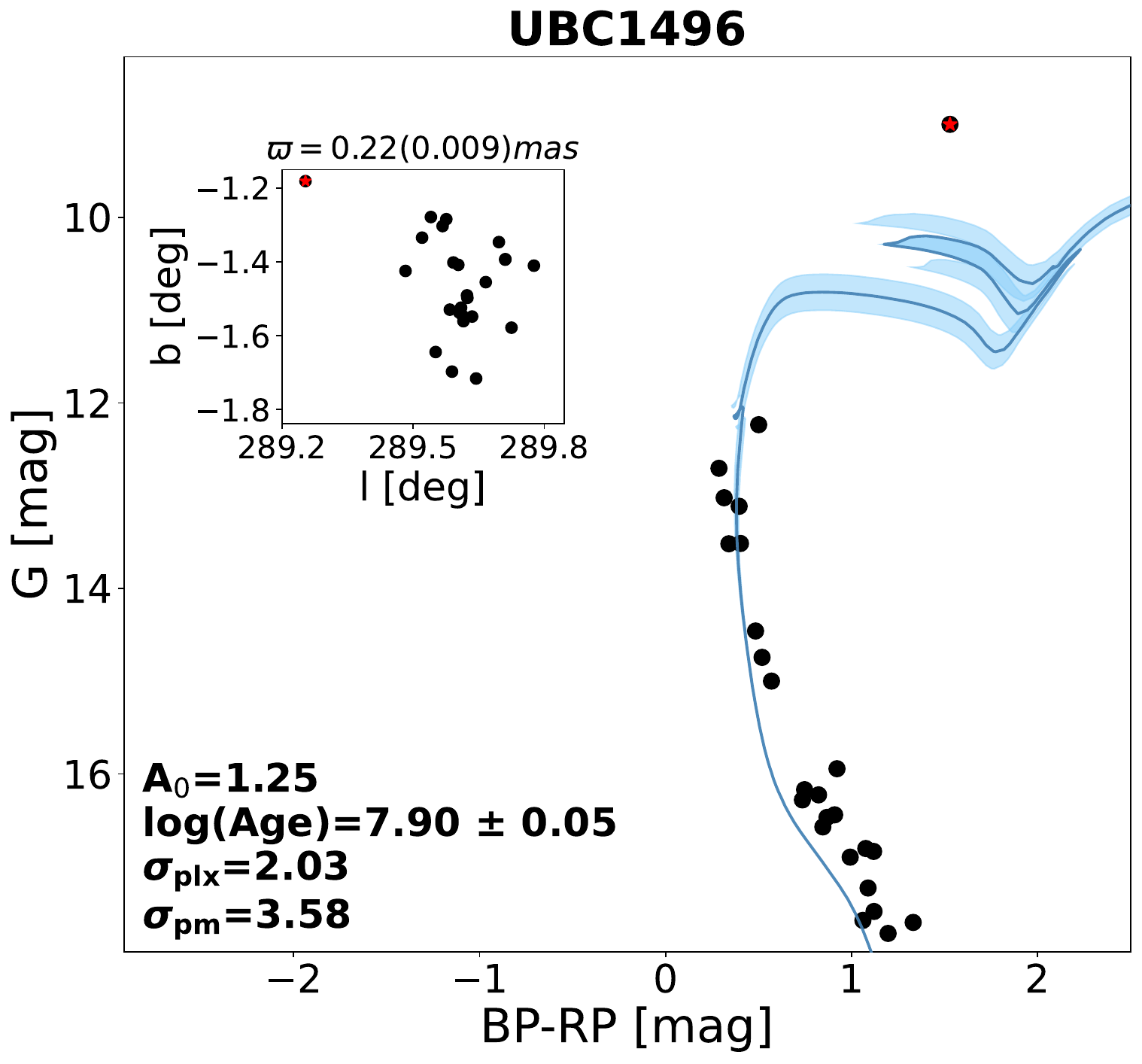}\\[0.4cm]

\includegraphics[width=0.23\linewidth]{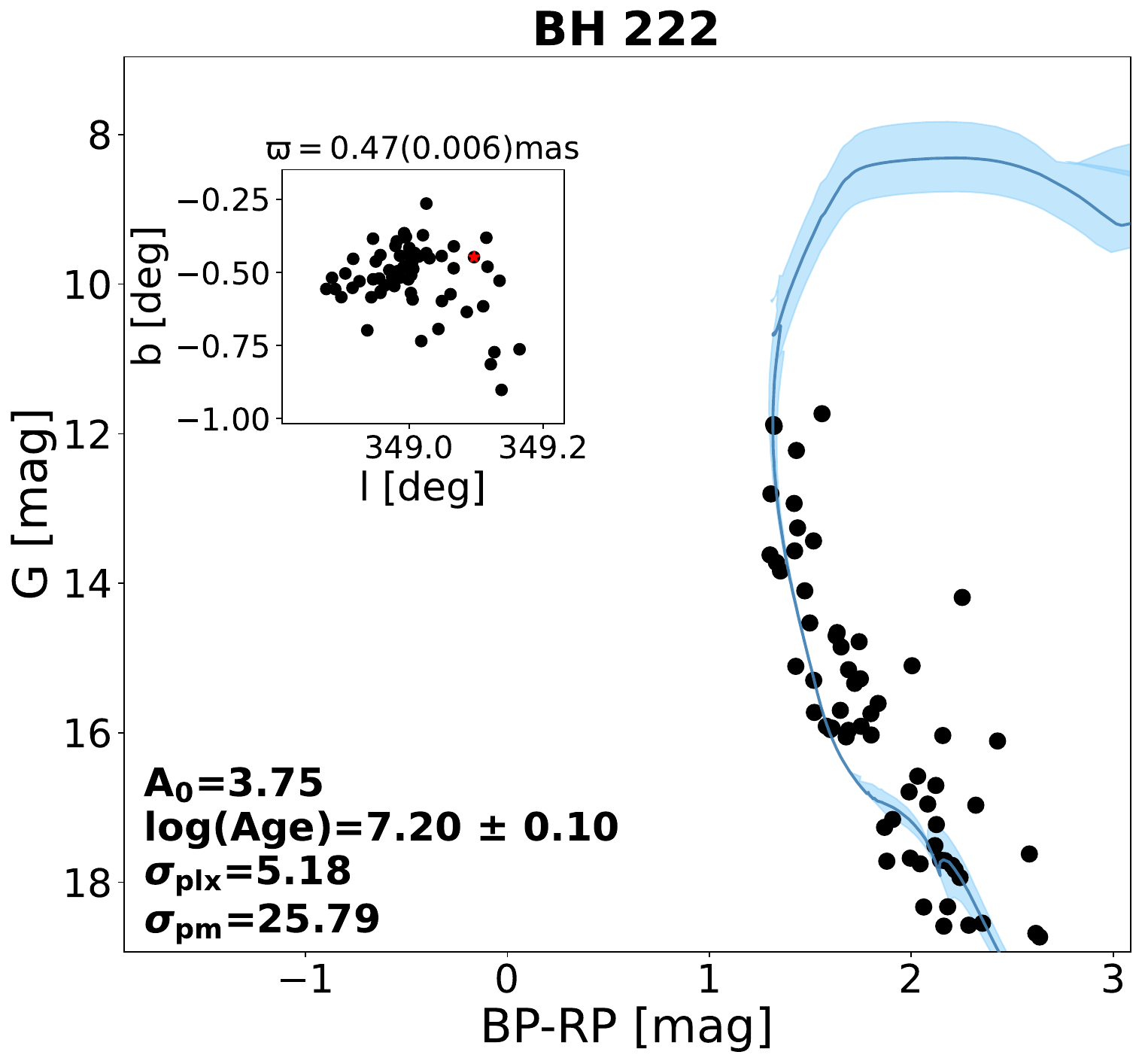}\hspace{0.02\linewidth}
\includegraphics[width=0.23\linewidth]{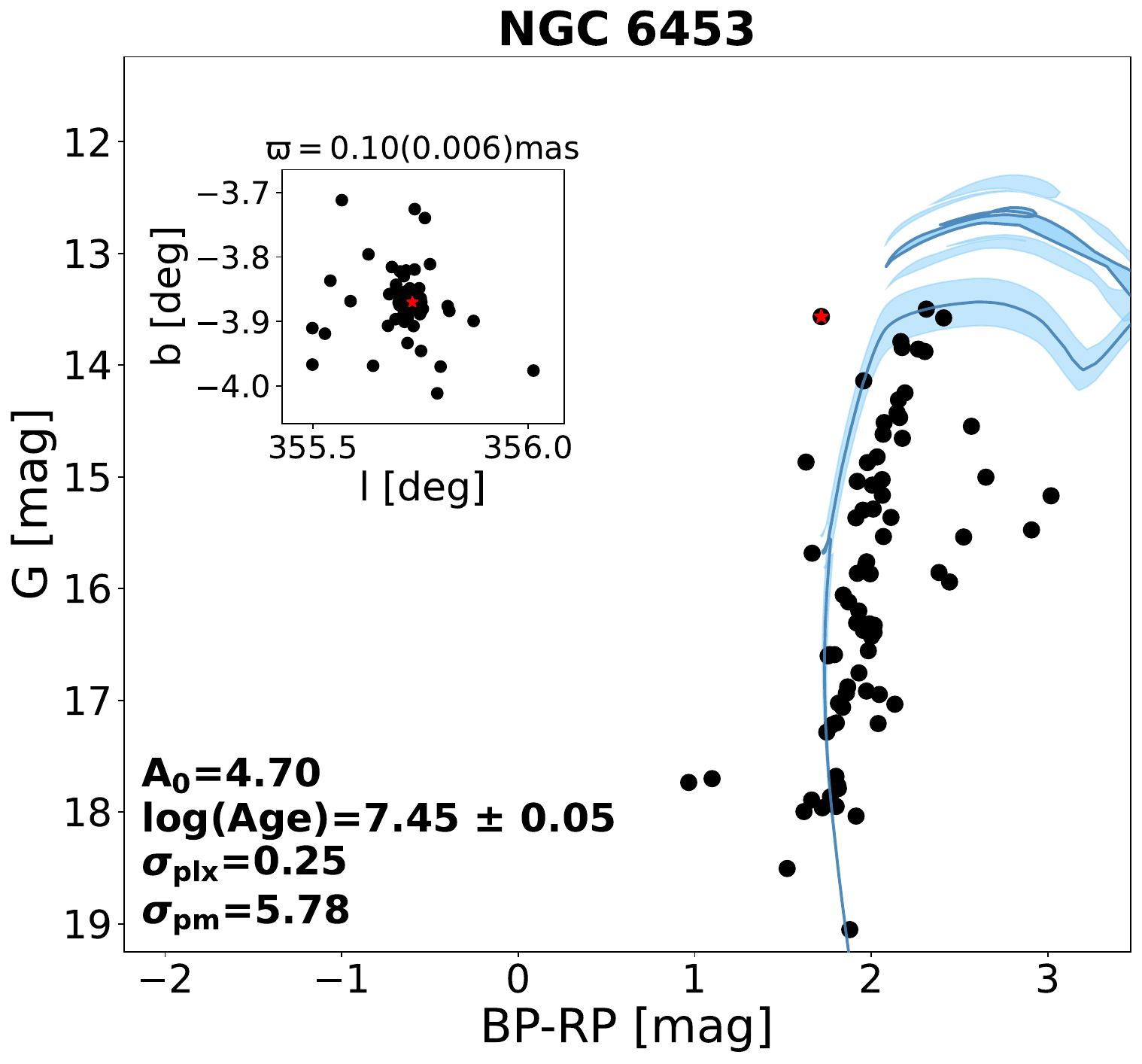}\\[0.4cm]

\caption{CMDs of the remaining rejected OC Cepheid samples, which also contain the extinction, age and distance parameters of the OCs, including 
Ruprecht 93, FSR 0172, HSC 2961, UBC 1496, BH 222, and NGC 6453.}
\end{figure*}

\newpage
\bibliography{references}{}

\begin{thebibliography}{}
\expandafter\ifx\csname natexlab\endcsname\relax\def\natexlab#1{#1}\fi
\providecommand{\url}[1]{\href{#1}{#1}}
\providecommand{\dodoi}[1]{doi:~\href{http://doi.org/#1}{\nolinkurl{#1}}}
\providecommand{\doeprint}[1]{\href{http://ascl.net/#1}{\nolinkurl{http://ascl.net/#1}}}
\providecommand{\doarXiv}[1]{\href{https://arxiv.org/abs/#1}{\nolinkurl{https://arxiv.org/abs/#1}}}

\bibitem[{{Alonso-Santiago} {et~al.}(2020){Alonso-Santiago}, {Negueruela}, {Marco}, {Tabernero}, \& {Castro}}]{Alonso-Santiago2020}
{Alonso-Santiago}, J., {Negueruela}, I., {Marco}, A., {Tabernero}, H.~M., \& {Castro}, N. 2020, \aap, 644, A136, \dodoi{10.1051/0004-6361/202038495}

\bibitem[{{Anderson} {et~al.}(2013){Anderson}, {Eyer}, \& {Mowlavi}}]{Anderson2013}
{Anderson}, R.~I., {Eyer}, L., \& {Mowlavi}, N. 2013, \mnras, 434, 2238, \dodoi{10.1093/mnras/stt1160}

\bibitem[{{Anderson} {et~al.}(2016){Anderson}, {Saio}, {Ekstr{\"o}m}, {Georgy}, \& {Meynet}}]{Anderson2016}
{Anderson}, R.~I., {Saio}, H., {Ekstr{\"o}m}, S., {Georgy}, C., \& {Meynet}, G. 2016, \aap, 591, A8, \dodoi{10.1051/0004-6361/201528031}

\bibitem[{{Bono} {et~al.}(2005){Bono}, {Marconi}, {Cassisi}, {Caputo}, {Gieren}, \& {Pietrzynski}}]{Bono2005}
{Bono}, G., {Marconi}, M., {Cassisi}, S., {et~al.} 2005, \apj, 621, 966, \dodoi{10.1086/427744}

\bibitem[{{Bressan} {et~al.}(2012){Bressan}, {Marigo}, {Girardi}, {Salasnich}, {Dal Cero}, {Rubele}, \& {Nanni}}]{Bressan2012}
{Bressan}, A., {Marigo}, P., {Girardi}, L., {et~al.} 2012, \mnras, 427, 127, \dodoi{10.1111/j.1365-2966.2012.21948.x}

\bibitem[{{Cantat-Gaudin} {et~al.}(2020){Cantat-Gaudin}, {Anders}, {Castro-Ginard}, {Jordi}, {Romero-G{\'o}mez}, {Soubiran}, {Casamiquela}, {Tarricq}, {Moitinho}, {Vallenari}, {Bragaglia}, {Krone-Martins}, \& {Kounkel}}]{CG2020}
{Cantat-Gaudin}, T., {Anders}, F., {Castro-Ginard}, A., {et~al.} 2020, \aap, 640, A1, \dodoi{10.1051/0004-6361/202038192}

\bibitem[{{Castro-Ginard} {et~al.}(2022){Castro-Ginard}, {Jordi}, {Luri}, {Cantat-Gaudin}, {Carrasco}, {Casamiquela}, {Anders}, {Balaguer-N{\'u}{\~n}ez}, \& {Badia}}]{Castro-Ginard2022}
{Castro-Ginard}, A., {Jordi}, C., {Luri}, X., {et~al.} 2022, \aap, 661, A118, \dodoi{10.1051/0004-6361/202142568}

\bibitem[{{Chen} {et~al.}(2019){Chen}, {Wang}, {Deng}, {de Grijs}, {Liu}, \& {Tian}}]{chen2019}
{Chen}, X., {Wang}, S., {Deng}, L., {et~al.} 2019, Nature Astronomy, 3, 320, \dodoi{10.1038/s41550-018-0686-7}

\bibitem[{{Chen} {et~al.}(2018){Chen}, {Wang}, {Deng}, {de Grijs}, \& {Yang}}]{chen2018}
{Chen}, X., {Wang}, S., {Deng}, L., {de Grijs}, R., \& {Yang}, M. 2018, \apjs, 237, 28, \dodoi{10.3847/1538-4365/aad32b}

\bibitem[{{Chen} {et~al.}(2020){Chen}, {Wang}, {Deng}, {de Grijs}, {Yang}, \& {Tian}}]{chen2020}
{Chen}, X., {Wang}, S., {Deng}, L., {et~al.} 2020, \apjs, 249, 18, \dodoi{10.3847/1538-4365/ab9cae}

\bibitem[{{Cruz Reyes} \& {Anderson}(2023)}]{mauricio2023}
{Cruz Reyes}, M., \& {Anderson}, R.~I. 2023, \aap, 672, A85, \dodoi{10.1051/0004-6361/202244775}

\bibitem[{{Cui} {et~al.}(2012){Cui}, {Zhao}, {Chu}, {Li}, {Li}, {Zhang}, {Su}, {Yao}, {Wang}, {Xing}, {Li}, {Zhu}, {Wang}, {Gu}, {Luo}, {Xu}, {Zhang}, {Liu}, {Zhang}, {Yang}, {Cao}, {Chen}, {Chen}, {Chen}, {Chen}, {Chu}, {Feng}, {Gong}, {Hou}, {Hu}, {Hu}, {Hu}, {Jia}, {Jiang}, {Jiang}, {Jiang}, {Jin}, {Li}, {Li}, {Li}, {Liu}, {Liu}, {Lu}, {Mao}, {Men}, {Qi}, {Qi}, {Shi}, {Tang}, {Tao}, {Wang}, {Wang}, {Wang}, {Wang}, {Wang}, {Wang}, {Wang}, {Wang}, {Wang}, {Wang}, {Wang}, {Wang}, {Xu}, {Xu}, {Yang}, {Yu}, {Yuan}, {Yuan}, {Zhai}, {Zhang}, {Zhang}, {Zhang}, {Zhao}, {Zhou}, {Zhou}, {Zhu}, \& {Zou}}]{Cui2012}
{Cui}, X.-Q., {Zhao}, Y.-H., {Chu}, Y.-Q., {et~al.} 2012, Research in Astronomy and Astrophysics, 12, 1197, \dodoi{10.1088/1674-4527/12/9/003}

\bibitem[{{De Somma} {et~al.}(2021){De Somma}, {Marconi}, {Cassisi}, {Ripepi}, {Pietrinferni}, {Molinaro}, {Leccia}, \& {Musella}}]{DeSomma2021}
{De Somma}, G., {Marconi}, M., {Cassisi}, S., {et~al.} 2021, \mnras, 508, 1473, \dodoi{10.1093/mnras/stab2611}

\bibitem[{{Deka} {et~al.}(2024){Deka}, {Bellinger}, {Kanbur}, {Deb}, {Bhardwaj}, {Randall}, {Kalici}, \& {Das}}]{Deka2024}
{Deka}, M., {Bellinger}, E.~P., {Kanbur}, S.~M., {et~al.} 2024, \mnras, 530, 5099, \dodoi{10.1093/mnras/stae1136}

\bibitem[{{Dias} {et~al.}(2021){Dias}, {Monteiro}, {Moitinho}, {L{\'e}pine}, {Carraro}, {Paunzen}, {Alessi}, \& {Villela}}]{dias2021}
{Dias}, W.~S., {Monteiro}, H., {Moitinho}, A., {et~al.} 2021, \mnras, 504, 356, \dodoi{10.1093/mnras/stab770}

\bibitem[{{Efremov}(1978)}]{Efremov1978}
{Efremov}, I.~N. 1978, \sovast, 22, 161

\bibitem[{{Espinoza-Arancibia} {et~al.}(2024){Espinoza-Arancibia}, {Pilecki}, {Pietrzy{\'n}ski}, {Smolec}, \& {Kervella}}]{Espinoza-Arancibia2024}
{Espinoza-Arancibia}, F., {Pilecki}, B., {Pietrzy{\'n}ski}, G., {Smolec}, R., \& {Kervella}, P. 2024, \aap, 682, A185, \dodoi{10.1051/0004-6361/202347804}

\bibitem[{{Evans} {et~al.}(1990){Evans}, {Szabados}, \& {Udalska}}]{Evans1990}
{Evans}, N.~R., {Szabados}, L., \& {Udalska}, J. 1990, \pasp, 102, 981, \dodoi{10.1086/132727}

\bibitem[{{Fabricius} {et~al.}(2021){Fabricius}, {Luri}, {Arenou}, {Babusiaux}, {Helmi}, {Muraveva}, {Reyl{\'e}}, {Spoto}, {Vallenari}, {Antoja}, {Balbinot}, {Barache}, {Bauchet}, {Bragaglia}, {Busonero}, {Cantat-Gaudin}, {Carrasco}, {Diakit{\'e}}, {Fabrizio}, {Figueras}, {Garcia-Gutierrez}, {Garofalo}, {Jordi}, {Kervella}, {Khanna}, {Leclerc}, {Licata}, {Lambert}, {Marrese}, {Masip}, {Ramos}, {Robichon}, {Robin}, {Romero-G{\'o}mez}, {Rubele}, \& {Weiler}}]{Fabricius2021}
{Fabricius}, C., {Luri}, X., {Arenou}, F., {et~al.} 2021, \aap, 649, A5, \dodoi{10.1051/0004-6361/202039834}

\bibitem[{{Freedman} {et~al.}(2001){Freedman}, {Madore}, {Gibson}, {Ferrarese}, {Kelson}, {Sakai}, {Mould}, {Kennicutt}, {Ford}, {Graham}, {Huchra}, {Hughes}, {Illingworth}, {Macri}, \& {Stetson}}]{Freedman2001}
{Freedman}, W.~L., {Madore}, B.~F., {Gibson}, B.~K., {et~al.} 2001, \apj, 553, 47, \dodoi{10.1086/320638}

\bibitem[{{Gaia Collaboration} {et~al.}(2018){Gaia Collaboration}, {Brown}, {Vallenari}, {Prusti}, {de Bruijne}, {Babusiaux}, {Bailer-Jones}, {Biermann}, {Evans}, {Eyer}, {Jansen}, {Jordi}, {Klioner}, {Lammers}, {Lindegren}, {Luri}, {Mignard}, {Panem}, {Pourbaix}, {Randich}, {Sartoretti}, {Siddiqui}, {Soubiran}, {van Leeuwen}, {Walton}, {Arenou}, {Bastian}, {Cropper}, {Drimmel}, {Katz}, {Lattanzi}, {Bakker}, {Cacciari}, {Casta{\~n}eda}, {Chaoul}, {Cheek}, {De Angeli}, {Fabricius}, {Guerra}, {Holl}, {Masana}, {Messineo}, {Mowlavi}, {Nienartowicz}, {Panuzzo}, {Portell}, {Riello}, {Seabroke}, {Tanga}, {Th{\'e}venin}, {Gracia-Abril}, {Comoretto}, {Garcia-Reinaldos}, {Teyssier}, {Altmann}, {Andrae}, {Audard}, {Bellas-Velidis}, {Benson}, {Berthier}, {Blomme}, {Burgess}, {Busso}, {Carry}, {Cellino}, {Clementini}, {Clotet}, {Creevey}, {Davidson}, {De Ridder}, {Delchambre}, {Dell'Oro}, {Ducourant}, {Fern{\'a}ndez-Hern{\'a}ndez}, {Fouesneau}, {Fr{\'e}mat}, {Galluccio}, {Garc{\'\i}a-Torres},
  {Gonz{\'a}lez-N{\'u}{\~n}ez}, {Gonz{\'a}lez-Vidal}, {Gosset}, {Guy}, {Halbwachs}, {Hambly}, {Harrison}, {Hern{\'a}ndez}, {Hestroffer}, {Hodgkin}, {Hutton}, {Jasniewicz}, {Jean-Antoine-Piccolo}, {Jordan}, {Korn}, {Krone-Martins}, {Lanzafame}, {Lebzelter}, {L{\"o}ffler}, {Manteiga}, {Marrese}, {Mart{\'\i}n-Fleitas}, {Moitinho}, {Mora}, {Muinonen}, {Osinde}, {Pancino}, {Pauwels}, {Petit}, {Recio-Blanco}, {Richards}, {Rimoldini}, {Robin}, {Sarro}, {Siopis}, {Smith}, {Sozzetti}, {S{\"u}veges}, {Torra}, {van Reeven}, {Abbas}, {Abreu Aramburu}, {Accart}, {Aerts}, {Altavilla}, {{\'A}lvarez}, {Alvarez}, {Alves}, {Anderson}, {Andrei}, {Anglada Varela}, {Antiche}, {Antoja}, {Arcay}, {Astraatmadja}, {Bach}, {Baker}, {Balaguer-N{\'u}{\~n}ez}, {Balm}, {Barache}, {Barata}, {Barbato}, {Barblan}, {Barklem}, {Barrado}, {Barros}, {Barstow}, {Bartholom{\'e} Mu{\~n}oz}, {Bassilana}, {Becciani}, {Bellazzini}, {Berihuete}, {Bertone}, {Bianchi}, {Bienaym{\'e}}, {Blanco-Cuaresma}, {Boch}, {Boeche}, {Bombrun}, {Borrachero},
  {Bossini}, {Bouquillon}, {Bourda}, {Bragaglia}, {Bramante}, {Breddels}, {Bressan}, {Brouillet}, {Br{\"u}semeister}, {Brugaletta}, {Bucciarelli}, {Burlacu}, {Busonero}, {Butkevich}, {Buzzi}, {Caffau}, {Cancelliere}, {Cannizzaro}, {Cantat-Gaudin}, {Carballo}, {Carlucci}, {Carrasco}, {Casamiquela}, {Castellani}, {Castro-Ginard}, {Charlot}, {Chemin}, {Chiavassa}, {Cocozza}, {Costigan}, {Cowell}, {Crifo}, {Crosta}, {Crowley}, {Cuypers}, {Dafonte}, {Damerdji}, {Dapergolas}, {David}, {David}, {de Laverny}, \& {De Luise}}]{gaia2018}
{Gaia Collaboration}, {Brown}, A.~G.~A., {Vallenari}, A., {et~al.} 2018, \aap, 616, A1, \dodoi{10.1051/0004-6361/201833051}

\bibitem[{{Gaia Collaboration} {et~al.}(2021){Gaia Collaboration}, {Brown}, {Vallenari}, {Prusti}, {de Bruijne}, {Babusiaux}, {Biermann}, {Creevey}, {Evans}, {Eyer}, {Hutton}, {Jansen}, {Jordi}, {Klioner}, {Lammers}, {Lindegren}, {Luri}, {Mignard}, {Panem}, {Pourbaix}, {Randich}, {Sartoretti}, {Soubiran}, {Walton}, {Arenou}, {Bailer-Jones}, {Bastian}, {Cropper}, {Drimmel}, {Katz}, {Lattanzi}, {van Leeuwen}, {Bakker}, {Cacciari}, {Casta{\~n}eda}, {De Angeli}, {Ducourant}, {Fabricius}, {Fouesneau}, {Fr{\'e}mat}, {Guerra}, {Guerrier}, {Guiraud}, {Jean-Antoine Piccolo}, {Masana}, {Messineo}, {Mowlavi}, {Nicolas}, {Nienartowicz}, {Pailler}, {Panuzzo}, {Riclet}, {Roux}, {Seabroke}, {Sordo}, {Tanga}, {Th{\'e}venin}, {Gracia-Abril}, {Portell}, {Teyssier}, {Altmann}, {Andrae}, {Bellas-Velidis}, {Benson}, {Berthier}, {Blomme}, {Brugaletta}, {Burgess}, {Busso}, {Carry}, {Cellino}, {Cheek}, {Clementini}, {Damerdji}, {Davidson}, {Delchambre}, {Dell'Oro}, {Fern{\'a}ndez-Hern{\'a}ndez}, {Galluccio}, {Garc{\'\i}a-Lario},
  {Garcia-Reinaldos}, {Gonz{\'a}lez-N{\'u}{\~n}ez}, {Gosset}, {Haigron}, {Halbwachs}, {Hambly}, {Harrison}, {Hatzidimitriou}, {Heiter}, {Hern{\'a}ndez}, {Hestroffer}, {Hodgkin}, {Holl}, {Jan{\ss}en}, {Jevardat de Fombelle}, {Jordan}, {Krone-Martins}, {Lanzafame}, {L{\"o}ffler}, {Lorca}, {Manteiga}, {Marchal}, {Marrese}, {Moitinho}, {Mora}, {Muinonen}, {Osborne}, {Pancino}, {Pauwels}, {Petit}, {Recio-Blanco}, {Richards}, {Riello}, {Rimoldini}, {Robin}, {Roegiers}, {Rybizki}, {Sarro}, {Siopis}, {Smith}, {Sozzetti}, {Ulla}, {Utrilla}, {van Leeuwen}, {van Reeven}, {Abbas}, {Abreu Aramburu}, {Accart}, {Aerts}, {Aguado}, {Ajaj}, {Altavilla}, {{\'A}lvarez}, {{\'A}lvarez Cid-Fuentes}, {Alves}, {Anderson}, {Anglada Varela}, {Antoja}, {Audard}, {Baines}, {Baker}, {Balaguer-N{\'u}{\~n}ez}, {Balbinot}, {Balog}, {Barache}, {Barbato}, {Barros}, {Barstow}, {Bartolom{\'e}}, {Bassilana}, {Bauchet}, {Baudesson-Stella}, {Becciani}, {Bellazzini}, {Bernet}, {Bertone}, {Bianchi}, {Blanco-Cuaresma}, {Boch}, {Bombrun}, {Bossini},
  {Bouquillon}, {Bragaglia}, {Bramante}, {Breedt}, {Bressan}, {Brouillet}, {Bucciarelli}, {Burlacu}, {Busonero}, {Butkevich}, {Buzzi}, {Caffau}, {Cancelliere}, {C{\'a}novas}, {Cantat-Gaudin}, {Carballo}, {Carlucci}, {Carnerero}, {Carrasco}, {Casamiquela}, {Castellani}, {Castro-Ginard}, {Castro Sampol}, {Chaoul}, {Charlot}, {Chemin}, {Chiavassa}, {Cioni}, {Comoretto}, {Cooper}, {Cornez}, {Cowell}, {Crifo}, {Crosta}, {Crowley}, {Dafonte}, {Dapergolas}, {David}, \& {David}}]{gaia2021}
---. 2021, \aap, 649, A1, \dodoi{10.1051/0004-6361/202039657}

\bibitem[{{Gaia Collaboration} {et~al.}(2023){Gaia Collaboration}, {Vallenari}, {Brown}, {Prusti}, {de Bruijne}, {Arenou}, {Babusiaux}, {Biermann}, {Creevey}, {Ducourant}, {Evans}, {Eyer}, {Guerra}, {Hutton}, {Jordi}, {Klioner}, {Lammers}, {Lindegren}, {Luri}, {Mignard}, {Panem}, {Pourbaix}, {Randich}, {Sartoretti}, {Soubiran}, {Tanga}, {Walton}, {Bailer-Jones}, {Bastian}, {Drimmel}, {Jansen}, {Katz}, {Lattanzi}, {van Leeuwen}, {Bakker}, {Cacciari}, {Casta{\~n}eda}, {De Angeli}, {Fabricius}, {Fouesneau}, {Fr{\'e}mat}, {Galluccio}, {Guerrier}, {Heiter}, {Masana}, {Messineo}, {Mowlavi}, {Nicolas}, {Nienartowicz}, {Pailler}, {Panuzzo}, {Riclet}, {Roux}, {Seabroke}, {Sordo}, {Th{\'e}venin}, {Gracia-Abril}, {Portell}, {Teyssier}, {Altmann}, {Andrae}, {Audard}, {Bellas-Velidis}, {Benson}, {Berthier}, {Blomme}, {Burgess}, {Busonero}, {Busso}, {C{\'a}novas}, {Carry}, {Cellino}, {Cheek}, {Clementini}, {Damerdji}, {Davidson}, {de Teodoro}, {Nu{\~n}ez Campos}, {Delchambre}, {Dell'Oro}, {Esquej},
  {Fern{\'a}ndez-Hern{\'a}ndez}, {Fraile}, {Garabato}, {Garc{\'\i}a-Lario}, {Gosset}, {Haigron}, {Halbwachs}, {Hambly}, {Harrison}, {Hern{\'a}ndez}, {Hestroffer}, {Hodgkin}, {Holl}, {Jan{\ss}en}, {Jevardat de Fombelle}, {Jordan}, {Krone-Martins}, {Lanzafame}, {L{\"o}ffler}, {Marchal}, {Marrese}, {Moitinho}, {Muinonen}, {Osborne}, {Pancino}, {Pauwels}, {Recio-Blanco}, {Reyl{\'e}}, {Riello}, {Rimoldini}, {Roegiers}, {Rybizki}, {Sarro}, {Siopis}, {Smith}, {Sozzetti}, {Utrilla}, {van Leeuwen}, {Abbas}, {{\'A}brah{\'a}m}, {Abreu Aramburu}, {Aerts}, {Aguado}, {Ajaj}, {Aldea-Montero}, {Altavilla}, {{\'A}lvarez}, {Alves}, {Anders}, {Anderson}, {Anglada Varela}, {Antoja}, {Baines}, {Baker}, {Balaguer-N{\'u}{\~n}ez}, {Balbinot}, {Balog}, {Barache}, {Barbato}, {Barros}, {Barstow}, {Bartolom{\'e}}, {Bassilana}, {Bauchet}, {Becciani}, {Bellazzini}, {Berihuete}, {Bernet}, {Bertone}, {Bianchi}, {Binnenfeld}, {Blanco-Cuaresma}, {Blazere}, {Boch}, {Bombrun}, {Bossini}, {Bouquillon}, {Bragaglia}, {Bramante}, {Breedt},
  {Bressan}, {Brouillet}, {Brugaletta}, {Bucciarelli}, {Burlacu}, {Butkevich}, {Buzzi}, {Caffau}, {Cancelliere}, {Cantat-Gaudin}, {Carballo}, {Carlucci}, {Carnerero}, {Carrasco}, {Casamiquela}, {Castellani}, {Castro-Ginard}, {Chaoul}, {Charlot}, {Chemin}, {Chiaramida}, {Chiavassa}, {Chornay}, {Comoretto}, {Contursi}, {Cooper}, {Cornez}, {Cowell}, {Crifo}, {Cropper}, {Crosta}, {Crowley}, {Dafonte}, {Dapergolas}, {David}, {David}, {de Laverny}, {De Luise}, \& {De March}}]{gaia2023}
{Gaia Collaboration}, {Vallenari}, A., {Brown}, A.~G.~A., {et~al.} 2023, \aap, 674, A1, \dodoi{10.1051/0004-6361/202243940}

\bibitem[{{Hao} {et~al.}(2022){Hao}, {Xu}, {Wu}, {Lin}, {Bian}, {Li}, \& {Liu}}]{Hao2022}
{Hao}, C.~J., {Xu}, Y., {Wu}, Z.~Y., {et~al.} 2022, \aap, 668, A13, \dodoi{10.1051/0004-6361/202244570}

\bibitem[{{Hayden} {et~al.}(2015){Hayden}, {Bovy}, {Holtzman}, {Nidever}, {Bird}, {Weinberg}, {Andrews}, {Majewski}, {Allende Prieto}, {Anders}, {Beers}, {Bizyaev}, {Chiappini}, {Cunha}, {Frinchaboy}, {Garc{\'\i}a-Her{\'n}andez}, {Garc{\'\i}a P{\'e}rez}, {Girardi}, {Harding}, {Hearty}, {Johnson}, {M{\'e}sz{\'a}ros}, {Minchev}, {O'Connell}, {Pan}, {Robin}, {Schiavon}, {Schneider}, {Schultheis}, {Shetrone}, {Skrutskie}, {Steinmetz}, {Smith}, {Wilson}, {Zamora}, \& {Zasowski}}]{Hayden2015}
{Hayden}, M.~R., {Bovy}, J., {Holtzman}, J.~A., {et~al.} 2015, \apj, 808, 132, \dodoi{10.1088/0004-637X/808/2/132}

\bibitem[{{He} {et~al.}(2023){He}, {Liu}, {Luo}, {Wang}, \& {Jiang}}]{he2023}
{He}, Z., {Liu}, X., {Luo}, Y., {Wang}, K., \& {Jiang}, Q. 2023, \apjs, 264, 8, \dodoi{10.3847/1538-4365/ac9af8}

\bibitem[{{He} {et~al.}(2022){He}, {Li}, {Zhong}, {Liu}, {Bai}, {Qin}, {Jiang}, {Zhang}, \& {Chen}}]{he2022}
{He}, Z., {Li}, C., {Zhong}, J., {et~al.} 2022, \apjs, 260, 8, \dodoi{10.3847/1538-4365/ac5cbb}

\bibitem[{{Heinze} {et~al.}(2018){Heinze}, {Tonry}, {Denneau}, {Flewelling}, {Stalder}, {Rest}, {Smith}, {Smartt}, \& {Weiland}}]{Heinze2018}
{Heinze}, A.~N., {Tonry}, J.~L., {Denneau}, L., {et~al.} 2018, \aj, 156, 241, \dodoi{10.3847/1538-3881/aae47f}

\bibitem[{{Hourihane} {et~al.}(2023){Hourihane}, {Fran{\c{c}}ois}, {Worley}, {Magrini}, {Gonneau}, {Casey}, {Gilmore}, {Randich}, {Sacco}, {Recio-Blanco}, {Korn}, {Allende Prieto}, {Smiljanic}, {Blomme}, {Bragaglia}, {Walton}, {Van Eck}, {Bensby}, {Lanzafame}, {Frasca}, {Franciosini}, {Damiani}, {Lind}, {Bergemann}, {Bonifacio}, {Hill}, {Lobel}, {Montes}, {Feuillet}, {Tautvai{\v{s}}ien{\.{e}}}, {Guiglion}, {Tabernero}, {Gonz{\'a}lez Hern{\'a}ndez}, {Gebran}, {Van der Swaelmen}, {Mikolaitis}, {Daflon}, {Merle}, {Morel}, {Lewis}, {Gonz{\'a}lez Solares}, {Murphy}, {Jeffries}, {Jackson}, {Feltzing}, {Prusti}, {Carraro}, {Biazzo}, {Prisinzano}, {Jofr{\'e}}, {Zaggia}, {Drazdauskas}, {Stonkut{\'e}}, {Marfil}, {Jim{\'e}nez-Esteban}, {Mahy}, {Guti{\'e}rrez Albarr{\'a}n}, {Berlanas}, {Santos}, {Morbidelli}, {Spina}, \& {Minkevi{\v{c}}i{\={u}}t{\.{e}}}}]{Hourihane2023}
{Hourihane}, A., {Fran{\c{c}}ois}, P., {Worley}, C.~C., {et~al.} 2023, \aap, 676, A129, \dodoi{10.1051/0004-6361/202345910}

\bibitem[{{Huang} {et~al.}(2010){Huang}, {Gies}, \& {McSwain}}]{Huang2010}
{Huang}, W., {Gies}, D.~R., \& {McSwain}, M.~V. 2010, \apj, 722, 605, \dodoi{10.1088/0004-637X/722/1/605}

\bibitem[{{Hunt} \& {Reffert}(2024)}]{huntemily2024}
{Hunt}, E.~L., \& {Reffert}, S. 2024, \aap, 686, A42, \dodoi{10.1051/0004-6361/202348662}

\bibitem[{{Inno} {et~al.}(2021){Inno}, {Rix}, {Stanek}, {Jayasinghe}, {Poggio}, {Drimmel}, \& {Rotundi}}]{inno2021}
{Inno}, L., {Rix}, H.-W., {Stanek}, K.~Z., {et~al.} 2021, \apj, 914, 127, \dodoi{10.3847/1538-4357/abf940}

\bibitem[{{Jayasinghe} {et~al.}(2018){Jayasinghe}, {Kochanek}, {Stanek}, {Shappee}, {Holoien}, {Thompson}, {Prieto}, {Dong}, {Pawlak}, {Shields}, {Pojmanski}, {Otero}, {Britt}, \& {Will}}]{Jayasinghe2018}
{Jayasinghe}, T., {Kochanek}, C.~S., {Stanek}, K.~Z., {et~al.} 2018, \mnras, 477, 3145, \dodoi{10.1093/mnras/sty838}

\bibitem[{{J{\"o}nsson} {et~al.}(2020){J{\"o}nsson}, {Holtzman}, {Allende Prieto}, {Cunha}, {Garc{\'\i}a-Hern{\'a}ndez}, {Hasselquist}, {Masseron}, {Osorio}, {Shetrone}, {Smith}, {Stringfellow}, {Bizyaev}, {Edvardsson}, {Majewski}, {M{\'e}sz{\'a}ros}, {Souto}, {Zamora}, {Beaton}, {Bovy}, {Donor}, {Pinsonneault}, {Poovelil}, \& {Sobeck}}]{Jonsson2020}
{J{\"o}nsson}, H., {Holtzman}, J.~A., {Allende Prieto}, C., {et~al.} 2020, \aj, 160, 120, \dodoi{10.3847/1538-3881/aba592}

\bibitem[{{Leavitt} \& {Pickering}(1912)}]{leavitt1912}
{Leavitt}, H.~S., \& {Pickering}, E.~C. 1912, Harvard College Observatory Circular, 173, 1

\bibitem[{{Lemasle} {et~al.}(2022){Lemasle}, {Lala}, {Kovtyukh}, {Hanke}, {Prudil}, {Bono}, {Braga}, {da Silva}, {Fabrizio}, {Fiorentino}, {Fran{\c{c}}ois}, {Grebel}, \& {Kniazev}}]{Lemasle2022}
{Lemasle}, B., {Lala}, H.~N., {Kovtyukh}, V., {et~al.} 2022, \aap, 668, A40, \dodoi{10.1051/0004-6361/202243273}

\bibitem[{{Lin} {et~al.}(2022){Lin}, {Xu}, {Hao}, {Liu}, {Li}, \& {Bian}}]{lin2022}
{Lin}, Z., {Xu}, Y., {Hao}, C., {et~al.} 2022, \apj, 938, 33, \dodoi{10.3847/1538-4357/ac9051}

\bibitem[{{Lindegren} {et~al.}(2021{\natexlab{a}}){Lindegren}, {Bastian}, {Biermann}, {Bombrun}, {de Torres}, {Gerlach}, {Geyer}, {Hern{\'a}ndez}, {Hilger}, {Hobbs}, {Klioner}, {Lammers}, {McMillan}, {Ramos-Lerate}, {Steidelm{\"u}ller}, {Stephenson}, \& {van Leeuwen}}]{Lindegren2021a}
{Lindegren}, L., {Bastian}, U., {Biermann}, M., {et~al.} 2021{\natexlab{a}}, \aap, 649, A4, \dodoi{10.1051/0004-6361/202039653}

\bibitem[{{Lindegren} {et~al.}(2021{\natexlab{b}}){Lindegren}, {Klioner}, {Hern{\'a}ndez}, {Bombrun}, {Ramos-Lerate}, {Steidelm{\"u}ller}, {Bastian}, {Biermann}, {de Torres}, {Gerlach}, {Geyer}, {Hilger}, {Hobbs}, {Lammers}, {McMillan}, {Stephenson}, {Casta{\~n}eda}, {Davidson}, {Fabricius}, {Gracia-Abril}, {Portell}, {Rowell}, {Teyssier}, {Torra}, {Bartolom{\'e}}, {Clotet}, {Garralda}, {Gonz{\'a}lez-Vidal}, {Torra}, {Abbas}, {Altmann}, {Anglada Varela}, {Balaguer-N{\'u}{\~n}ez}, {Balog}, {Barache}, {Becciani}, {Bernet}, {Bertone}, {Bianchi}, {Bouquillon}, {Brown}, {Bucciarelli}, {Busonero}, {Butkevich}, {Buzzi}, {Cancelliere}, {Carlucci}, {Charlot}, {Cioni}, {Crosta}, {Crowley}, {del Peloso}, {del Pozo}, {Drimmel}, {Esquej}, {Fienga}, {Fraile}, {Gai}, {Garcia-Reinaldos}, {Guerra}, {Hambly}, {Hauser}, {Jan{\ss}en}, {Jordan}, {Kostrzewa-Rutkowska}, {Lattanzi}, {Liao}, {Licata}, {Lister}, {L{\"o}ffler}, {Marchant}, {Masip}, {Mignard}, {Mints}, {Molina}, {Mora}, {Morbidelli}, {Murphy}, {Pagani}, {Panuzzo},
  {Pe{\~n}alosa Esteller}, {Poggio}, {Re Fiorentin}, {Riva}, {Sagrist{\`a} Sell{\'e}s}, {Sanchez Gimenez}, {Sarasso}, {Sciacca}, {Siddiqui}, {Smart}, {Souami}, {Spagna}, {Steele}, {Taris}, {Utrilla}, {van Reeven}, \& {Vecchiato}}]{Lindegren2021b}
{Lindegren}, L., {Klioner}, S.~A., {Hern{\'a}ndez}, J., {et~al.} 2021{\natexlab{b}}, \aap, 649, A2, \dodoi{10.1051/0004-6361/202039709}

\bibitem[{{Lohr} {et~al.}(2018){Lohr}, {Negueruela}, {Tabernero}, {Clark}, {Lewis}, \& {Roche}}]{Lohr2018}
{Lohr}, M.~E., {Negueruela}, I., {Tabernero}, H.~M., {et~al.} 2018, \mnras, 478, 3825, \dodoi{10.1093/mnras/sty1280}

\bibitem[{{Madore}(1982)}]{Madore1982}
{Madore}, B.~F. 1982, \apj, 253, 575, \dodoi{10.1086/159659}

\bibitem[{{Madore} {et~al.}(2017){Madore}, {Freedman}, \& {Moak}}]{Madore2017}
{Madore}, B.~F., {Freedman}, W.~L., \& {Moak}, S. 2017, \apj, 842, 42, \dodoi{10.3847/1538-4357/aa6e4d}

\bibitem[{{Ma{\'\i}z Apell{\'a}niz} {et~al.}(2021){Ma{\'\i}z Apell{\'a}niz}, {Pantaleoni Gonz{\'a}lez}, \& {Barb{\'a}}}]{MaizApellaniz2021}
{Ma{\'\i}z Apell{\'a}niz}, J., {Pantaleoni Gonz{\'a}lez}, M., \& {Barb{\'a}}, R.~H. 2021, \aap, 649, A13, \dodoi{10.1051/0004-6361/202140418}

\bibitem[{{Majaess} {et~al.}(2024){Majaess}, {Turner}, {Minniti}, {Alonso-Garcia}, \& {Saito}}]{Majaess2024}
{Majaess}, D., {Turner}, D.~G., {Minniti}, D., {Alonso-Garcia}, J., \& {Saito}, R.~K. 2024, \pasp, 136, 094202, \dodoi{10.1088/1538-3873/ad7405}

\bibitem[{{Mermilliod} {et~al.}(2009){Mermilliod}, {Mayor}, \& {Udry}}]{Mermilliod2009}
{Mermilliod}, J.~C., {Mayor}, M., \& {Udry}, S. 2009, \aap, 498, 949, \dodoi{10.1051/0004-6361/200810244}

\bibitem[{{Minniti} {et~al.}(2010){Minniti}, {Lucas}, {Emerson}, {Saito}, {Hempel}, {Pietrukowicz}, {Ahumada}, {Alonso}, {Alonso-Garcia}, {Arias}, {Bandyopadhyay}, {Barb{\'a}}, {Barbuy}, {Bedin}, {Bica}, {Borissova}, {Bronfman}, {Carraro}, {Catelan}, {Clari{\'a}}, {Cross}, {de Grijs}, {D{\'e}k{\'a}ny}, {Drew}, {Fari{\~n}a}, {Feinstein}, {Fern{\'a}ndez Laj{\'u}s}, {Gamen}, {Geisler}, {Gieren}, {Goldman}, {Gonzalez}, {Gunthardt}, {Gurovich}, {Hambly}, {Irwin}, {Ivanov}, {Jord{\'a}n}, {Kerins}, {Kinemuchi}, {Kurtev}, {L{\'o}pez-Corredoira}, {Maccarone}, {Masetti}, {Merlo}, {Messineo}, {Mirabel}, {Monaco}, {Morelli}, {Padilla}, {Palma}, {Parisi}, {Pignata}, {Rejkuba}, {Roman-Lopes}, {Sale}, {Schreiber}, {Schr{\"o}der}, {Smith}, {Sodr{\'e}}, {Soto}, {Tamura}, {Tappert}, {Thompson}, {Toledo}, {Zoccali}, \& {Pietrzynski}}]{Minniti2010}
{Minniti}, D., {Lucas}, P.~W., {Emerson}, J.~P., {et~al.} 2010, \na, 15, 433, \dodoi{10.1016/j.newast.2009.12.002}

\bibitem[{{Pietrukowicz} {et~al.}(2021){Pietrukowicz}, {Soszy{\'n}ski}, \& {Udalski}}]{pietrukowicz2021}
{Pietrukowicz}, P., {Soszy{\'n}ski}, I., \& {Udalski}, A. 2021, \actaa, 71, 205, \dodoi{10.32023/0001-5237/71.3.2}

\bibitem[{{Pilecki}(2024)}]{Pilecki2024}
{Pilecki}, B. 2024, \apjl, 970, L14, \dodoi{10.3847/2041-8213/ad5b54}

\bibitem[{{Riello} {et~al.}(2021){Riello}, {De Angeli}, {Evans}, {Montegriffo}, {Carrasco}, {Busso}, {Palaversa}, {Burgess}, {Diener}, {Davidson}, {Rowell}, {Fabricius}, {Jordi}, {Bellazzini}, {Pancino}, {Harrison}, {Cacciari}, {van Leeuwen}, {Hambly}, {Hodgkin}, {Osborne}, {Altavilla}, {Barstow}, {Brown}, {Castellani}, {Cowell}, {De Luise}, {Gilmore}, {Giuffrida}, {Hidalgo}, {Holland}, {Marinoni}, {Pagani}, {Piersimoni}, {Pulone}, {Ragaini}, {Rainer}, {Richards}, {Sanna}, {Walton}, {Weiler}, \& {Yoldas}}]{Riello2021}
{Riello}, M., {De Angeli}, F., {Evans}, D.~W., {et~al.} 2021, \aap, 649, A3, \dodoi{10.1051/0004-6361/202039587}

\bibitem[{{Riess} {et~al.}(2021){Riess}, {Casertano}, {Yuan}, {Bowers}, {Macri}, {Zinn}, \& {Scolnic}}]{Riess2021}
{Riess}, A.~G., {Casertano}, S., {Yuan}, W., {et~al.} 2021, \apjl, 908, L6, \dodoi{10.3847/2041-8213/abdbaf}

\bibitem[{{Riess} {et~al.}(2019){Riess}, {Casertano}, {Yuan}, {Macri}, \& {Scolnic}}]{riess2019}
{Riess}, A.~G., {Casertano}, S., {Yuan}, W., {Macri}, L.~M., \& {Scolnic}, D. 2019, \apj, 876, 85, \dodoi{10.3847/1538-4357/ab1422}

\bibitem[{{Riess} {et~al.}(2022){Riess}, {Breuval}, {Yuan}, {Casertano}, {Macri}, {Bowers}, {Scolnic}, {Cantat-Gaudin}, {Anderson}, \& {Cruz Reyes}}]{Riess2022}
{Riess}, A.~G., {Breuval}, L., {Yuan}, W., {et~al.} 2022, \apj, 938, 36, \dodoi{10.3847/1538-4357/ac8f24}

\bibitem[{{Ripepi} {et~al.}(2019){Ripepi}, {Molinaro}, {Musella}, {Marconi}, {Leccia}, \& {Eyer}}]{Ripepi2019}
{Ripepi}, V., {Molinaro}, R., {Musella}, I., {et~al.} 2019, \aap, 625, A14, \dodoi{10.1051/0004-6361/201834506}

\bibitem[{{Ripepi} {et~al.}(2022){Ripepi}, {Catanzaro}, {Clementini}, {De Somma}, {Drimmel}, {Leccia}, {Marconi}, {Molinaro}, {Musella}, \& {Poggio}}]{Ripepi2022}
{Ripepi}, V., {Catanzaro}, G., {Clementini}, G., {et~al.} 2022, \aap, 659, A167, \dodoi{10.1051/0004-6361/202142649}

\bibitem[{{Ripepi} {et~al.}(2023){Ripepi}, {Clementini}, {Molinaro}, {Leccia}, {Plachy}, {Moln{\'a}r}, {Rimoldini}, {Musella}, {Marconi}, {Garofalo}, {Audard}, {Holl}, {Evans}, {Jevardat de Fombelle}, {Lecoeur-Taibi}, {Marchal}, {Mowlavi}, {Muraveva}, {Nienartowicz}, {Sartoretti}, {Szabados}, \& {Eyer}}]{Ripepi2023}
{Ripepi}, V., {Clementini}, G., {Molinaro}, R., {et~al.} 2023, \aap, 674, A17, \dodoi{10.1051/0004-6361/202243990}

\bibitem[{{Sandage} \& {Tammann}(2006)}]{Sandage2006}
{Sandage}, A., \& {Tammann}, G.~A. 2006, \araa, 44, 93, \dodoi{10.1146/annurev.astro.43.072103.150612}

\bibitem[{{Shetye} {et~al.}(2024){Shetye}, {Viviani}, {Anderson}, {Mowlavi}, {Eyer}, {Evans}, \& {Szabados}}]{Shetye2024}
{Shetye}, S.~S., {Viviani}, G., {Anderson}, R.~I., {et~al.} 2024, \aap, 690, A284, \dodoi{10.1051/0004-6361/202450185}

\bibitem[{{Skowron} {et~al.}(2019){Skowron}, {Skowron}, {Mr{\'o}z}, {Udalski}, {Pietrukowicz}, {Soszy{\'n}ski}, {Szyma{\'n}ski}, {Poleski}, {Koz{\l}owski}, {Ulaczyk}, {Rybicki}, {Iwanek}, {. Wrona}, \& {Gromadzki}}]{skowron2019}
{Skowron}, D.~M., {Skowron}, J., {Mr{\'o}z}, P., {et~al.} 2019, \actaa, 69, 305, \dodoi{10.32023/0001-5237/69.4.1}

\bibitem[{{Soszy{\'n}ski} {et~al.}(2017){Soszy{\'n}ski}, {Udalski}, {Szyma{\'n}ski}, {Wyrzykowski}, {Ulaczyk}, {Poleski}, {Pietrukowicz}, {Koz{\l}owski}, {Skowron}, {Skowron}, {Mr{\'o}z}, {Pawlak}, {Rybicki}, \& {Jacyszyn-Dobrzeniecka}}]{Soszynski2017}
{Soszy{\'n}ski}, I., {Udalski}, A., {Szyma{\'n}ski}, M.~K., {et~al.} 2017, \actaa, 67, 297, \dodoi{10.32023/0001-5237/67.4.1}

\bibitem[{{Taylor}(2005)}]{Taylor2005}
{Taylor}, M.~B. 2005, in Astronomical Society of the Pacific Conference Series, Vol. 347, Astronomical Data Analysis Software and Systems XIV, ed. P.~{Shopbell}, M.~{Britton}, \& R.~{Ebert}, 29

\bibitem[{{Turner} {et~al.}(2006){Turner}, {Abdel-Sabour Abdel-Latif}, \& {Berdnikov}}]{Turner2006}
{Turner}, D.~G., {Abdel-Sabour Abdel-Latif}, M., \& {Berdnikov}, L.~N. 2006, \pasp, 118, 410, \dodoi{10.1086/499501}

\bibitem[{{Udalski} {et~al.}(2018){Udalski}, {Soszy{\'n}ski}, {Pietrukowicz}, {Szyma{\'n}ski}, {Skowron}, {Skowron}, {Mr{\'o}z}, {Poleski}, {Koz{\l}owski}, {Ulaczyk}, {Rybicki}, {Iwanek}, \& {Wrona}}]{Udalski2018}
{Udalski}, A., {Soszy{\'n}ski}, I., {Pietrukowicz}, P., {et~al.} 2018, \actaa, 68, 315, \dodoi{10.32023/0001-5237/68.4.1}

\bibitem[{{Vasiliev} \& {Baumgardt}(2021)}]{Vasiliev2021}
{Vasiliev}, E., \& {Baumgardt}, H. 2021, \mnras, 505, 5978, \dodoi{10.1093/mnras/stab1475}

\bibitem[{{Wang} {et~al.}(2024){Wang}, {Xu}, {Lin}, {Hao}, {Liu}, \& {Li}}]{wang2024}
{Wang}, H., {Xu}, Y., {Lin}, Z., {et~al.} 2024, \aj, 168, 34, \dodoi{10.3847/1538-3881/ad50d3}

\bibitem[{{Wright} {et~al.}(2010){Wright}, {Eisenhardt}, {Mainzer}, {Ressler}, {Cutri}, {Jarrett}, {Kirkpatrick}, {Padgett}, {McMillan}, {Skrutskie}, {Stanford}, {Cohen}, {Walker}, {Mather}, {Leisawitz}, {Gautier}, {McLean}, {Benford}, {Lonsdale}, {Blain}, {Mendez}, {Irace}, {Duval}, {Liu}, {Royer}, {Heinrichsen}, {Howard}, {Shannon}, {Kendall}, {Walsh}, {Larsen}, {Cardon}, {Schick}, {Schwalm}, {Abid}, {Fabinsky}, {Naes}, \& {Tsai}}]{Wright2010}
{Wright}, E.~L., {Eisenhardt}, P. R.~M., {Mainzer}, A.~K., {et~al.} 2010, \aj, 140, 1868, \dodoi{10.1088/0004-6256/140/6/1868}

\bibitem[{{Zhou} \& {Chen}(2021)}]{zc2021}
{Zhou}, X., \& {Chen}, X. 2021, \mnras, 504, 4768, \dodoi{10.1093/mnras/stab1209}

\bibitem[{{Zinn}(2021)}]{Zinn2021}
{Zinn}, J.~C. 2021, \aj, 161, 214, \dodoi{10.3847/1538-3881/abe936}

\bibitem[{{Zorec} \& {Royer}(2012)}]{Zorec2012}
{Zorec}, J., \& {Royer}, F. 2012, \aap, 537, A120, \dodoi{10.1051/0004-6361/201117691}

\end{thebibliography}
\bibliographystyle{aasjournal}


\end{CJK*}
\end{document}